\documentclass[longauth]{aa} % for the long lists of affiliations 
\usepackage[version=3]{mhchem}   %molecules
\usepackage[nottoc, notlof, notlot]{tocbibind}   %biblio
\usepackage{enumitem}
\usepackage{graphicx}
\usepackage{xcolor}
\usepackage{colortbl}
\usepackage{lscape}
\usepackage{soul}
\usepackage{hyperref}
\usepackage{float}
\usepackage{subfig}

\usepackage{txfonts}
\newcommand*{\mb}[1]{{ \color{red}#1\color{black}}}

\newcommand{\mf}{CH$_3$OCHO}
\newcommand{\hii}{H\scriptsize{II}}
\newcommand{\uchii}{UC-H\scriptsize{II}}
\newcommand{\lbol}{L$_{\mathrm{bol}}$}
\newcommand{\Msun}{\ensuremath{\mathrm M_{\odot}}}
\newcommand{\Lsun}{\ensuremath{\mathrm L_{\odot}}}

\defcitealias{motte2022}{Paper~I}
\defcitealias{ginsburg2022}{Paper~II}
\defcitealias{pouteau2022}{Paper~III}
\defcitealias{brouillet2022}{Paper~IV}
\defcitealias{nony2023}{Paper~V}
\defcitealias{pouteau2023}{Paper~VI}
\defcitealias{cunningham2023}{Paper~VII}
\defcitealias{diaz2023}{Paper~VIII}
\defcitealias{towner2024}{Paper~IX}
\defcitealias{armante2024}{Paper~X}
\defcitealias{louvet2024}{Paper~XII}

\begin{document}

   \title{ALMA-IMF XI: The sample of hot core candidates}
   \subtitle{A rich population of young high-mass proto-stars unveiled by the emission of methyl formate}

   \author{M. Bonfand\inst{1,2} , T. Csengeri\inst{2}, S. Bontemps\inst{2}, N. Brouillet\inst{2}, F. Motte\inst{3}, F. Louvet\inst{3}, A. Ginsburg\inst{4}, N. Cunningham\inst{3}, R. Galván-Madrid\inst{5}, F. Herpin\inst{2}, F. Wyrowski\inst{6}, M. Valeille-Manet\inst{2}, A. M. Stutz\inst{7,8}, J. Di Francesco\inst{9}, A. Gusdorf\inst{10,11}, M. Fernández-López\inst{12}, B. Lefloch\inst{2}, H-L. Liu\inst{7,13}, P. Sanhueza\inst{14,15}, R. H. Álvarez-Gutiérrez\inst{7}, F. Olguin\inst{16}, T. Nony\inst{5}, A. Lopez-Sepulcre\inst{3,17}, P. Dell'Ova\inst{10,11}, Y. Pouteau\inst{3}, D. Jeff\inst{4}, H.-R. V. Chen\inst{16}, M. Armante\inst{10,11}, A. Towner\inst{18}, L. Bronfman\inst{19}, and N. Kessler\inst{2}}

              \institute{\inst{1} Departments of Astronomy and Chemistry, University of Virginia, Charlottesville, VA 22904, USA \\
              \email{pgu4gb@virginia.edu}  \\
              \inst{2} Laboratoire d’Astrophysique de Bordeaux, Univ. Bordeaux, CNRS, B18N, allée Geoffroy Saint-Hilaire, 33615 Pessac, France \\
              \inst{3} Univ. Grenoble Alpes, CNRS, IPAG, 38000 Grenoble, France \\
              \inst{4} Department of Astronomy, University of Florida, PO Box 112055, USA \\
              \inst{5} Instituto de Radioastronomía y Astrofísica, Universidad Nacional Autónoma de México, Morelia, Michoacán 58089, México \\
              \inst{6} Max-Planck Institut für Radioastronomie, Auf dem Hügel 69, 53121 Bonn, Germany \\
              \inst{7} Departamento de Astronomía, Universidad de Concepción, Casilla 160-C, 4030000 Concepción, Chile \\
              \inst{8} Max-Planck-Institute for Astronomy, Königstuhl 17, 69117 Heidelberg, Germany \\
              \inst{9} Herzberg Astronomy and Astrophysics Research Centre, National Research Council of Canada, 5071 West Saanich Road, Victoria, BC V9E 2E7 Canada \\
              \inst{10} Laboratoire de Physique de l’École Normale Supérieure, ENS, Univ. PSL, CNRS, Sorbonne Université, Université de Paris, Paris, France \\
              \inst{11} Observatoire de Paris, PSL University, Sorbonne Université, LERMA, 75014, Paris, France \\
              \inst{12} Instituto Argentino de Radioastronomía (CCT-La Plata, CONICET; CICPBA), C.C. No. 5, 1894, Villa Elisa, Buenos Aires, Argentina \\
              \inst{13} Department of Astronomy, Yunnan University, Kunming, 650091, PR China \\
              \inst{14} National Astronomical Observatory of Japan, National Institutes of Natural Sciences, 2-21-1 Osawa, Mitaka, Tokyo 181-8588, Japan \\
              \inst{15} Department of Astronomical Science, SOKENDAI (The Graduate University for Advanced Studies), 2-21-1 Osawa, Mitaka, Tokyo 181-8588, Japan \\
              \inst{16} Institute of Astronomy, National Tsing Hua University, Hsinchu 30013, Taiwan \\
              \inst{17} Institut de Radioastronomie Millimétrique (IRAM), 300 rue de la Piscine, 38406 Saint-Martin-D’Hères, France \\
              \inst{18} Steward Observatory, University of Arizona, 933 North Cherry Avenue, Tucson, AZ 85721, USA \\
              \inst{19} Departamento de Astronomía, Universidad de Chile, Casilla 36-D, Santiago, Chile \\
             }
       
   \authorrunning{M. Bonfand et al.}

   \date{Received 1 September 2023; accepted 22 February 2024}

% 5 token are mandatory
 
  \abstract
  % leave it empty if necessary  
   {The star formation process leads to an increased chemical complexity in the interstellar medium. Sites associated with high-mass star and cluster formation exhibit a so-called hot core phase, characterized by high temperatures and column densities of complex organic molecules.}
   {We aim to systematically search for and identify a sample of hot cores towards the 15 Galactic protoclusters of the ALMA-IMF Large Program and investigate their statistical properties.}
   {We built a comprehensive census of hot core candidates towards the ALMA-IMF protoclusters based on the detection of two {\mf} emission lines at 216.1\,GHz. We used the source extraction algorithm \textsl{GExt2D} to identify peaks of methyl formate (\mf) emission that is a complex species commonly observed towards sites of star formation.  We performed a cross-matching with the catalog of thermal dust continuum sources from the ALMA-IMF 1.3\,mm continuum data to infer their physical properties.}
   {We built up a catalog of 76 hot core candidates with masses ranging from $\sim$0.2\,$\Msun$ to $\sim$80\,$\Msun$, of which 56 are new detections. A large majority of these objects, identified in methyl formate emission are compact, rather circular, with deconvolved FWHM sizes of $\sim$\,2300\,au on average. The central sources of two target fields show more extended methyl formate emission, that is rather circular, with deconvolved FWHM sizes of $\sim$ 6700\,au and 13400\,au. About 30\% of our sample of methyl formate sources have core masses above 8\,$\Msun$ within sizes ranging from $\sim$\,1000\,au to 13400\,au, which well correspond to archetypical hot cores. The origin of the \mf\ emission toward the lower-mass cores can be explained by a mixture of contribution from shocks, or may correspond to objects in a more evolved state, \textit{i.e.} beyond the hot core stage. We find that the fraction of hot core candidates increases with the core mass, suggesting that the brightest dust cores are all in the hot core phase.} 
   {Our results suggest that most of these compact methyl formate sources are readily explained by simple symmetric models, while collective effects from radiative heating and shocks from compact protoclusters are needed to explain the observed extended {\mf} emission. The large fraction of hot core candidates towards the most massive cores suggests that they rapidly enter the hot core phase and feedback effects from the forming protostar(s) impact their environment on short time-scales.}

%Extended emission of methyl formate ($>$ 5000\,au), with a more complex morphology (i.e. not circular), is observed towards three regions.

   \keywords{stars: formation -- stars: massive -- stars: formation -- stars: protostars -- ISM: abundances -- ISM: molecules -- radio lines: ISM -- Line: formation -- Line: profiles} 
   %\keywords{ }
   %giant planet formation --
            %    $\kappa$-mechanism --
            %    stability of gas spheres
            %   }

   \maketitle

%
%-------------------------------------------------------------------
\section{Introduction}
%--------------------------------------------------------------------
\label{section-intro}

Star formation plays a key role in building the complex inventory of interstellar chemical species in various astronomical sources, which in turn serve as powerful diagnostic tools to study their surrounding environment \citep[see e.g.,][and references therein]{jorgensen2020, ceccarelli2022}. Through the observation of molecular emission lines, it is possible to investigate the still poorly constrained physical conditions and chemical processes that connect the different stages of star formation. In comparison to low-mass stars, the formation process of high-mass stars ($M_{\star}$ $>$ 8\,$\Msun$) is still less well-understood \citep{tan2013, motte2018}. The early evolutionary stage of high-mass star formation is expected to be short. For example, \citet{motte2007} estimate a pre-stellar phase of $< 10^4$~yr based on the core population in Cygnus-X, \citet{bonfand2017} estimated a lifetime of 6 $\times$ 10$^4$~yr for the hot core phase in the Galactic center molecular cloud Sgr B2(N), and \citet{csengeri2014} estimate $\sim7.5\times10^4$~yr for the phase prior to the emergence of strong infrared emission, corresponding to stars of type B0 or earlier, based on the statistics of massive clumps uncovered by the ATLASGAL survey. In addition, both mechanical and radiative feedback effects from already formed (proto)stars in a clustered environment  complicate the physical and chemical structure of high-mass star-forming regions further. As a consequence, the evolutionary sequence for high-mass star formation remains inadequately tested. Nevertheless, different observational signatures can be used to characterize the deeply embedded protostar, such as hot molecular cores, hyper-, and ultra-compact {\hii} regions that are exclusively associated with sites of high-mass star and cluster formation. Hyper-, and ultra-compact {\hii} regions are characterised by free-free emission from ionised gas pinpointing a (proto)stellar mass $>$8-15\,$\Msun$ \citep{hosokawa2009}. Free-free emission may also arise from an ionising jet component \citep[for a review see e.g.][]{Anglada2018}. Hot molecular cores (HCs) are identified based on association with a variety of complex organic molecules (COMs\footnote{Complex organic molecules are carbon-bearing molecules that are composed of at least six atoms \citep{herbst2009}.}), relatively high excitation temperatures ($>$100\,K), high gas densities ($n_{\mathrm H_2}$ = 10$^5$ -- 10$^8$ cm$^{-3}$), compact sizes ($<$ 0.1 pc), high bolometric luminosities ($>$10$^{4}$ $\Lsun$) and large core masses (10 -- 1000\,$\Msun$) \citep[see, e.g.,][]{kurtz2000,cesaroni2005, bonfand2019}. 

The exact origin of COMs is still strongly debated, i.e. grain-surface \citep[see e.g.,][]{garrod2006, garrod2013} vs. gas-phase production \citep[see, e.g.,][]{Charnley1992, balucani2015,balucani2018,vasyunin2013}. Though, over the past decades, they have been detected and studied in great detail towards several prominent hot cores, such as the well known galactic center source SgrB2(N) \citep[][]{belloche2013, belloche2016, bonfand2017, belloche2019} and the nearby star-forming region Orion\,KL \citep{brouillet2015, cernicharo2016, tercero2018}, where many of the first detections of interstellar molecules at radio and (sub)millimeter wavelengths were made \citep[see][and references therein]{mcguire2022}. COMs have also been recognised towards the low-mass counterparts of hot cores, so-called hot corinos \citep{bottinelli2004,ceccarelli2004}, that are Class 0 protostars, such as NGC 1333-IRAS 2A and -IRAS 4A \citep{taquet2015}, and IRAS 16293-2422 \citep{jorgensen2012, richard2013}. Regardless of where COMs are detected, their spectra carry information on the chemical and physical properties of their envelopes, their morphologies and probably their evolutionary stages \citep[see, e.g.,][]{allen2018, bonfand2019, jorgensen2020, gieser2021}. Investigating the chemical composition of star-forming cores in different environments and at different evolutionary stages is crucial for understanding the formation and early evolution of high-mass stars as well as the pathways for the chemical enrichment of the star-forming gas.

Here we analyse observational data from the ALMA-IMF Large Program: \textit{ALMA transforms our view of the origin of stellar masses} \citep[][hereafter \citetalias{motte2022} and \citetalias{ginsburg2022}, respectively]{motte2022, ginsburg2022} that uncovers a large population of star forming cores over various evolutionary stages and Galactic environments. ALMA-IMF is a survey of 15 massive nearby Galactic protoclusters that aims to statistically investigate the properties of a large sample of star-forming cores to understand the link between the core mass function and the initial mass function \citep[][hereafter \citetalias{pouteau2022} \citetalias{pouteau2023} \citetalias{nony2023}, respectively]{pouteau2022, pouteau2023, nony2023}. The 15 target regions were identified based on the ATLASGAL survey \citep[][]{schuller2009, csengeri2014}, and the catalog of \citet[][]{csengeri2017} describing the 200 brightest clumps of the survey. They were selected to probe massive protoclusters at different evolutionary stages within a distance of 2 - 5.5~kpc. \citetalias{motte2022} gives an overview of the selected targets, where the ALMA-IMF protoclusters were classified into three types of regions, based on the amount of dense gas in the cloud which has potentially been impacted by {\hii} region(s): i) young protoclusters devoid of internal ionizing sources, ii) intermediate protoclusters, that harbor a few HC- or {\uchii} regions as small, localized bubbles of ionized gas, or iii) evolved protoclusters, that contain bright and extended {\hii} regions and hence gas removal has started. Some of the targeted clouds host several well-known high-mass star-forming regions associated with strong radio continuum emission originating from {\uchii} regions, such as: G008.67 \citep{hernandez2014}, G010.62 \citep{liu2019, law2021}, G012.80 \citep{immer2014}, G333.60 \citep{lo2015}, W51-E \citep{mehringer1994,zhang1998,ginsburg2016a,rivilla2017b}, and W51-IRS2 \citep{henkel2013}. Other regions are known to harbor some of the brightest hot cores in the Galactic plane, G327.29 \citep[][and references therein]{wyrowski2008, bisschop2013}, G351.77 \citep{leurini2008}, and the W51e1/e2 hot cores of the W51-E protocluster \citep{zhang1997, ginsburg2017}. G328.25 is a well-studied hot core precursor, isolated down to $\sim$400\,au scales \citep{csengeri2018,csengeri2019, bouscasse2022}. Finally, \citet[][hereafter, \citetalias{brouillet2022}]{brouillet2022} identified eight hot cores towards the young protocluster W43-MM1, studied as part of the pilot project (2013.1.01365.S), which served as a preparation for the ALMA-IMF Large Program. 

With a $\sim$6.7 GHz non-continuous bandwidth, the ALMA-IMF data have already started to reveal the rich molecular content of several young star-forming cores. From a first-look analysis of the data, we showed in \citetalias{motte2022} that emission lines of COMs are detected over multiple spectral windows of the observational setup, suggesting that the dataset can be efficiently used to investigate the hot core phenomenon. Among the detected COMs within the ALMA-IMF band, we focus here on methyl formate (\mf), commonly detected towards both low- and high-mass star-forming regions, with a broad range of column densities. For instance, \citet{coletta2020} investigated IRAM--30m data obtained in three bands (3, 2, and 0.9\,mm) towards 39 star-forming regions,  and derived column densities for methyl formate ranging from $\sim$ 4$\times$10$^{15}$ up to 4$\times$10$^{18}$\,cm$^{-2}$.

In the current chemical models of hot cores, \mf\ is formed at early times during the star formation process, primarily through solid-phase radical-addition reactions that occur around 20--40 K \citep[see, e.g,][]{garrod2006, garrod2022}. Experimental studies lead by \citet[][]{ishibashi2021} showed that methyl formate can also be formed efficiently on water ice at 10\,K, via the photolysis of methanol. Then, radiative heating from the central protostar leads to the thermal sublimation of water ices from the grain surfaces. \mf\ is released into the gas phase when the temperature reaches $\sim$ 120 K \citep{garrod2022} and significant thermal desorption still occurs up to $\sim$160 K \citep{bonfand2019, garrod2022}. Recently, \citet{bouscasse2022, busch2022} and \citet[][]{bouscasse2024} found increased abundances of several O-bearing COMs, including \mf\ at lower temperatures of $\lesssim$ 100\,K towards Sgr\,B2(N1), the cold extended envelope of G328.25, and other infrared quiet massive clumps, suggesting that other desorption processes are at work below the thermal desorption temperature. One possible explanation proposed by \citet{busch2022} would be a partial thermal desorption of molecules from the outer, CO-rich layers of the ice mantles, at the end of the cold collapse. Given its low binding energy, CO would desorb at much lower temperatures (20–30 K). As a result, COMs that are also abundant in these layers may be able to co-desorb at temperatures $<$100\,K. Once the upper layers, which are rich in CO, had desorbed along with some COMs, COMs would still be present in the water-rich layers beneath, to be released at higher temperatures when water-ice desorbs. \citet{burke2015} undertook detailed experimental studies  showing that methyl formate may also desorb from the ices as a pure desorption feature and therefore in typical hot core conditions it would desorb at lower temperatures, starting at 77 K, or 108 K for mixed ices (i.e. methyl formate:\ce{H2O} ices).

Methyl formate has also been observed in the cold gas phase towards prestellar cores and other cold environments \citep[][]{bacman2012, cernicharo2012, vastel2014}, suggesting that low-temperature mechanisms are needed to explain the presence of \mf\ in the gas phase. The UV-driven photo desorption of surface molecules was shown to have only a limited ability to desorb molecules at visual extinction values $>$ 1 under the assumption of the standard interstellar radiation field and cosmic-ray (CR) ionisation rate \citep{jin2020}. On the other hand, chemical desorption \citep[\textit{i.e.} desorption induced by the release of chemical energy upon formation of a molecule,][]{garrod2007} is able to drive substantial COM desorption at low temperatures. \citet{balucani2015} showed that \mf \ may also efficiently form via the gas-phase oxidation of \ce{CH3OCH2}. This reaction does not have an activation barrier and it is triggered by a series of gas-phase reactions following the non-thermal desorption \citep[\textit{i.e.} cosmic ray-induced heating of grains and/or chemical desorption][respectively]{hasegawa1993, garrod2007} of solid-phase methanol, such that it may be efficient even at low temperatures. Finally, several O-bearing COMs, including methyl formate, have been detected in accretion shocks towards both high-mass \citep{csengeri2018,csengeri2019} and low-mass objects \citep{Imai2022}. In addition, methyl formate has also been detected towards shocks related to outflow activity by \citet{palau2017}. In these cases, sputtering may play a role in breaking the grains and liberating \mf\ into the gas phase.
  
In the present paper, we aim to systematically identify intermediate- to high-mass protostars associated with emission from \mf\ towards the 15 ALMA-IMF protoclusters. Our goal is to provide a catalog of hot core candidates from various cloud environments that are undergoing different dynamical events (e.g., gas inflow, protostellar outflows and expanding {\hii} regions). In Sect.\,\ref{section-obs} we present the observational data and the continuum core catalog used for our analysis. The method to identify and extract the hot core candidates from the ALMA-IMF data is described in Sect.\,\ref{section-MF}, while the resulting catalog of hot core candidates is presented in Sect.\,\ref{section-MFcat}. In Sect.\,\ref{section-MFproperties} we derive the physical properties of the hot core candidates, while the chemical origin of the methyl formate emission, as well as the exact nature of the sources is discussed in Sect.\,\ref{section-discussion}. Finally, our results are summarized in Sect.\,\ref{section-ccl}. Additional material, such as the spectra extracted towards the hot core candidates, the continuum maps, the H$_{41 \alpha}$ maps, as well as detailed explanations on the methods to estimate the free-free contamination are given in the Appendix \ref{appendix-spectra} to \ref{appendix-freefree}.

%--------------------------------------------------------------------
\section{Observations and core catalogs}
%--------------------------------------------------------------------
\label{section-obs}

The ALMA-IMF Large Program (2017.1.01355.L, PIs: Motte, Ginsburg, Louvet, Sanhueza) was undertaken to image 15 of the most massive Galactic protoclusters over the same physical scale, sensitivity, and spectral coverage, allowing us a homogeneous characterization of these star-forming regions. The overview of the scientific goals of the ALMA-IMF program, and the target selection is described in \citetalias{motte2022}; the detailed description of the observing setup, data reduction pipeline, and the subsequent data quality assessment is detailed in \citetalias{ginsburg2022}. The data reduction of the ALMA-IMF spectral windows is described in \citet[][hereafter, \citetalias{cunningham2023}]{cunningham2023}.

%-------------------------------
\subsection{Spectral line datacubes}
%-------------------------------
\label{section-linecubes}

The ALMA-IMF dataset consists of 15 mosaics covering a field of view of 1\,pc$^2$ to 8\,pc$^2$ obtained with the ALMA 12-m array. Table \ref{TAB-targetlist} provides an overview of the 15 targeted protoclusters, with the cube centers, the rest velocities ($V_{\mathrm{LSR}}$) of the protoclusters, their distances to the Sun and their evolutionary stages. The full spectral setup is composed of 12 spectral windows (spw): eight at 1.3 mm (ALMA band 6, hereafter B6) and four at 3 mm (ALMA band 3, hereafter B3), which represent a $\sim$ 6.7 GHz non-continuous bandwidth per protocluster. The detailed characteristics of these 12 spw are given in Table 2 of \citetalias{motte2022}, including an overview of the main spectral lines they cover. In \citetalias{cunningham2023} we provide the full spectral line data products for the 15 protoclusters. They were produced using the custom ALMA-IMF imaging pipeline\footnote{\href{https://github.com/ALMA-IMF/reduction}{https://github.com/ALMA-IMF/reduction}} originally developed to process the continuum data as described in \citetalias{ginsburg2022}, and subsequently adapted to process the spectral line datacubes as described in \citetalias{cunningham2023}. In short, up to two different ALMA 12-m array configurations were combined in the uv-plane for each field, and corrected for system temperature and spectral data normalization (see also Section 2 of \citetalias{ginsburg2022} for more details). Then, the pipeline performs a line cleaning with parameters optimized for each field, and applies the Jorsater-van-Moorsel \citep[“JvM”,][]{jorsater1995} correction. The deconvolved datacubes have a constant beam over all the channels. Finally, we use the STATCONT software \citep[][]{sanchez2018} with the sigma-clipping algorithm to systematically remove the continuum emission in the image plane and produce datacubes containing only spectral line emission.
 
%======================
% TABLE: LIST ALMA-IMF FIELDS
%======================
\begin{table*}[!t]
\begin{center}
  \caption{\label{TAB-targetlist} List of the 15 massive protoclusters targeted by ALMA-IMF.} 
  \setlength{\tabcolsep}{1.6mm}
  \begin{tabular}{lrrllcc}
    \hline
    \hline 
    \multicolumn{1}{c}{Field} & \multicolumn{2}{c}{Cube center [IRCS - J2000]} & \multicolumn{1}{c}{$V_{\mathrm{LSR}}$} & \multicolumn{1}{c}{$d$} & \multicolumn{1}{c}{Evolutionary} & \multicolumn{1}{c}{$M_{\mathrm{cloud}}$} \\
              & \multicolumn{1}{c}{RA[h:m:s]} & \multicolumn{1}{c}{DEC[$^{\circ}$:$\arcmin$:$\arcsec$]} &  \multicolumn{1}{c}{[km.s$^{-1}$]} & \multicolumn{1}{c}{[kpc]} & \multicolumn{1}{c}{stage} & \multicolumn{1}{c}{$\times$ 10$^3 \Msun$} \\
    \hline
                  G008.67  &  18:06:21.12 & $-$21:37:16.7 & +37.6 & 3.4 & I & 3.1 \\
                  G010.62  &  18:10:28.80 & $-$19:55:48.3 & $-$2.0 & 5.0 & E & 6.7\\
                  G012.80  &  18:14:13.37 & $-$17:55:45.2 & +37.0 & 2.4 & E & 4.6 \\
                  G327.29  &  15:53:08.13 & $-$54:37:08.6 & $-$45.0 & 2.5 & Y & 5.1 \\
                  G328.25  &  15:57:59.68 & $-$53:57:59.8 & $-$43.0 & 2.5 & Y & 2.5 \\
                  G333.60  &  16:22:09.36 & $-$50:05:59.2 & $-$47.0 & 4.2 & E & 12.0 \\
                  G337.92  &  16:41:10.62 & $-$47:08:02.9 & $-$40.0& 2.7 & Y & 2.5 \\
                  G338.93  &  16:40:34.42 & $-$45:41:40.6 & $-$62.0 & 3.9 & Y & 7.1 \\
                  G351.77  &  17:26:42.62 & $-$36:09:20.5 & $-$3.0 & 2.0 & I & 2.5 \\
                  G353.41  &  17:30:26.28 & $-$34:41:49.7 & $-$17.0 & 2.0 & I & 2.5 \\
                  W43-MM1  &  18:47:47.00 & $-$01:54:26.0 & +97.0 & 5.5 & Y & 13.4 \\
                  W43-MM2  &  18:47:36.61 & $-$02:00:51.7 & +97.0 &  5.5 & Y & 11.6 \\
                  W43-MM3  &  18:47:41.46 & $-$02:00:28.2 & +97.0 & 5.5 & I & 5.2 \\
                  W51-E    &  19:23:44.18 & +14:30:28.9 & +55.0 & 5.4 & I & 32.7 \\
                  W51-IRS2 &  19:23:39.81 & +14:31:02.9 & +55.0 & 5.4 & E & 20.6 \\
    \hline
\end{tabular}
\end{center}
\vspace{-4mm}
\tablefoot{The central positions of the mosaics are taken from the \mf\ cube headers. The rest velocities ($V_{\mathrm{LSR}}$), distances to the Sun ($d$), evolutionary stages (Young, Intermediate, Evolved), and cloud mass ($M_{\mathrm{cloud}}$) computed from the 870 $\mu$m integrated flux, are from \citetalias{motte2022}.}
\end{table*}
% ======================

\begin{figure}[!t]
\begin{center}
       \includegraphics[width=\hsize]{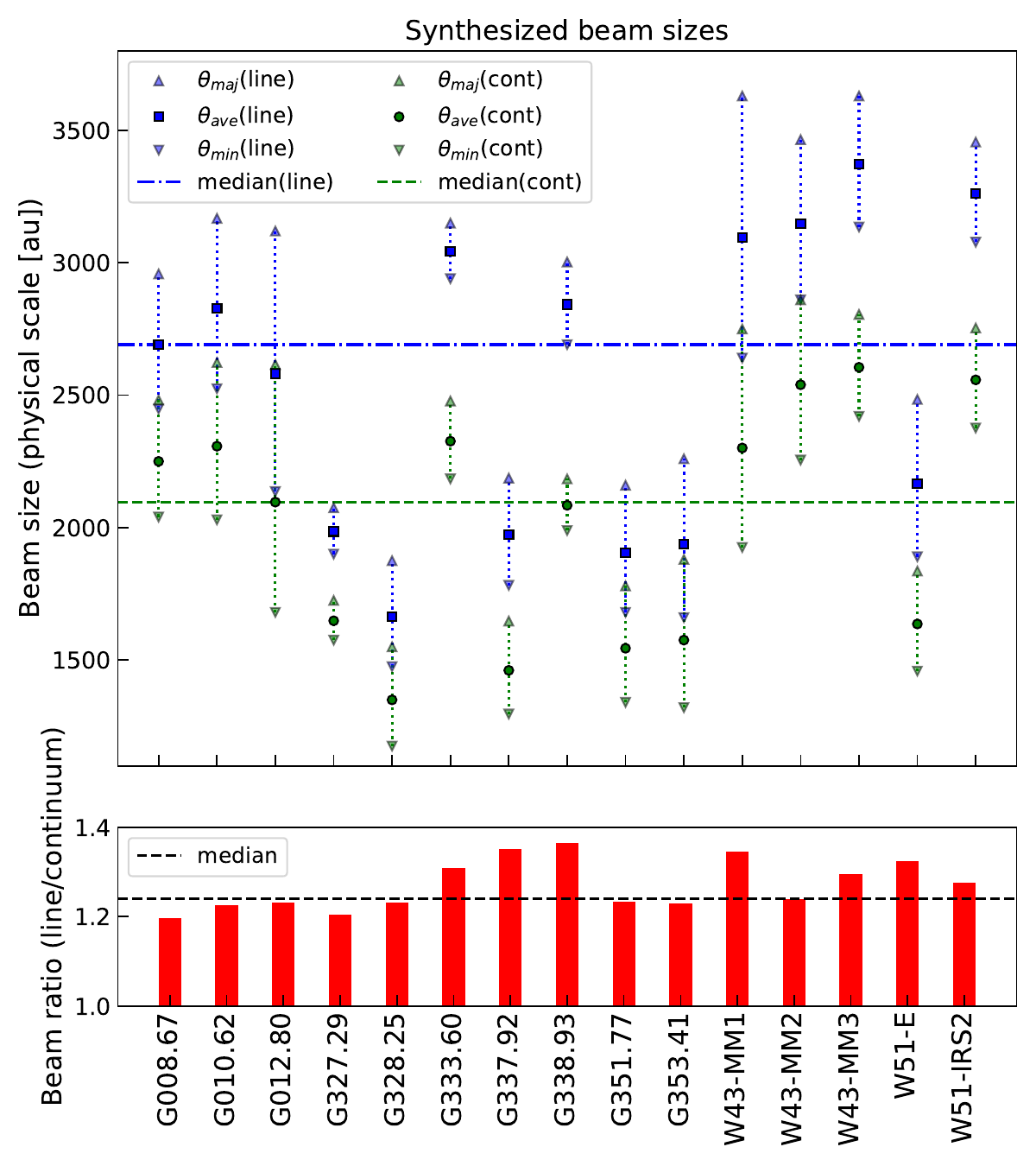} \\
  \caption{\label{FIG-beam-size} \textbf{(Top panel)} Comparison of line cube (blue) and continuum map (green) synthesized beam sizes for the 15 ALMA-IMF fields. For each field, the average beam size is defined as $\theta_{\mathrm{ave}}$ = $\sqrt{\theta_{\mathrm{maj}} \times \theta_{\mathrm{min}}}$, where $\theta_{\mathrm{maj}}$ and $\theta_{\mathrm{min}}$ are expressed in physical scale (au). The horizontal blue and green dashed lines show the median beam sizes for the 15 line cubes and continuum maps, respectively. \textbf{(Bottom panel)} Ratio of the average line cube to continuum map beam size for the 15 ALMA-IMF fields. The horizontal dashed line shows the median.}
  \end{center}
\end{figure}

 In the present paper, we focus our analysis on the 234\,MHz-wide spw centered on 216.2 GHz, at 1.3 mm (B6--spw0), that contains four strong emission lines of methyl formate, as well as DCO$^+$ (3-2), and OC$^{33}$S (18-17), with a spectral resolution of 0.17\,km\,s$^{-1}$ (\textit{i.e.} 122\,kHz). The angular resolution of the observations was chosen to achieve a physical resolution of about 2500\,au for each individual protocluster considering their different distances. The resulting angular resolution of the B6--spw0 line cubes, using a robust weighting of 0, ranges from $\sim$ 0.4$\arcsec$ to 1.1$\arcsec$, depending on the distance of the protocluster. The synthesized beams of the datacubes, given by the geometric mean of the major and minor axes ($\theta_{\mathrm{ave}} = \sqrt{\theta_{\mathrm{maj}} \times \theta_{\mathrm{min}}}$), are shown in Fig. \ref{FIG-beam-size} and listed in Table \ref{TAB-linecubes}. The flux densities, $S$, measured per beam (in Jy\,beam$^{-1}$) in the datacubes are converted to effective brightness temperatures ($T_{\mathrm B}$, in K) as follows: 
 \begin{equation}
 T_{\mathrm B}\,[\mathrm K] = S \,[\mathrm{Jy\,beam^{-1}}] \times \frac{c^2}{2 k_{\mathrm{b}} \nu^2} \times \frac{1}{\Omega_{\mathrm{beam}}}, 
 \end{equation}
 where $c$ is the speed of light, $k_{\mathrm{b}}$ is the Boltzmann constant, $\nu$ the central frequency of the considered spw (see Table \ref{TAB-linecubes}), and $\Omega_{\mathrm{beam}}$ the beam solid angle of the line cubes given by $\Omega_{\mathrm{beam}}$ = $\theta_{\mathrm{ave}}^2 \times \frac{\pi}{4 ln(2)}$. Finally, in order to estimate the noise in a homogeneous manner, we use the line cubes prior to the correction for the primary beam response. For each field, we measure the rms noise within a polygon that is defined as a region devoid of emission. The rms noise levels estimated in this way are given in Table \ref{TAB-linecubes} in units of mJy per clean beam and K. 

The ALMA-IMF spectral coverage includes other potential tracers of heated gas, such as high $E_{\rm up}/k$ transitions of \ce{CH3OH}, and \ce{CH3CN} lines. However, several of their transitions exhibit a considerably more extended morphology and hence provide a potentially more confused view of hot cores compared to that of the selected spectrally well-resolved {\mf} lines (see \citetalias{brouillet2022}). A more detailed comparison of these tracers will be subject for further studies.

%======================
% FIGURE: COMPARE BEAM SIZE LINE CUBE vs. CONT CORE
%======================
%======================

%======================
% TABLE: SPW PROPERTIES
%======================
\begin{table*}[!t]
\begin{center}
  \caption{\label{TAB-linecubes} Observational characteristics of the B6-spw0 line cubes, 1.3 mm and 3mm continuum maps used in the present study.} 
  \begin{tabular}{l|lrrl|lrrlrr}
    \hline
    \hline
                                      & \multicolumn{4}{c|}{Line cubes}   & \multicolumn{6}{c}{Continuum maps} \\
                                      & \multicolumn{4}{c|}{}   & \multicolumn{3}{c}{1.3 mm}  & \multicolumn{3}{c}{3 mm} \\                                      
    \hline                                 
    \multicolumn{1}{c|}{Protocluster}  & \multicolumn{1}{c}{$\theta_{maj} \times \theta_{min}$} & PA & \multicolumn{2}{c|}{rms} & \multicolumn{1}{c}{$\theta_{maj} \times \theta_{min}$} & \multicolumn{1}{c}{PA}  & $\nu_{\mathrm{1.3mm}}$ & \multicolumn{1}{c}{$\theta_{maj} \times \theta_{min}$} & \multicolumn{1}{c}{PA}  & $\nu_{\mathrm{3mm}}$ \\ 
    \multicolumn{1}{c|}{name}          & \multicolumn{1}{c}{[$\arcsec$ $\times$ $\arcsec$]} & [deg] & [mJy beam$^{-1}$] & [K] & \multicolumn{1}{c}{[$\arcsec$ $\times$ $\arcsec$]} & [deg] & \multicolumn{1}{c}{[GHz]}  & \multicolumn{1}{c}{[$\arcsec$ $\times$ $\arcsec$]} & [deg] & \multicolumn{1}{c}{[GHz]}     \\
    \hline
                  G008.67  &  0.87$\times$0.72 & -82 &   8.6 & 0.36 &  0.73$\times$0.60 & -82 & 228.732 &  0.51$\times$0.40 & +72 &  100.526  \\ %4.3 &  0.01      
                  G010.62  &  0.64$\times$0.51 & -74 & 2.4 & 0.19 & 0.53$\times$0.41 & -78 & 229.268  & 0.39$\times$0.32 & -80 & 100.704 \\% 1.6 &  0.05   \\
                  G012.80  &  1.30$\times$0.89 & +77 &  13.2 & 0.30 & 1.09$\times$0.70 & +75 & 229.080 & 1.48$\times$1.26 & +89 &  100.680 \\ %6.0 &  0.01   \\
                  G327.29  &  0.83$\times$0.76 & -53 & 10.0 & 0.41 & 0.69$\times$0.63 & -41 & 229.507 & 0.43$\times$0.37 & +70 &   101.776  \\ %4.8 &  0.03   \\
                  G328.25  &  0.75$\times$0.59 & -13 & 16.7 & 0.99 & 0.62$\times$0.47 & -112 & 227.575   & 0.62$\times$0.44 & -83 &  101.500  \\  %15.0 &  0.01   \\
                  G333.60  &  0.75$\times$0.70 & -37 & 3.4 & 0.17 & 0.59$\times$0.52 & -33 & 229.062  & 0.46$\times$0.44 & +50 & 100.756 \\ % 1.3 &  0.18   \\    
                  G337.92  &  0.81$\times$0.66 & -51 & 4.2 & 0.21 & 0.61$\times$0.48 & -56 & 227.503  &  0.46$\times$0.41 & +78 &   101.602  \\ % 1.5 &  0.16 \\
                  G338.93  &  0.77$\times$0.69 & +80 & 4.0 & 0.20 & 0.56$\times$0.51 & -85 & 229.226 &  0.41$\times$0.39 & +84 &  100.882 \\ %  1.5 &  0.20   \\
                  G351.77  &  1.08$\times$0.84 & +88 & 15.1 & 0.44 & 0.89$\times$0.67 & +87 & 227.991  &  1.52$\times$1.30 & +89 &  100.228 \\ % 6.6 &  0.03   \\
                  G353.41  &  1.13$\times$0.83 & +86 & 15.3 & 0.43 & 0.94$\times$0.66 & +85 & 229.431  &  1.46$\times$1.27 & +77 &  100.547 \\ % 7.2 &  0.03   \\
                  W43-MM1  &  0.66$\times$0.48 & -81 &  2.2 & 0.18 & 0.50$\times$0.35 & -77 & 229.680 & 0.56$\times$0.34 & -73 &  99.795 \\ % 1.9 &  0.26   \\
                  W43-MM2  &  0.63$\times$0.52 & -80 &  2.1 & 0.17 & 0.52$\times$0.41 & -76 & 227.597  &  0.31$\times$0.24 & -72 & 101.017 \\ % 1.2 &  0.40   \\
                  W43-MM3  &  0.66$\times$0.57 & +86 &  2.3 & 0.16 & 0.51$\times$0.44 & +90 & 228.931 & 0.41$\times$0.29 & -83 &   100.911   \\ %1.0 &  0.53   \\
                  W51-E    &  0.46$\times$0.35 & +30 &  2.1 & 0.34 & 0.34$\times$0.27 & +26 & 228.918 & 0.29$\times$0.27 & +71 & 101.426 \\ %  1.2 &  0.75  \\
                  W51-IRS2 &  0.64$\times$0.57 & -19 &  2.3 & 0.16 & 0.51$\times$0.44 & -26 & 228.530  & 0.29$\times$0.27 & -57 &  101.263  \\ % 1.6 &  0.39   \\
    \hline 
\end{tabular}
\end{center}
\vspace{-4mm}
\tablefoot{The synthesized beam sizes are read from the header of the line cubes and continuum maps (see \citetalias{motte2022}, \citetalias{ginsburg2022}, and \citetalias{cunningham2023}). The central frequencies of the continuum maps, $\nu_{1.3 mm}$ and $\nu_{3 mm}$, are from \citetalias{louvet2024}. The rms noise level is measured in the JvM-corrected cubes in units of mJy per clean beam, for channels that are free of strong emission, and within a subregion that is free of thermal emission. See \citetalias{cunningham2023} for more details on how to recover the noise in units of Jy per dirty beam.}
\end{table*}
% ======================

%-------------------------------
\subsection{Continuum maps and core catalogs}
%-------------------------------
\label{section-corecat}

The first data release of the ALMA-IMF continuum images at 1.3 mm and 3 mm, along with a complete description of the data reduction and imaging process, are presented in \citetalias{ginsburg2022}. The exact central frequency of the 1.3 mm and the 3 mm continuum maps, along with the average synthesized beam sizes are given for each field in Table \ref{TAB-linecubes}. Figure \ref{FIG-beam-size} shows that the average synthesized beam size of the line cubes is systematically larger than that of the continuum maps at 1.3 mm, with a median ratio (line cube over continuum map beam) of $\sim$1.24, and a difference ranging from 20\% to 36\%, depending on the protocluster.

\citet[][hereafter, \citetalias{louvet2024}]{louvet2024} present the catalogs of dust continuum cores extracted from the continuum images at 1.3 mm, computed using maps that consider only the line-free channels (also referred to as \textit{cleanest} maps). Two sets of \textit{cleanest} continuum maps were used for the source extraction: the continuum maps at their native angular resolution (1400 -- 2700\,au) also referred to as unsmoothed data, and the continuum maps that were all smoothed to the same physical resolution of 2700 au, that implies a reduced angular resolution compared to the Briggs 0 weighted gridding of the spw used here. For the current analysis we focus exclusively on the unsmoothed continuum data, thus benefiting from the original angular resolution of the data. In \citetalias{louvet2024}, the multi-scale source and filament extraction method \textsl{getsf} \citep[][]{menshchikov2021} was used to separate the compact source-like peaks from their backgrounds, using spatial decomposition before extracting sources, that are defined as relatively round emission peaks, significantly stronger than the local surrounding fluctuations of background and noise. In total 807 compact continuum cores were extracted from the 15 ALMA-IMF protoclusters using \textsl{getsf}, including 140 sources that are largely contaminated by free-free emission, according to the spectral index calculations presented in \citetalias{louvet2024}. The core catalogs can be found on the ALMA-IMF large program website \footnote{\href{https://www.almaimf.com/}{https://www.almaimf.com/}}, and in \citetalias{louvet2024}.

%(i.e. $\alpha <$ 2)

%--------------------------------------------------------------------
\section{Identification of hot core candidates}
%--------------------------------------------------------------------
\label{section-MF}

We present here a simple approach, independent from the continuum core identification, to extract hot core candidates towards the 15 massive protoclusters, based on the spatial distribution of a single COM, methyl formate (\mf). A deeper search for hot cores using other spectral lines from the complete ALMA-IMF dataset will be presented in a forthcoming paper.

%------------------
\subsection{\mf\ integrated intensity (moment 0) maps} 
\label{section-CH3OCHO}

\begin{figure}[!h]
   \begin{center}
    \includegraphics[width=\hsize]{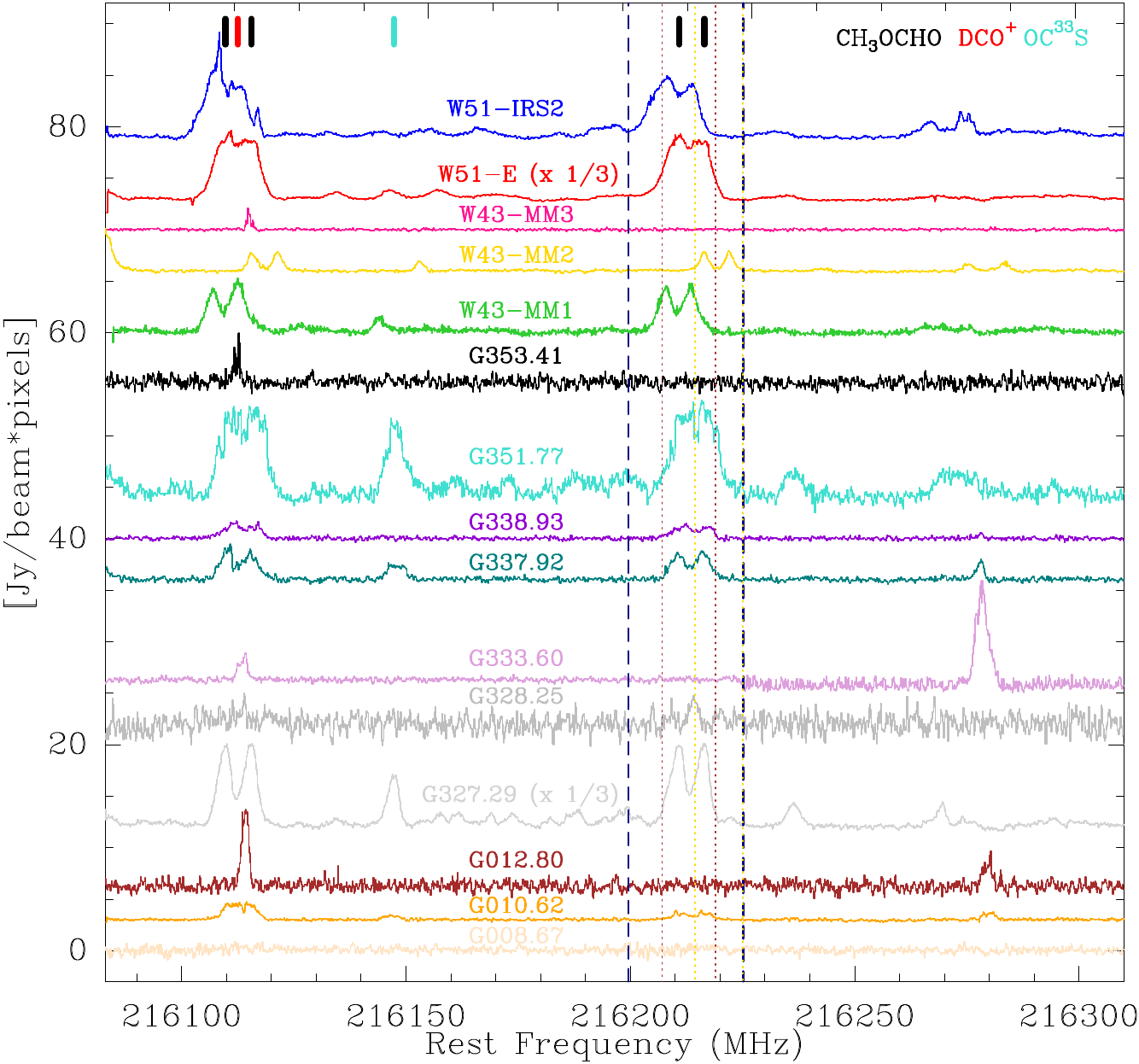} 
    \caption{\label{FIG-integrated-spectra} Continuum-subtracted spectra integrated over the full field of view of the 15 ALMA-IMF B6-spw0 line cubes. The value in parentheses (if any), indicates the scaling factor applied to the spectrum. The dark blue vertical dashed lines show the channel range used to compute the  moment 0  maps of methyl formate for all protocluster, except for G012.80 and W43-MM2, for which tighter velocity ranges, shown in brown and yellow dotted lines, respectively, were used. The vertical colored bars on top of the plot show the rest frequencies of the corresponding species indicated in the top right corner.} 
   \end{center}
\end{figure}

The ALMA-IMF spectral setup covers four strong transitions of \mf\ in its B6-spw0 at 216.2 GHz (see the exact rest frequencies listed in Table\,\ref{TAB-transitions}). The four transitions share the same upper level energy, $E_{\mathrm{up}}$/$k$ = 109 K, so they most likely trace the same region within the source envelope and also exhibit similar line profiles. Figure \ref{FIG-integrated-spectra} shows the spectra observed between 216.08 GHz and 216.32 GHz (i.e. 234\,MHz wide), spatially averaged over the full field of view of the 15 ALMA-IMF fields. The four transitions of \mf\ are gathered into two pairs of lines. The spectral resolution of 0.17\,km\,s$^{-1}$ is sufficient to resolve the lines with at least 11 channels, considering the Full Width at Half Maximum (FWHM) of the lines ranging between $\sim$ 2 and 6 km\,s$^{-1}$, depending on the protocluster. However, in each pair, the two transitions are separated by $\sim$5.7 km s$^{-1}$, such that depending on the linewidth of each \mf\ transition, they may be partially blended. Except in the case of G327.29, G351.77, and W51-E, the averaged spectrum shows a relatively low contamination from other molecules, such that \mf\ lines are easy to identify. 

In most cases, the two \mf \ pairs have similar shapes and intensities. However, in the cases of G010.62, G012.80, G333.60, W43-MM1, W43-MM3, W51-E and W51-IRS2, the first pair of \mf\ lines, centered at 216.113 GHz, is strongly contaminated by the DCO$^+$ (3--2) line (see Table \ref{TAB-transitions}). Furthermore, most fields exhibit complex spectra, with multiple velocity features, which may come either from multiple sources detected in the field with different $V_{\mathrm{LSR}}$ (see last column of Table\,\ref{TAB-MFcat}), or resulting from multiple velocity components of \mf\,spatially centered on the same core but slightly shifted in velocity. Therefore, we create moment 0 maps of methyl formate by integrating the spectral intensity over a broad velocity range of $\sim$ 35\,km\,s$^{-1}$ (\textit{i.e.} 206 channels), that covers the \mf\ pair of lines that is not contaminated by DCO$^+$ (see vertical dashed lines in Fig. \ref{FIG-integrated-spectra}). This velocity range was selected as the best compromise to take into account that different sources may have different $V_{\mathrm{LSR}}$ ($>$10\,km\,s$^{-1}$ dispersion in the core $V_{\mathrm{LSR}}$, see Fig. \ref{FIG-integrated-spectra} of \citetalias{cunningham2023}, and also Sect.\,\ref{section-vlsr}), and excluding emission from other species. In the case of G012.80 and W43-MM2 we use a custom, tighter, velocity range of $\sim$15 km s$^{-1}$ (\textit{i.e.} 88 channels) to increase the signal-to-noise ratio (S/N) of the very faint \mf\ emission lines. 

\begin{figure*}[h!]
\resizebox{\hsize}{!}
   {\begin{tabular}{cc}  
       \includegraphics[width=\hsize]{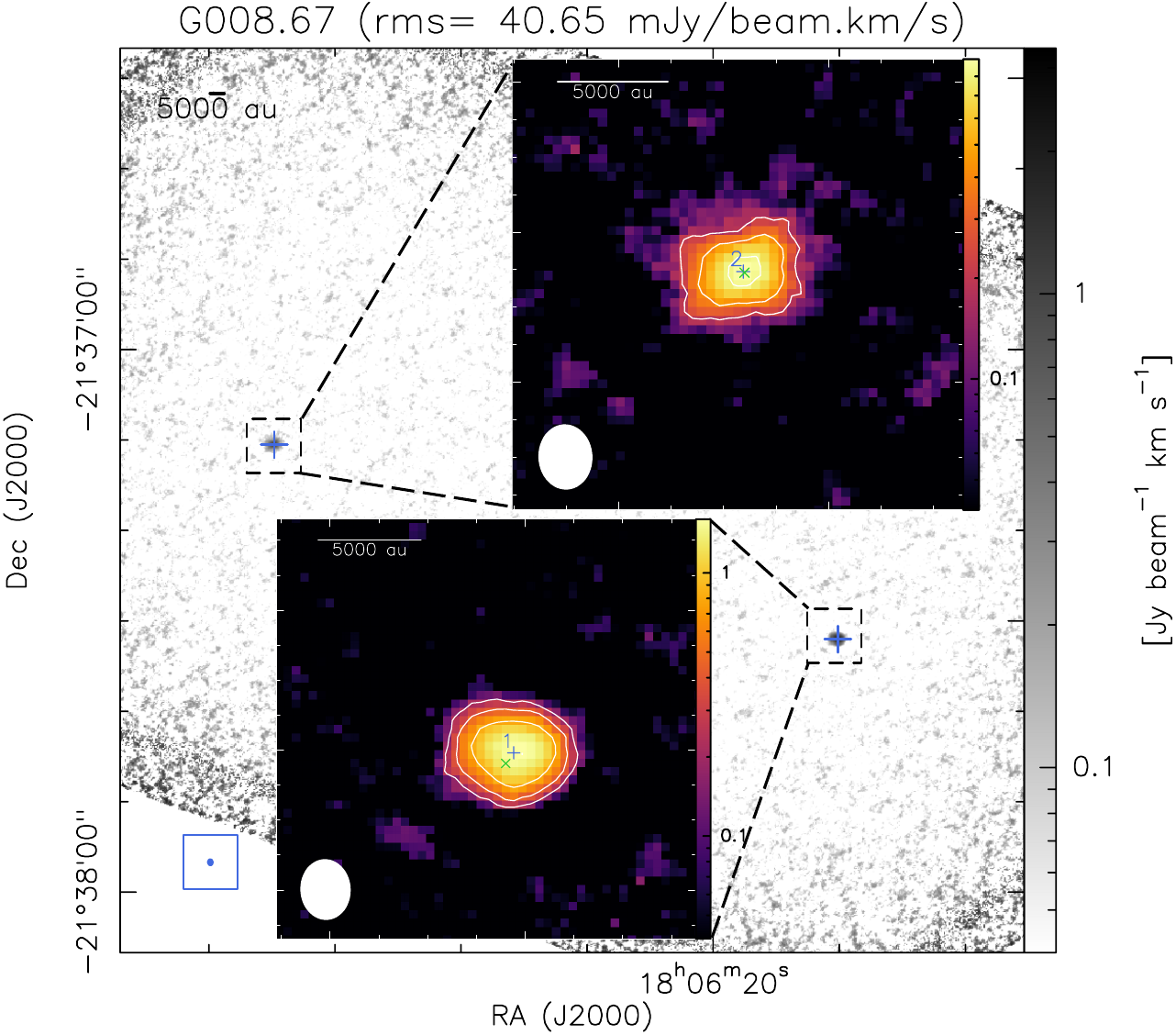}  &
       \includegraphics[width=\hsize]{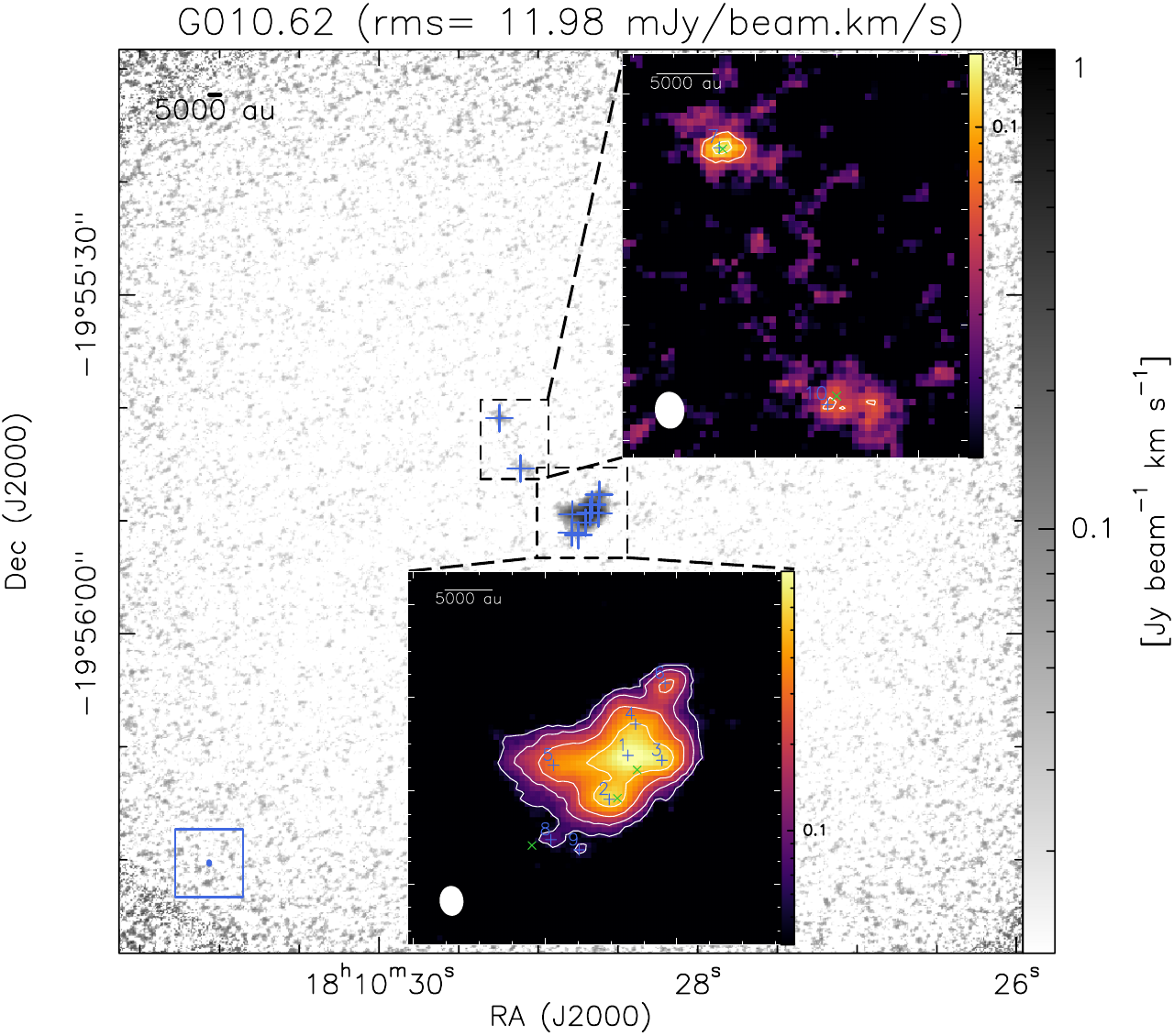}  \\
       \includegraphics[width=\hsize]{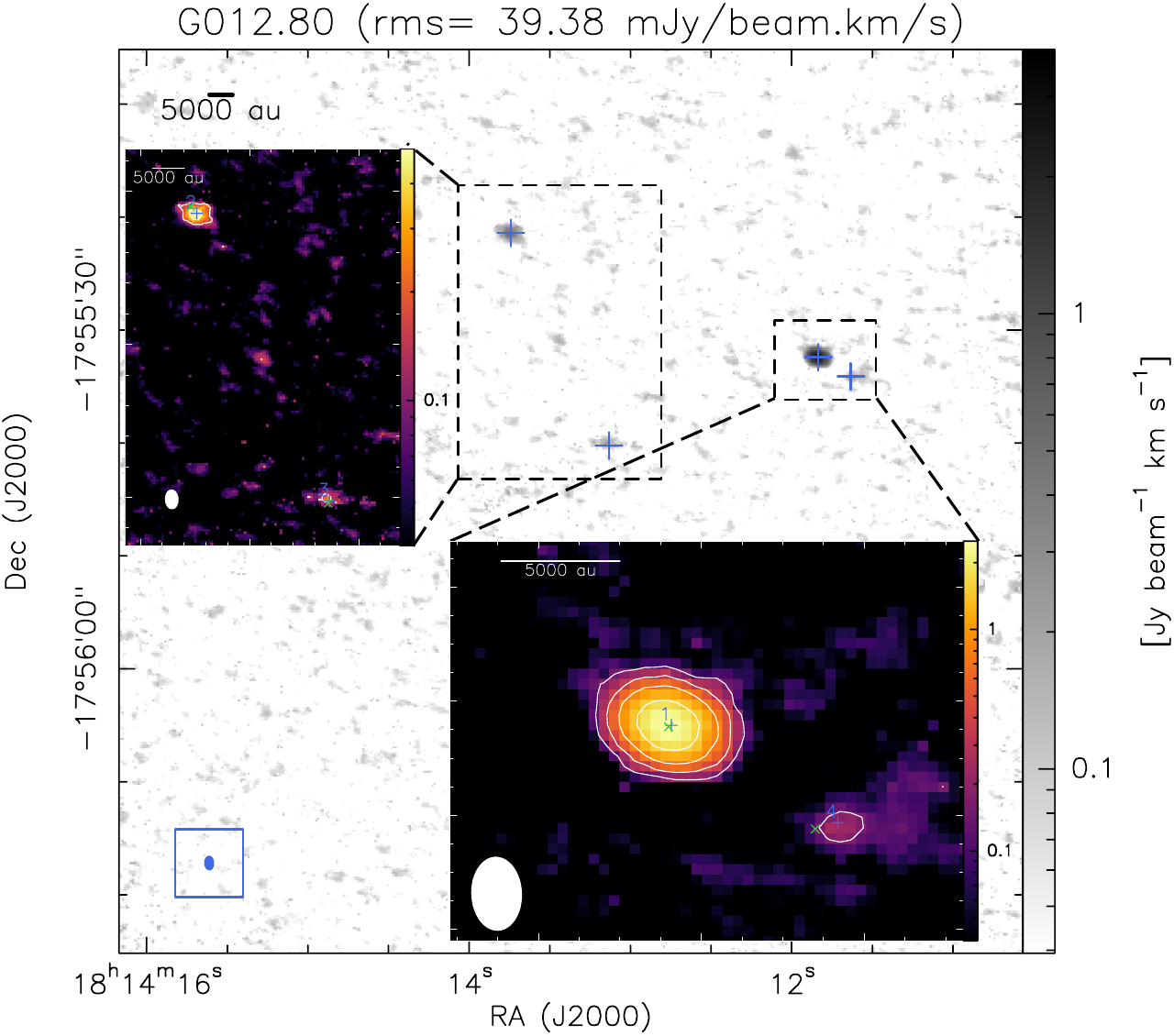}  &
       \includegraphics[width=\hsize]{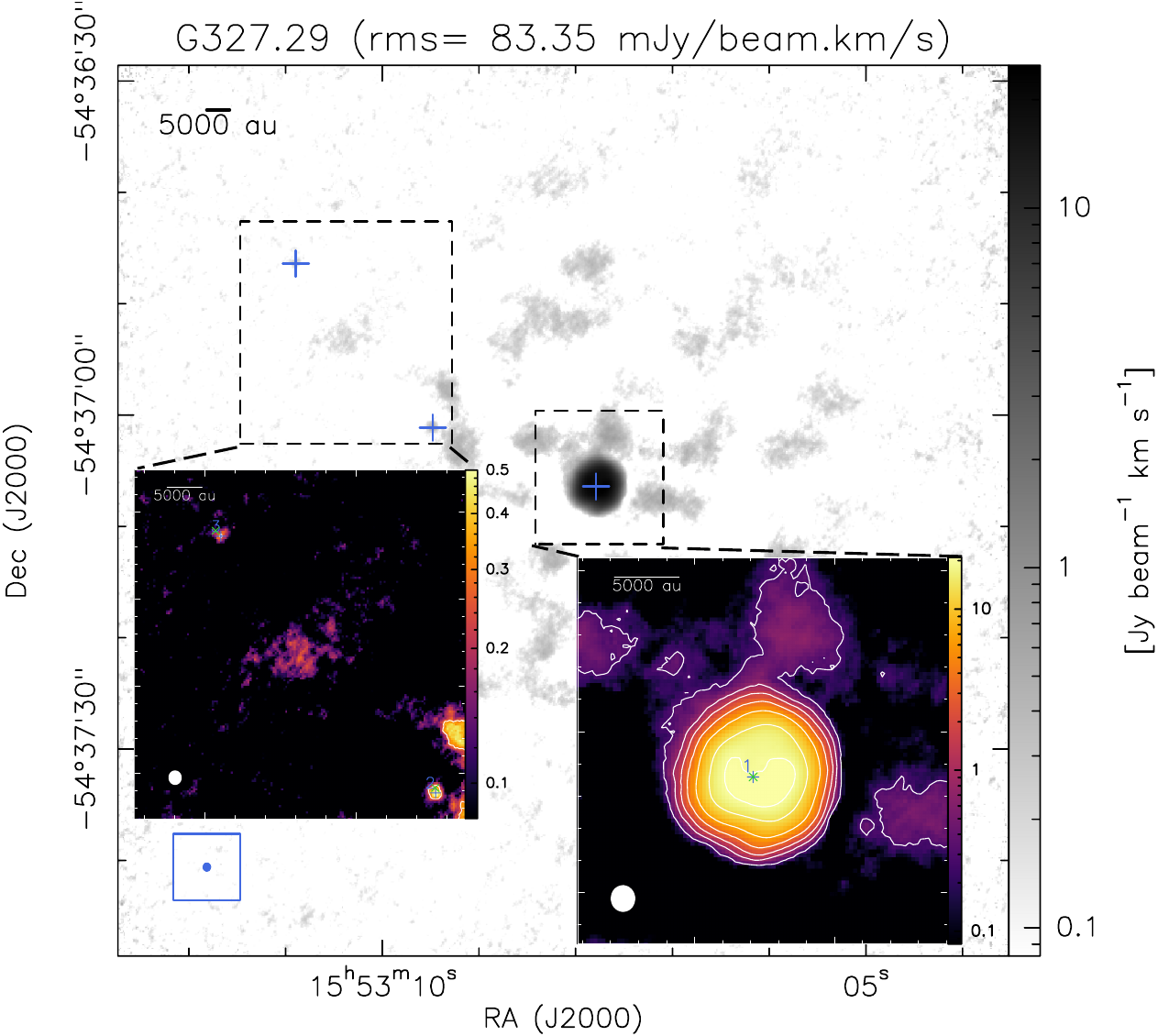}  \\ 
    \end{tabular}}
        \caption{\label{FIG-mom0-maps} Moment 0 maps of methyl formate obtained towards the 15 ALMA-IMF protoclusters as described in Sect.\,\ref{section-CH3OCHO}. Contours start at 5\,$\sigma$ (the 1\,$\sigma$ rms noise level is indicated on top of each panel) and double in value thereafter. In each panel the blue crosses show the peak positions of the methyl formate sources extracted with \textsl{GExt2D}, while the green crosses show their associated continuum cores from the \textsl{getsf} unsmoothed catalog (\citetalias{louvet2024}). The blue or white ellipses in each panel represent the synthesized beam of the line cubes. The figure continues on the next page.}
    \end{figure*}

\begin{figure*}[h!]
    \resizebox{\hsize}{!}
   {\begin{tabular}{cc} 
       \includegraphics[width=\hsize]{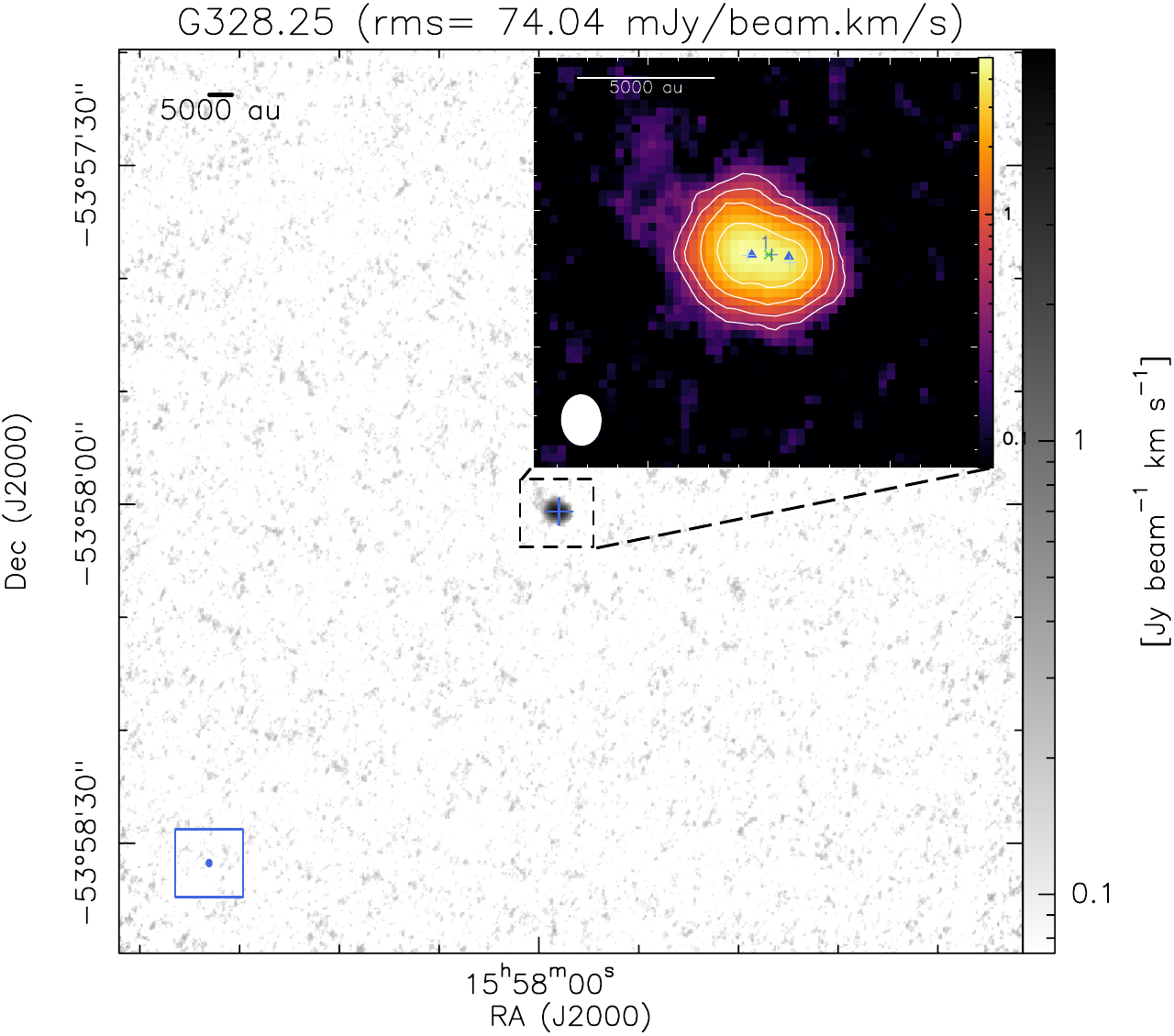} &
       \includegraphics[width=\hsize]{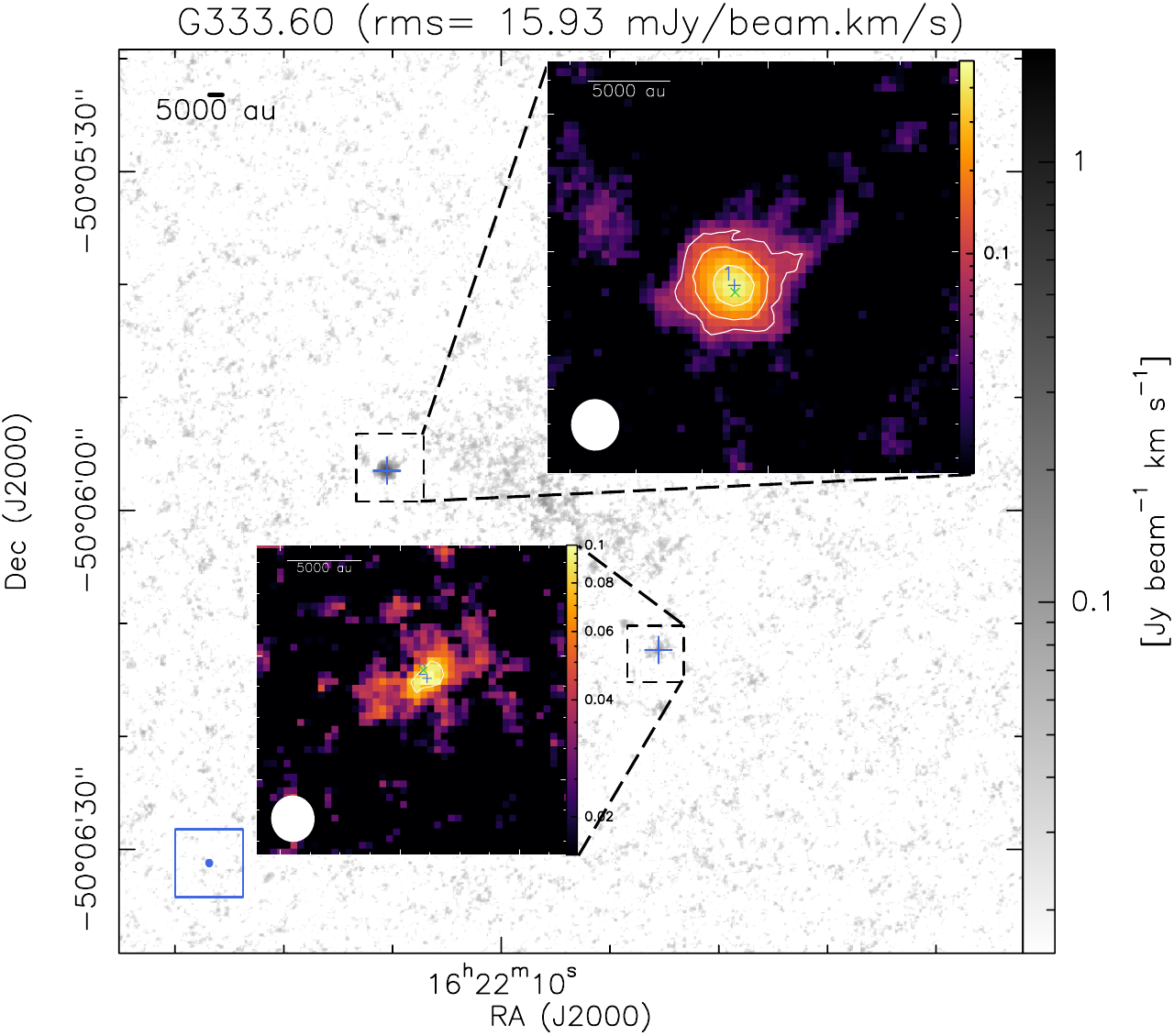}  \\
       \includegraphics[width=\hsize]{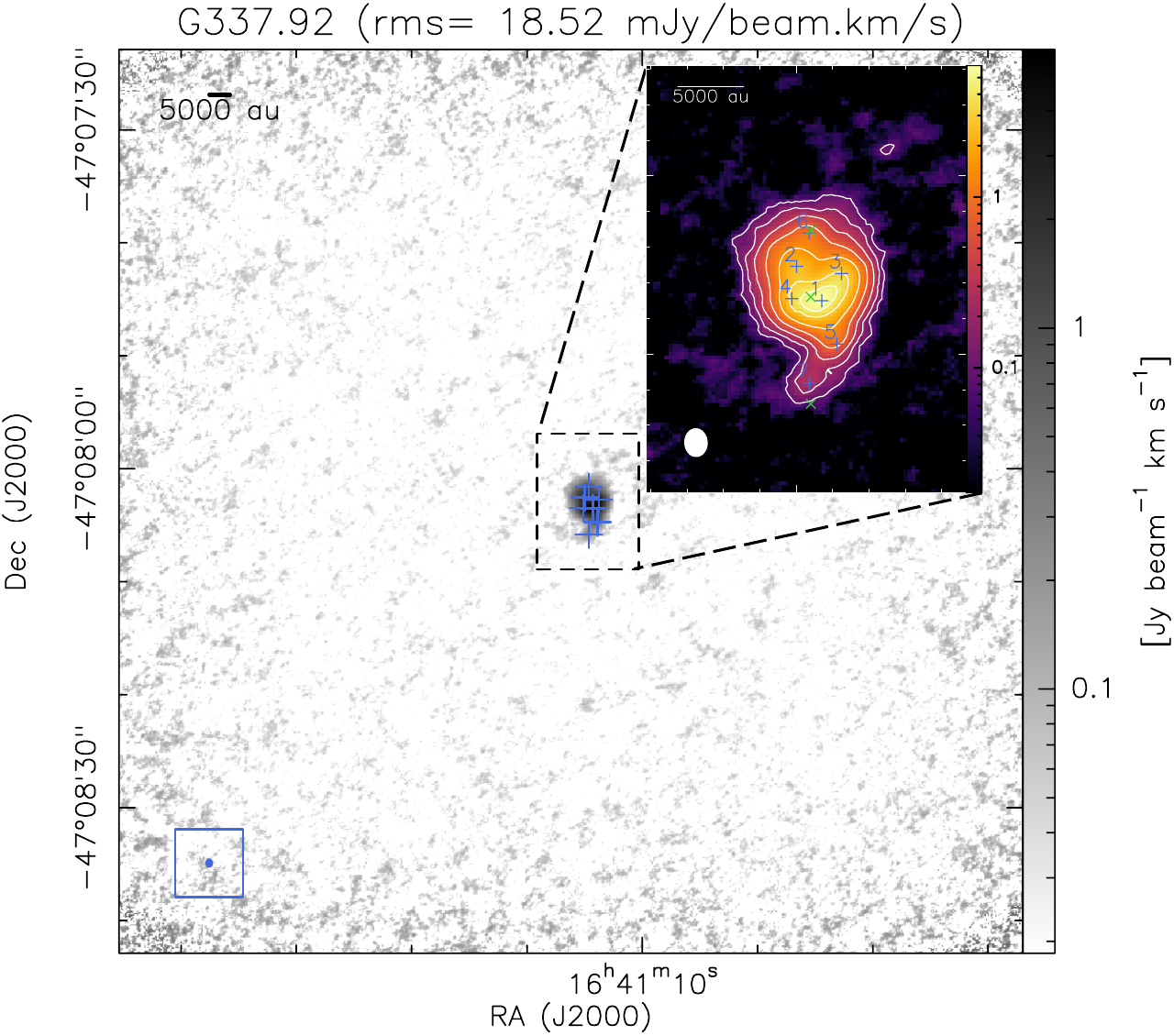}  &
       \includegraphics[width=\hsize]{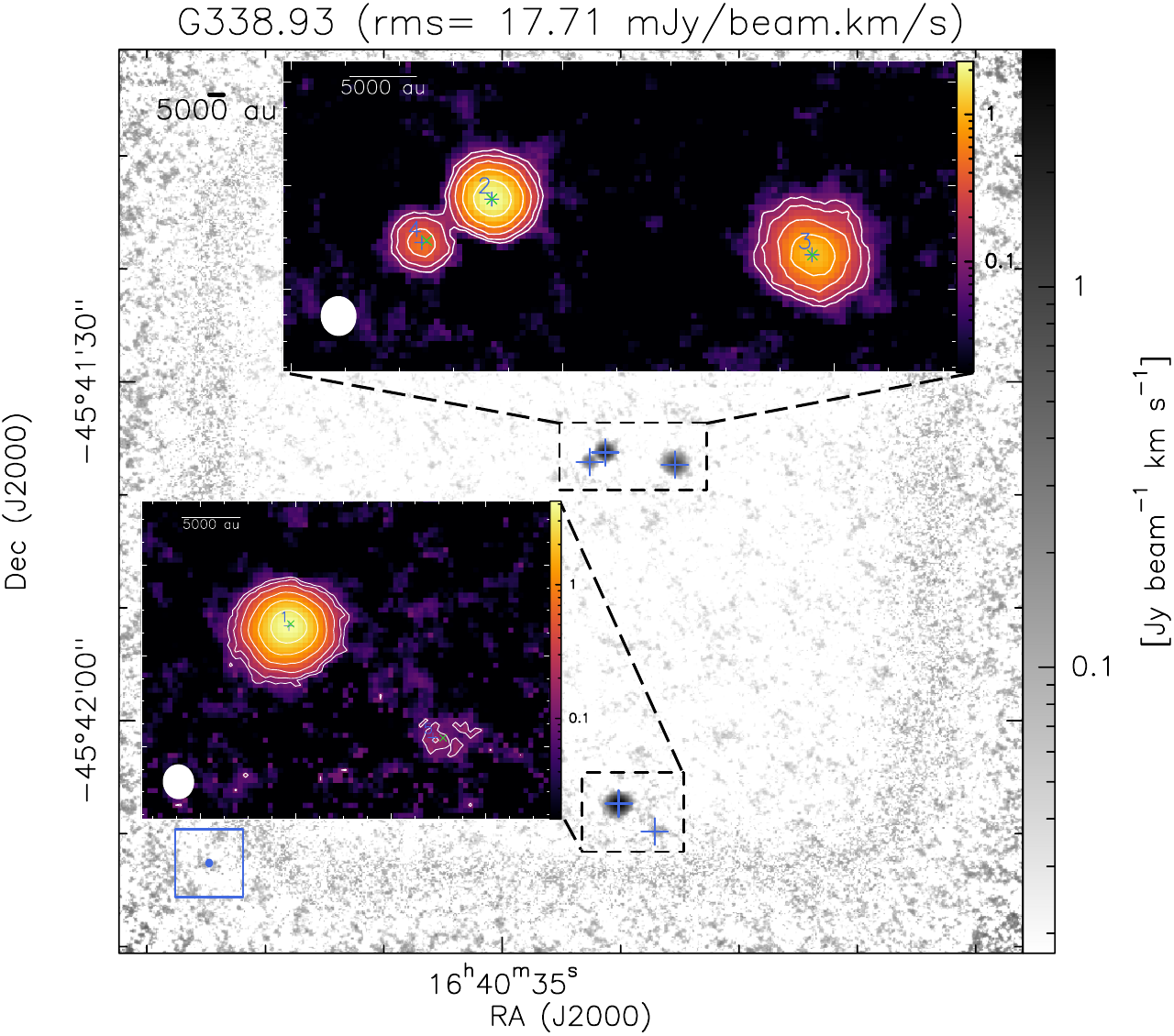}  \\
    \end{tabular}}
        \caption{\label{FIG-mom0-maps2} Same as Fig.\,\ref{FIG-mom0-maps}. In the case of G328.25, the two light blue crosses show the peak position of methyl formate initially extracted by \textsl{GExt2D}, that correspond to the position of the accretion shocks identified by \citet{csengeri2018}, of which the positions are marked with blue triangles. The figure continues on the next page.}
    \end{figure*}

\begin{figure*}[h!]
    \resizebox{\hsize}{!}
   {\begin{tabular}{cc} 
       \includegraphics[width=\hsize]{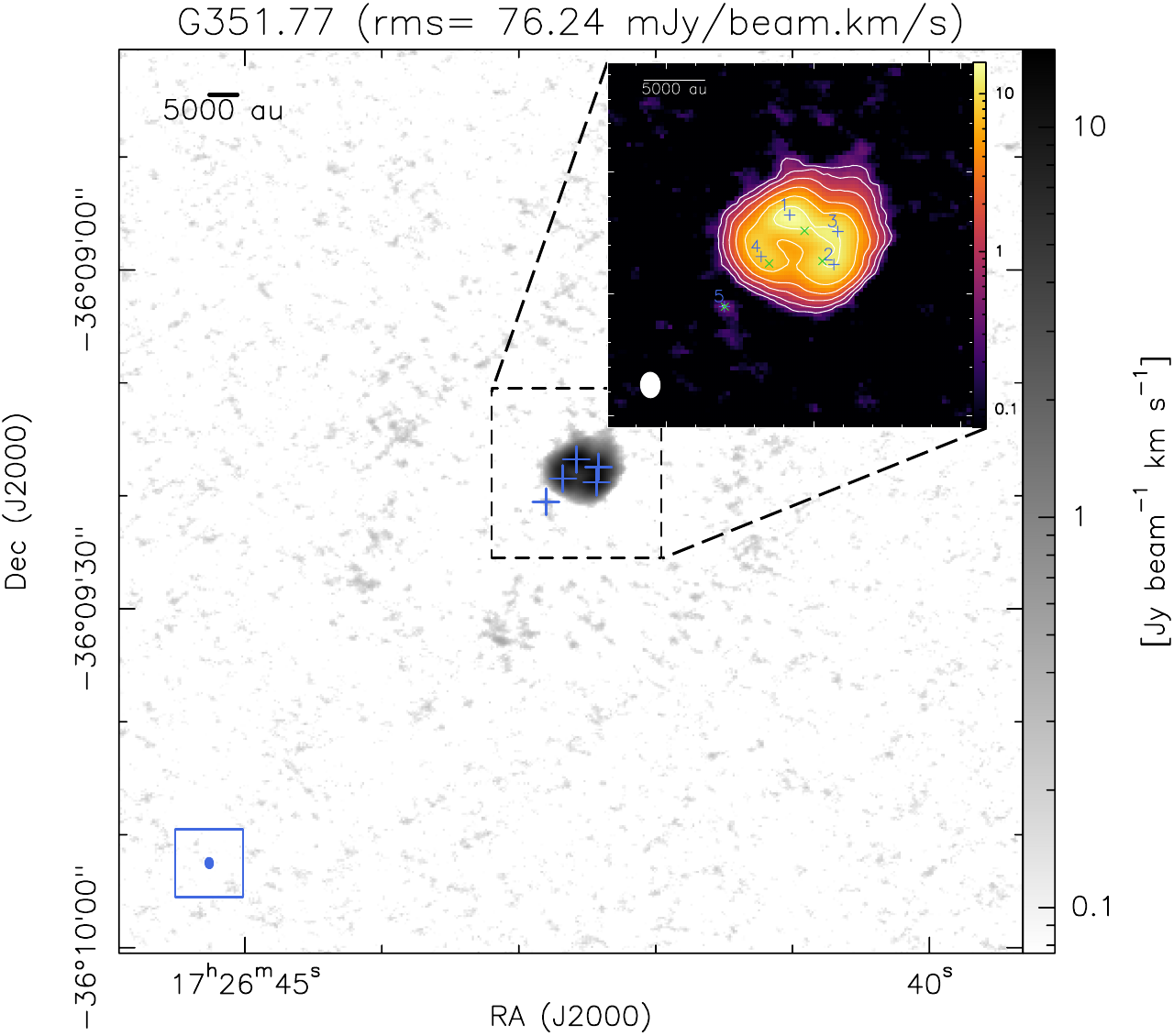} &
       \includegraphics[width=\hsize]{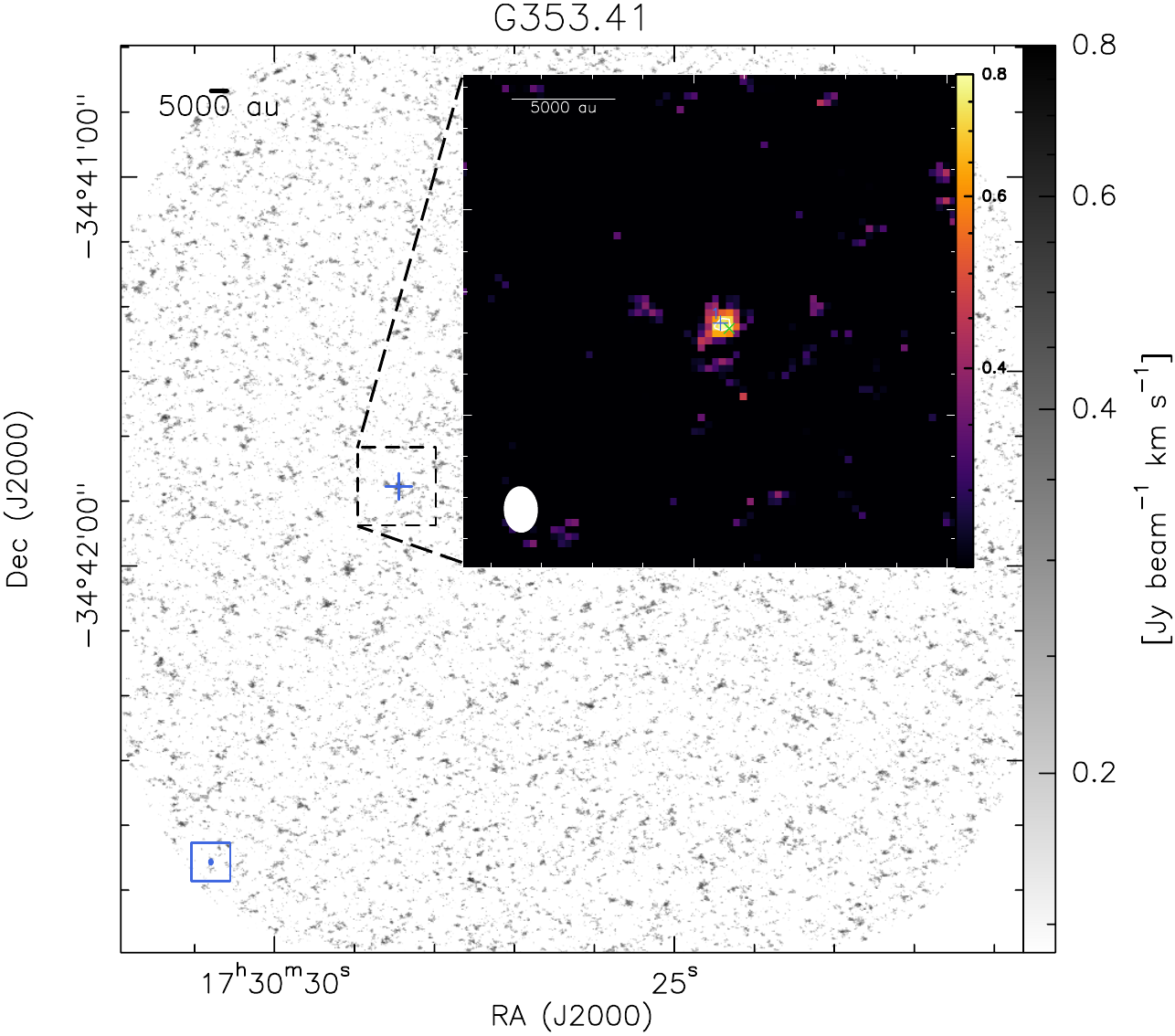}  \\
       \includegraphics[width=\hsize]{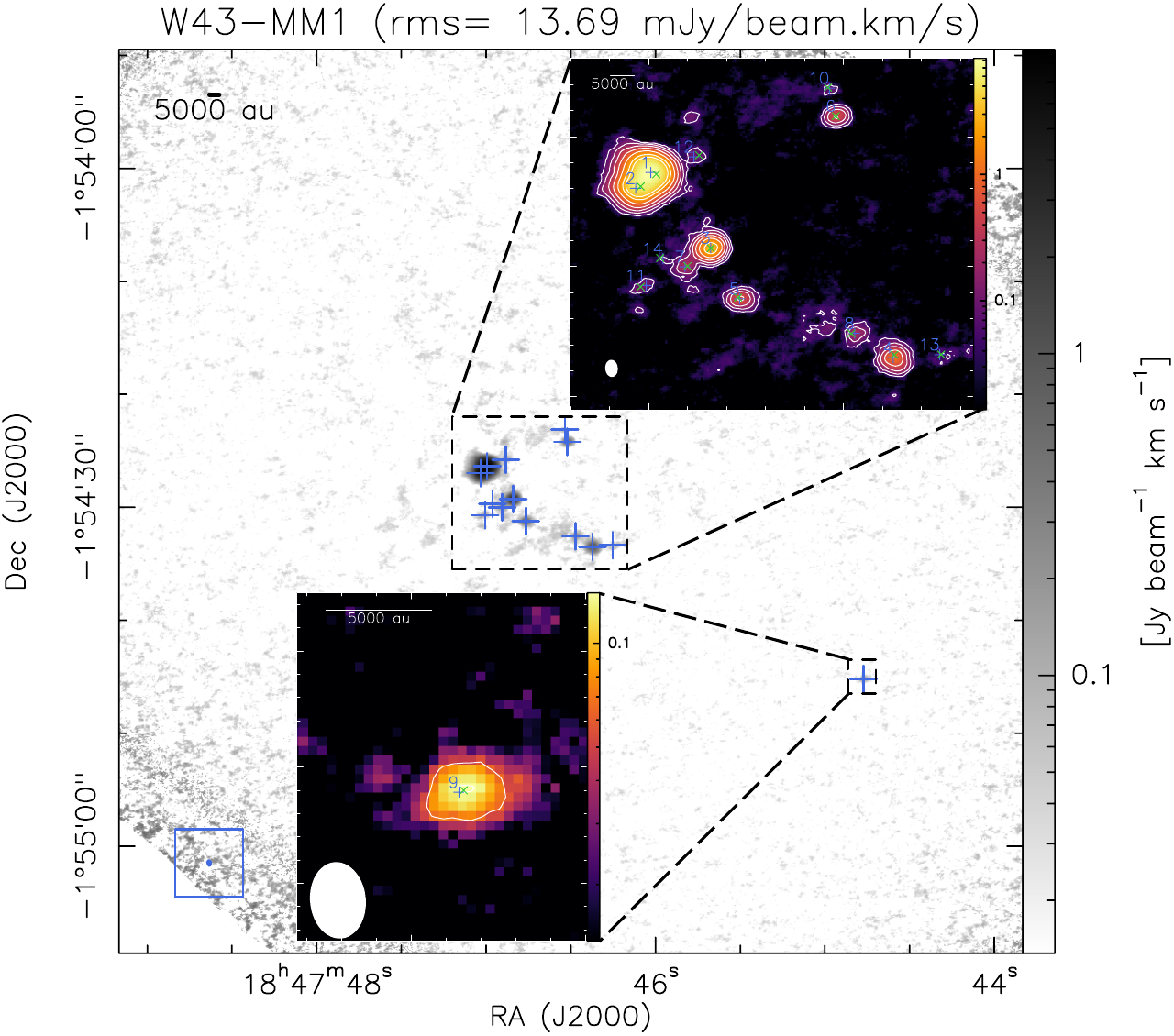}  &
       \includegraphics[width=\hsize]{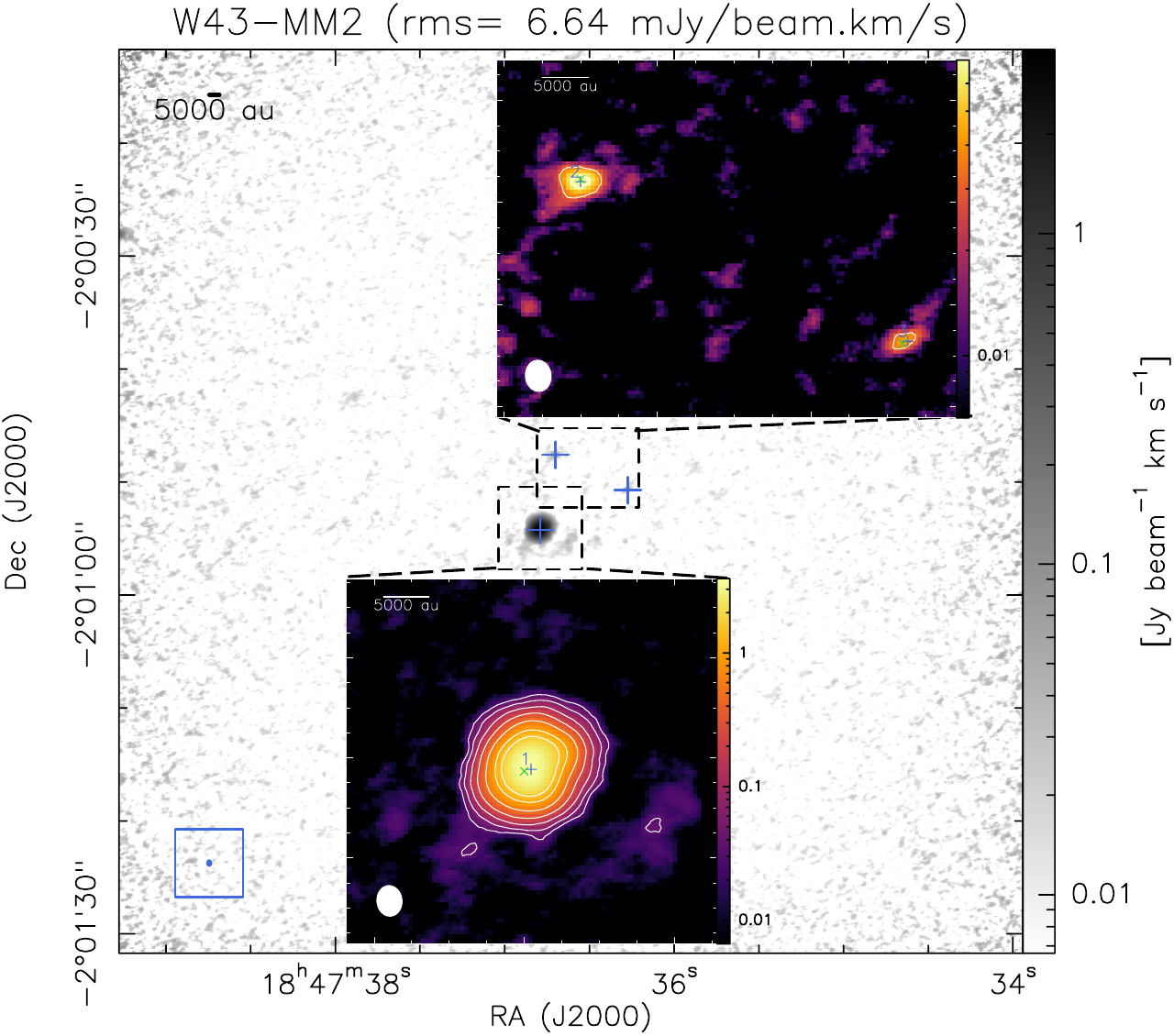}  \\
    \end{tabular}}
        \caption{\label{FIG-mom0-maps3} Same as Fig.\,\ref{FIG-mom0-maps}. The figure continues on the next page.}
    \end{figure*}

\begin{figure*}[h!]
    \resizebox{\hsize}{!}
   {\begin{tabular}{cc}  
       \includegraphics[width=\hsize]{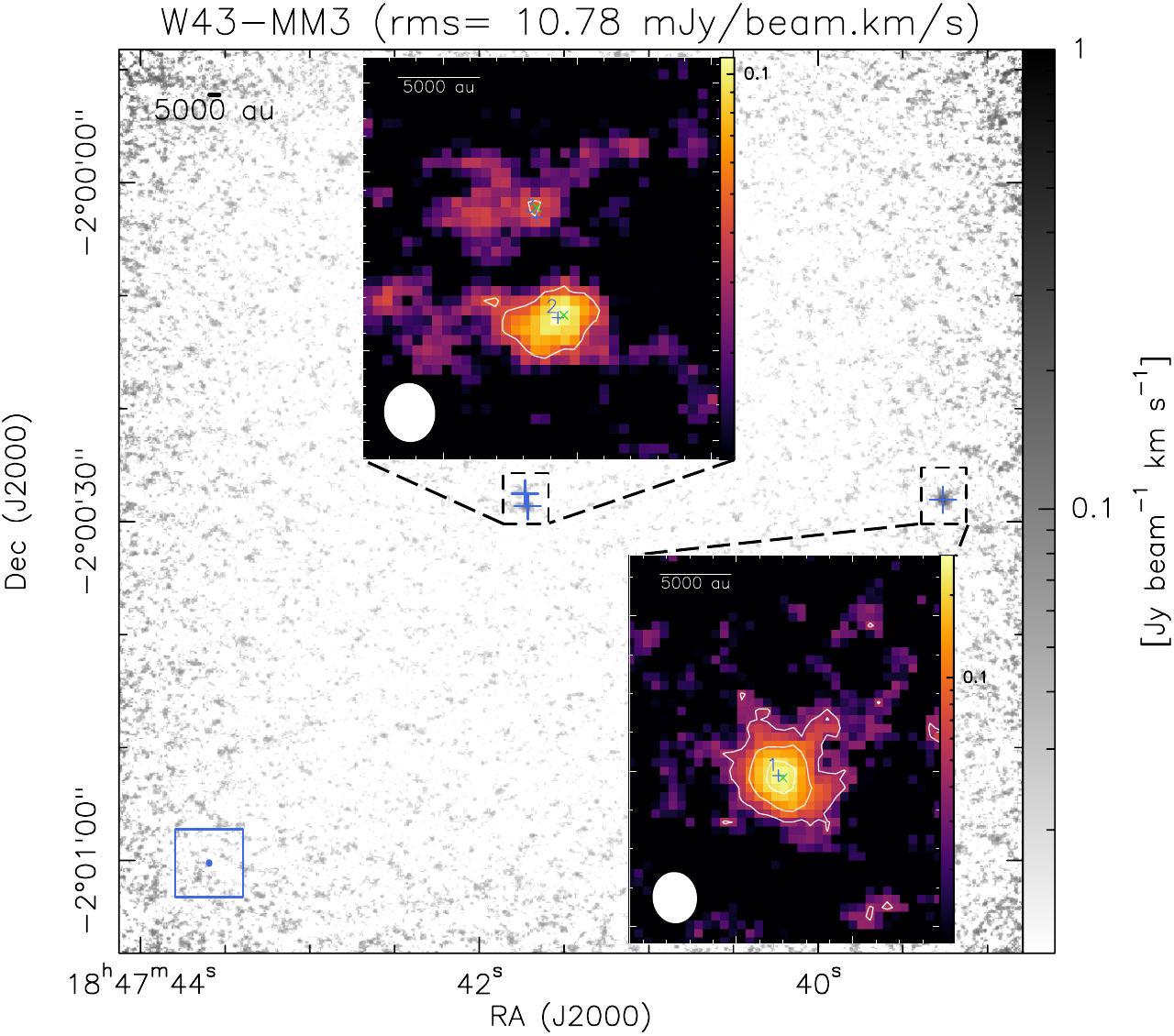}  &
       \includegraphics[width=\hsize]{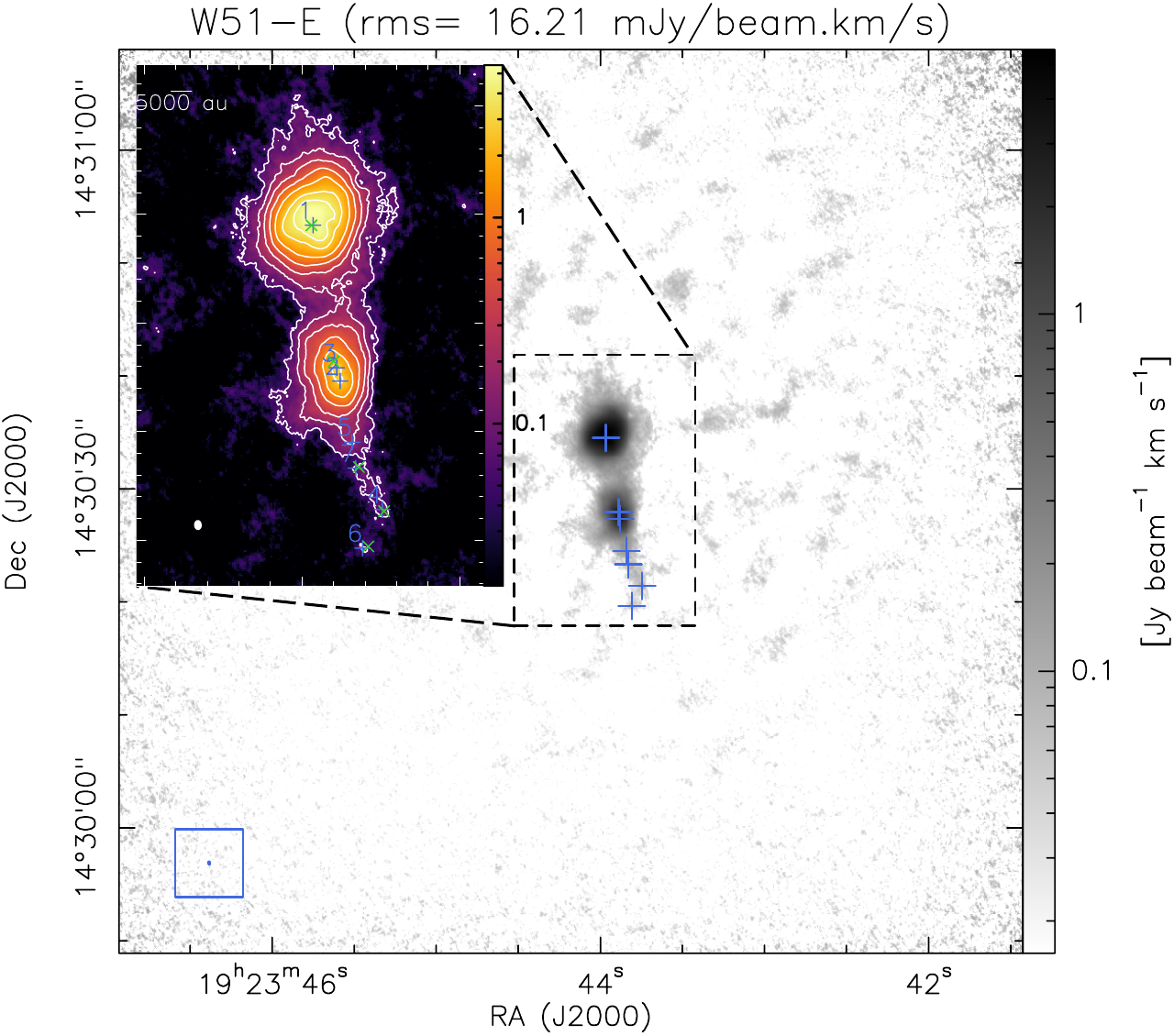}  \\
       \includegraphics[width=\hsize]{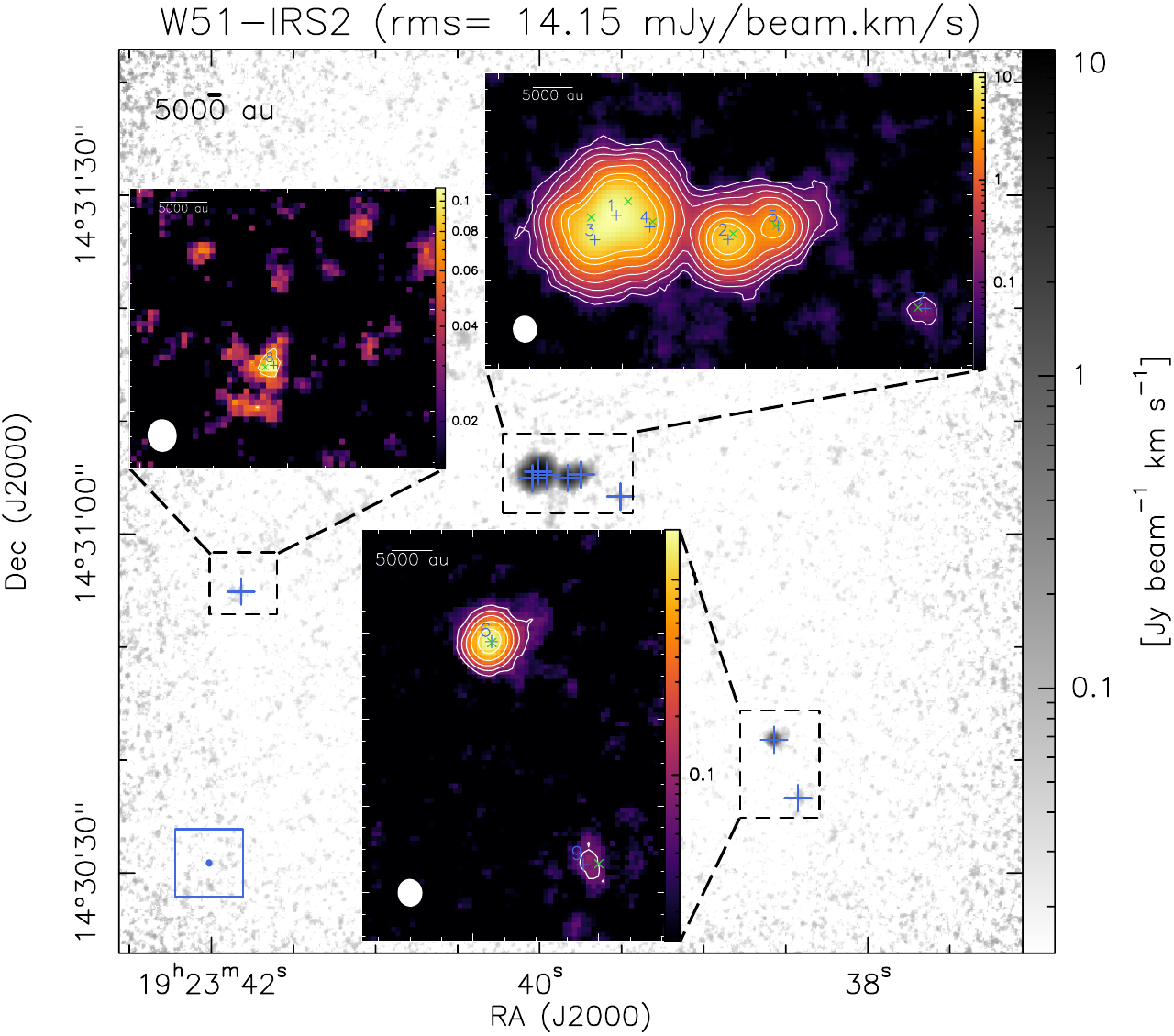}  \\
    \end{tabular}}
        \caption{\label{FIG-mom0-maps4} Same as Fig.\,\ref{FIG-mom0-maps}.}
    \end{figure*}

Figures \ref{FIG-mom0-maps}--\ref{FIG-mom0-maps4} display the moment 0 maps of the methyl formate line pair 2 and shows that the emission from \mf\ traces a diversity of structures across the 15 ALMA-IMF protoclusters. We can mainly distinguish two types of structures: 
\begin{itemize}
\item extended structures ($>$5000 au) that may contain one or more sources, this is the case of five ALMA-IMF protoclusters: G010.62, G327.29, G337.92, G351.77, and W51-E, two of which are young, two are intermediate, and one is evolved according to \citetalias{motte2022}. In the case of G010.62, G337.92, and G351.77, the methyl formate emission exhibits a more complex spatial structure that is not axisymmetric (\textit{i.e.} not circular).
\item The other ten protoclusters harbor individual objects, with rather compact, elliptical or circular emission, with an extent of a few thousands au, that may be clustered or isolated.
\end{itemize}

%======================
% FIGURE: INTEGRATED SPECTRA
%======================
%======================

%======================
% TABLE: TRANSITIONS 
%======================
\begin{table*}[!t]
\begin{center}
  \caption{\label{TAB-transitions} Properties of the investigated transitions.} 
  \setlength{\tabcolsep}{1.6mm}
  \begin{tabular}{ccrcc}
    \hline
    \hline
    Species       & Freq         & E$_{\mathrm{up}}/k$ &  A$_{ij}$  & J$_{\mathrm{up}}$(K$_{\mathrm{a}}$,K$_{\mathrm{c}})$ – J$_{\mathrm{low}}$(K$_{\mathrm{a}}$,K$_{\mathrm{c}}$) \\
                  & [MHz]        & [K]     &  [s$^{-1}$] &  \\
    \hline
    \multicolumn{5}{c}{Line pair 1} \\
    \hline
    \mf,\,vt=0  & 216109.780   & 109.3 & 1.49$\times$10$^{-4}$ &  19(2, 18) -- 18(2, 17) E \\ %vb=1
    DCO$^+$,v=0      & 216112.582   & 20.7 & 7.66$\times$10$^{-4}$   &  3 -- 2 \\    
    \mf,\,vt=0  & 216115.572   & 109.3 & 1.49$\times$10$^{-4}$ & 19(2, 18) -- 18(2, 17) A  \\ %vb=0
     \hline
    \multicolumn{5}{c}{Line pair 2} \\
    \hline
    \mf,\,vt=0  & 216210.906   & 109.3 & 1.49$\times$10$^{-4}$  & 19(1, 18) -- 18(1, 17) E  \\ %vb=2
    \mf,\,vt=0  & 216216.539   & 109.3 & 1.49$\times$10$^{-4}$  &  19(1, 18) -- 18(1, 17) A \\ %vb=0
    \hline
\end{tabular}
\end{center}
\vspace{-4mm}
\tablefoot{The spectroscopic predictions (frequencies, upper energy levels and Einstein coefficients) are taken from the JPL catalog \citep{pearson2010}.}
\end{table*}
% ======================

%======================
% FIGURE: moment 0 MAPS
%======================

%======================   

%------------------
\subsection{Source extraction}
%------------------
\label{section-MFextraction}

Given the large dataset used for this analysis, with varying dynamic range and morphology across the different fields, the method used for the source extraction must be as homogeneous and automatic as possible. Therefore, in order to extract in a systematic way compact and centrally peaked methyl formate sources from the 15 moment 0 maps, we use the source extraction algorithm \textsl{GExt2D} \citep[][]{bontemps2024}, which is based on a Gaussian fitting of the strongest curvature points in intensity maps and optimised for compact source identification, similar to the \textsl{CutEX} algorithm of \citet{molinari2011}. The source extraction and characterization is made in two steps:
\begin{itemize}
\item in a first step, \textsl{GExt2D} computes the second derivative of the \mf\ moment 0 map and looks for local maxima in the curvature map, that indicates the presence of compact sources, of which it extracts the coordinates of the central position.
\item In the second step, source sizes (FWHM) and the peak values of the integrated intensity maps (Jy beam$^{-1}$ km s$^{-1}$) are measured for each individual source by fitting 2D Gaussians to its central position, in the primary-beam corrected \mf\ moment 0 map. 
\end{itemize}
In order to facilitate the source detection in the first step, we use the moment 0 maps prior to the correction for the primary beam response, which exhibit a homogeneous noise level in the entire field. However, since we cover some of the brightest Galactic protoclusters, some maps are affected by dynamic range limitations. This is particularly an issue for the G327.29 protocluster (see Fig. \ref{FIG-mom0-maps}) and leads to a significantly larger average noise over the map, due to the central, brightest source being surrounded by strong sidelobes. In order to prevent \textsl{GExt2D} from detecting spurious sources (\textit{i.e.} bright emission associated with strong sidelobes), we have manually identified in each map a region in which the noise is the most representative of the whole field, which is different from the polygon we used to measure the rms noise level in the line cubes in Sect. \ref{section-linecubes}. The source extraction starts with the strongest fluctuation in the map and proceeds to fainter fluctuations, finding local maxima down to noise-dominated curvature values. To be ultimately selected, a peak must be significant both in curvature and intensity. We set the detection threshold to a signal-to-noise ratio of 2.5, that is related to the local noise fluctuation in the curvature map. The detection thus stops when it reaches a S/N\,=\,2.5 in curvature for a single pixel. We note that for the faintest sources, an offset of 1-2 pixels with respect to the real peak of emission may occur, which can be explained by an inhomogeneous noise distribution in the image or because of the background subtraction.

In order to remove spurious sources from our catalog, we visually inspected the single-pixel spectra extracted towards the peak position of all the sources identified with \textsl{GExt2D}. As some spectra may show strong fluctuations due to inhomogeneous noise or inaccurate continuum subtraction, only the sources for which the two \mf\ line pairs are detected above the 3\,$\sigma$ noise level given in Table\,\ref{TAB-linecubes} are considered as robust detections and are used in the rest of our analysis. Their spectra are showed in Figs.\,\ref{FIG-MF-spectra}--\ref{FIG-MF-spectra4}.

In the case of G327.29 and G351.77, a closer look at the spectra extracted towards the individual methyl formate sources, in particular G327.29--MF1, G351.77--MF1, MF2 and MF3, shows that the velocity range used for the moment 0 maps is marginally contaminated by  emission from other spectral lines. Using a narrower velocity range for the moment~0 maps for these sources gives, however, consistent parameters for the peak position and deconvolved source size. The indicated velocity range is, however necessary to extract all methyl formate emission observed towards the fainter sources G351.77--MF5, G327.29--MF1 and MF2. For this reason, for the rest of our analysis we use the same velocity range of 35\,km\,s$^{-1}$ for G327.29 and G351.77 as for the other regions.

%------------------
\subsection{Fraction of channels containing emission}
%------------------
\label{section-line-density}

We use the spectra shown in Figs.\,\ref{FIG-MF-spectra}--\ref{FIG-MF-spectra4} to assess the spectral line richness of each methyl formate source. To do so, we count the number of channels that contain emission above the 3$\sigma$ noise level, using the $rms$ values listed in Table\,\ref{TAB-linecubes}. The percentage of channels containing emission above \mb{3$\sigma$} in the spectrum observed towards each methyl formate source is shown in Table\,\ref{TAB-MFcat}. These values range between 1 and 77\%, where the sources with the highest percentage of channels containing emission above the threshold are expected to be the richest in emission lines. This percentage is well correlated with the peak intensity measured in the mehtyl formate moment 0 maps. However, because of the sensitivity limitation of the dataset, we may miss fainter emission lines from more compact sources (see also discussion in Sect.\,\ref{section-discussion-nature-sources}). For this reason, the fraction of channels containing emission in B6-spw0 is not used as an additional quantitative criterion to classify potential hot cores in the rest of the paper.

%---------------------------------
\subsection{$V_{\mathrm{LSR}}$ estimates}
%---------------------------------
\label{section-vlsr}

Using the position of the methyl formate sources identified with the \textsl{GExt2D} algorithm, we extracted single-pixel spectra to fit the \mf\ lines. We derive the $V_{\mathrm{LSR}}$ of each methyl formate source by fitting a single component, 1D-Gaussian to each of the three methyl formate lines that are not contaminated by DCO$^+$ emission (see Table\,\ref{TAB-transitions}). The average $V_{\mathrm{LSR}}$ for each methyl formate source are provided in Table\,\ref{TAB-MFcat}. We find that in most cases, the average centroid $V_{\mathrm{LSR}}$ of the methyl formate sources are consistent with the protocluster $V_{\mathrm{LSR}}$ given in Table\,\ref{TAB-targetlist}, with velocity offsets $V_{\mathrm{off}} <$ 5\,km~s$^{-1}$, where $V_{\mathrm{off}}$ = $| V_{\mathrm{LSR}}$ (MF) -- $V_{\mathrm{LSR}}$ (protocluster)$|$. In the case of G333.60, W43-MM2, W51-E, and W51-IRS2, however, the velocity offset of some methyl formate sources is $>$ 5\,km~s$^{-1}$, and may be up to $\sim$ 9\,km~s$^{-1}$. 

Using the fits from single DCN ($J=3-2$) line observed towards the whole sample of continuum cores spectra in \citetalias{cunningham2023}, we found no obvious correlation between the spread of the core $V_{\mathrm{LSR}}$ and the evolutionary stage of the protocluster. \\

%---------------------------------
\section{The catalog of hot core candidates}
%---------------------------------
\label{section-MFcat}

Hereafter, we define a hot core candidate as a peak of methyl formate emission extracted from the moment 0 maps with the \textsl{GExt2D} algorithm. In the following subsections we present the catalog of hot core candidates, including new detections, and we discuss in more details the identification of hot core candidates in regions with compact and extended \mf\ emission.

%---------------------------------
\subsection{Statistics of hot core candidates}
%---------------------------------
\label{section-MFstat}

 All the 15 ALMA-IMF protoclusters, including the youngest ones, exhibit some emission in the investigated \mf \ transitions and harbor at least one potential hot core candidate (see Figs\,\ref{FIG-mom0-maps}--\ref{FIG-mom0-maps4}). Overall, we find a total of 76 methyl formate sources, which is about an order of magnitude less cores compared to the number of purely dust continuum cores, from the \textsl{getsf} unsmoothed catalog (\citetalias{louvet2024}, see also Sect.\,\ref{section-corecat}). The full list of methyl formate sources is given in Table \ref{TAB-MFcat}, with their coordinates and peak values measured in the \mf\ moment 0 maps with \textsl{GExt2D}. Important characteristics of the hot core candidates (FWHM sizes and total gas masses) are derived and discussed in Sect.\,\ref{section-MFproperties}.
 
\begin{figure}[!t]
\begin{center}
       \includegraphics[width=\hsize]{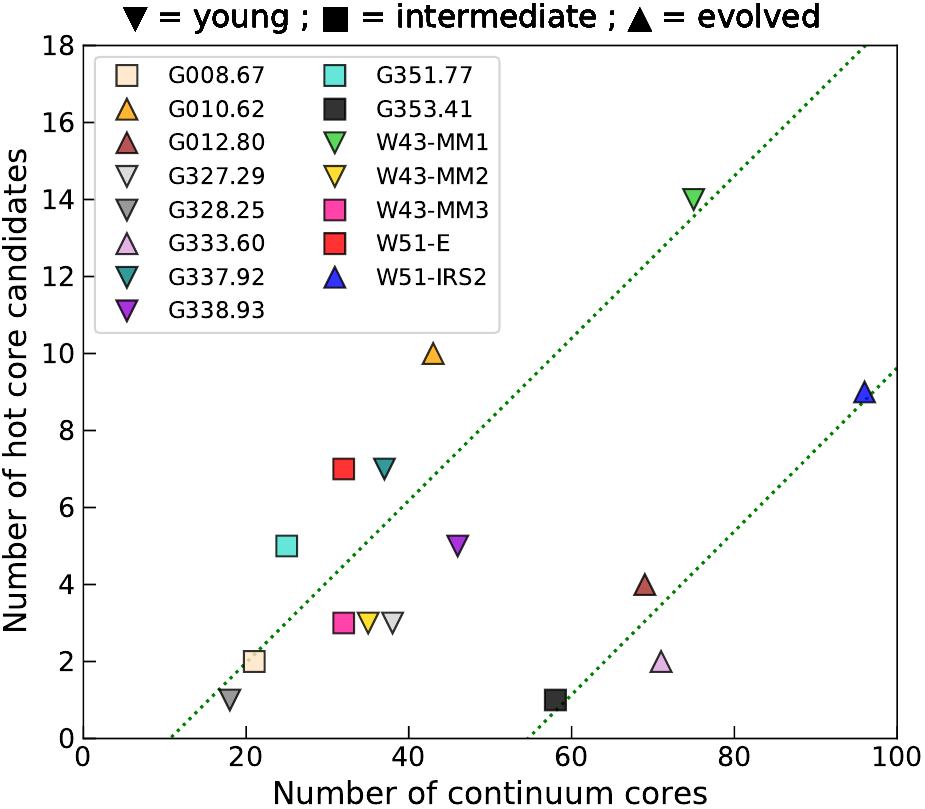} \\
  \caption{\label{FIG-nb-MFs} Number of hot core candidates per ALMA-IMF protocluster, as a function of the number of dust continuum cores from the unsmoothed core catalog presented in \citetalias{louvet2024}, excluding free-free sources. The different symbols represent the different evolutionary stages of the protoclusters: young, intermediate and evolved, as indicated on top of the figure. The dashed lines show linear fits to the two distinct groups of protoclusters.}
  \end{center}
\end{figure}

In Fig.~\ref{FIG-nb-MFs} we show the number of compact methyl formate sources per region, as a function of the number of dust continuum cores from the \textsl{getsf} unsmoothed catalog presented in \citetalias{louvet2024}, excluding free-free sources. We distinguish two groups of sources, one with the three evolved protoclusters, G012.80, G333.92, and W51-IRS2, as well as the intermediate region, G353.41, and the other one with the remaining 11 protoclusters. In both groups there is an increasing trend of the number of hot core candidates as a function of the the number of continuum cores. The region with the largest number of hot core candidates is the young protocluster W43-MM1, with as many as 14 compact methyl formate sources in a single field. The young protocluster G328.25 and the intermediate one G353.41 both harbor only a single hot core candidate. Their particular cases are further discussed in Sects.\,\ref{section-MF-compact} and \ref{section-discussion-G353}. 

%======================
% FIGURE: NB HCs VS NB CONT CORES
%======================
%======================

\begin{figure}[!t]
\begin{center}
       \includegraphics[width=\hsize]{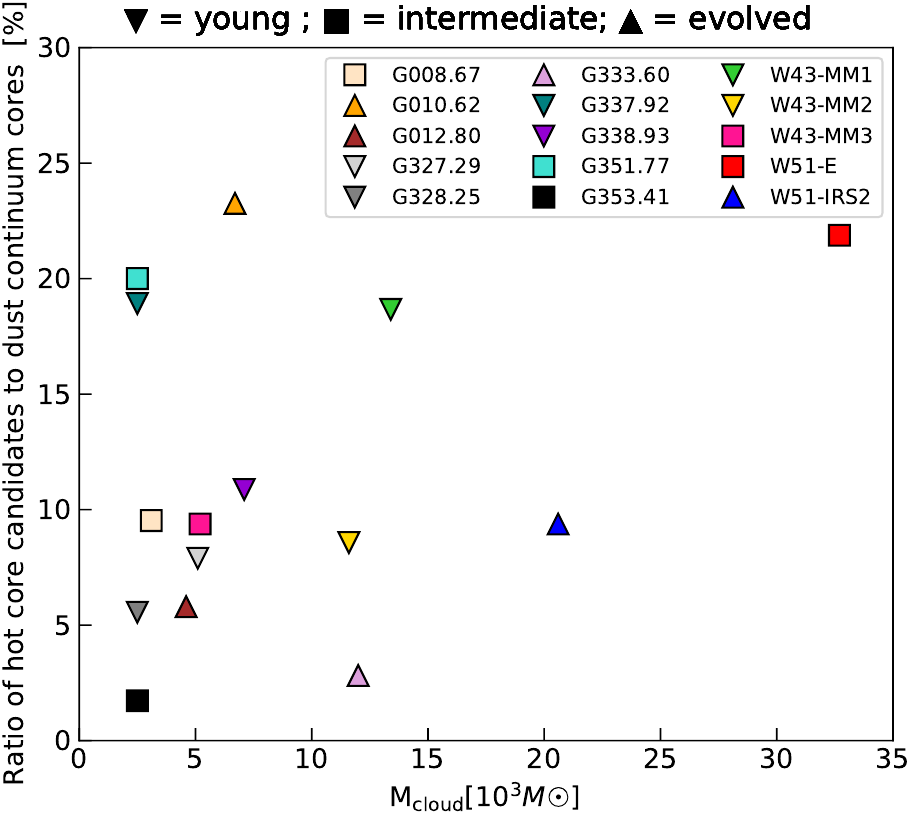}  \\
  \caption{\label{FIG-frac-MF-Mclump} Ratio of the number of hot core candidates to the number of dust continuum cores from the unsmoothed core catalog presented in \citetalias{louvet2024}, excluding free-free sources, as a function of the mass of the protocluster ($M_{\rm cloud}$). The different symbols represent the different evolutionary stages of the ALMA-IMF protoclusters: young, intermediate and evolved, as indicated on top of the figure.}
\end{center}
\end{figure}

In Fig.~\,\ref{FIG-frac-MF-Mclump} we show for each ALMA-IMF protocluster, the ratio of the number of hot core candidates to the number of dust continuum cores, as a function of the mass of the protocluster, $M_{\rm cloud}$. It shows that in all cases, the number of hot core candidates per region never represents more than 25\% the number of dust continuum cores. Furthermore, no clear trend emerges, neither as a function of clump mass, nor of the evolutionary stage of the protocluster. Young, intermediate, and evolved protoclusters do not exhibit any clear difference, suggesting that the methyl formate source properties are independent of the evolutionary stage of their hosting clumps.

%======================
% FIGURE: FRACTION HC/CONT VS MClump
%======================
%======================

%---------------------------------
\subsection{Hot core candidates detected in regions with compact \mf\ emission}
%---------------------------------
\label{section-MF-compact}

For nine out of the 15 ALMA-IMF protoclusters, the source identification is relatively straightforward since they mainly harbor individual objects, with rather compact, elliptical or circular emission, with an extent of a few thousands au. 

In particular, G008.67 harbors two individual, elliptical, compact sources. Towards G012.80 we identified four individual, rather elliptical sources, two of which are well resolved, and two are compact sources. We identified two faint methyl formate sources towards G333.60 that is one of the most evolved regions in our sample. G338.93 is a young region that harbors 5 isolated, circular, compact sources. G353.41 is a more evolved region that is very bright in the continuum at 1.3 mm, and strongly affected by ionized gas coming from {\uchii} regions (see Fig. 2 of \citetalias{motte2022}). This region is a remarkable outlier of the ALMA-IMF sample as it hosts only one weak \mf\ source, despite the fact that it hosts a large number of continuum cores, with 57 sources identified in the \textsl{getsf} unsmoothed core catalog (see also Sect.\,\ref{section-discussion-G353}). The largest number of methyl formate sources, 14, is found towards the young protocluster W43-MM1, where most of the sources are resolved and appear as isolated sources. We identified three individual compact methyl formate sources towards both W43-MM2 and W43-MM3, of which the larger ones are rather circular. 

The case of G328.25 is somewhat particular because \citet{csengeri2019} show extended \mf\ emission associated with accretion shocks (see the blue triangles in Fig. \ref{FIG-mom0-maps2}), that are resolved at an angular resolution of 0.23$\arcsec$ ($\sim$ 575\,au at the distance of G328.25). These two distinct peaks have also been identified and extracted with the \textsl{GExt2D} algorithm from the ALMA-IMF \mf\ moment~0  map (see the light blue crosses in Fig. \ref{FIG-mom0-maps2}), where the emission is marginally extended in \mf\ at an angular resolution of 0.67\arcsec ($\sim$ 1675\,au). Based on an unbiased spectral line survey obtained with the APEX telescope towards G328.25, \citet{bouscasse2022} analysed the molecular composition of this region and extracted the excitation conditions for several species. Based on the properties of COMs, they suggest that this source corresponds to an emerging hot core. We thus report the peak positions of the \mf\ emission in Table\,\ref{TAB-MFcat} (as G328.25--shock1 and shock2), but we consider this source to be a single core, at the peak position of the continuum core.

%These positions correspond to the location of accretion shocks identified by \citet{csengeri2018, csengeri2019}.

%---------------------------------
\subsection{Hot core candidates detected in regions with extended \mf\ emission}
%---------------------------------
\label{section-MF-extended}

The other six ALMA-IMF protoclusters exhibit both compact sources and extended emission of methyl formate. G327.29 and W51-E, are dominated by a central bright source, while the four other protoclusters exhibit extended, non axisymmetric emission. 

The central source of G327.29 is dominated by extremely bright emission in methyl formate, in fact both the methyl formate and the continuum emission features are similar, circularly symmetric, except towards its central position (see Fig\,\ref{FIG-mom0-maps}), where an arc-like emission feature suggests that the lower part of the circle is brighter. Such features could be explained by intrinsic inhomogeneity in the \mf\ emitting gas, but also by dust opacity. With a 2D Gaussian fit to the \mf \ emission, we measure an extent of 2.7 \arcsec\ (deconvolved FWHM), which corresponds to a size of $\sim$6800 au at the distance of G327.29, and is $>$3 times larger that the synthesized beam of the line datacube. This size is considerably larger than most of the other methyl formate sources that are typically compact sources. For simplicity, we consider the bright source seen in methyl formate towards G327.29 to be a single, individual core (G327.29--MF1) associated with the peak position of the continuum emission, which is consistent with the results of \citet{gibb2000, bisschop2013, wyrowski2008}. Two additional, individual, fainter methyl formate sources are detected towards G327.29, well offset from the central source. 

We find another source similar to G327.29--MF1 that is in the W51-E protocluster, W51-E--MF1, also known in the litterature as W51-e2. This central source is dominated by very bright circular emission, extended up to 2.5$\arcsec$, which corresponds to $\sim$ 13400\,au at the distance of the protocluster, and is $>$6 times larger that the synthesized beam of the line datacube. In this case, assuming a single source associated with the peak of the continuum emission is consistent with the results presented by \citet{ginsburg2017} and \citet{goddi2020} who argue that this source is powered by a single central massive star. In addition to the central source, two methyl formate sources have been identified in the bright emission South of the main one, which is elongated in the North-South direction. We could also identify in the same direction, four additional fainter, clustered sources.

%2\rlap{.}{$\arcsec$}5

Towards the W51-IRS2 protocluster, nine methyl formate sources have been identified that are rather circular. Five of them are particularly bright and clustered in the center of the field. These sources could easily be identified by our source-extraction algorithm, and they indeed correspond to the same peaks seen in methyl formate moment 0 maps obtained at higher angular resolution  of 0.2\arcsec\ by \citet{ginsburg2017} (see their Fig.\,4).

We find that three regions, G010.62, G351.77, and G337.92, exhibit extended \mf\ emission with a complex clustered structure. In this case, the \mf\ peaks are surrounded by non axisimmetric extended emission, and we report here only the peak positions extracted by \textsl{GExt2D}. In the case of G351.77 and G337.92, we identified four and seven individual sources, respectively. In the case of G010.62, that is a well known {\uchii} region, we detected two isolated methyl formate sources, and eight more sources in a clustered blob in the center of the field. The nature of these sources, associated with the {\uchii} region, is further discussed in Sect.\,\ref{section-discussion-nature-sources}.

%---------------------------------
\subsection{Newly discovered hot core candidates}
%---------------------------------
\label{section-MFnew}

In this section we discuss the compact methyl formate sources identified with our analysis that were not qualified before as hot cores in the literature, and are thus newly discovered hot core candidates based on the ALMA-IMF Large Program. Overall, we find 56 sources that could be considered as new hot core candidates, which represents more than two third (76\%) of the ALMA-IMF methyl formate source sample.

\paragraph{G008.67} harbors two compact methyl formate sources, of which G008.67--MF2, coincides with the compact hot core identified from \ce{CH3CN} observations conducted with the SMA at about 3$\arcsec$ resolution, which corresponds to $>$10000\,au at the distance of G008.67 \citep{hernandez2014}. G008.67--MF1 is a new detection. 

\paragraph{G012.80} harbors four compact methyl formate sources. Our hot core candidate G012.80--MF2, corresponds to the W33-Main North region, while G012.80--MF1 and G012.80--MF4 coincide with W33-Main West source in the SMA continuum map at 345 GHz from \citet{immer2014} (see their Fig. 7 at 2.3$\arcsec$ resolution, which corresponds to 5500~au at the distance of G012.80). While they discuss the nature of these regions, these sources have not been qualified as hot cores. We thus consider four new hot core candidates towards G012.80. 

\paragraph{G333.60} is known as a bright and extended {\hii} region \citep{lo2015}, for which we are not aware of dedicated observations to search for hot core emission at high angular resolution. We identified two faint methyl formate sources in the \mf\ moment 0 maps obtained towards G333.60, which are new detections. 

\paragraph{G338.93} harbors five compact methyl formate sources which have never been reported as hot cores before, to the best of our knowledge. We thus consider them as five new detections.

\paragraph{G351.77} has been previously recognised as a bright hot core by several authors, such as \citet{leurini2008, liu2020, taniguchi2023}. Thanks to our improved angular resolution, we could split the bright emission in the central part of G351.77 and thus report four new detections in this region. \citet{beuther2017} resolves the small-scale structure of the G351.77 hot core down to 0.06$\arcsec$ angular resolution and find indication for multiplicity at such small scales.

\paragraph{G353.41} was recently covered by the ATOMS survey \citep[][at 1.6$\arcsec$, which corresponds to 3200\,au at the distance of the  protocluster]{liu2020}, however no hot core detection has been reported towards G353.41. Our hot core candidate is therefore a new detection in this region. 

\paragraph{W43-MM1, W43-MM2, and W43-MM3} constitute a mini starburst region. We find 14 compact methyl formate sources towards W43-MM1, eight of which correspond to the positions identified by \citetalias{brouillet2022} (see also second column of Table\,\ref{TAB-MFcat}). One of our sources is outside their investigated field of view, and five are new detections. In this case, our approach using only the \mf\ emission is more sensitive compared to their method, relying on line density estimates within a broader bandwidth ($\sim$ 2 GHz) towards the peak positions of the continuum cores. Together with the six sources detected towards W43-MM2 and W43-MM3, we detect 11 new hot core candidates in the W43 protocluster.

\paragraph{G327.29} harbors a well known central hot core. In addition, we identified two other fainter sources, G327.29--MF2 and G327.29--MF3, well offset from the central source. Their positions coincide with the continuum peaks SMM2 and SMM4 identified by \citet{leurini2017} in the SABOCA continuum emission map at 350~$\mu$m (see their Fig. 3). Since these sources were not qualified as hot cores by \citet{leurini2017}, we consider them as new detections.

\paragraph{W51-E and W51-IRS2} have been previously studied and recognized as hosting several bright hot cores \citep[see, e.g.,][and references therein]{ginsburg2017}. Only the fainter methyl formate sources extracted from the moment 0 maps are considered as new detections. It represents four sources towards W51-E (MF4 -- MF7), and another four towards W51-IRS2 (MF6--MF9).

\paragraph{G337.92} has not been the subject detailed high angular-resolution studies on its chemical content before the ALMA-IMF program, and thus the seven individual methyl formate sources are considered as new detections.

\paragraph{G010.62} is another prominent hot core in the Galactic plane. Our hot core candidates G010.62--MF3 and G010.62--MF5 correspond to the well resolved individual objects MF1 and MF2 from \citep{law2021} based on ALMA observations at higher angular resolution compared to that of ALMA-IMF. G010.62--MF3 and G010.62--MF2 correspond to source 1 and 2 from \citet{taniguchi2023}. Furthermore, some of our remaining hot core candidates correspond to well identified peaks in \ce{CH3OH} in \citet{law2021}, although, they have not been identified and discussed as hot cores. Overall, we propose seven sources to be new detections in this region. 

%======================
% TABLE: catalog HC candidates         
%======================
\begin{table*}[h!]
  \caption{\label{TAB-MFcat} Catalog of the sources extracted from the moment 0 maps of methyl formate, using the \textsl{GExt2D} algorithm.} 
  \setlength{\tabcolsep}{0.4mm}
  \resizebox{\textwidth}{!}{\begin{tabular}{llllrrrrrrrrcl}
    \hline
    \hline
   \multicolumn{1}{c}{ID}  & \multicolumn{1}{c}{Name} & \multicolumn{1}{c}{RA$^{(a)}$} & \multicolumn{1}{c}{Dec$^{(a)}$} & \multicolumn{1}{c}{$S^{\mathrm{peak} (b)}_{\mathrm{MF}}$} & \multicolumn{1}{c}{S/N$^{(c)}$} & \multicolumn{1}{c}{$\theta_{\mathrm{maj}} \times \theta_{\mathrm{min}}^{(d)}$} &   \multicolumn{1}{c}{PA$^{(d)}$} & \multicolumn{1}{c}{$\theta_{\mathrm{maj}}^{\mathrm{dec}} \times \theta_{\mathrm{min}}^{\mathrm{dec} (e)}$} &  \multicolumn{1}{c}{PA$^{\mathrm{dec} (e)}$} & \multicolumn{1}{c}{FWHM$_{\mathrm{MF}}^{\mathrm{dec} (f)}$} & \multicolumn{1}{c}{$V_{\mathrm{LSR}}^{(g)}$} & \multicolumn{1}{c}{\%channels$^{(h)}$} & \multicolumn{1}{c}{Tentative} \\  
\multicolumn{2}{c}{ }  & \multicolumn{1}{c}{[h:m:s]}  & \multicolumn{1}{c}{[$^{\circ}$:$\arcmin$:$\arcsec$]} & \multicolumn{2}{c}{[mJy beam$^{-1}$}  & \multicolumn{1}{c}{[$\arcsec$ $\times$ $\arcsec$]}  & \multicolumn{1}{c}{[deg]} & \multicolumn{1}{c}{[$\arcsec$ $\times$ $\arcsec$]}  & \multicolumn{1}{c}{[deg]} & \multicolumn{1}{c}{[au]} & \multicolumn{1}{c}{[km s$^{-1}$]} & \multicolumn{1}{c}{[\%]} & \multicolumn{1}{c}{classification$^{(i)}$} \\
     \multicolumn{2}{c}{ }  & \multicolumn{2}{c}{ } & \multicolumn{2}{c}{km s$^{-1}$]} &  &  &   &  &  & &  & \\
         \hline
\multicolumn{14}{c}{G008.67} \\
\hline
1 & G008.67--MF1 & 18:06:19.01 & -21:37:32.0 & 1696.2 & 35.9 & 1.11 $\times$ 0.88 & 85.2 & \_ & \_ & 1346.4 & 34.2$\pm$0.1 & 13 & HC$^*$ \\
2 & G008.67--MF2 & 18:06:23.48 & -21:37:10.5 & 850.9 & 25.6 & 1.12 $\times$ 0.87 & 107.8 & \_  & \_ & 1346.4 & 41.7$\pm$0.1 & 8 & \\
\hline
\multicolumn{14}{c}{G010.62} \\
\hline
1 & G010.62--MF1 & 18:10:28.67 & -19:55:49.2 & 745.7 & 28.3 & 1.12 $\times$ 0.87 & 93.9 & 0.99 $\times$ 0.59 & -83.2 & 3811.5 & -0.8$\pm$0.1 & 18 &  \\
2 & G010.62--MF2 & 18:10:28.70 & -19:55:50.2 & 524.4 & 19.0 & 0.93 $\times$ 0.83 & 78.5 & 0.76 $\times$ 0.54 & -88.8 & 3212.5 & -3.2$\pm$0.1 & 16 & \\
3 & G010.62--MF3 & 18:10:28.62 & -19:55:49.3 & 391.6 & 12.7 & 0.77 $\times$ 0.69 & 126.5 & 0.57 $\times$ 0.28 & -64.9 & 2019.6 & 1.6$\pm$0.1 & 26 & HC$^*$ \\
4 & G010.62--MF4 & 18:10:28.66 & -19:55:48.6 & 372.6 & 12.3 & 1.01 $\times$ 0.71 & 86.0 & 0.87 $\times$ 0.33 & -89.6 & 2653.2 &  -0.9$\pm$0.1 & 9 & \\
5 & G010.62--MF5 & 18:10:28.78 & -19:55:49.5 & 321.9 & 14.7 & 1.34 $\times$ 0.96 & 93.0 & 1.24 $\times$ 0.73 & -85.0 & 4727.2 & -4.7$\pm$0.1 & 6 & \\
6 & G010.62--MF6 & 18:10:28.61 & -19:55:47.7 & 216.9 & 12.4 & 0.84 $\times$ 0.76 & 119.1 & 0.66 $\times$ 0.41 & -67.7 & 2613.6 & -4.1$\pm$0.1 & 6 & \\
7 & G010.62--MF7 & 18:10:29.24 & -19:55:40.9 & 117.0 & 10.8 & 0.67 $\times$ 0.49 & 97.2 & 0.44 $\times$ 0.40 & -78.9 & 2098.7 & -0.2$\pm$0.6 & 7 & \\
8 & G010.62--MF8 & 18:10:28.79 & -19:55:51.0 & 50.6 & 4.8 & 0.62 $\times$ 0.60 & 95.5 & \_ & \_ & 1415.6 & -3.4$\pm$0.2 & 4 & \\
9 & G010.62--MF9 & 18:10:28.74 & -19:55:51.3 & 46.6 & 3.9 & 0.64 $\times$ 0.46 & 17.5 & \_ & \_ & 1415.6 & -4.3$\pm$0.2 & 4 & \\
10 & G010.62--MF10 & 18:10:29.11 & -19:55:45.4 & 46.5 & 6.1 & 0.62 $\times$ 0.62 & -0.7 & \_ & \_ & 1415.6 & -6.2$\pm$0.1 & 3 & \\
\hline
\multicolumn{14}{c}{G012.80} \\
\hline
1 & G012.80--MF1 & 18:14:11.83 & -17:55:32.4 & 2439.2 & 107.0 & 1.37 $\times$ 1.04 & 73.3 & 1.04 $\times$ 0.77 & 75.1 & 2160.0 & 37.8$\pm$0.3 & 12 & HC$^*$ \\
2 & G012.80--MF2 & 18:14:13.74 & -17:55:21.4 & 427.2 & 22.8 & 1.55 $\times$ 1.18 & 73.9 & 1.27 $\times$ 0.54 & 75.2 & 1994.4 & 36.8$\pm$0.4 & 5 & HC$^*$ \\
3 & G012.80--MF3 & 18:14:13.13 & -17:55:40.2 & 282.6 & 17.3 & 2.13 $\times$ 1.42 & 62.4 & 1.93 $\times$ 0.61 & 66.1 & 2611.2 & 36.9$\pm$0.2 & 3 &  \\
4 & G012.80--MF4 & 18:14:11.63 & -17:55:34.1 & 162.7 & 6.1 & 1.40 $\times$ 0.93 & 90.4 & 1.06 $\times$ 0.89 & 84.2 & 2340.0 & 37.7$\pm$0.3 & 1 & \\
\hline
\multicolumn{14}{c}{G327.29} \\
\hline
1 & G327.29--MF1 & 15:53:07.79 & -54:37:06.4 & 20680.0 & \_ & 2.85 $\times$ 2.71 & 70.1 & 2.73 $\times$ 2.59 & 74.1 & 6665.0 & -43.5$\pm$0.3 & 77 & HC \\
2 & G327.29--MF2 & 15:53:09.48 & -54:37:01.1 & 455.6 & 9.6 & 0.82 $\times$ 0.77 & -3.8 & \_ & \_ & 992.5 & -46.7$\pm$0.2 & 5  & HC$^*$ \\
3 & G327.29--MF3 & 15:53:10.89 & -54:36:46.4 & 260.0 & 10.0 & 1.00 $\times$ 0.77 & 97.4 & 0.64 $\times$ 0.27 & -76.9 & 1047.5 & -45.2$\pm$0.4 & 1 & \\
\hline
\multicolumn{14}{c}{G328.25} \\
\hline
1 & G328.25--MF1 & 15:57:59.80 & -53:58:00.7 & 2596.9 & \_ & 1.42 $\times$ 1.02 & -74.1 & 1.23 $\times$ 0.80 & -68.2 & 2490.0 & -40.0$\pm$0.1 & 14 & HC$^*$ \\
\_ & G328.25--shock1$^{(*)}$ & 15:57:59.83  & -53:58:00.7 &  &  &  &  &  &  & &   &  & shock \\
\_ & G328.25--shock2$^{(*)}$ & 15:57:59.76 & -53:58:00.8 &  &  &  &  &  &  &  &   &  & shock \\
\hline
\multicolumn{14}{c}{G333.60} \\
\hline
1 & G333.60--MF1 & 16:22:11.05 & -50:05:56.5 & 406.7 & 21.5 & 0.94 $\times$ 0.82 & 49.4 & 0.56 $\times$ 0.43 & 47.4 & 2091.6 & -52.7$\pm$0.1 & 14 & HC$^*$ \\
2 & G333.60--MF2 &16:22:08.55 & -50:06:12.4 & 79.8 & 6.4 & 0.77 $\times$ 0.51 & 125.5 & 0.32 $\times$ 0.53 & -51.4 & 1755.6 & -46.1$\pm$0.7 & 4 & HC$^*$ \\
\hline
\multicolumn{14}{c}{G337.92} \\
\hline
1 & G337.92--MF1 & 16:41:10.42 & -47:08:03.5 & 4899.7 & 50.1 & 1.31 $\times$ 0.96 & 116.0 & 1.12 $\times$ 0.53 & -61.1 & 2095.2 & -40.4$\pm$0.1 & 62 & HC \\
2 & G337.92--MF2 & 16:41:10.49 & -47:08:02.5 & 2124.8 & 28.5 & 1.15 $\times$ 1.08 & 23.4 & 0.87 $\times$ 0.80 & -32.3 & 2270.7 & -40.4$\pm$0.1 & 18 & \\
3 & G337.92--MF3 & 16:41:10.37 & -47:08:02.7 & 1946.7 & 24.6 & 1.06 $\times$ 0.96 & 6.4 & 0.77 $\times$ 0.60 & -23.9 & 1852.2 & -38.2$\pm$0.1 & 22 & HC$^*$ \\
4 & G337.92--MF4 & 16:41:10.51 & -47:08:03.4 & 1623.9 & 15.6 & 0.98 $\times$ 0.81 & 71.9 & 0.64 $\times$ 0.35 & -86.2 & 1282.5 & -42.8$\pm$0.6 & 54 & \\
5 & G337.92--MF5 & 16:41:10.38 & -47:08:04.7 & 806.0 & 6.8 & 1.10 $\times$ 0.85 & -216.0 & 0.88 $\times$ 0.29 & -40.6 & 1385.1 & -41.0$\pm$0.1 & 6 & \\
6 & G337.92--MF6 & 16:41:10.46 & -47:08:01.6 & 612.0 & 12.0 & 1.12 $\times$ 0.75 & 72.8 & 0.83 $\times$ 0.21 & 81.9 & 1136.7 & -38.7$\pm$0.1 & 10 & \\
7 & G337.92--MF7 & 16:41:10.46 & -47:08:05.8 & 223.2 & 6.7 & 0.79 $\times$ 0.75 & 40.9 & \_ & \_ & 988.2 & -40.6$\pm$0.1 & 4 & \\
\hline
\multicolumn{14}{c}{G338.93} \\
\hline
1 & G338.93--MF1 & 16:40:34.01 & -45:42:07.3 & 3977.6 & 57.3 & 1.05 $\times$ 0.97 & 92.3 & 0.79 $\times$ 0.59 & 87.3 & 2675.4 & -63.7$\pm$0.1 & 25  & HC$^*$ \\
2 & G338.93--MF2 & 16:40:34.13 & -45:41:36.3 & 2229.0 & 67.3 & 0.89 $\times$ 0.84 & 66.4 & 0.56 $\times$ 0.34 & 74.1 & 1727.7 & -61.2$\pm$0.3 & 17 & HC \\
3 & G338.93--MF3 & 16:40:33.54 & -45:41:37.3 & 1046.0 & 51.1 & 1.12 $\times$ 0.99 & 53.6 & 0.87 $\times$ 0.63 & 61.2 & 2913.3 & -61.4$\pm$0.1 & 15 & HC$^*$ \\
4 & G338.93--MF4 & 16:40:34.25 & -45:41:37.1 & 493.7 & 31.4 & 0.83 $\times$ 0.73 & 229.6 & \_ & \_ & 1419.6 & -60.0$\pm$0.2 & 10 & HC$^*$ \\
5 & G338.93--MF5 & 16:40:33.71 & -45:42:09.8 & 121.5 & 4.8 & 1.00 $\times$ 0.67 & 64.7 & 0.72 $\times$ 0.35 & 67.4 & 1977.3 & -63.5$\pm$0.1 & 4  & \\
\hline
\multicolumn{14}{c}{G351.77} \\
\hline
1 & G351.77--MF1 & 17:26:42.58 & -36:09:16.7 & 14392.7 & 27.9 & 2.05 $\times$ 1.54 & 92.9 & 1.87 $\times$ 1.10 & -88.0 & 2882.0 & -7.4$\pm$0.1 & 54 & HC$^*$ \\
2 & G351.77--MF2 & 17:26:42.43 & -36:09:18.8 & 9804.6 & 34.6 & 2.27 $\times$ 1.72 & 112.5 & 2.09 $\times$ 1.36 & -71.4 & 3388.0 & -2.4$\pm$0.1 & 55 & HC \\
3 & G351.77--MF3 & 17:26:42.41 & -36:09:17.4 & 8855.3 & 23.7 & 1.95 $\times$ 1.52 & 128.6 & 1.71 $\times$ 1.14 & -59.4 & 2800.0 & -1.9$\pm$0.1 & 45 &  HC \\
4 & G351.77--MF4 & 17:26:42.67 & -36:09:18.5 & 8059.5 & 24.6 & 1.79 $\times$ 1.19 & 61.7 & 1.56 $\times$ 0.58 & 66.7 & 1904.0 & -6.0$\pm$0.1 & 21 &  \\
5 & G351.77--MF5 & 17:26:42.80 & -36:09:20.5 & 302.1 & 2.4 & 0.76 $\times$ 0.53 & 61.8 & 0.91 $\times$ 0.40 & 77.9 & 1212.0 & -4.8$\pm$0.3 & 2 &  \\
\hline
\multicolumn{14}{c}{G353.41} \\
\hline
1 & G353.41--MF1 & 17:30:28.44 & -34:41:47.7 & 360.1 & 6.1 & 1.09 $\times$ 0.74 & -228.0 & 0.72 $\times$ 0.56 & -69.7 & 1288.0 & -19.1$\pm$3.6 & 3 & \\
\hline
\end{tabular}}
\tablefoot{The peak position$^{(a)}$, peak intensity$^{(b)}$, signal-to-noise ratio$^{(c)}$, major and minor axes$^{(d)}$ as well as position angle$^{(d)}$ of the 2D Gaussian, are derived using \textsl{GExt2D}, except for the brightest source of G327.29, G328.25 and W51-E, where the peak position of the methyl formate emission is set as the position of the brightest compact continuum core. In the case of G328.25, G328--shock1 and G328--shock2$^{(*)}$ indicate the peak positions of the methyl formate emission initially extracted by  \textsl{GExt2D}, that correspond to accretion shocks \citet[see Fig.\,\ref{FIG-mom0-maps2}, as well as][]{csengeri2018}. The major and minor axes$^{(e)}$ and position angle$^{(e)}$ deconvolved from the line cube beam size as explained in Appendix\,\ref{appendix-deconvolve-source-size}. The mean deconvolved source size$^{(f)}$ of the methyl formate emission is computed at the distance of each protocluster. When the deconvolved source size falls below the minimum size set for each protocluster (see Sect.\,\ref{section-MFsizes}), then the deconvolved major and minor axes, as well as the position angle values are left blank, and the mean deconvolved source size of the methyl formate emission (FWHM$^{\mathrm{dec}_{\mathrm{MF}}}$) is set to half the synthesized beam size of the line cube}. The rest velocity$^{(g)}$ of the source is derived from the fits to the three \mf\ lines that are not contaminated by DCO$^+$ and the uncertainty represents the standard deviation. Percentage of the total number of channels$^{(h)}$ per spw that contain emission above the 3$\sigma$ noise level (Sect.\ref{section-line-density}). The last column$^{(i)}$ indicates the methyl formate sources tentatively classified as hot cores (HC) based on their mass $>$ 8\,$\Msun$. The sources with their lowest estimated mass $<$\,8\,$\Msun$ are marked with a star (HC$^*$).The table continues on the next page.
\end{table*}

\begin{table*}[h!]
\ContinuedFloat
  \caption{continued.} 
  \setlength{\tabcolsep}{0.4mm}
   \resizebox{\textwidth}{!}{\begin{tabular}{llllrrrrrrrrcl}
    \hline
    \hline
   \multicolumn{1}{c}{ID}  & \multicolumn{1}{c}{Name} & \multicolumn{1}{c}{RA$^{(a)}$} & \multicolumn{1}{c}{Dec$^{(a)}$} & \multicolumn{1}{c}{$S^{\mathrm{peak} (b)}_{\mathrm{MF}}$} & \multicolumn{1}{c}{S/N$^{(c)}$} & \multicolumn{1}{c}{$\theta_{\mathrm{maj}} \times \theta_{\mathrm{min}}^{(d)}$} &   \multicolumn{1}{c}{PA$^{(d)}$} & \multicolumn{1}{c}{$\theta_{\mathrm{maj}}^{\mathrm{dec}} \times \theta_{\mathrm{min}}^{\mathrm{dec} (e)}$} &  \multicolumn{1}{c}{PA$^{\mathrm{dec} (e)}$} & \multicolumn{1}{c}{FWHM$_{\mathrm{MF}}^{\mathrm{dec} (f)}$} & \multicolumn{1}{c}{$V_{\mathrm{LSR}}^{(g)}$} & \multicolumn{1}{c}{\%channels$^{(h)}$} & \multicolumn{1}{c}{Tentative} \\  
\multicolumn{2}{c}{ }  & \multicolumn{1}{c}{[h:m:s]}  & \multicolumn{1}{c}{[$^{\circ}$:$\arcmin$:$\arcsec$]} & \multicolumn{2}{c}{[mJy beam$^{-1}$}  & \multicolumn{1}{c}{[$\arcsec$ $\times$ $\arcsec$]}  & \multicolumn{1}{c}{[deg]} & \multicolumn{1}{c}{[$\arcsec$ $\times$ $\arcsec$]}  & \multicolumn{1}{c}{[deg]} & \multicolumn{1}{c}{[au]} & \multicolumn{1}{c}{[km s$^{-1}$]} & \multicolumn{1}{c}{[\%]} & \multicolumn{1}{c}{classification} \\
     \multicolumn{2}{c}{ }  & \multicolumn{2}{c}{ } & \multicolumn{2}{c}{km s$^{-1}$]} &  &  &   &  &  & &  & \\
         \hline
         \multicolumn{14}{c}{W43-MM1} \\
\hline
1 & W43-MM1--MF1(\#4) & 18:47:46.99 & -01:54:26.4 & 8661.5 & 76.3 & 1.09 $\times$ 0.95 & 66.2 & 0.96 $\times$ 0.71 & 79.2 & 4576.0 & 101.8$\pm$0.1 & 76 & HC \\
2 & W43-MM1--MF2(\#1) & 18:47:47.03 & -01:54:27.0 & 4822.0 & 33.5 & 0.91 $\times$ 0.78 & 58.6 & 0.75 $\times$ 0.48 & 77.7 & 3327.5 & 99.6$\pm$0.1 & 53 & HC  \\
3 & W43-MM1--MF3(\#2) & 18:47:46.84 & -01:54:29.3 & 3043.8 & 74.4 & 0.70 $\times$ 0.59 & 99.6 & 0.51 $\times$ 0.29 & -80.8 & 2123.0 & 99.5$\pm$0.2 & 70 & HC \\
4 & W43-MM1--MF4(\#3) & 18:47:46.37 & -01:54:33.5 & 1000.3 & 31.8 & 0.73 $\times$ 0.65 & 74.4 & \_ & \_ & 1545.5 & 97.2$\pm$0.1 & 36 & HC  \\
5 & W43-MM1--MF5(\#5) & 18:47:46.76 & -01:54:31.2 & 575.8 & 41.9 & 0.72 $\times$ 0.53 & 91.6 & 0.54 $\times$ 0.37 & -85.0 & 2502.5 & 98.9$\pm$1.0 & 22 & HC$^*$  \\
6 & W43-MM1--MF6(\#11) & 18:47:46.51 & -01:54:24.2 & 554.4 & 40.8 & 0.67 $\times$ 0.50 & 97.9 & 0.47 $\times$ 0.42 & -81.6 & 2453.0 & 93.9$\pm$0.2 & 34 & \\
7 & W43-MM1--MF7(\#10) & 18:47:46.90 & -01:54:30.0 & 212.6 & 12.8 & 0.84 $\times$ 0.56 & 136.5 & 0.65 $\times$ 0.24 & -55.3 & 2205.5 & 100.6$\pm$0.2 & 19 & \\
8 & W43-MM1--MF8(\#9)  & 18:47:46.47 & -01:54:32.6 & 412.3 & 14.1 & 0.67 $\times$ 0.56 & 124.8 & 0.45 $\times$ 0.31  & -71.2 & 2084.5 & 96.2$\pm$0.3 & 20 & HC$^*$  \\ 
9 & W43-MM1--MF9 & 18:47:44.77 & -01:54:45.2 & 128.5 & 12.1 & 0.64 $\times$ 0.44 & 102.1 & 0.49 $\times$ 0.42 & -79.4 & 2519.0 & 95.2$\pm$0.1 & 13 & \\
10 & W43-MM1--MF10 & 18:47:46.53 & -01:54:23.1 & 90.9 & 7.0 & 0.62 $\times$ 0.41 & 89.4 & 0.50 $\times$ 0.39 & -86.0 & 2475.0 & 97.1$\pm$0.2  & 17 & HC$^*$  \\
11 & W43-MM1--MF11 & 18:47:47.00 & -01:54:30.7 & 89.3 & 8.1 & 0.64 $\times$ 0.45 & 92.2 & 0.42 $\times$ 0.47 & -84.4 & 2475.0 & 100.1$\pm$0.3 & 15 & HC$^*$  \\
12 & W43-MM1--MF12 & 18:47:46.88 & -01:54:25.8 & 88.2 & 3.6 & 0.52 $\times$ 0.42 & 216.8 & \_ & \_ & 1545.5 & 99.4$\pm$0.1 & 11 &  \\
13 & W43-MM1--MF13 & 18:47:46.25 & -01:54:33.4 & 50.0 & 4.7 & 0.90 $\times$ 0.60 & 108.0 & 0.76 $\times$ 0.26 & -74.8 & 2458.5 & 97.2$\pm$0.1 & 11  &  \\
14 & W43-MM1--MF14 & 18:47:46.96 & -01:54:29.7 & 49.0 & 2.6 & 0.56 $\times$ 0.37 & 89.8 & 0.54 $\times$ 0.29 & -85.3 & 2194.5 & 100.9$\pm$0.3 & 16 & \\ 
\hline
\multicolumn{14}{c}{W43-MM2} \\
\hline
1 & W43-MM2--MF1 & 18:47:36.79 & -02:00:54.2 & 4459.6 & 80.9 & 1.04 $\times$ 0.94 & -33.2 & 0.88 $\times$ 0.73 & -49.4 & 4444.0 & 88.6$\pm$0.1 & 62 & HC \\
2 & W43-MM2--MF2 & 18:47:36.70 & -02:00:47.6 & 71.3 & 16.4 & 0.82 $\times$ 0.55 & 93.9 & 0.64 $\times$ 0.30 & -84.5 & 2420.0 & 89.9$\pm$0.7 & 3 & \\
3 & W43-MM2--MF3 & 18:47:36.27 & -02:00:50.7 & 46.3 & 6.3 & 0.63 $\times$ 0.42 & 110.9 & 0.46 $\times$ 0.36 & -72.9 & 2255.0 & 91.0$\pm$0.3 & 1 & \\
\hline
\multicolumn{14}{c}{W43-MM3} \\
\hline
1 & W43-MM3--MF1 & 18:47:39.26 & -02:00:28.1 & 175.8 & 16.1 & 0.86 $\times$ 0.78 & 29.1 & 0.61 $\times$ 0.48 & 51.6 & 2981.0 & 94.8$\pm$1.7 & 13 &  \\
2 & W43-MM3--MF2 & 18:47:41.71 & -02:00:28.6 & 76.7 & 16.8 & 0.78 $\times$ 0.64 & 103.1 & \_ & \_ & 1688.5 & 92.8$\pm$0.3 & 10 & HC$^*$  \\
3 & W43-MM3--MF3 & 18:47:41.73 & -02:00:27.5 & 23.0 & 4.9 & 0.87 $\times$ 0.58 & 98.4 & 0.65 $\times$ 0.30 & -84.0 & 2464.0 & 93.2$\pm$0.3 & 2 & \\
\hline
\multicolumn{14}{c}{W51-E} \\
\hline
1 & W51-E--MF1 & 19:23:43.97 & 14:30:34.5 & 4215.0 & \_ & 2.74 $\times$ 2.32 & 23.9 & 2.71 $\times$ 2.27 & 24.1 & 13435.2 & 56.3$\pm$0.3 & 73 & HC \\
2 & W51-E--MF2 & 19:23:43.87 & 14:30:27.3 & 2119.5 & 40.9 & 1.14 $\times$ 0.95 & 114.6 & 1.04 $\times$ 0.89 & -66.9 & 5211.0 & 54.9$\pm$0.8 & 65  & HC$^*$ \\
3 & W51-E--MF3 & 19:23:43.88 & 14:30:27.9 & 1794.7 & 26.5 & 0.88 $\times$ 0.78 & 314.7 & 0.76 $\times$ 0.69 & -33.1 & 3936.6 & 60.0$\pm$0.5 & 72 & HC \\
4 & W51-E--MF4 & 19:23:43.74 & 14:30:21.4 & 81.5 & 10.7 & 0.67 $\times$ 0.45 & 24.5 & 0.57 $\times$ 0.08 & 25.9 & 1188.0 & 62.7$\pm$0.1 & 7 & \\
5 & W51-E--MF5 & 19:23:43.84 & 14:30:24.5 & 73.6 & 8.6 & 0.83 $\times$ 0.74 & -36.8 & 0.71 $\times$ 0.63 & -18.3 & 3634.2 & 58.1$\pm$0.5 & 9  & HC$^*$ \\
6 & W51-E--MF6 & 19:23:43.80 & 14:30:19.6 & 40.9 & 10.2 & 0.82 $\times$ 0.66 & 102.0 & 0.69 $\times$ 0.55 & -86.4 & 3337.2 & 58.4$\pm$0.1 & 3 & \\
7 & W51-E--MF7 & 19:23:43.82 & 14:30:23.3 & 31.9 & 3.1 & 0.75 $\times$ 0.56 & 81.1 & 0.63 $\times$ 0.39 & 70.6 & 2683.8 & 61.6$\pm$0.8 & 4 & HC$^*$ \\
\hline
\multicolumn{14}{c}{W51-IRS2} \\
\hline
1 & W51-IRS2--MF1 & 19:23:40.00 & 14:31:05.5 & 10097.2 & 57.0 & 1.37 $\times$ 1.13 & 138.3 & 1.24 $\times$ 0.94 & -39.0 & 5837.4 & 58.3$\pm$0.3 & 73 & HC \\
2 & W51-IRS2--MF2 & 19:23:39.82 & 14:31:05.0 & 5162.1 & 47.8 & 0.96 $\times$ 0.89 & 57.2 & 0.72 $\times$ 0.68 & 38.5 & 3796.2 & 61.1$\pm$0.1 & 73 & HC \\
3 & W51-IRS2--MF3 & 19:23:40.04 & 14:31:04.9 & 3049.0 & 15.8 & 1.25 $\times$ 1.01 & 69.9 & 1.08 $\times$ 0.84 & 69.7 & 5151.5 & 56.8$\pm$0.5 & 72 & HC \\
4 & W51-IRS2--MF4 & 19:23:39.95 & 14:31:05.2 & 2933.2 & 15.4 & 1.1 $\times$ 0.83 & 2.8 & 0.94 $\times$ 0.54 & 0.5 & 3882.6 & 58.8$\pm$0.1 & 48 & HC \\
5 & W51-IRS2--MF5 & 19:23:39.74 & 14:31:05.3 & 2716.2 & 29.5 & 0.89 $\times$ 0.83 & -17.4 & 0.68 $\times$ 0.54 & -18.0 & 3288.6 & 63.0$\pm$0.4 & 40 & HC \\
6 & W51-IRS2--MF6 & 19:23:38.57 & 14:30:41.8 & 1720.5 & 91.2 & 0.71 $\times$ 0.64 & -23.6 & \_ & \_ & 1630.8 & 62.6$\pm$0.1 & 23  & HC \\
7 & W51-IRS2--MF7 & 19:23:39.50 & 14:31:03.3 & 105.3 & 8.5 & 0.65 $\times$ 0.59 & -130.1 & \_ & \_ & 1630.8 & 63.7$\pm$0.9 & 7 & \\
8 & W51-IRS2--MF8 & 19:23:41.81 & 14:30:54.9 & 84.7 & 5.4 & 0.71 $\times$ 0.47 & -13.6 & 0.43 $\times$ 0.43 & -14.8 & 2305.7 & 55.2$\pm$0.6 & 9 & \\
9 & W51-IRS2--MF9 & 19:23:38.42 & 14:30:36.6 & 52.4 & 5.0 & 1.04 $\times$ 0.85 & 15.1 & 0.86 $\times$ 0.59 & 9.3 & 3850.2 & 60.8$\pm$0.2 & 5 & \\
\hline
\end{tabular}}
\tablefoot{The sources previously identified towards W43-MM1 in \citetalias{brouillet2022} are indicated in parentheses in the second column after the source name.}
\end{table*}
% ======================

%--------------------------------------------------------------------
\section{Properties of the methyl formate sources}
%--------------------------------------------------------------------
\label{section-MFproperties}

In the following subsections, we investigate the physical properties of the 76 sources identified and extracted from the \mf \ moment 0 maps using the \textsl{GExt2D} algorithm. \\

%------------------
\subsection{Continuum emission associated with methyl formate compact sources}
%------------------
\label{section-contassociation}

\begin{figure}[!t]
   \begin{center}
       \includegraphics[width=\hsize]{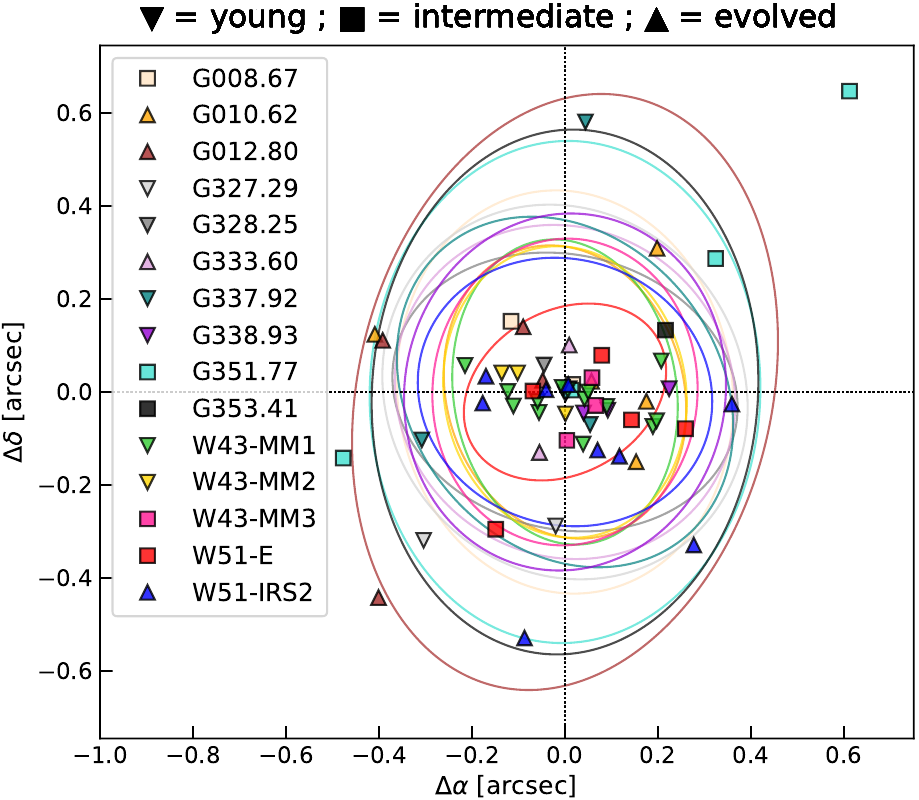}  \\
  \caption{\label{FIG-distance-cont-MF} Two-dimensional distribution of the angular offsets between the peak position of each \mf \ source extracted with \textsl{GExt2D} and its associated continuum core extracted with \textsl{getsf}. The different symbols represent the different evolutionary stages of the protoclusters: young, intermediate, or evolved, as shown on top of the plot. The dashed vertical and horizontal lines indicate the x = 0 and y = 0 axes, respectively. The ellipses represent the synthesized beam sizes (FWHM) of the 15 ALMA-IMF B6-spw0 line cubes. The diameter of the corresponding ellipse (\textit{i.e.} $\theta_{\mathrm{ave}}$) is used as a criterion to associate \mf \ sources with continuum cores towards each protocluster. The methyl formate sources that are not associated with any compact continuum core are not shown in this figure.}   
  \end{center}
\end{figure}

In order to characterize the physical properties of hot core candidates, we use the thermal dust continuum emission associated with these sources. To this regard, we cross-matched our catalog of methyl formate sources with that of the continuum cores from \citetalias{louvet2024} (see Sect. \ref{section-corecat}). We associate a methyl formate source to a continuum core if the angular offset between their respective peak positions is smaller than the diameter (FWHM) of the synthesized beam of the \mf \ line datacube. Figure \ref{FIG-distance-cont-MF} shows the angular offsets computed between each methyl formate source and its closest continuum core from the \textsl{getsf} unsmoothed catalog. We find that a large majority of the methyl formate sources have a good positional correspondence to that of the continuum cores, with in total 84\% of the methyl formate sources having a counterpart in the unsmoothed continuum core catalog. These continuum cores are indicated with green crosses in the moment 0 maps of methyl formate shown in Figs.\,\ref{FIG-mom0-maps}--\ref{FIG-mom0-maps4}. On the other hand, 12 methyl formate sources (i.e. $\sim$ 16\% of the total sample) do not coincide with any compact continuum core from the \textsl{getsf} unsmoothed catalog. These sources are found towards four protoclusters, G010.62, G337.92, G351.77, and W51-E, which are two intermediate regions, a young, and an evolved one. In Figs. \,\ref{FIG-cont-mom0-maps}--\ref{FIG-cont-mom0-maps4} we compare the 1.3\,mm continuum emission with the contours of the methyl formate integrated intensity. For the four protoclusters listed above, we find that the methyl formate compact sources that are not associated with any compact continuum core are exclusively found in regions of extended methyl formate emission, and also coincide with extended continuum emission at 1.3 mm. These cases are addressed in more details below.

%======================
% FIGURE: CHECK DISTANCES ASSOCIATED CONT SOURCES
%======================
%======================

In the intermediate stage protocluster W51-E, two methyl formate sources, W51-E--MF2 and W51-E--MF5, located in the extended North-South emission, could not be associated to any compact continuum core based on our position-match criterion. They fall, however, on the extended continuum emission that exhibits some fluctuations in the vicinity of the methyl formate sources (see Fig.\,\ref{FIG-cont-mom0-maps4}). It is likely that both the complexity of the emission and a lower background to core emission contrast hinders the identification of their continuum counterpart.

In the other intermediate evolutionary stage region, G351.77, the overall continuum emission at 1.3\,mm is extended in the West-East direction (see Fig.\,\ref{FIG-cont-mom0-maps3}), and does not resemble the shape of the \mf\ emission. While G351.77--MF2 and G351.77--MF4 have a compact continuum core nearby, the brightest continuum core is somewhat in between G351.77--MF1 and G351.77--MF3. Our position-matching criterion associates the continuum core to G351.77--MF1, while G351.77--MF3 cannot be associated to any continuum core. Chemical segregation, blending of unresolved sources, or again the low contrast between the peak and the background could lead to such positional shifts between the continuum and the \mf\ emission.  

 Towards the central part of the young protocluster G337.92 (see also Sect.\,\ref{section-MFstat}) the \mf\ emission exhibits an extended blob. Only sources G337.92--MF1, MF6 and MF7 seem to be associated with continuum peaks at 1.3\,mm (see Fig.\,\ref{FIG-cont-mom0-maps2}). The other four sources G337.92--MF2, MF3, MF4, and MF5 do not closely coincide with any continuum peak and cannot be associated with any compact continuum core using our position-matching criterion. It is possible that these \mf\ peaks correspond to inhomogeneities in extended emission heated by a single central source, or source blending prevents a firm association to continuum cores.

A similar case is observed towards the evolved region G010.62, where the \mf\ spatial distribution is not symmetric, and exhibit a complex morphology that does not show a close correlation with the distribution of the 1.3~mm continuum emission (see Fig.\,\ref{FIG-cont-mom0-maps}). This extended \mf\ emission is unlikely to be attributed to a single source due to its spatial extent (see Sect.\,\ref{section-MF-extended}), and sources G010.62--MF3, MF4, MF5, MF6, and MF9 do not find any continuum counterpart in the \textsl{getsf} unsmoothed continuum core catalog. 

For the four ALMA-IMF regions mentioned above, where the methyl formate sources lie in the extended 1.3\,mm continuum emission but cannot be associated with compact continuum cores, it is possible that the source extraction algorithm fails to disentangle and decompose the compact continuum cores on the top of a bright and extended background. The \textsl{getsf} definition of sources is the following \citep[see also Sects.\,1 and 3.2.2 of][]{menshchikov2021}: sources are the relatively round emission peaks that are significantly stronger than the local surrounding fluctuations (of background and noise), indicating the presence of the physical objects in space that produced the observed emission. If a structure is too elongated or has a very complex shape, it is unlikely to be identified as a compact source. The nature of the 12 methyl formate sources listed above that could not be associated with a compact continuum core at 1.3,\ mm is further discussed in Sect. \ref{section-discussion-nature-sources}.

Table \ref{TAB-cont-cat-getsf} lists the peak positions, peak ($S^{\mathrm{peak}}$) and integrated fluxes ($S^{\mathrm{int}}$) measured in both the continuum maps at 1.3\,mm and 3\,mm, as well as the source sizes (FWHM) of all the continuum cores associated to methyl formate sources. For the 12 methyl formate sources that are not associated to compact continuum cores, their flux is measured within the beam size in the 1.3\,mm continuum emission maps at the peak position of the \mf\ emission. The flux is then corrected by subtracting the background emission estimated at this position during the source extraction process (see Sect.\,\ref{section-corecat}). Since no emission size is fitted for these sources, we use the average beam size of the continuum maps, $\theta_{\mathrm{ave}}^{\mathrm{cont}}$, as the continuum source size (\textit{i.e.} FWHM$_{\mathrm{cont}}$ = $\theta_{\mathrm{ave}}^{\mathrm{cont}}$), such that in this case $S^{\mathrm{peak}}$ = $S^{\mathrm{int}}$. The resulting values are listed in Table\,\ref{TAB-cont-cat-getsf}. The methyl formate sources that are not associated to compact continuum cores are marked with a $*$ in the first column.

%------------------
\subsection{Free-free contamination}
%------------------
\label{section-freefree}

Reaching a certain stage in their evolution, high-mass (proto)stars develop ionising radiation that leads to the emergence of HC-{\hii} and {\uchii} regions. Such sources exhibit free-free emission that may contribute to the observed continuum emission at 3 mm, and potentially even at 1.3 mm. The relative contribution of emission from ionised gas versus that of thermal dust continuum emission, however, depends on several factors, such as the source size of the ionising emission and its optical depth. Since the ALMA-IMF fields cover massive protoclusters in a range of evolutionary stages, the contamination from free-free emission cannot be ignored for the total gas mass estimates for several sources. 

The ALMA-IMF dataset covers the H$_{41 \alpha}$ recombination line at 92.0 GHz, which originates from ionized gas coming from {\hii} regions (see e.g, Fig.\,2 of \citetalias{motte2022}), and we refer for a detailed analysis to \citet{galvanmadrid2024}. Using this information we identify 17 methyl formate sources that lie in intermediate and evolved regions containing free-free emission, these are G008.67, G010.62, G012.80, G333.60, W51-E and W51-IRS2 (see Figs.\,\ref{FIG-ellipses-h41alpha}, \ref{FIG-ellipses-h41alpha2}, and \ref{FIG-ellipses-h41alpha4}). For these regions, in order to determine the contribution of free-free emission to the 1.3 mm flux densities, we rely on the dual band approach of ALMA-IMF and exploit the dust continuum emission at 1.3~mm, and 3~mm, like done in \citetalias{pouteau2022} and \citetalias{louvet2024}. First the 3 mm integrated fluxes are rescaled to the 1.3 mm sizes to allow a direct comparison of these fluxes as described in \citetalias{pouteau2022}. Then we compute the theoretical flux ratio expected for thermal dust emission ($\gamma^{\mathrm{dust}}_{\mathrm{th}}$) as explained in Appendix \ref{appendix-freefree}. Figure \ref{FIG-freefree-contribution} shows the flux ratio ($S^{\mathrm{int}}_{\mathrm{1.3mm}}$/$S^{\mathrm{int}}_{\mathrm{3mm}}$) measured towards the 17 sources potentially affected by free-free emission, compared to the theoretical ratio computed assuming dust temperatures ranging from 50\,K to 150\,K (see Sect.\,\ref{section-MFTd}) and a dust emissivity exponent $\alpha$ ranging from 3.2 to 3.8 (green shaded area). For each source with a flux ratio $<$ $\gamma^{\mathrm{dust}}_{\mathrm{th}}$, a correction factor (frac$_{ff}$) must be applied to both its peak and integrated flux measured at both 1.3 mm and 3 mm to take into account the free-free contribution, as described in Appendix \ref{appendix-freefree}. These correction factors are listed in the last column of Table \ref{appendix-cont-core-cat}. The correction factor indicates the fraction of the flux initially measured that is due to free-free emission for each continuum core. We note that the 1.3 mm continuum emission measured towards G010.62--MF1 and G010.62--MF2 shows in both cases a level of free-free contamination, frac$_{ff}$, of 100\%. It suggests that their millimeter continuum emission is entirely due to ionised gas, which calls into question the nature of these two sources, which we further discuss in Sect. \ref{section-discussion-nature-sources}.

%======================
% FIGURE: FREEFREE CONTRIBUTION
%======================
\begin{figure}[!t]
   \begin{center}
       \includegraphics[width=\hsize]{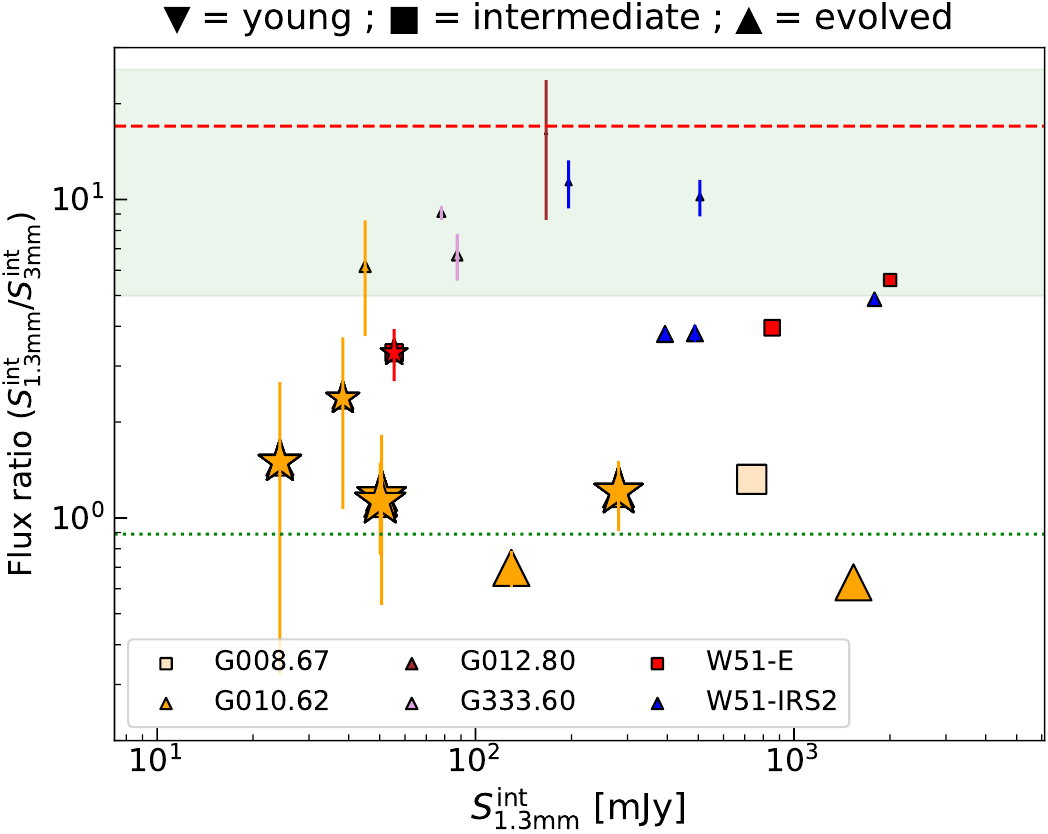}  \\
  \caption{\label{FIG-freefree-contribution} Flux ratio ($S_{\mathrm{1.3mm}}^{\mathrm{int}}$/$S_{\mathrm{3mm}}^{\mathrm{int}}$) measured towards the methyl formate sources of which the position coincides with the H$_{41 \alpha}$ emission in six ALMA-IMF protoclusters (see Figs.\,\ref{FIG-ellipses-h41alpha}--\ref{FIG-ellipses-h41alpha4}). The size of each marker is proportional to the level of free-free contamination (\textit{e.g} the biggest markers correspond to G010.62--MF1 and G010.62--MF2 for which 100\% of the flux measured at 1.3 mm is expected to be due to free-free emission). The green shaded area shows the theoretical ratio expected for dust cores, computed for a dust emissivity exponent $\alpha$ ranging from 2 to 4, for optically thin and thick dust emission, respectively. The red dashed line shows the theoretical ratio obtained using $\alpha$ = 3.5 (as in \citetalias{pouteau2022}), while the green dotted line shows the theoretical ratio expected for optically thin {\hii} regions, corresponding to $\alpha$ = -0.1.}
  \end{center}
\end{figure}
%======================

%---------------------------------
\subsection{Source size}
%---------------------------------
\label{section-MFsizes}

% We made a deconvolution considering the ellipticity of the sources and of the synthesized beam of the line cubes.

\begin{figure}[!h]
   \begin{center}
     \includegraphics[width=\hsize]{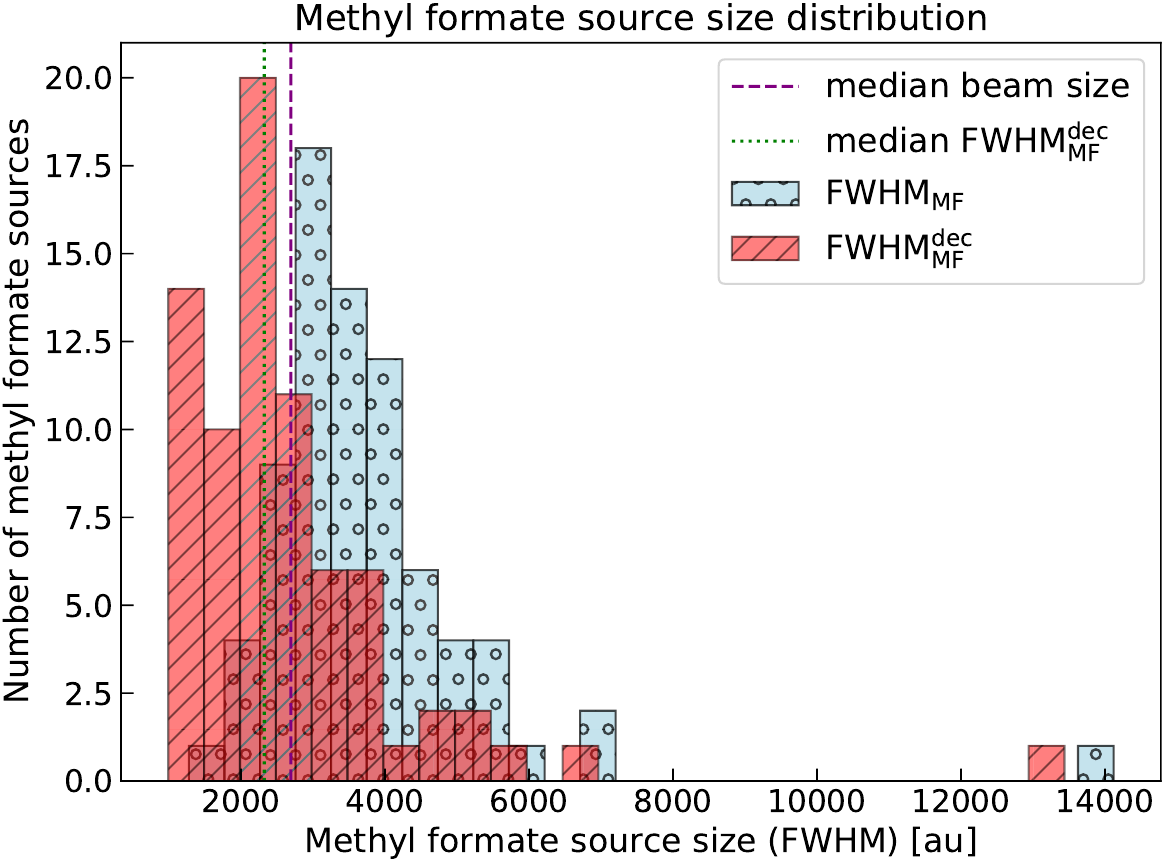} \\
    \caption{\label{FIG-histo-size} Methyl formate source size distribution derived using \textsl{GExt2D} fitting 2D Gaussians to the \mf \ moment 0 maps. The non-deconvolved (FWHM$_{\mathrm{MF}}$) and deconvolved sizes (FWHM$_{\mathrm{MF}}^{\mathrm{dec}}$) are shown in blue and red, respectively. The green dotted line shows the median of the deconvolved source sizes, while the dashed purple line shows the median beam size of the line cubes.}
   \end{center}
\end{figure}

We estimate the size of the methyl formate sources from the FWHMs of the 2D Gaussian fitting to the \mf\ moment 0 maps using \textsl{GExt2D}, as described in Sect. \ref{section-MFextraction}. The resulting minor ($\theta_{\mathrm{min}}$) and major axes ($\theta_{\mathrm{maj}}$) are deconvolved from the synthesized beam size of the line cube, considering the ellipticity of the sources and of the synthesized beam, as described in Appendix \ref{appendix-deconvolve-source-size}. We have set a minimum deconvolved size for each region to half the synthesized beam of the line cube, in order to limit deconvolution effects that may give excessively small and thus unrealistic sizes. The sizes before ($\theta_{\mathrm{maj}} \times \theta_{\mathrm{min}}$) and after deconvolution ($\theta_{\mathrm{maj}}^{\mathrm{dec}} \times \theta_{\mathrm{min}}^{\mathrm{dec}}$) are listed for each methyl formate source in Table \ref{TAB-MFcat}, along with physical sizes at the distance of the respective protocluster (FWHM$^{\mathrm{dec}}_{\mathrm{MF}}$ in au). Figure \ref{FIG-histo-size} shows the distribution of the physical sizes before (FWHM$_{\mathrm{MF}}$) and after (FWHM$^{\mathrm{dec}}_{\mathrm{MF}}$) beam deconvolution. The methyl formate sources exhibit deconvolved source sizes ranging from $\sim$ 990\,au to 13400\,au, with a median size of about 2300\,au. The two outliers of the distribution correspond to W51-E--MF1 and G327.29--MF1. The majority of the sources are marginally resolved, with a handful of sources staying unresolved (\textit{i.e.} FWHM$^{\mathrm{dec}}_{\mathrm{MF}}$ < median beam size of the line cubes).

\begin{figure}[!ht]
  \begin{center}
       \includegraphics[width=\hsize]{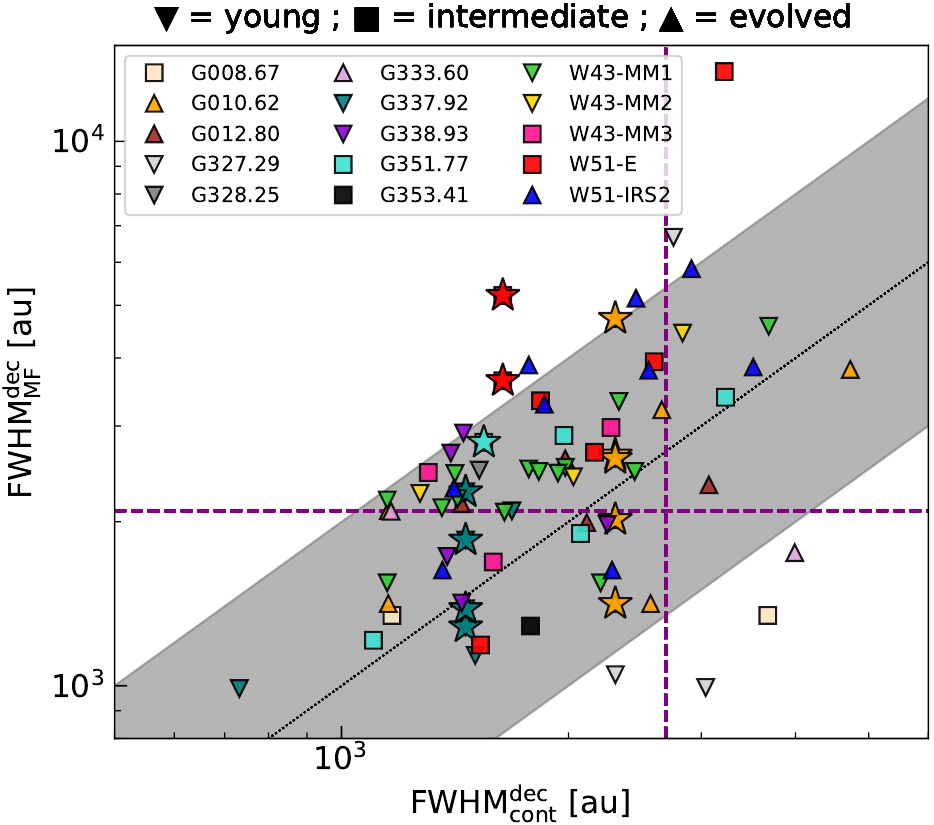}  \\
  \caption{\label{FIG-size-cont-MF} Methyl formate source sizes derived with \textsl{GExt2D} compared to the associated continuum core sizes taken from the unsmoothed core catalog of \citetalias{louvet2024}. All sizes are deconvolved from the beam sizes. The black dotted line shows the one to one ratio, while the grey shaded area shows where the FWHM$^{\mathrm{dec}}_{\mathrm{MF}}$ sizes are within a factor two from their associated FWHM$^{\mathrm{dec}}_{\mathrm{cont}}$ sizes. The different symbols represent the different evolutionary stages of the protoclusters: young, intermediate, or evolved, as shown on top of the figure. The methyl formate sources that are not associated with compact continuum cores are shown with stars. For these sources we assumed FWHM$^{\mathrm{dec}}_{\mathrm{cont}}$ = $\theta^{\mathrm{cont}}_{\mathrm{ave}}$. The vertical and horizontal dashed lines show the median beam sizes of the line cubes and the continuum maps, respectively.} 
  \end{center}
\end{figure}

In Figure\,\ref{FIG-size-cont-MF} we compare the methyl formate deconvolved source sizes to that of their associated continuum cores. While about 74\% of the methyl formate sources are found to be more extended than their associated continuum core, overall, for $\sim$87\% of the sources, both their methyl formate and continuum emission deconvolved source sizes agree within a factor of two (grey shaded area). 

In Figs.\,\ref{FIG-ellipses-h41alpha}--\ref{FIG-ellipses-h41alpha4} the deconvolved source sizes of the methyl formate sources are outlined with blue ellipses for each ALMA-IMF protocluster, compared to that of their associated continuum cores shown green ellipses.

%======================
% FIGURE: HISTOS SIZE 
%======================
%=====================

%======================
% FIGURE: COMPARE CONT vs. HC SIZES
%======================
%======================

%----------------------------
\subsection{Temperature estimates}
%----------------------------
\label{section-MFTd}

In order to obtain mass estimates of the cores from the thermal dust continuum emission (see Sect.\,\ref{section-masses}), the dust temperature, $T_{\mathrm{d}}$, is a critical parameter. Since for the current analysis we rely only on the \mf\ lines, we need to adopt an estimate of the temperature that best characterize the methyl formate sources. \mf\ has a lower binding energy \citep[4210 K,][]{burke2015} compared to water (4815 K, Jin et al. in prep.), such that it is trapped in water ices until the temperature exceeds $\sim$120\,K. If the observed methyl formate emission originates only from thermal desorption, \mf\ is released into the gas phase via co-desorption with water above 120\,K. At that point we expect a rise in \mf\ abundance within the thermal sublimation radius, which corresponds to the extent of the heated gas traced by \mf. Significant thermal desorption still occurs up to $\sim$160 K \citep{bonfand2019, garrod2022}. However, as mentioned already in Sect.\,\ref{section-intro}, {\mf} has already been observed in the gas phase below the thermal desorption temperature \citep[e.g.][]{busch2022,bouscasse2024}, and shocks from accretion-ejection processes \citep{palau2017, csengeri2019} can also lead to enhancements of some gas-phase COMs, including methyl formate. 

Both gas- and dust-based temperature estimates have been previously performed for the W43 protocluster from the ALMA-IMF data (see \citealp{motte2018nat} and \citetalias{pouteau2022}). Dust-based temperature estimates using \textsl{Herschel} and APEX data with the resolution-improving PPMAP method \citep[Point Process MAPping procedure,][]{marsh2015} provided temperatures below 65\,K for our sample of methyl formate sources in W43-MM2 and W43-MM3. For the W43-MM1 region, \citet{motte2018nat} derived dust temperatures of 21--93\,K for the 14 continuum cores associated to methyl formate sources, while gas-based temperature estimates in \citetalias{brouillet2022} suggest excitation temperatures of 120--160\,K using \ce{CH3CN} lines detected towards the seven most massive hot cores. Discrepancies between the dust and gas based temperature estimates may suggest strong temperature gradients towards our compact methyl formate sources and hence the adopted temperatures may be subject to significant uncertainties. For the cold continuum sources we use here dust-based temperature estimates made using PPMAP by \citet[][]{dellova2024} that allows us to probe the dust temperature at scales larger than 2.5$\arcsec$. These temperature values are, however, not adequate for hot core sources that have deeply embedded internal heating sources on smaller scales.

A few other ALMA-IMF protoclusters have dedicated studies at the spatial resolution of individual cores (see Sect.\,\ref{section-MFstat}). \citet{taniguchi2023} derived excitation temperatures of $\sim$200\,K towards G010.62, from the analysis of \ce{CH3CN} lines observed at 0.3$\arcsec$ resolution (i.e. $\sim$1500\,au at the distance of the protocluster). \citet{law2021} report  higher temperatures, up to 400\,K from the analysis of \ce{CH3OH} transitions (see their Fig 6.). These results were obtained from ALMA data at very high angular resolution, 0.14$\arcsec$, which corresponds to a physical scale of $\sim$700\,au at the distance of the protocluster, much smaller than the deconvolved FWHM sizes we derived from the methyl formate emission (\textit{i.e.} 1400--3800\,au), such that we expect this temperature to be diluted at the resolution of the ALMA-IMF data. 

Rotational temperatures of 100\,K and 165\,K have been derived based on the analysis of \mf\ and \ce{CH3OH} lines, respectively, detected towards G351.77 in the ATOMS survey \citep{liu2021}. Furthermore, several 6.7\,GHz class II methanol masers have been detected towards G351.77 \citep[see, e.g,][]{beuther2009}, which suggests gas temperatures $\geq$100\,K \citep{sobolev1997, cragg2005}. Similar to the case of the central bright source of G327.29, which also exhibit a 6.7\,GHz class II methanol maser \cite[see, e.g.,][]{wyrowski2008}. \\

Based on the results listed above, we adopt a canonical dust temperature of 100 $\pm$ 50\,K for all methyl formate sources, that takes into account the discrepancies in the temperature estimates previously made towards some of the ALMA-IMF protoclusters. There are six exceptions to this assumption where a higher temperature is warranted. In particular, the central bright emission observed in both continuum and COMs towards W51-E has been investigated in detail by \citet{ginsburg2017}, who report a peak excitation temperature $>$350\,K based on the analysis of \ce{CH3OH} emission lines (see their Fig 6) detected in their 0.3$\arcsec$ resolution data, which corresponds to 1800~au at the distance of the protocluster. We assume that this emission mostly comes from the three main, brightest methyl formate sources, W51-E--MF1, MF2, and MF3, for which we adopt a higher dust temperature of 300$\pm$100\,K. 

In the case of W51-IRS2, the bright emission seen towards the Northern cores seems to be dominated by the methyl formate sources we have identified as W51-IRS2--MF1 and W51-IRS2--MF3 \citep[see Fig. 4 of][]{ginsburg2017}. Similar to the W51-E main sources, we adopt a higher dust temperature of 300$\pm$100\,K for these two objects. This is consistent with the detection of several ammonia (\ce{NH3}) masers in this region, which suggests temperatures as high as 300\,K \citep{henkel2013}.

Finally, the central source of G327.29 is somewhat similar to the extreme methyl formate sources in the W51 regions, in terms of its spatial extent and brightness, and it is also associated with several 6.7 Class II methanol masers. Vibrationally excited state transitions of COMs further suggest more elevated temperatures \citep[$T_{\mathrm{rot}} >$ 180\,K, see][]{gibb2000}, and hence we also adopt here 300$\pm$100\,K for the central G327.29--MF1 source.

%-------------------------
\subsection{Mass estimates}
%-------------------------
\label{section-masses}

Masses are computed from the 1.3 mm flux density from the \textsl{getsf} unsmoothed catalog from \citetalias{louvet2024}. We take into account potential contamination for free-free emission (Sect.\,\ref{section-freefree}) and use source specific dust temperature estimates (Sect.\,\ref{section-MFTd}). A previous analysis of the ALMA-IMF data has shown that the most massive objects may reach high densities, e.g., up to 2$\times$10$^{8}$ cm$^{-3}$ in the W43 protocluster (\citetalias{pouteau2022}), and thus the dust thermal continuum emission may become optically thick (see Table \ref{TAB-cont-cat-getsf}). In order to take into account dust opacities in the mass estimates we use the following equation \citep[][]{motte2018nat, pouteau2022}: 
\begin{equation}
\label{physical-profiles-eq3}
M_{\mathrm{core}} = -\frac{\Omega_{\mathrm{beam}} \times d^2}{\kappa_{\mathrm{1.3mm}}} \frac{S^{\mathrm{int}}_{\mathrm{1.3mm}}}{S^{\mathrm{peak}}_{\mathrm{1.3mm}}} ln \left( 1 - \frac{S^{\mathrm{peak}}_{\mathrm{1.3mm}}}{\Omega_{\mathrm{beam}} \times B_{\mathrm{1.3mm}}(\mathrm{T_d})} \right) \,,
\end{equation}
where the 1.3 mm peak and integrated flux, $S^{\mathrm{peak}}_{\mathrm{1.3mm}}$ and $S^{\mathrm{int}}_{\mathrm{1.3mm}}$, respectively, are corrected for the free-free contamination (see Sect. \ref{section-freefree}). Following \citetalias{pouteau2022}, we adopted a dust opacity coefficient per unit of mass $\kappa_{1.3mm}$ = 0.01 cm$^{2}$ g$^{-1}$ (assuming a gas-to-dust ratio of 100), which is adapted to dense cores \citep[see][]{ossenkopf1994}. The distance of the source to the Sun, $d$ is given in Table \ref{TAB-targetlist}. The solid angle of the continuum beam is given by $\Omega_{beam} = \frac{\pi}{4\ln 2} \times \theta_{\mathrm{ave}}^{\mathrm{cont}}\,^2$. Finally, $B_{1.3mm}(T_d)$ is the Planck function at the dust temperature $T_d$ (see Sect. \ref{section-MFTd}). 

We list in Table \ref{TAB-cont-cat-getsf} the mass estimates obtained for all methyl formate sources, with dust temperatures ranging from 50~K to 150~K (or 200~K to 400~K for the most extreme sources, see Sect.\,\ref{section-MFTd}). A factor 3 of difference in the assumed dust temperature leads to at most a factor 4.5 of difference in the mass estimates. 

%ICI
%To test the impact of using a single temperature for all cores, we also computed masses using temperatures from a uniform random sampling between 50\,K and 150\,K for all sources, except the most extreme ones for which we used a temperature range between 200\,K and 400\,K. We display one realisation with a magenta line in Fig.\,\ref{FIG-histo-mass-distribution}. We find that using a single temperature, $T_{\mathrm{d}}$ = 100\,K or 300\,K, and a random temperature distribution leads to comparable mass histograms.

\begin{figure}[!t]
\begin{center}
       \includegraphics[width=\hsize]{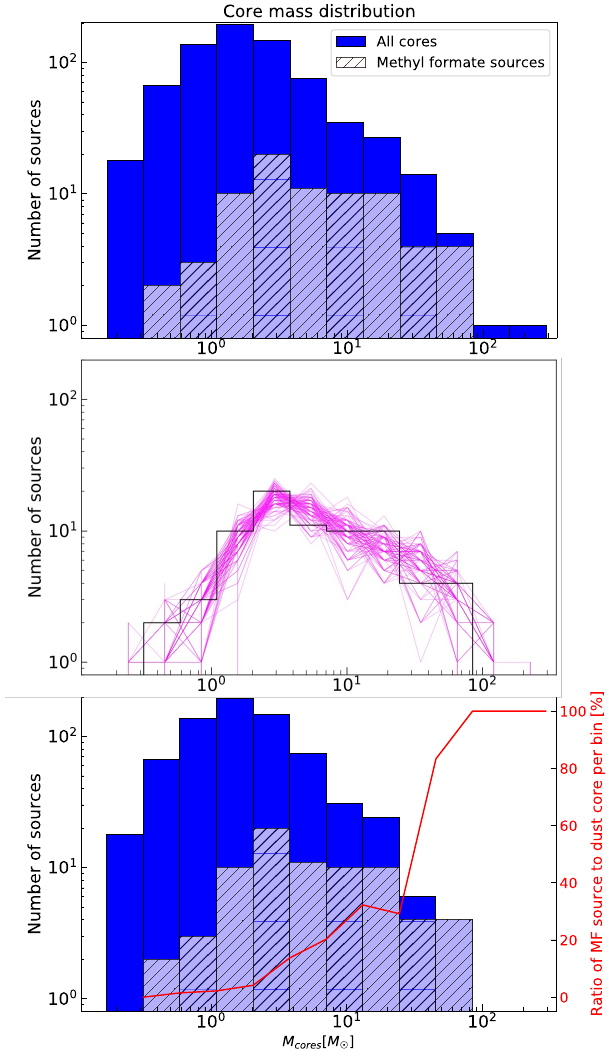}  
  \caption{\label{FIG-histo-mass-distribution} Mass distribution of the methyl formate sources (black hatched histogram) computed using $T_{\mathrm{d}}$ = 100\,K for all sources except the six most extreme ones for which we used 300\,K instead. The magenta lines in the middle panel show in comparison the mass distributions obtained for 100 realisations of randomly sampled temperatures between 50 and 150\,K (or 200 and 400\,K for the six most extreme sources). The blue histogram in the top panel shows the mass distribution of all cores, \textit{i.e.} all compact dust continuum cores (without free-free contaminated sources) plus all the methyl formate sources. In the bottom panel we removed from the blue histogram 22 compact dust continuum cores for which the PPMAP method does not provide robust temperature estimates. The red line shows the ratio of methyl formate sources to the total number of cores per bin.}
  \end{center}
\end{figure}

Figure\,\ref{FIG-histo-mass-distribution} shows the distribution of masses of the 76 methyl formate sources computed using a dust temperature of 100 K for all sources, except the most extreme ones for which we used 300 K (see Sect.\,\ref{section-MFTd}). In the top panel of Fig.\,\ref{FIG-histo-mass-distribution} we also show in blue the mass distribution of all the ALMA-IMF cores, \textit{i.e} the methyl formate sources, plus the compact dust continuum cores from the \textsl{getsf} unsmoothed catalog (\citetalias{louvet2024}), for which the masses were computed using dust temperatures ranging from $\sim$19 -- 73\,K provided by the PPMAP temperature maps built for each protocluster (\citealp[][]{dellova2024}, see Sect.\,\ref{section-MFTd}). While the methyl formate sources range in mass from $\sim$ 0.2\,$\Msun$ to $\sim$80\,$\Msun$, with a median mass of $\sim$ 3.8\,$\Msun$, the dust continuum cores without methyl formate emission reach masses as high as $\sim$293\,$\Msun$, albeit with a lower median mass of 1.6\,$\Msun$. In the bottom panel of Fig.\,\ref{FIG-histo-mass-distribution} we also show a  mass distribution where we removed 22 dust continuum cores that spatially coincide with extended methyl formate emission, and hence their temperatures estimates could have more significant uncertainties. The PPMAP temperature maps have a 2.5$\arcsec$ angular resolution \citep[][]{dellova2024}, that is insufficient to properly trace the temperature of compact heating sources close to other internally heated sources. These sources are found towards the young protoclusters G327.29, G338.93, W43-MM1, and W43-MM2, the intermediate ones G351.77 and W51-E, and finally the evolved protocluster W51-IRS2 (see the green triangles in Figs.\,\ref{FIG-cont-mom0-maps}--\ref{FIG-cont-mom0-maps4}).

We investigate the fraction of continuum cores associated with compact methyl formate emission with respect to the total population of dust cores. The red line in the bottom panel of Fig.\,\ref{FIG-histo-mass-distribution} shows that the fraction of compact continuum cores that are associated with methyl formate emission is higher for the most massive cores. Among all the continuum cores that have masses above 8\,$\Msun$, about 41\% of are associated with compact methyl formate emission. This ratio increases to 90\% if we consider the cores with masses $>$39\,$\Msun$. Clearly, uncertainty in the temperature estimates for the most massive cores impacts this fraction. To mitigate this, we used 1000 realisations of randomly sampled temperatures from a uniform distribution between 50 and 150\,K for all methyl formate sources, and between 200 and 400\,K for the six most extreme sources. The middle panel of Fig.\,\ref{FIG-histo-mass-distribution} shows 100 of the 1000 realisations for the comparison. We find that 38--48\% of the dust continuum cores with masses above 8\,$\Msun$ are associated with methyl formate emission, and this ratio increases to 90\% for the cores with masses that range between $\sim$ 30 and 40\,$\Msun$.

%======================
% FIGURE: HISTO MASS DISTRIBUTION
%======================
%======================

%--------------------------------------------------------------------
\section{Discussion}
%--------------------------------------------------------------------
\label{section-discussion}

%------------------------------
\subsection{Chemical origin of the \mf\ emission}
%------------------------------
\label{section-discussion-origin-ch3ocho}

 \mf\ is expected to form at early times during the star formation process, mainly on the surface of cold interstellar dust grains (20--40 K), through radical-addition reactions \citep[see, e.g,][]{garrod2006, garrod2022}. In the classical picture of hot core related chemistry, where we consider hot cores as chemically enhanced regions radiatively heated by a central high-mass, still accreting protostar, the chemical species frozen out onto dust grain ice mantles co-desorb with water into the gas phase when the dust temperature reaches $\sim$ 120~K \citep{garrod2022}. As mentioned in Sect.\,\ref{section-intro}, other mechanisms may also be responsible for the presence of \mf\ in the gas phase, in particular at lower temperatures, such as chemical desorption, and grain sputtering due to shocks related to accretion and outflow activity. 

\begin{figure}[!t]
   \begin{center}
       \includegraphics[width=\hsize]{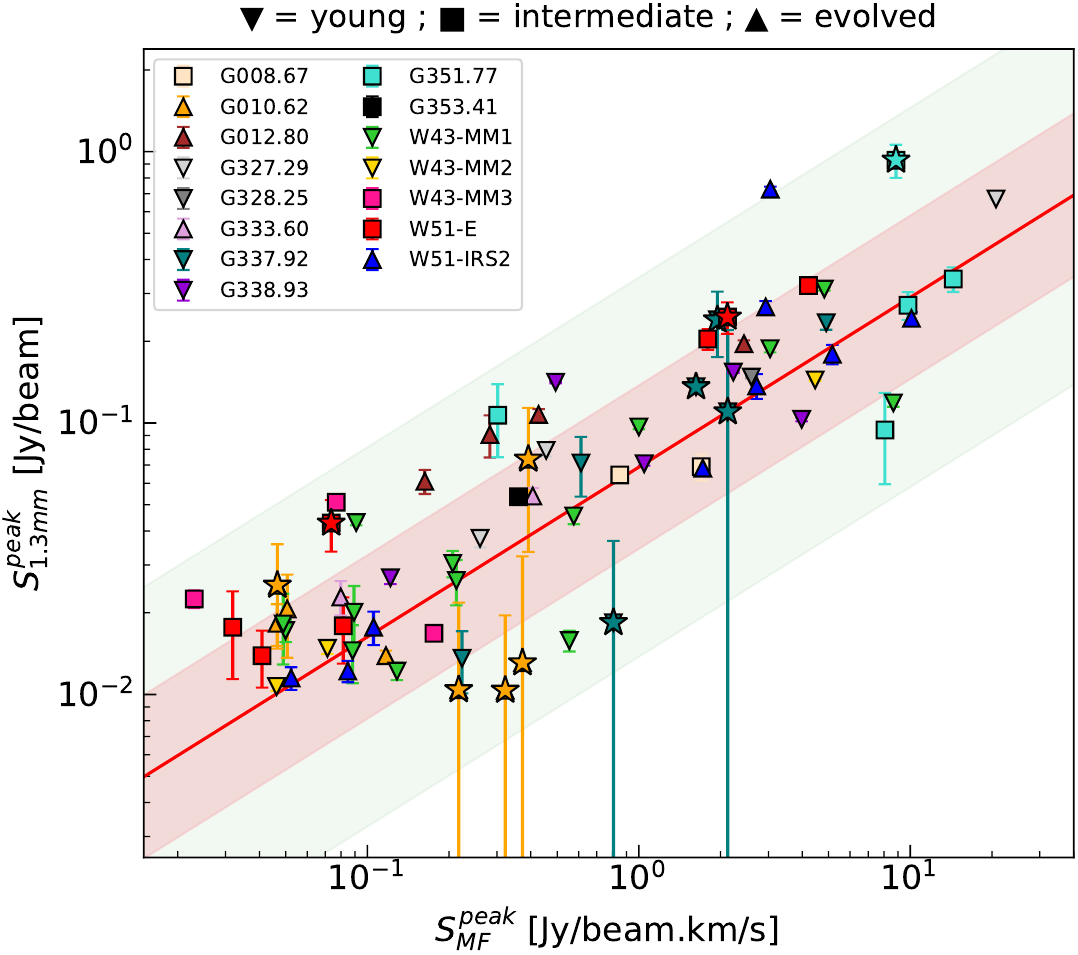}  \\
 \caption{\label{FIG-flux-ratio} Peak continuum flux density at 1.3\,mm vs. peak flux extracted in the moment 0 maps of \mf\ by \textsl{GExt2D}. The different symbols represent the different evolutionary stages of the protoclusters: young, intermediate and evolved, as indicated on top of the figure, while the stars shows the methyl formate sources that are not associated with any compact dust continuum core. The red line shows a weighted linear fit, while the red an green shaded regions show a factor of two and five deviation from this fit, respectively.}
  \end{center}
\end{figure}

In Fig.\,\ref{FIG-flux-ratio} we compare the 1.3\,mm continuum peak flux density to the peak flux measured in the moment 0 maps of methyl formate. We find that the continuum and methyl formate intensity appears to be relatively well correlated, with a Pearson's coefficient $\rho$\,=\,0.7 and assorted with a $P_{\mathrm{value}} <$ 0.001. About 70\% of the methyl formate sources have fluxes within a factor two from the expected value given by a weighted linear fit. This correlation suggests that the methyl formate emission for this sample is likely to share similar chemical origin, assuming that the methyl formate flux densities are directly related to the \mf\,gas-phase abundance. For the rest of the sources, larger discrepancies, up to a factor five, are observed (green shaded area). In particular, a group of 26 sources with low peak continuum flux densities, $I^{\mathrm{peak}}_{\mathrm{1.3 mm}}$ = 10 -- 20 mJy\,beam$^{-1}$, turn out to have methyl formate fluxes, $I^{\mathrm{peak}}_{\mathrm{MF}}$, spread over more than one order of magnitude. For these 26 sources, that are among the faintest methyl formate emitting sources and the faintest continuum sources, the methyl formate detected in the gas phase towards them could thus have a different chemical origin than the rest of the sources.

%======================
% FIGURE: Flux ratio MF vs. 1.3mm continuum
%======================
%======================

%-----------------------------
\subsection{Heating sources of hot cores?}
%------------------------------
\label{section-discussion-heating}

%\timea{The FWHM size contains about 66\% of the area under a 1D Gaussian. Taking FWHM/2 as a radius, therefore, does not consider all the emission. There are various approaches to overcome this, but if you take the FWHM as a radius, one gets $>$ 95\% of the emission. This is my argument to use the FWHM as a radius.}

Analytical expressions to describe the temperature distribution of a spherically symmetric, centrally illuminated core have been proposed in the literature \citep{Goldreich1974, Wilner1995}. Here we use the relation from \citet{RowanRobinson1980,Wolfire1986} and \citet{Wilner1995} for a bolometric luminosity (\lbol) range between 1\,$\Lsun$ and $6\times10^6$\,$\Lsun$:

\begin{equation}
    T_{\mathrm d}(R)=37 \,\mathrm{K} \times \Big( \frac{L}{L_{\odot}}\Big)^{0.25} \times \Big(\frac{R}{100 \,\mathrm{au}}\Big)^{-0.4}
\end{equation}

\begin{figure}[!t]
\begin{center}
       \includegraphics[width=\hsize]{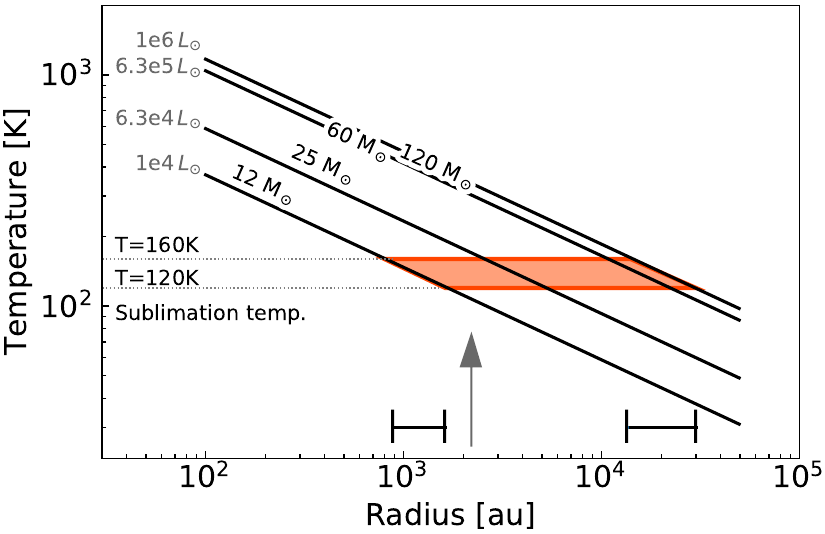}  \\
  \caption{\label{FIG-luminosity-size} Temperature profiles as a function of radius for centrally illuminated dusty cores from 12 $\Msun$ to 120 $\Msun$, following \citet{Wilner1995}. The red shaded area shows the interval that corresponds to the sublimation temperature of the ices (120--160 K). The two horizontal black lines represent the range of radii expected for a heating source with a luminosity of 10$^4$\,$\Lsun$ (12\,$\Msun$) and 10$^6$\,$\Lsun$ (120\,$\Msun$). The gray arrow indicates the median size of the methyl formate sources in our sample.} 
  \end{center}
\end{figure}

If we assume that the observed \mf\ emission traces the thermal sublimation radius with $T_{\rm d}$ = 120--160\,K (see Sect.\,\ref{section-MFTd}), we can use the measured \mf\ emission sizes to estimate the bolometric luminosity of the heating source and infer the type of the embedded (proto)star. Fig.\,\ref{FIG-luminosity-size} shows that for a heating source with a luminosity of 10$^4$\,$\Lsun$, the minimum radius is about 800--1600 au for \mf\ sublimation at 120--160 K. The smallest deconvolved FWHM sizes for our sample of \mf\ sources are about 990 au. Taking the measured deconvolved FWHM as a radius here, this would correspond to heating sources with \lbol\,$\sim$\,10$^4$\,$\Lsun$. This suggests that the population of methyl formate sources could be explained by deeply embedded heating sources corresponding to emerging B0 or earlier type ZAMS stars. On the other hand, the most extended, circularly symmetric \mf\ sources (the central sources of G327.29 and W51-E) reaching about 13400 au sizes could also be consistent with a very luminous single embedded source, because a  10$^6$\,$\Lsun$ heating source would have its \mf\ sublimation radius at 14000 to 30000 au. This size range corresponds well to the largest deconvolved FWHM sizes of \mf\ emitting regions of the ALMA-IMF sample. The population of circularly symmetric isolated \mf\ emission could thus be explained by individual heating sources with \lbol \,$\sim$10$^4$-10$^6$\,$\Lsun$. Protostellar evolutionary tracks show that when an emerging protostar reaches about $\sim$10$^4$\,$\Lsun$, the total luminosity is principally dominated by the protostellar luminosity \citep{hosokawa2009}. Although in these models the ZAMS luminosity and the corresponding mass are highly dependent on the accretion rate, we can tentatively convert these luminosities to stellar masses bearing in mind that our understanding of stars with M$_\star >$60\,$\Msun$ is very limited. Stellar evolutionary models predict a ZAMS stellar mass of 12\,$\Msun$ for a stellar luminosity of 10$^4$\,$\Lsun$ \citep{Ekstrom2012}, while a ZAMS stellar mass of 70\,$\Msun$ to a luminosity of 10$^6$\,$\Lsun$. Stellar models have, however a scatter between 60\,$\Lsun$ and 100--120\,$\Msun$ for the high luminosity range \citep{Meynet2000,Martins2005}. Should the methyl formate emission trace gas beyond the sublimation radius, i.e. corresponding to lower temperatures as suggested by \citet{busch2022,bouscasse2022} and \citet[][]{bouscasse2024}, the inferred luminosities are overestimated by our assumptions here, and consequently also the mass of the central protostar.

Overall, the inferred luminosities likely probe embedded heating sources that could well correspond to high-mass protostars, i.e. precursors of O and early B stars. The most extended sources could be explained by a single luminous protostar with a current stellar mass of about 70--120\,$\Msun$ for the central objects of G327.29 and W51-E, although multiplicity cannot be excluded.

%======================
% FIGURE: T PROFILE L
%======================
%======================

%------------------------------
\subsection{Nature of the methyl formate sources}
%------------------------------
\label{section-discussion-nature-sources}

To further investigate the nature of the methyl formate sources, we show in Fig.\,\ref{FIG-mass-to-size} the mass-to-size distribution of our sample of hot core candidates, where the mass corresponds to the core mass inferred from the dust thermal emission (see Sect. \ref{section-masses}) and the size is derived from the \mf \ emission (see Sect. \ref{section-MFsizes}). Assuming a 30--100\% efficiency for the core mass being converted to stellar mass \citep[see e.g.,][]{louvet2014,konyves2015}, cores with total gas mass above 8--28\,$\Msun$ are expected to form high-mass stars, and thus are excellent candidates for hot cores and their precursors.In total, 38 sources have their highest estimated mass (\textit{i.e.} for $T_{\mathrm{d}}$\,=\,50 or 200\,K) above 8~$\Msun$, which represents about 50\% of our sample. This number drops to 22\% (\textit{i.e.} 17 sources) if we consider only a 30\% efficiency for converting core mass into stellar mass. The fraction of sources with masses above 8\,$\Msun$ decreases to 24\% (\textit{i.e.} 18 sources), if we consider the lowest estimated mass (\textit{i.e.} for $T_{\mathrm{d}}$\,=\,150 or 400\,K). This suggests that the majority of the methyl formate sources correspond to low- and intermediate-mass cores.

The 38 methyl formate sources with masses $>$ 8~$\Msun$ in Fig.\,\ref{FIG-mass-to-size} span a range of deconvolved sizes of over one order of magnitude, from $\sim$ 1300 au to 13400 au. This range of masses and sizes correspond to archetypical hot cores (see Table \ref{TAB-cores-corinos}) and are $>$10 times larger compared to hot corinos. These sources are indicated in the last column of Table\,\ref{TAB-MFcat} as HC, or HC$^*$ for the ones of which the lowest estimated mass is $<$ 8~$\Msun$. The latest could be interpreted as hot cores around intermediate mass objects \citep[see, e.g,][]{fuente2014}. Four of the 38 sources, G010.62--MF3, G337.92--MF3, W51-E--MF2 and G351--MF3, are not associated with any 1.3mm compact continuum core from the \textsl{getsf} unsmoothed catalog (see the sources marked with stars in Fig.\,\ref{FIG-mass-to-size}). We also note that five ALMA-IMF protoclusters do not contain any emerging hot cores, which are 3 evolved clusters, G010.62, G012.80, and G333.60, and two intermediate ones, G353.41, and W43-MM3, where the identified methyl formate sources have low associated core masses. 

The rest of the compact methyl formate sources (\textit{i.e.} 53 sources, which corresponds to $\sim$70\% of the sample), have masses $<$ 8\,$\Msun$. Such core masses would correspond to intermediate mass and low-mass cores corresponding to hot corinos, i.e. chemically active Class 0/I sources, and  intermediate mass objects (see Table \ref{TAB-cores-corinos}). However, these objects have sizes ranging from $\sim$1000\,au to 4700\,au that is still a factor of at least 3 and up to 15 larger than the sizes typically found for hot corinos (see Table \ref{TAB-cores-corinos}) and would require sources $>$10$^4$\,$\Lsun$ as a heating source. The nature of these sources is, therefore, unclear. Based on the sensitivity of the dataset, we estimated in \citetalias{motte2022} that a hot core, like Orion-KL, with spectral lines as bright as $\sim$20--30 K in a 2000 au beam \citep[e.g.][]{brouillet2015, pagani2017} would be detected in ALMA-IMF, but our sensitivity is insufficient to detect hot corinos like IRAS 16293 \citep{cazaux2003} or IRAS 2A \citep{bottinelli2007} and IRAS 4B with compact sizes of at most 100 au \citep{bottinelli2004}. Hence these detections cannot correspond to the compact, radiatively heated hot gas phase of a hot corino population surrounding low-mass protostars. Instead, such cores could correspond to already formed intermediate to high-mass  protostars with a stellar mass of 5-12 $\Msun$, where most of the core material has been accreted already.  

An alternative explanation is that a significant contribution to the observed extended \mf\ emission could originate from spatially extended shocks either due to ejection or infall of material. In this scenario the cores would accrete more material from their surroundings. Spatially resolved ALMA observations confirm this scenario for the principal hot core precursor of the G328.25 protocluster, where the extended \mf\ emission, here decomposed into two peak positions, correspond to accretion shocks \citep[see also][]{csengeri2018}. \\

\begin{figure}[!t]
\begin{center}
       \includegraphics[width=\hsize]{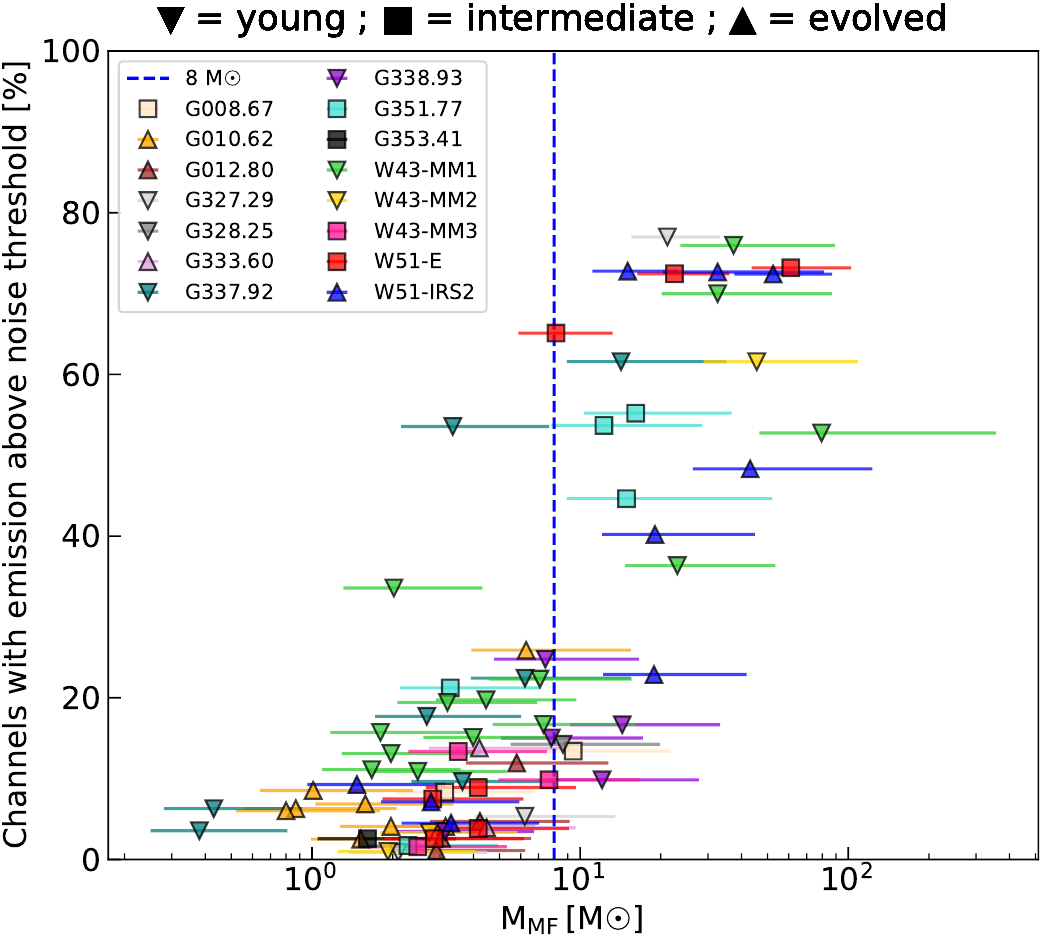}  \\
  \caption{\label{FIG-line-density} Number of channels that contain emission above the 3$\sigma$ noise level in the observed spectrum (Table\,\ref{TAB-MFcat}) as a function of the masses of the methyl formate sources. The different symbols represent the different evolutionary stages of the protoclusters: young, intermediate and evolved, as indicated on top of the figure, while the stars shows the methyl formate sources that are not associated with any compact dust continuum core. The horizontal blue dashed line shows the mass threshold of 8\,$\Msun$.}
  \end{center}
\end{figure}

\begin{figure}[!t]
\begin{center}
       \includegraphics[width=\hsize]{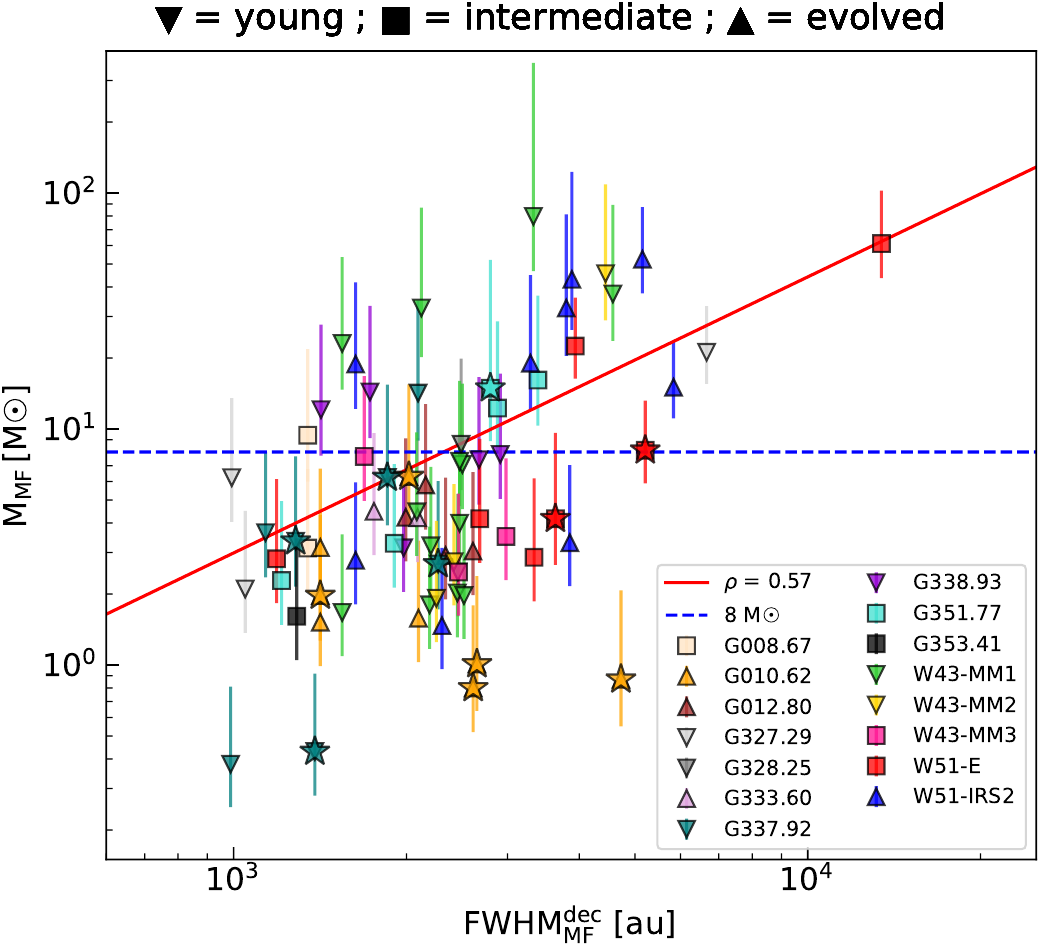}  \\
  \caption{\label{FIG-mass-to-size} Mass-to-size plot for the 76 methyl formate sources extracted from the moment 0 maps (Figs.\,\ref{FIG-mom0-maps}--\ref{FIG-mom0-maps4}). The different symbols represent the different evolutionary stages of the protoclusters: young, intermediate and evolved, as indicated on top of the figure, while the stars shows the methyl formate sources that are not associated with any compact dust continuum core. The error bars show the range of masses obtained considering the uncertainties on $T_{\mathrm d}$ (300$\pm$ 100K for the most extreme objects and 100$\pm$ 50K for the rest, see Sect.\,\ref{section-MFTd}). The weighted fit to the datapoints (red line) shows that the most massive cores tend to harbor a more extended emission of methyl formate.}   
  \end{center}
\end{figure}

We discuss in Fig.\,\ref{FIG-line-density} the comparison of the source line richness as a function of core mass. We show that there is a general trend of more massive cores harboring spectra with more channels with emission above the noise threshold, \textit{i.e.} being more line rich. All but one sources with more than 36\% channels containing emission in their spectra are identified as hot cores, \textit{i.e.} with masses $>$ 8\,$\Msun$. For the rest of the methyl formate sources, their spectra are found to contain 9--25\% of channels with emission. Figure\,\ref{FIG-line-density} shows that there is no clear threshold between hot cores and lower-mass objects for the sources that have around 20\% of their channels containing emission. For this reason, in the current paper we base our source classification on the mass threshold only, while the trends with line richness will be further discussed in a subsequent paper Csengeri et al. (in prep.). \\

%======================
% FIGURE: SPETCRAL LINE DENSITY
%======================
%======================

As seen in Sect.\,\ref{section-freefree}, 19 compact \mf\ sources are associated with extended free-free emission (see Figs.\,\ref{FIG-ellipses-h41alpha}, \ref{FIG-ellipses-h41alpha2}, and \ref{FIG-ellipses-h41alpha4}), calling into question the nature of several sources. In particular, towards the G010.62 protocluster, we have estimated that $\sim$9 -- 100\% of the flux measured at 1.3 mm towards the methyl formate sources is due to free-free emission. These sources are spatially coincident with the bright central {\hii} region suggesting that these sources are more evolved than the two other more isolated sources in the same field (G010.62--MF7 and G010.62--MF10). In particular, the 1.3~mm continuum emission measured towards the sources G010.62--MF1 and G010.62--MF2 is estimated to be 100\% due to free-free emission, suggesting that they are entirely surrounded by ionised gas. Visual inspection of their spectra using the ALMA-IMF spw7 in B6 (centered at 232.45\,GHz, with a bandwidth of $\sim$2\,GHz), has revealed plenty of strong molecular lines confirming the presence of hot gas rich in COMs, despite the presence of ionized gas. Since the emission of molecular gas is necessarily associated with the presence of dust, this may suggest that our method somewhat overestimates the free-free contribution towards these sources. In fact, for {\uchii} regions the emission is often optically thick, while we assumed optically thin emission, and the emitting region for the ionised gas could also be smaller compared to the dust continuum emission. Alternatively, a complex mixture of ionised and molecular gas along the line of sight could also explain our observations. Emission of COMs in the vicinity, or in partial overlap with {\uchii} regions has already been observed, for instance towards the Galactic center cloud Sgr B2(N) \citep[see][]{bonfand2017}. It is also possible that in this field, the different observed structures (i.e. methyl formate extended emission and {\hii} region) are not part of the same spatially coherent structure, i.e. the {\hii} region shell is just expanding, therefore the molecular emission comes from the outside. \\

Overall, we propose that we see the \mf\ emission arising from different population of sources, with the most massive cores corresponding to hot cores where radiative heating has liberated \mf\ into the gas phase. The origin of \mf\ emission for the lower mass cores is uncertain, and could be explained either by a continuous accretion and a shock origin related to infall and accretion processes, or by having the cores in a more evolved stage where most of the core mass has already been accreted. Cores associated with free-free emission warrant a more precise understanding of the amount of free-free contribution, and represent the latest stages of high-mass star formation where the (proto)star already ionises its surroundings, and the molecular core material is being exhausted. \\

%======================
% FIGURE: MASS_HC vs. SIZE_HC
%======================
%======================

%======================
% TABLE: COMPARE PROPERTIES HOT CORES/CORINOS
%======================
\begin{table}[!t]
\begin{center}
  \caption{\label{TAB-cores-corinos} Summary of the main properties of archetypical hot cores/corinos.} 
  \setlength{\tabcolsep}{1.6mm}
  \begin{tabular}{ccc}
    \hline
    \hline
    \multicolumn{1}{c}{Source type} & \multicolumn{1}{c}{Size}  & \multicolumn{1}{c}{$M$} \\
                                    & \multicolumn{1}{c}{[au]} & \multicolumn{1}{c}{[$\Msun$]} \\
    \hline
    Hot cores & $>$ 2000 &  8 -- 1000    \\
    Hot corinos & $<300$ &   $<$ 8     \\
    Our sources & 990 -- 13400 &   0.25 -- 263  \\
   \hline
\end{tabular}
\end{center}
\vspace{-4mm}
\tablefoot{See e.g. \citet{cesaroni1998, cesaroni2005, widicus2017, sanchezmonge2014, bonfand2017, bonfand2019} for hot core properties, and \citet{ceccarelli2000, ceccarelli2007, lee2022, okoda2022, chahine2022} for hot corinos.}
\end{table}
% ======================

%------------------------------
\subsection{Emergence and life time of hot cores}
%------------------------------
\label{section-lifetime}

Using the relative fraction of hot cores versus the total number of continuum cores, we can provide a rough estimate for the time-scale required for the emergence of hot cores. Similar time-scale estimates have been done using source counts to compare hot core to {\hii} region timescales in \citet{wilner2001}. We rely on the assumption that the \mf\ emission  originates from the same mechanism over the sample (see, however Sect. \ref{section-discussion-origin-ch3ocho}), and that all objects following the same evolutionary path will develop into radiatively heated hot cores. This is unlikely to hold  for the entire sample of \mf\ sources, especially the lower core mass population, therefore, we consider only the highest mass cores that are the most robust hot core candidates. Figure \ref{FIG-histo-mass-distribution} shows that the relative fraction of dust cores with and without hot core emission increases with the core mass, which supports the picture where all massive cores go through the hot core stage. We notice that above a mass threshold of $\sim$8\,$\Msun$, this fraction increases from $\sim$ 20\% to 100\% increasing with core mass. We assume a time-scale for the protostellar phase of $3\times10^5$~yr based on \citep{Duarte2013} and also used in \citet{csengeri2014}. 
For a core mass range between 30 and 50~$M_\odot$, the fraction of hot cores to the total number of dust cores rapidly increases to 30--80\%. Adopting a constant $3\times10^5$~yr time-scale for the protostellar phase over this mass range, we find that the time scales for the hot core phase is between 0.9$\times10^5$~yr  and 2.4$\times10^5$~yr. Uncertainties only due to the mass estimates and the protostellar life-time itself can lead to variations of a factor of a few. Since we do not have a full statistics of the more evolved HC-{\uchii} stages, this estimate does not account for a potential overlap in the hot core and HC-{\uchii} phase. 

%If a fraction of 50\% for a core mass above 50\,$\Msun$ are hot cores, and adopting this time-scale for the protostellar phase of $3\times10^5$~yr, we find that massive cores spend at least 1.5$\times10^5$~yr in the hot core phase.

%------------------------------
\subsection{What type of clouds host massive hot cores?}
%------------------------------
\label{section-discussion-nb-HCs-per-protocluster}

Figure \ref{FIG-frac-MF-Mclump} shows that the highest ratios of methyl formate sources to continuum cores ($\leq$25\%) are found towards the five following protoclusters: G351.77, G337.92, G010.62, W43-MM1, and W51-E, with no obvious correlation with their evolutionary stage. They all contain at least one massive (\textit{i.e.} $>$8\,$\Msun$) hot core, except for G010.62, of which the brightest central region is strongly contaminated by free-free emission. In particular, 100\% of the continuum flux measured at 1.3 mm towards the two brightest methyl formate sources, G010.62--MF1 and MF2 have been attributed to free-free emission (see Sect.\,\ref{section-discussion-nature-sources}), preventing us from computing mass estimates for these sources. Furthermore, we found five methyl formate sources without associated compact continuum core in this region, which may lead to inaccurate extraction of the flux from the continuum maps, and thus inaccurate mass estimates (see Sect.\,\ref{section-contassociation} and Table\,\ref{TAB-cont-cat-getsf}). In addition to G010.62, we find four other ALMA-IMF protoclusters that do not contain any massive hot cores, the two evolved clusters G012.80 and G333.60, and two intermediate ones G353.41, and W43-MM3. 

\begin{figure}[!t]
\begin{center}
       \includegraphics[width=\hsize]{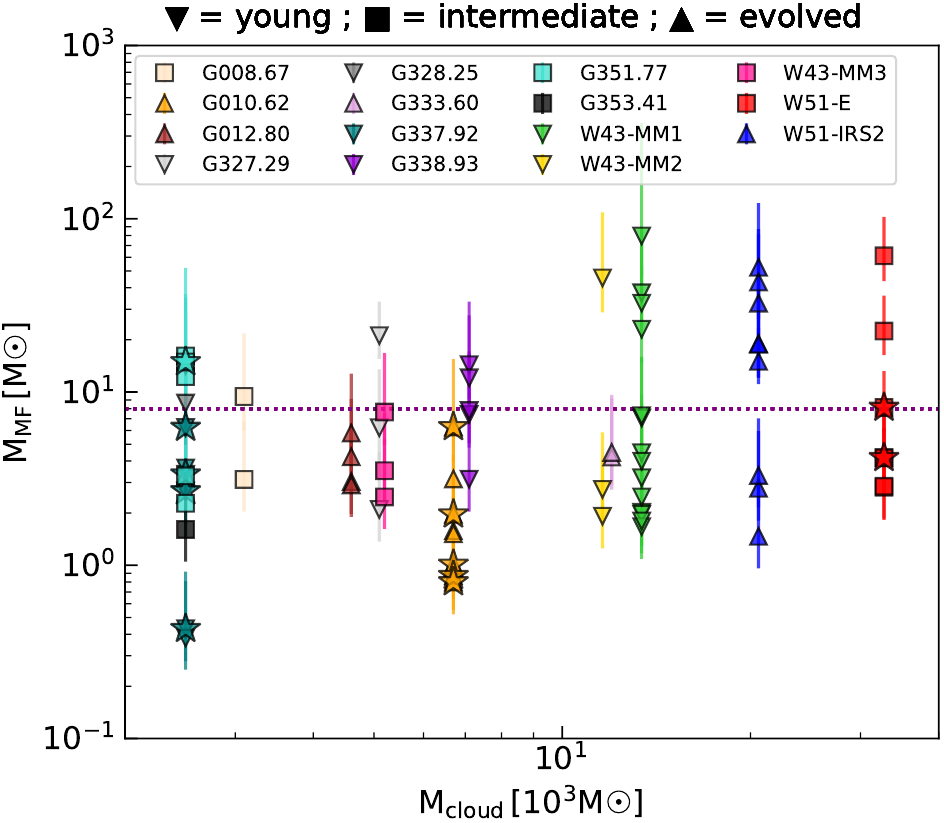}  \\
  \caption{\label{FIG-mHC-Mclump} Mass distribution of the methyl formate sources as a function of their parent cloud mass. The different symbols represent the different evolutionary stages of the protoclusters: young, intermediate and evolved, as indicated on top of the figure, while the stars shows the methyl formate sources that are not associated with any compact dust continuum core. The error bars show the range of masses obtained considering the uncertainties on $T_{\mathrm d}$ (300$\pm$ 100K for the most extreme objects and 100$\pm$ 50K for the rest, see Sect.\,\ref{section-MFTd}). The horizontal dotted line shows the 8\,$\Msun$ mass threshold.}
  \end{center}
\end{figure}

Figure \ref{FIG-mHC-Mclump} highlights the hot cores with masses above 8\,$\Msun$, which are found towards ten ALMA-IMF protoclusters, with no obvious correlation with their evolutionary stage. The ten clouds span a range of mass from 2.5$\times$10$^3$\,$\Msun$ to 32.7$\times$10$^3$\,$\Msun$, suggesting that the presence of massive hot cores also does not depend on the total mass of the protocluster. 

%======================
% FIGURE: MASS_HC vs. MASS_CLUMP
%======================
%======================

%------------------------------
\subsection{G353.41: a chemically inactive  massive protocluster?}
%------------------------------
\label{section-discussion-G353}

G353.41 is an intermediate evolutionary stage protocluster that is very bright in the continuum at 1.3~mm, and strongly affected by ionized gas coming from {\uchii} regions (see Fig.~2 of \citetalias{motte2022}). This region is a remarkable outlier of the ALMA-IMF hot core sample as it hosts only one weak, low-mass ($<$2\,$\Msun$) compact methyl formate source. Despite the fact that it hosts as many as 57 compact dust continuum cores, these are mainly low-to intermediate mass cores, with masses that range from 0.2 to 9.8\,$\Msun$ (see \citetalias{louvet2024}). A visual inspection of the ALMA-IMF spectra in B6 spw7 (centered at 232.45\,GHz, with a bandwidth of $\sim$2 GHz) extracted towards the three more massive cores (\textit{i.e.} $>$7\,$\Msun$) did not reveal any strong molecular lines, confirming the lack of hot gas rich in COMs. It suggests that despite being massive, and qualifying for the ALMA-IMF selection, this protocluster is likely to form only low- to intermediate-mass stars, and lacks strong emission from COMs. It is yet unclear what physical conditions can explain this characteristics, a more complete investigation of the overall energetics of the protocluster is needed to discuss the potential origin of its chemically poor stage.

%--------------------------------------------------------------------
\section{Summary}
%--------------------------------------------------------------------
\label{section-ccl}

We investigated the spatial distribution of methyl formate emission towards 15 massive protoclusters targeted by the ALMA-IMF Large Program. Methyl formate is a complex species commonly detected in star-forming regions, that we used to search for hot core candidates in our dataset. We computed and analysed moment 0 maps combining two strong transitions of methyl formate at 216.2 GHz, with $E_{\rm up}/k$ = 109\,K. We used a source-extraction algorithm to extract and characterize the cores that exhibit methyl formate emission. We cross-matched our catalog of methyl formate sources with that of the compact continuum cores of the ALMA-IMF dataset in order to derive and compare their physical properties and constrain their nature. Our main findings are summarized below: 

   \begin{enumerate}
      \item We find that in most cases, methyl formate traces compact sources, that may be clustered or isolated, with sizes (FWHM) between $\sim$ 1000 and 6000\,au, and a median extent of $\sim$2300\,au. There are two outliers in the two young protoclusters, G327.29 and W51-E, with more extended emission of methyl formate above 6000\,au and reaching 13400\,au. 
      \item We built a catalog of 76 compact methyl formate sources, which is about an order of magnitude less compared to the number of purely dust continuum cores extracted from the ALMA-IMF dataset. We identified 56 of these methyl formate sources (\textit{i.e.} 76\% of the sample) as new detections. 
      \item We find an increasing trend in the number of methyl formate sources as a function of the number of continuum cores per protocluster, but the fraction of hot core candidates to dust continuum cores never exceeds 25\% per protocluster. We find no clear correlation between the number of methyl formate sources and the clump mass, nor the evolutionary stage of the protocluster, suggesting that the number of hot core candidates is independent from the properties of their host protocluster.
      \item Assuming a mean dust temperature of 100\,K, we estimated core masses ranging from $\sim$ 0.2\,$\Msun$ to 80\,$\Msun$ for the sources that exhibit methyl formate emission. We found that the relative fraction of compact continuum cores with and without methyl formate emission increases with the core mass, which supports the picture where all massive cores go through the hot core stage.  
      \item We showed that the population of hot core candidates could be explained by deeply embedded heating sources corresponding to high-mass protostars, i.e. emerging B0 or earlier type ZAMS stars, with \lbol \,$\sim$10$^4$-10$^6$\,$\Lsun$. The most extended sources, \textit{i.e.} the central sources of G327.29 and W51-E reaching about 13400\,au sizes, can also be explained by a single very luminous embedded source with 10$^6$\,$\Lsun$, although multiplicity at smaller scales cannot be excluded. 
      \item We found that about 50\% of the methyl formate sources have masses above 8\,$\Msun$ and span a range of sizes (FWHM) from 1300 au to 13400 au, which correspond well to archetypical hot cores. The rest of the compact methyl formate sources, have masses $<$8\,$\Msun$ and sizes ranging from 1000\,au to 4700\,au, that correspond to intermediate mass objects, where a significant contribution from spatially extended shocks could also explain the observed \mf\ emission. 
      \item Considering the fraction of cores with hot core emission and assuming a time-scale for the protostellar phase of $3\times10^5$~yr, we found that massive cores spend at least 1.5$\times10^5$~yr in the hot core phase.
      \item We found that all protoclusters harbor at least one methyl formate source and ten out of 15 contain massive hot cores, suggesting that the presence of massive hot cores does not depend neither on the total mass of the protocluster nor on its evolutionary stage. We found one outlier, G353.41 that clearly lacks strong emission from COMs.
   \end{enumerate}

Overall, we propose that the methyl formate emission arises from different population of objects over the ALMA-IMF target regions, with the most massive cores corresponding to centrally heated hot cores, where \mf\ is released into the gas phase via thermal desorption, and the origin of \mf\ emission for the lower-mass cores could be  explained by either shocks related to infall and accretion processes, or by having the cores in a more evolved stage, where most of the core mass has already been accreted to form intermediate to high-mass protostars.

\begin{acknowledgements}

We thank the referee for the careful reading of the manuscript and providing useful comments. This paper makes use of the ALMA data ADS/JAO.ALMA2017.1.01355.L. ALMA is a partnership of ESO (representing its member states), NSF (USA) and NINS (Japan), together with NRC (Canada), MOST and ASIAA (Taiwan), and KASI (Republic of Korea), in cooperation with the Republic of Chile. The Joint ALMA Observatory is operated by ESO, AUI/NRAO and NAOJ. This project has received funding from the European Research Council (ERC) via the ERC Synergy Grant ECOGAL (grant 855130), from the French Agence Nationale de la Recherche (ANR) through the project COSMHIC (ANR-20-CE31-0009), and the French Programme National de Physique Stellaire and Physique et Chimie du Milieu Interstellaire (PNPS and PCMI) of CNRS/INSU (with INC/INP/IN2P3). The project leading to this publication has received support from ORP, that is funded by the European Union’s Horizon 2020 research and innovation programme under grant agreement No 101004719 [ORP]. M.B. thanks R. T. Garrod for the fruitful discussions on the chemical origin of methyl formate. M.B. thanks J. Pety for providing the equations for the source size deconvolution. M.B. is currently a postdoctoral fellow in the University of Virginia’s VICO collaboration and is funded by grants from the NASA Astrophysics Theory Program (grant number 80NSSC18K0558) and the NSF Astronomy \& Astrophysics program (grant number 2206516). T.Cs. has received financial support from the French State in the framework of the IdEx Université de Bordeaux Investments for the future Program. S.B. acknowledges support by the French Agence Nationale de la Recherche (ANR) through the project GENESIS (ANR-16-CE92-0035-01). F.M., N.C., and Y.P. acknowledge the COSMHIC ANR and the ECOGAL ERC. A.G. acknowledges support from the National Science Foundation under grant AST-2008101. R.G.-M. acknowledges support from UNAM-PAPIIT project IN104319 and from CONACyT Ciencia de Frontera project ID: 86372. A.S gratefully acknowledges support by the Fondecyt Regular (projectcode 1220610), and ANID BASAL projects ACE210002 and FB210003. R.A. gratefully acknowledges support from ANID Beca Doctorado Nacional 21200897. A.L.S. acknowledges funding from the European Research Council (ERC) under the European Union’s Horizon 2020 research and innovation programme, for the Project “The Dawn of Organic Chemistry ” (DOC), grant agreement No 741002. L.B. gratefully acknowledges support by the ANID BASAL project  FB210003.

\end{acknowledgements}

%--------------------------------------------------------------------
% BIBLIO
%--------------------------------------------------------------------
\bibliographystyle{aa}
\bibliography{biblio_draft}

%--------------------------------------------------------------------
% APPENDIX
%--------------------------------------------------------------------
\begin{appendix}

\onecolumn

%------------------------------------------------------------------
\section{\mf\ spectra}
%------------------------------------------------------------------
\label{appendix-spectra}

\begin{figure*}[h!]
   \resizebox{\hsize}{!}
  {\begin{tabular}{cc}  
       \includegraphics[width=\hsize]{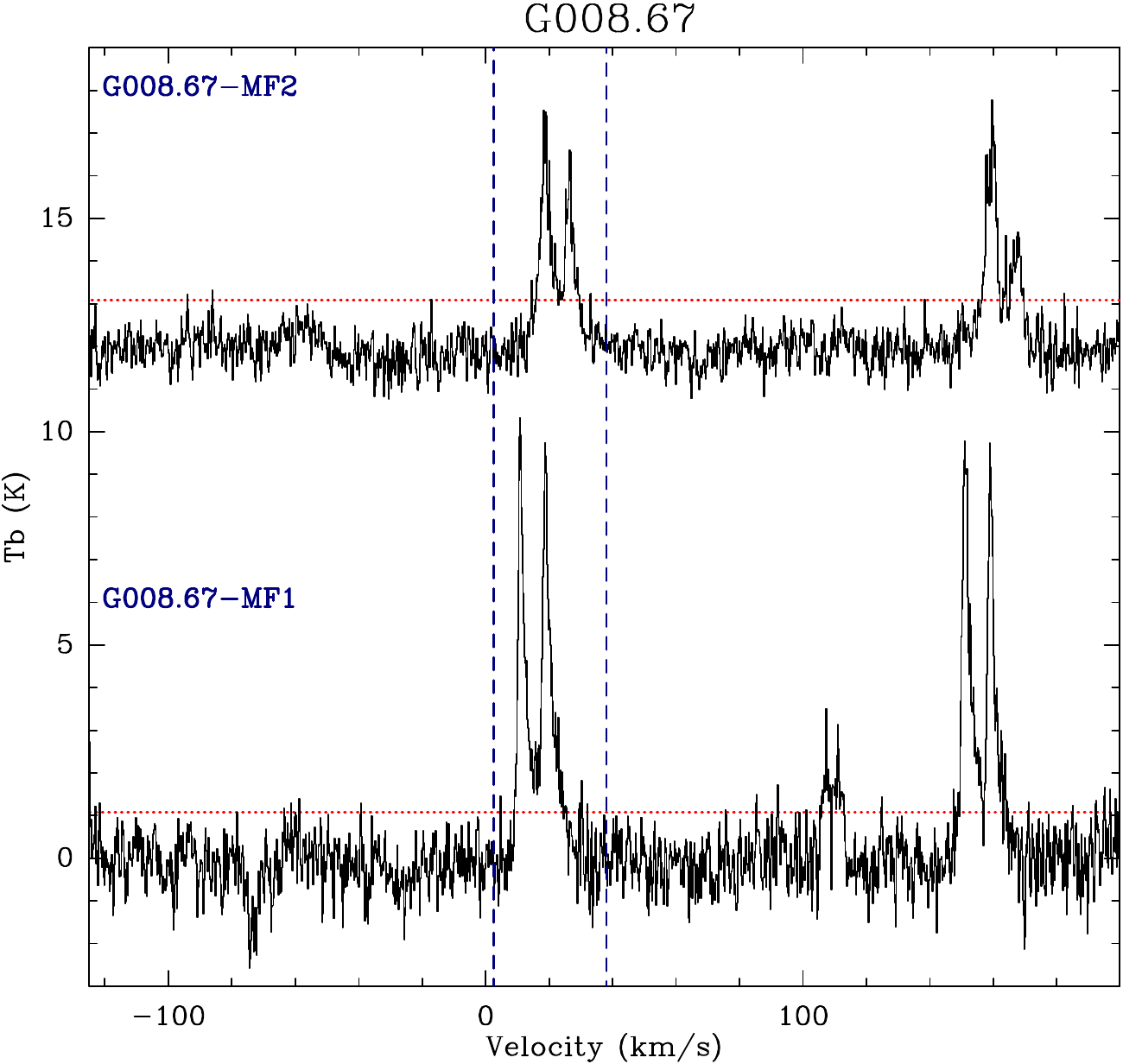}  &
       \includegraphics[width=\hsize]{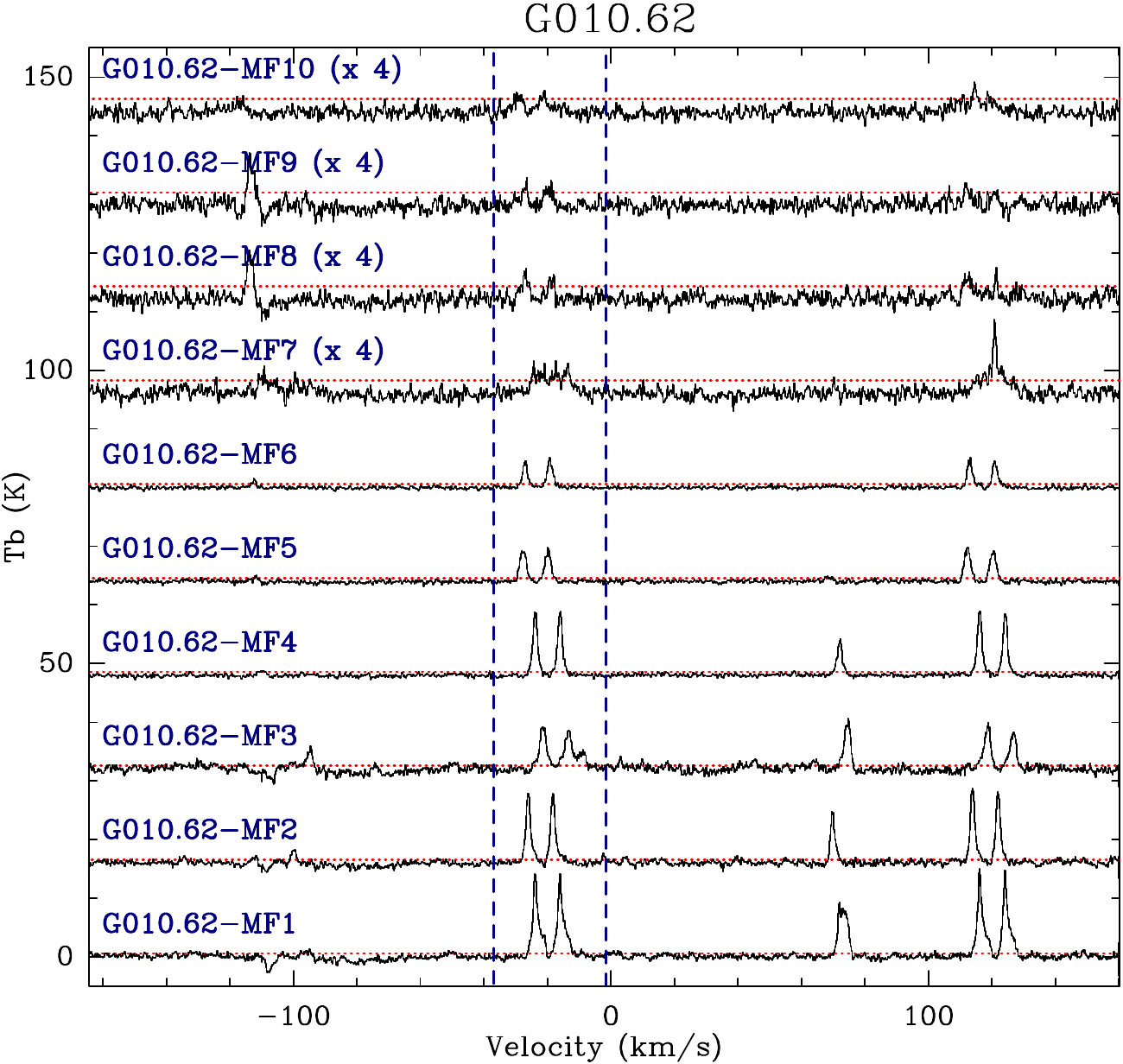}  \\
       \includegraphics[width=\hsize]{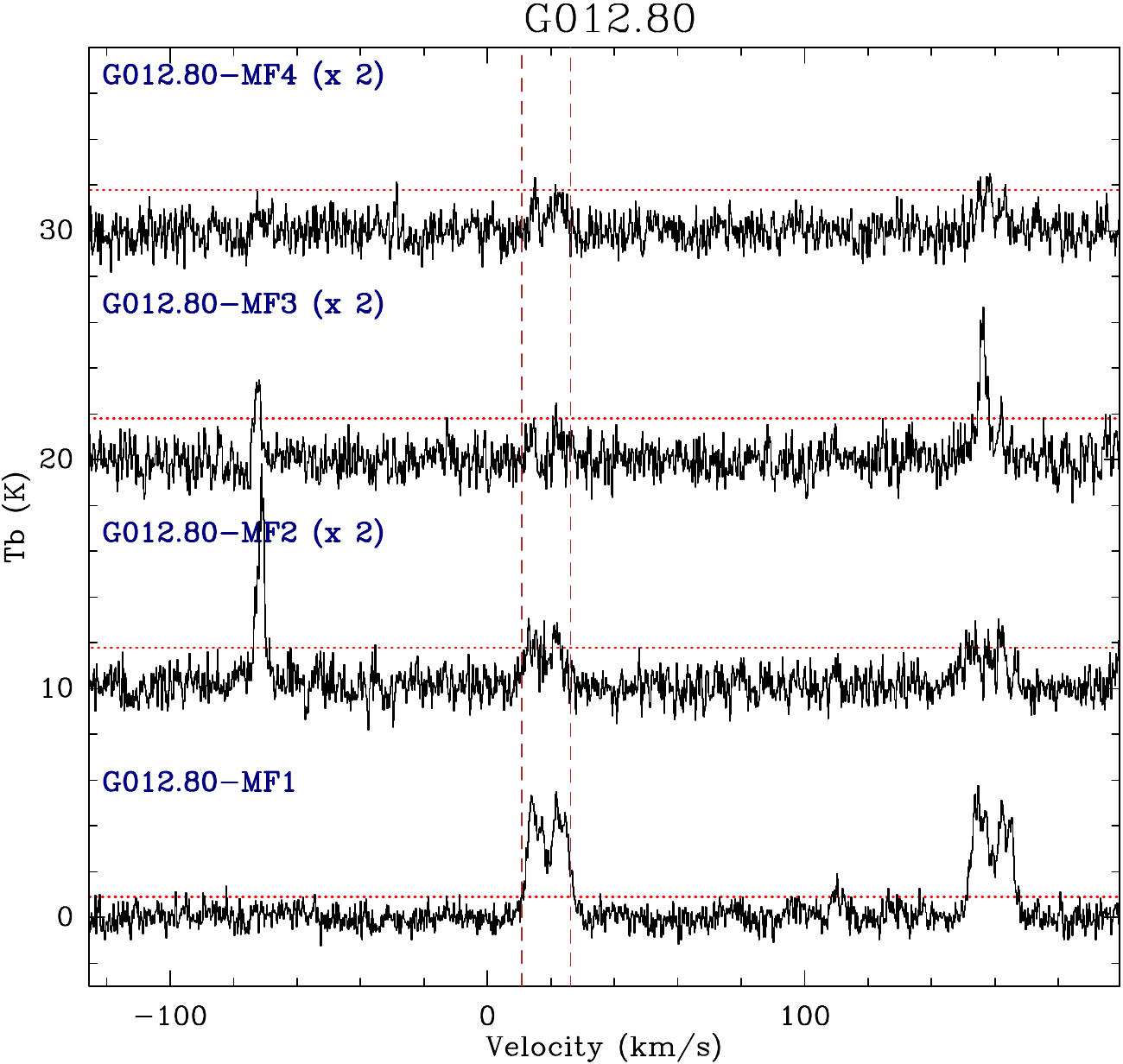}  &
       \includegraphics[width=\hsize]{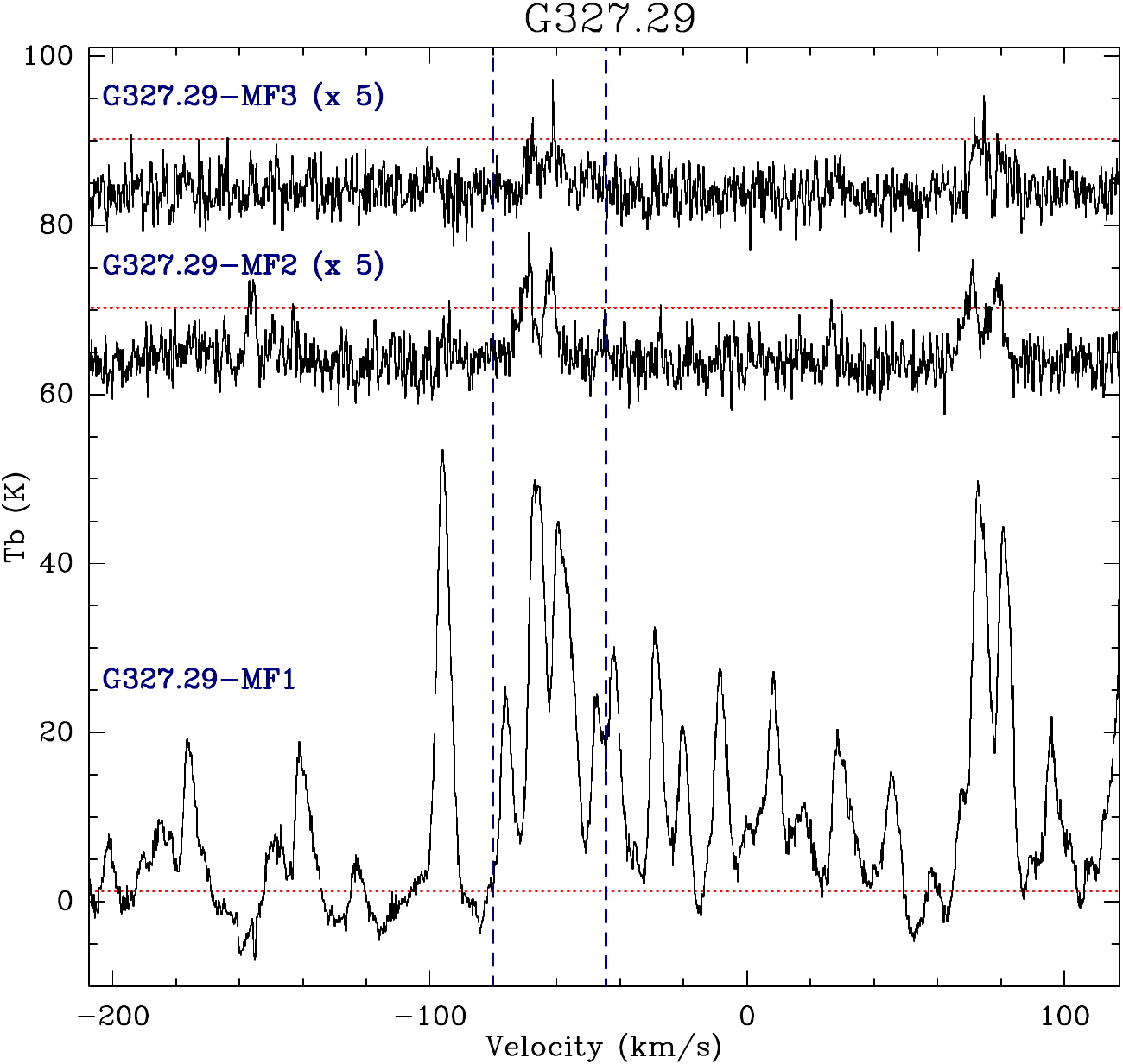}  \\
       \end{tabular}}
  \caption{\label{FIG-MF-spectra} Single-pixel continuum-subtracted spectra extracted towards the peak position of the compact methyl formate sources in the 15 ALMA-IMF protoclusters. The spectra are shifted along the y axis and the value in parentheses (if any), indicates the scaling factor applied to the spectrum. The vertical dashed lines show the velocity range used to compute the methyl formate moment 0 maps as in Fig.\,\ref{FIG-integrated-spectra}. The red horizontal dotted lines show the 3$\sigma$ threshold, using the rms noise level measured in the line cubes (see Table\,\ref{TAB-linecubes}). The figure continues on the next page.} 
\end{figure*}

\begin{figure*}[t!]
   \resizebox{\hsize}{!}
  {\begin{tabular}{cc} 
       \includegraphics[width=\hsize]{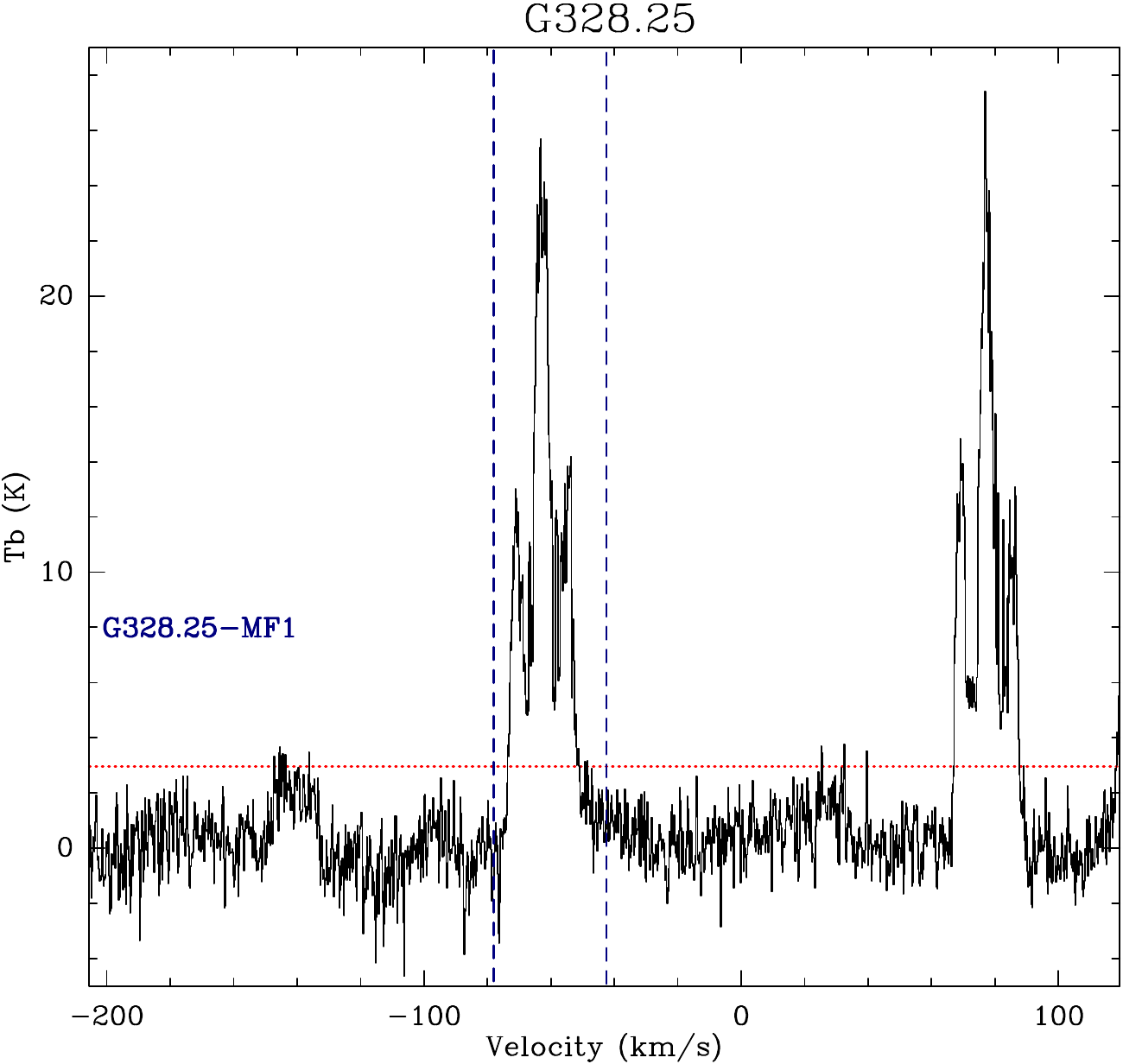}  &
       \includegraphics[width=\hsize]{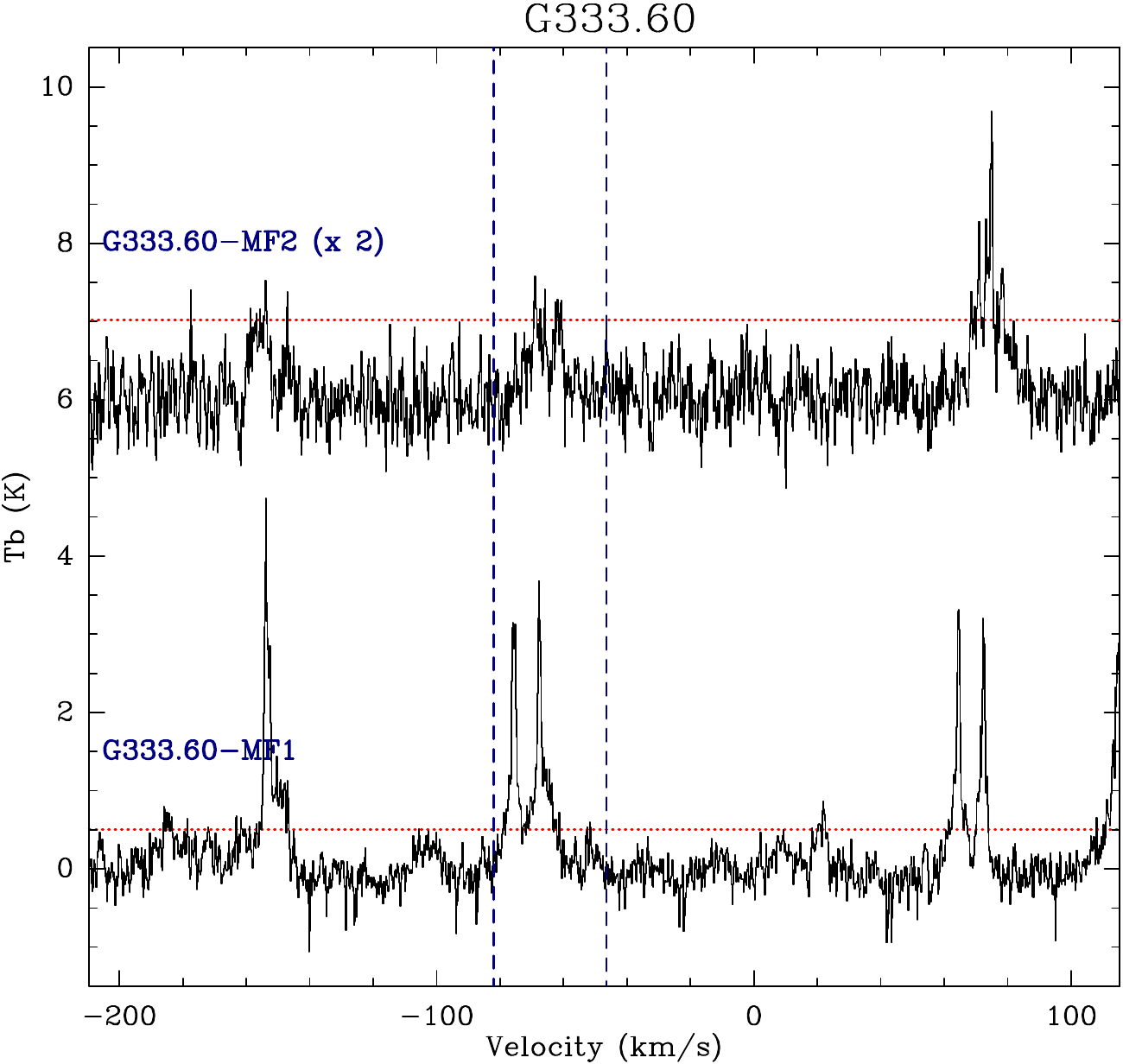}  \\
       \includegraphics[width=\hsize]{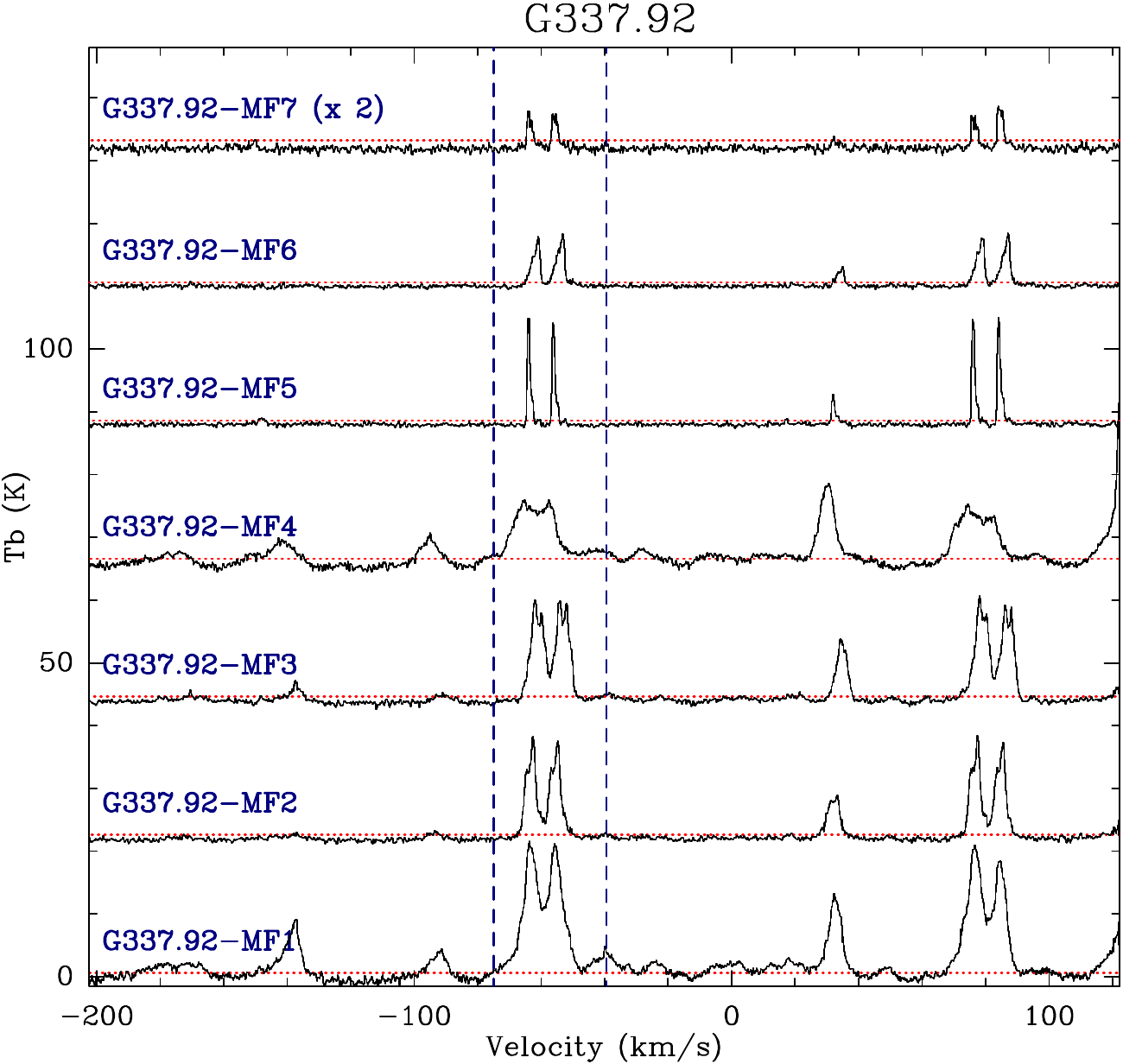}  &
       \includegraphics[width=\hsize]{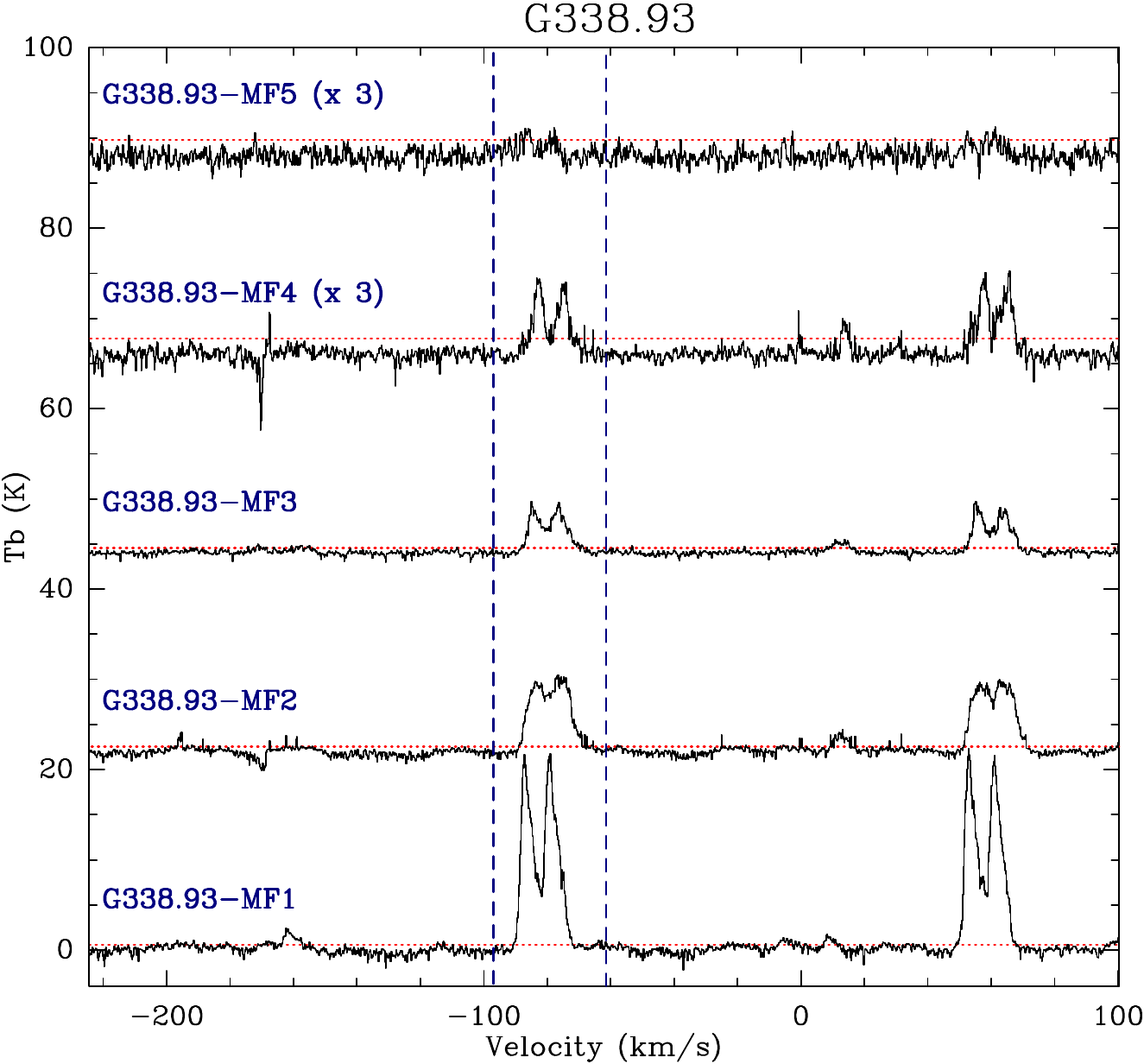}  \\
       \end{tabular}}
       \caption{\label{FIG-MF-spectra2} Same as Fig.\,\ref{FIG-MF-spectra}. The figure continues on the next page.} 
\end{figure*}

\begin{figure*}[t!]
   \resizebox{\hsize}{!}
  {\begin{tabular}{cc} 
       \includegraphics[width=\hsize]{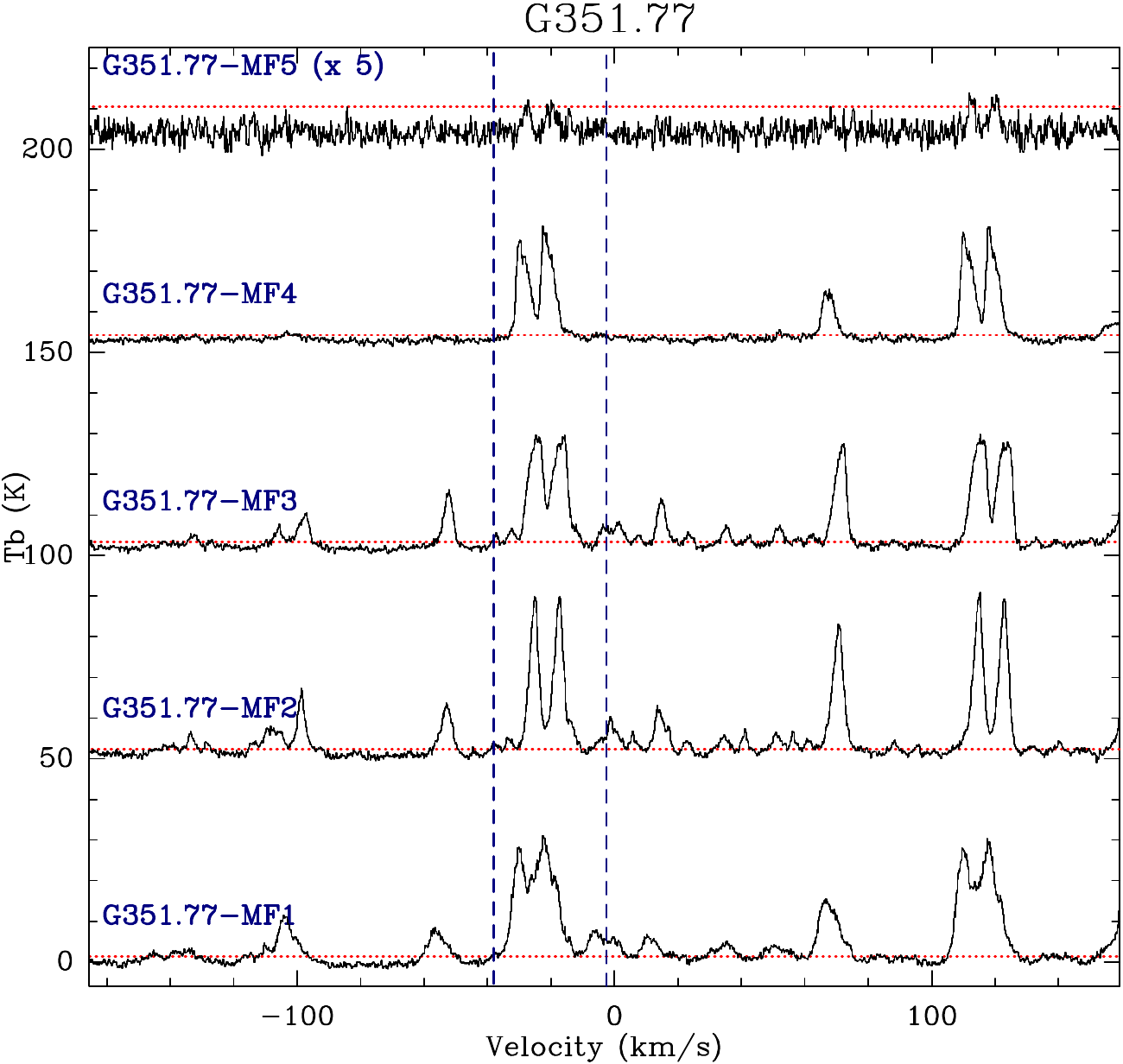}  &
       \includegraphics[width=\hsize]{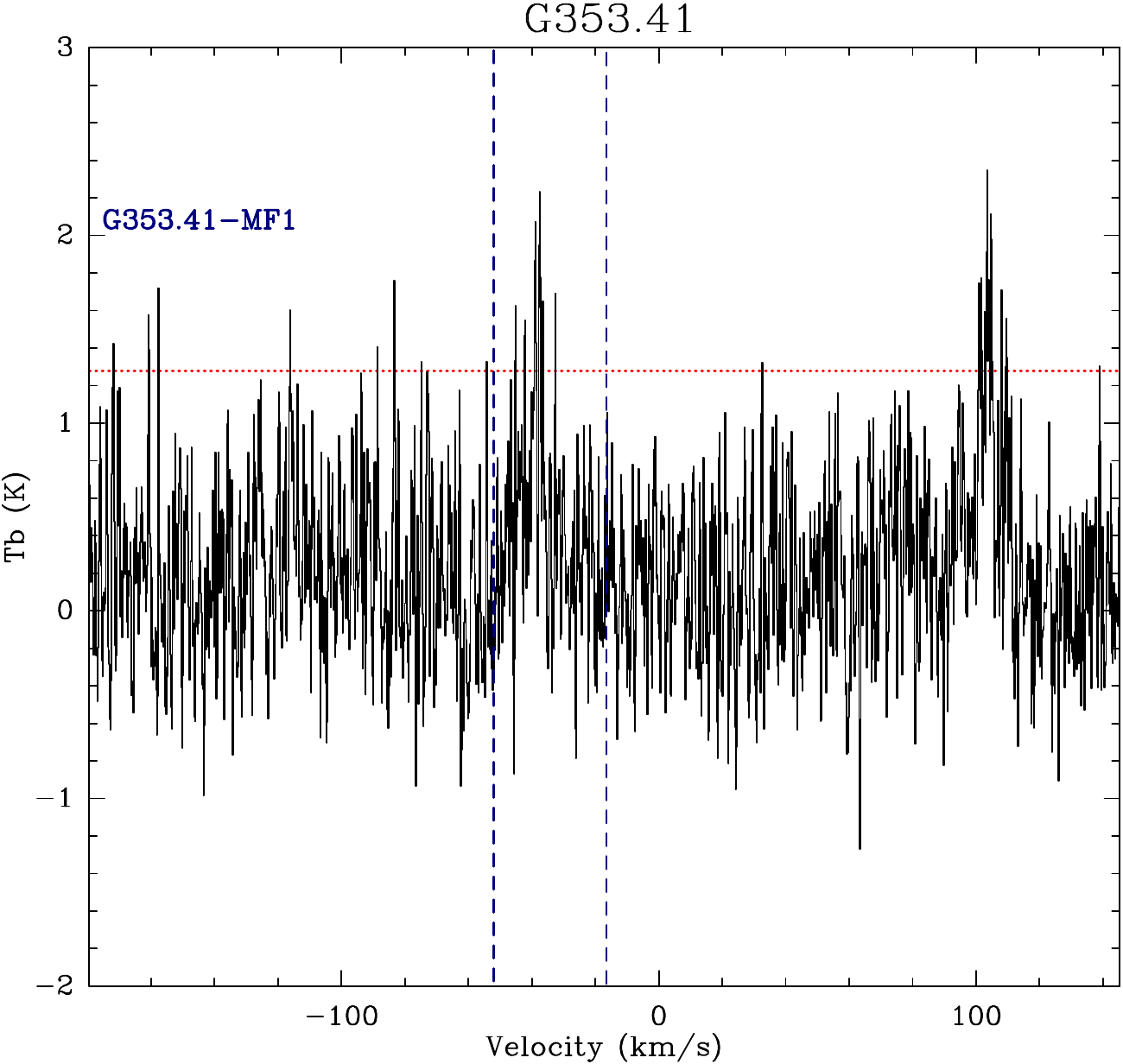}  \\
       \includegraphics[width=\hsize]{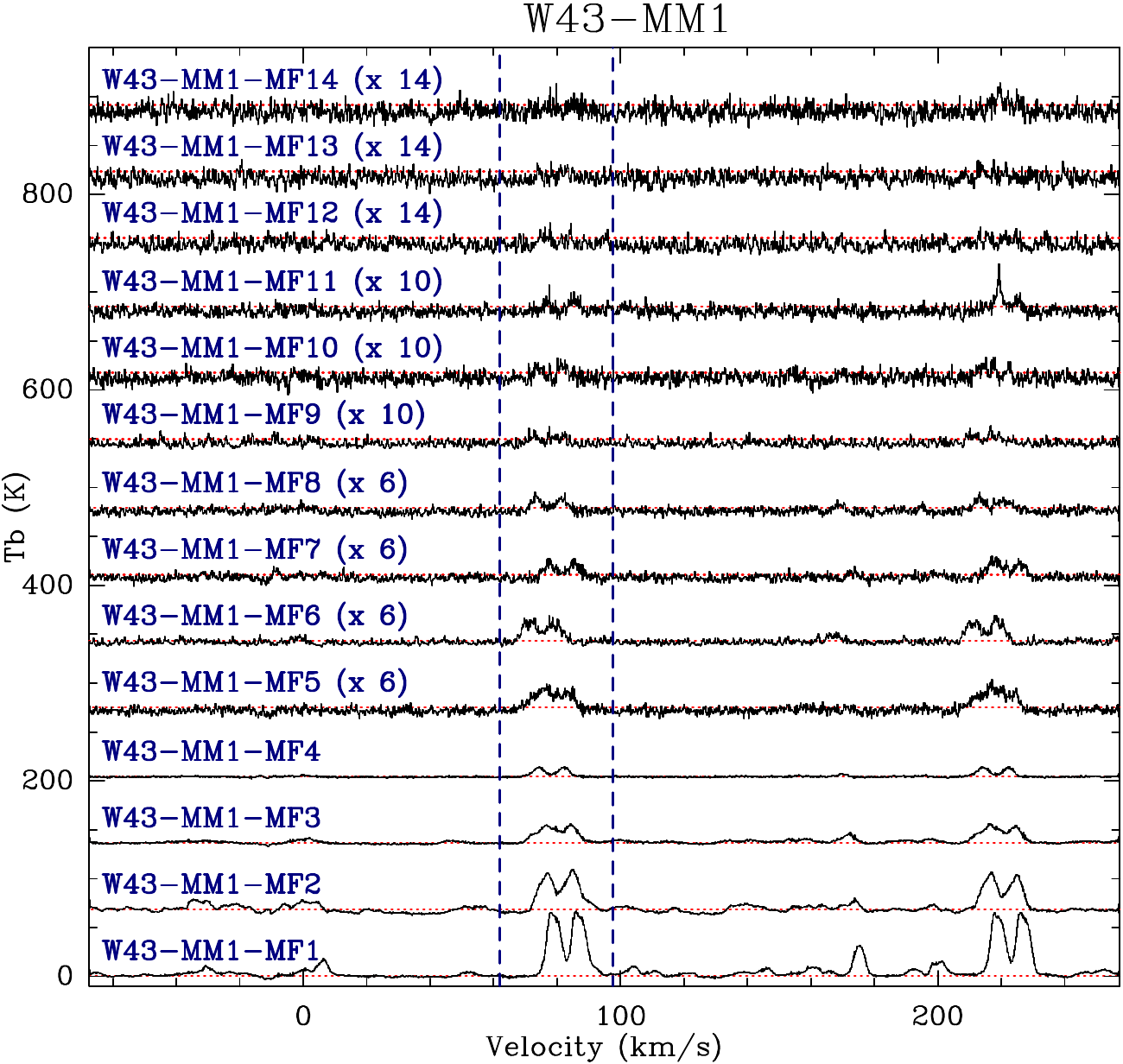}  &
       \includegraphics[width=\hsize]{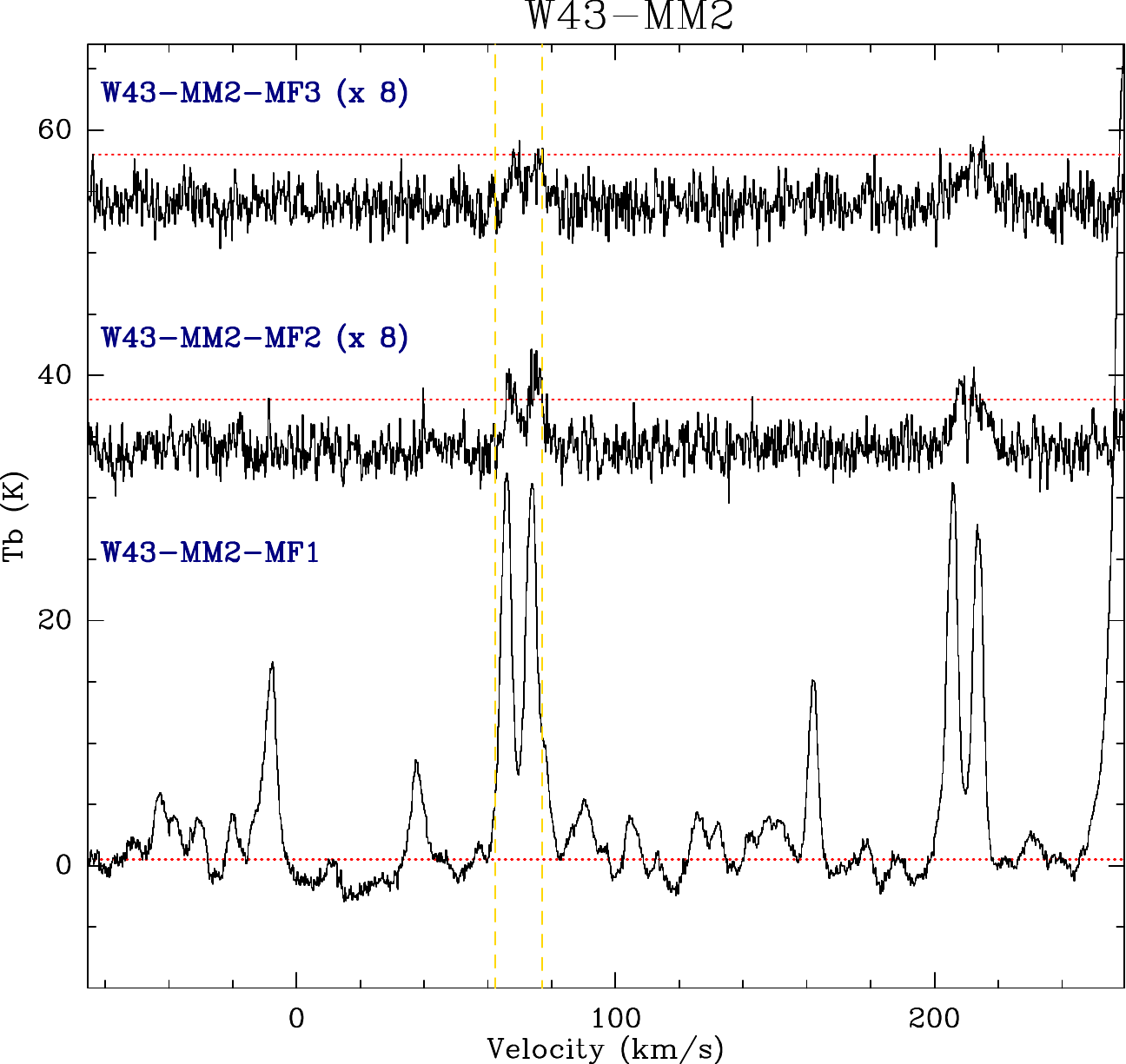}  \\
       \end{tabular}}
       \caption{\label{FIG-MF-spectra3} Same as Fig.\,\ref{FIG-MF-spectra}. The figure continues on the next page.}
\end{figure*}

\begin{figure*}[t!]
   \resizebox{\hsize}{!}
  {\begin{tabular}{cc} 
       \includegraphics[width=\hsize]{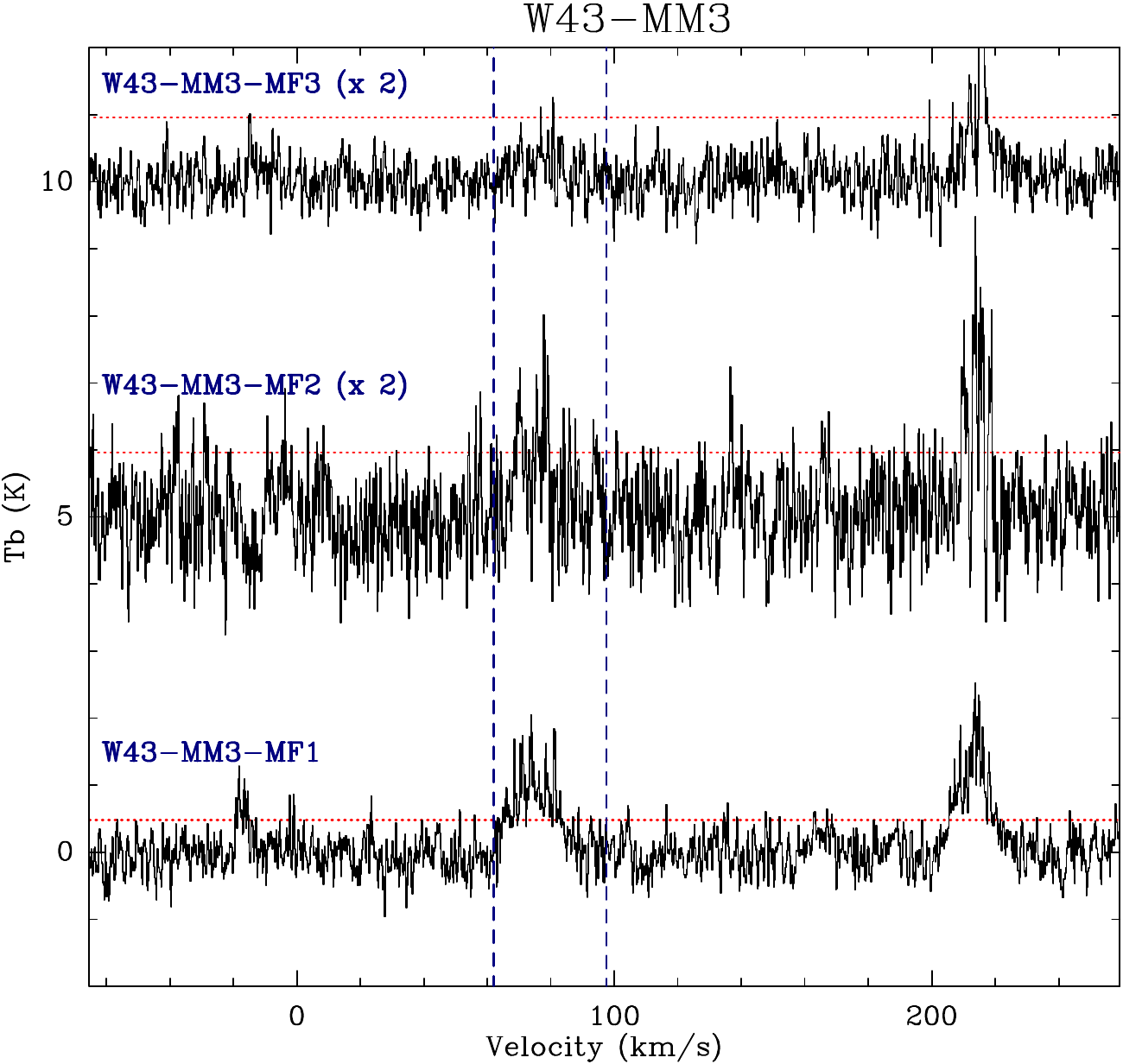}  &
       \includegraphics[width=\hsize]{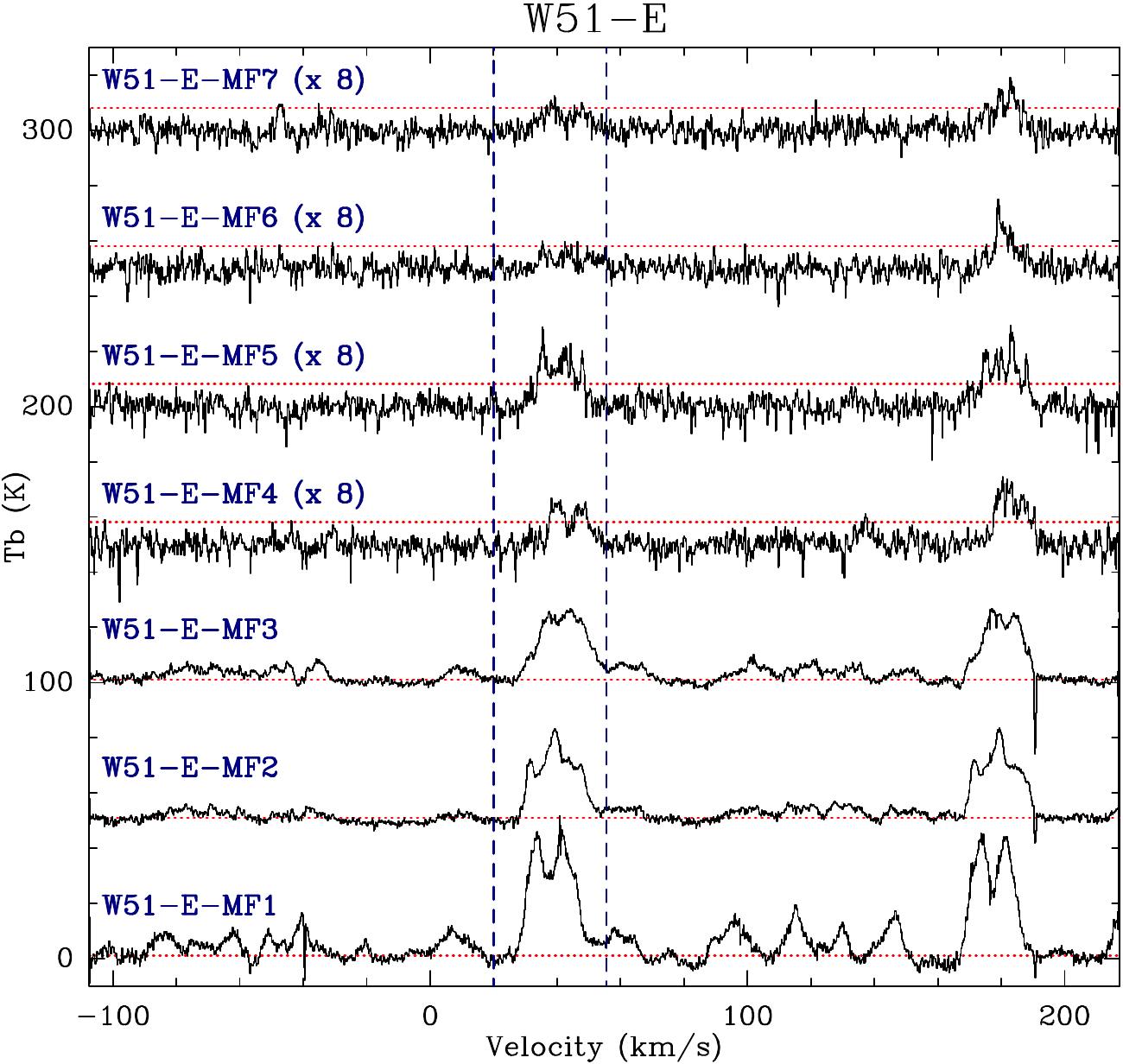}  \\
       \includegraphics[width=\hsize]{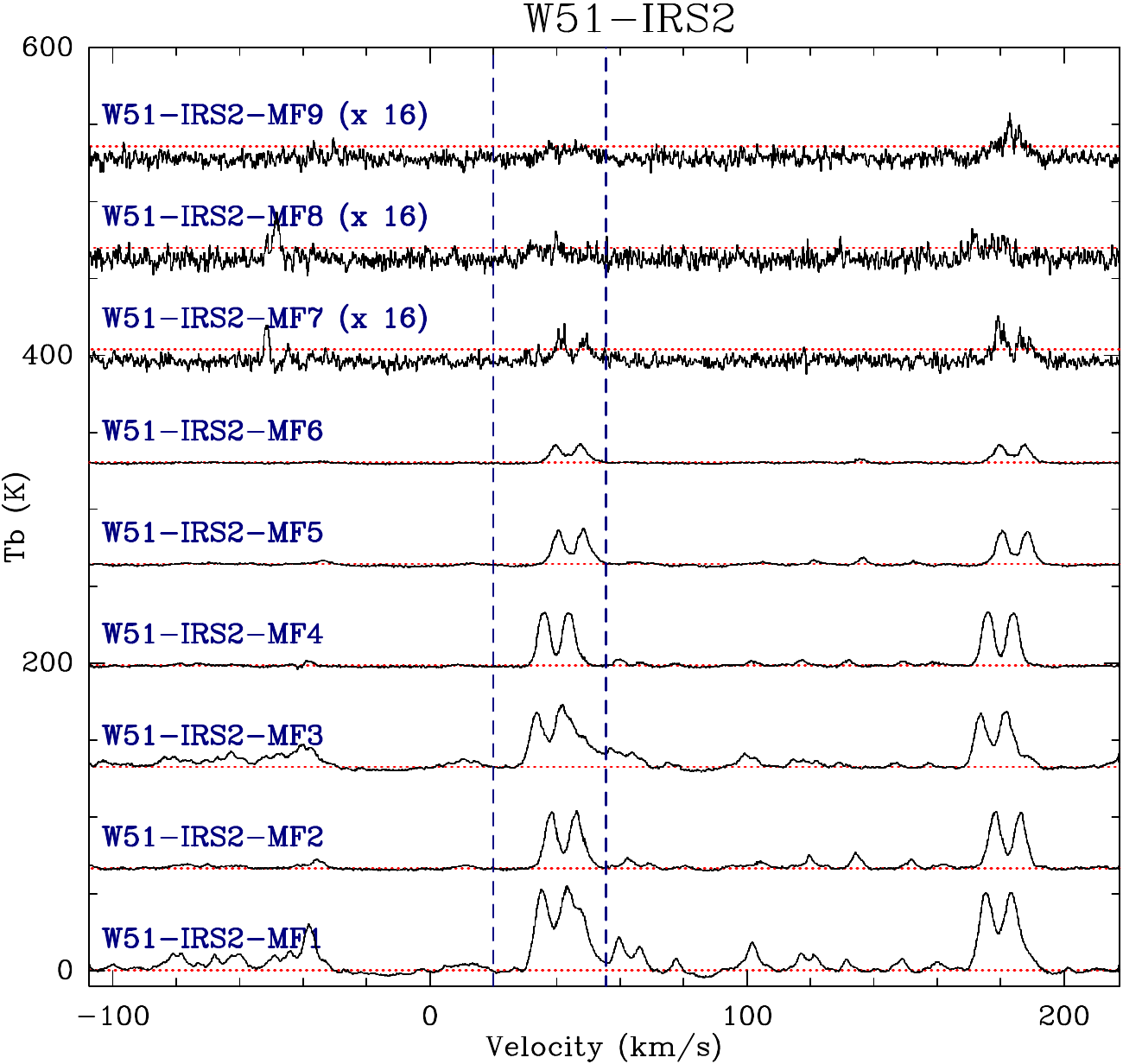}  \\
       \end{tabular}}
       \caption{\label{FIG-MF-spectra4} Same as Fig.\,\ref{FIG-MF-spectra}.} 
\end{figure*}

In Figs.\,\ref{FIG-MF-spectra}--\ref{FIG-MF-spectra4} we present the single-pixel continuum-subtracted spectra extracted from the ALMA-IMF B6-spw0 line cubes towards the position of all methyl formate sources in each of the 15 protoclusters.

%======================
% FIGURE: SPECTRA
%======================

\clearpage

%------------------------------------------------------------------
\section{Continuum emission compared with \mf\ spatial distribution}
%------------------------------------------------------------------
\label{appendix-cont-mom0-maps}

\begin{figure*}[hbt!]
   \resizebox{\hsize}{!}
   {\begin{tabular}{cc}  
       \includegraphics[width=\hsize]{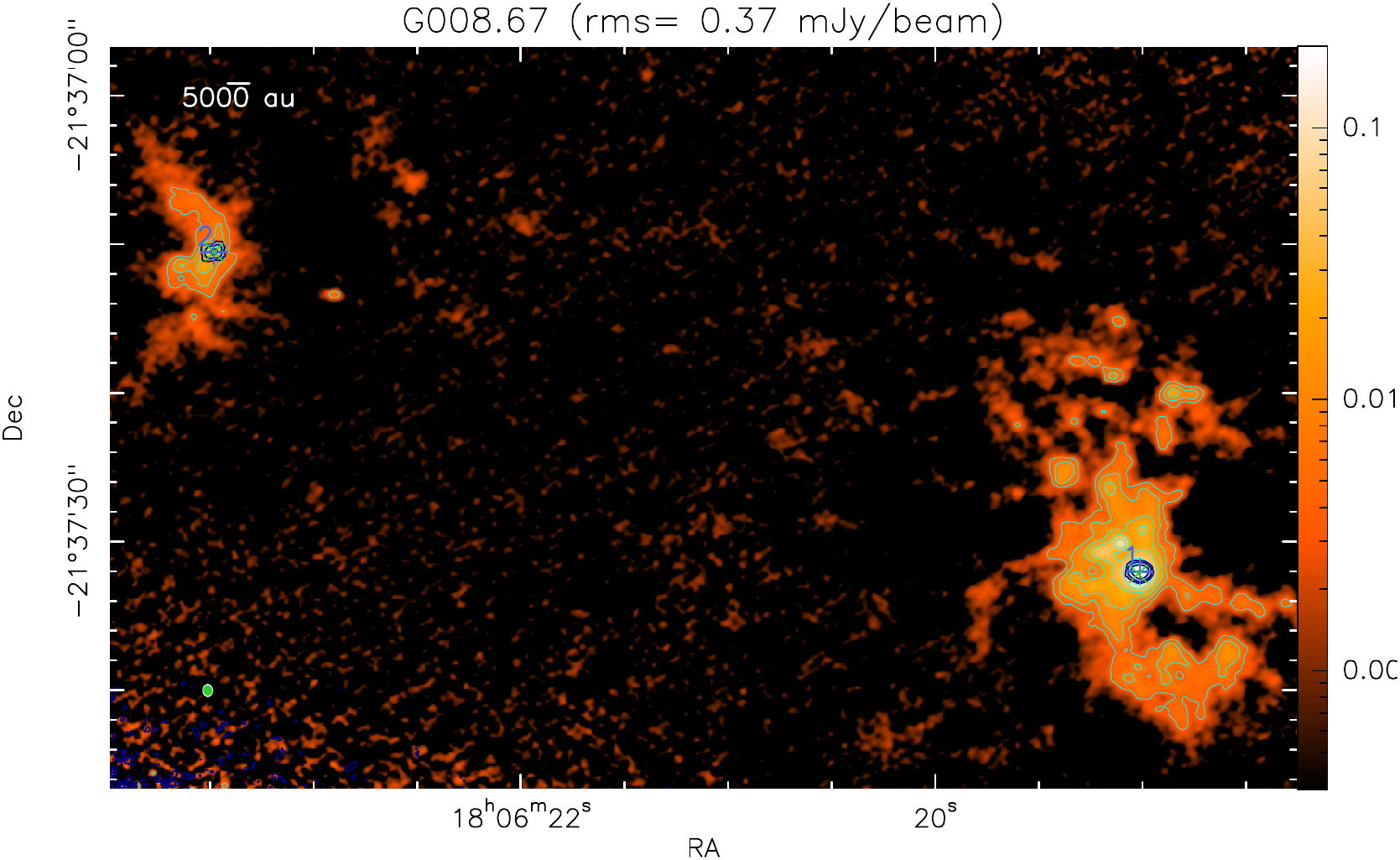}  &
       \includegraphics[width=\hsize]{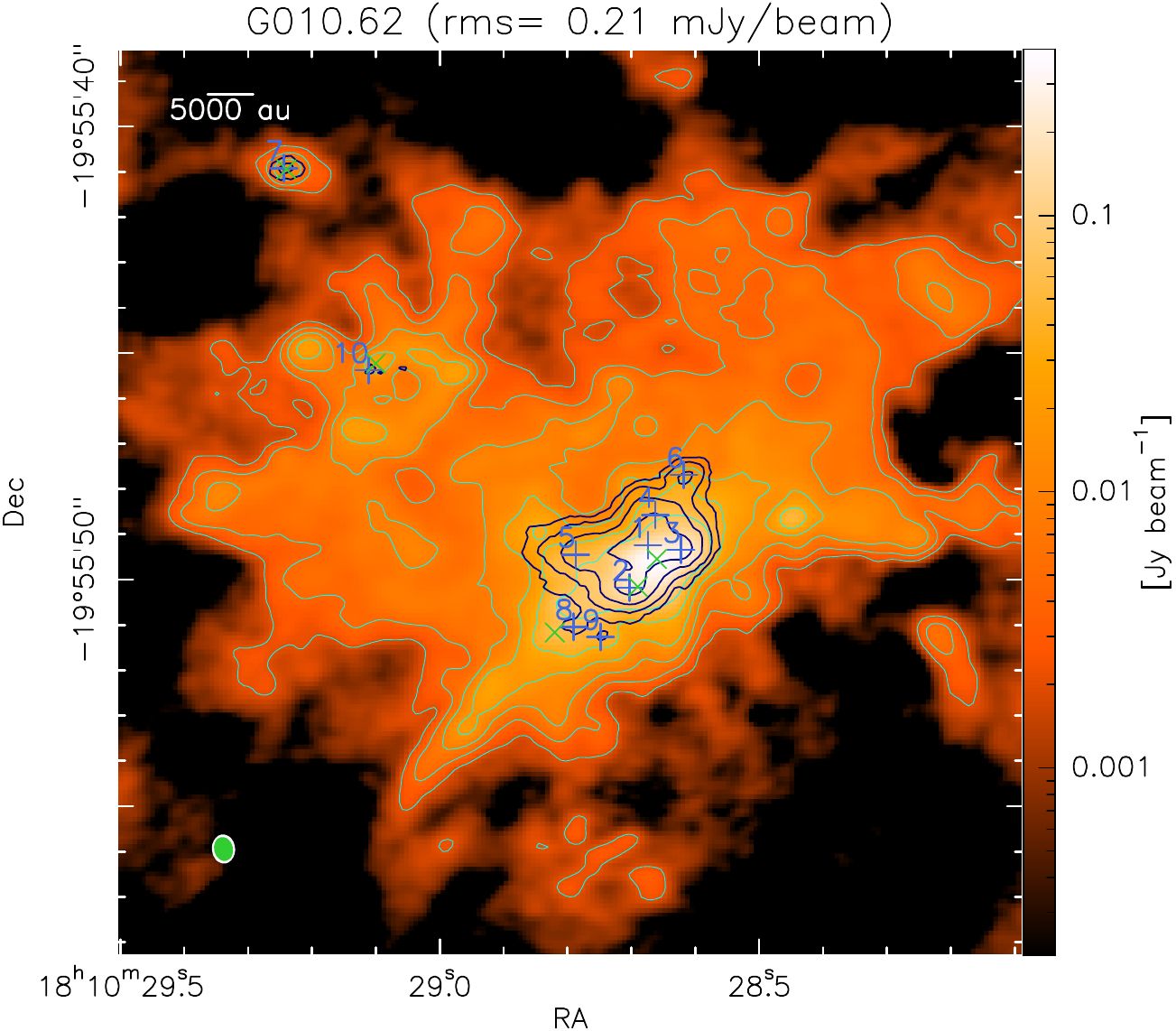}  \\
       \includegraphics[width=\hsize]{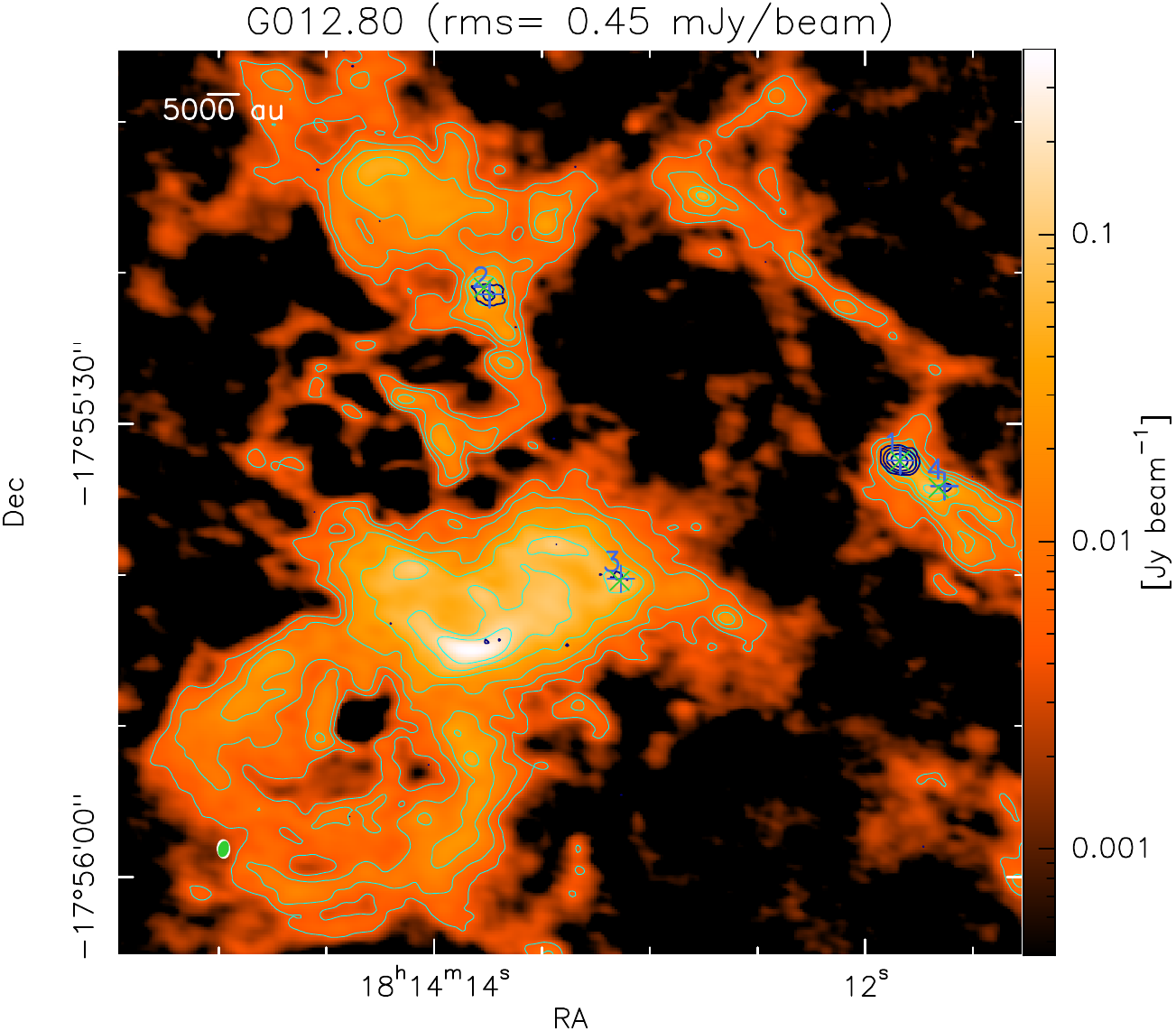}  &
       \includegraphics[width=\hsize]{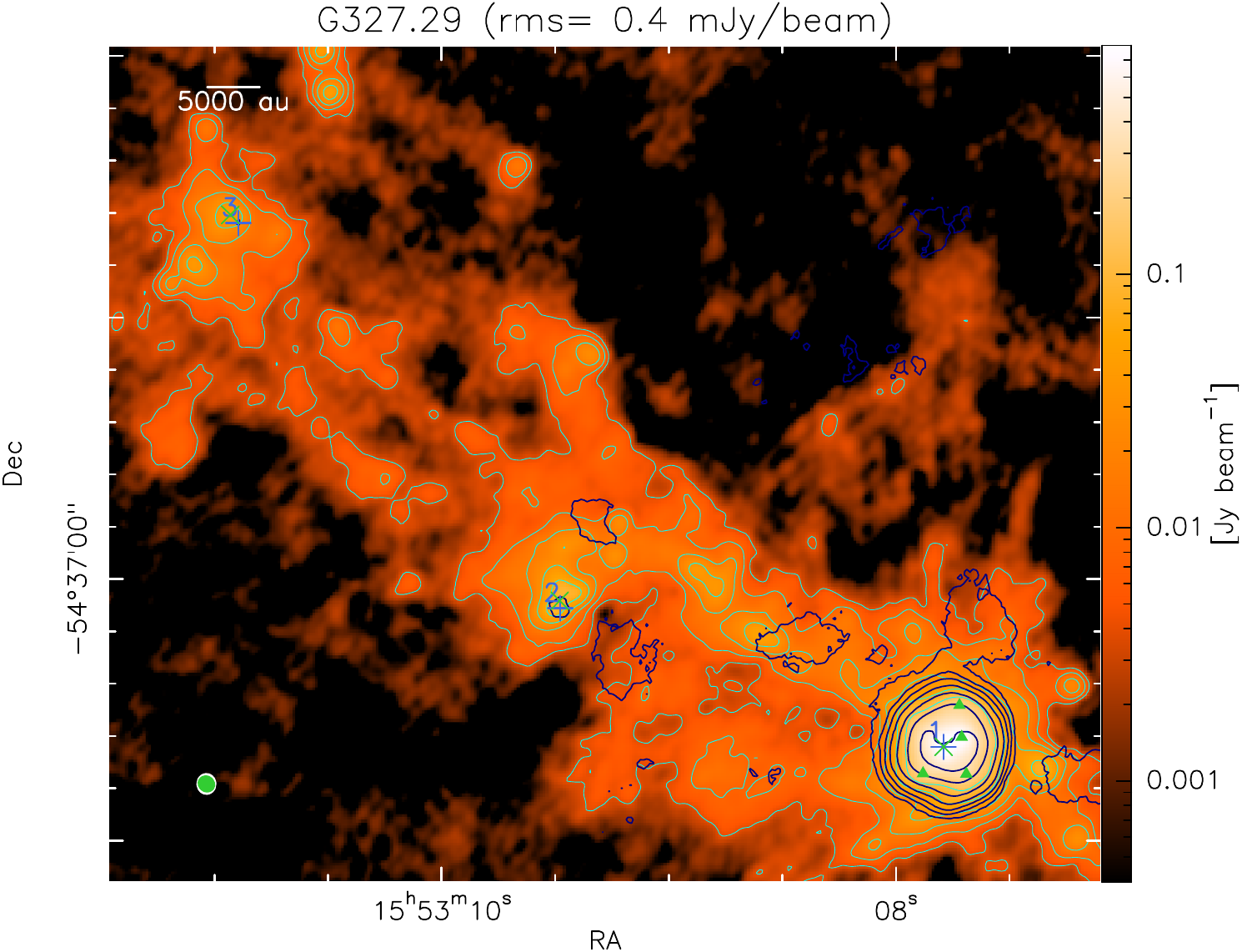}  \\
       \end{tabular}}
  \caption{\label{FIG-cont-mom0-maps} Continuum emission maps obtained at 1.3 mm towards the 15 ALMA-IMF protoclusters (background image). The contours, shown in light blue, start at 5\,$\sigma$ and double in value thereafter. The 1\,$\sigma$ rms noise level measured in each continuum map is indicated on top of each panel. The dark blue contours show the methyl formate integrated emission as in Fig.\,\ref{FIG-mom0-maps}. The blue crosses show the peak positions of the methyl formate sources extracted with \textsl{GExt2D}, while the green crosses show their associated continuum cores from the \textsl{getsf} unsmoothed catalog (\citetalias{louvet2024}). The green triangles show the continuum cores that coincide with extended methyl formate emission but that are not associated with hot core candidates, such that they have been removed from the lower panel of Fig.\,\ref{FIG-histo-mass-distribution}. The green and white ellipses represent the synthesized beam sizes of the continuum maps and the line cubes, respectively. The figure continues on the next page.}
  \end{figure*}

\begin{figure*}[hbt!]
   \resizebox{\hsize}{!}
   {\begin{tabular}{cc}
       \includegraphics[width=\hsize]{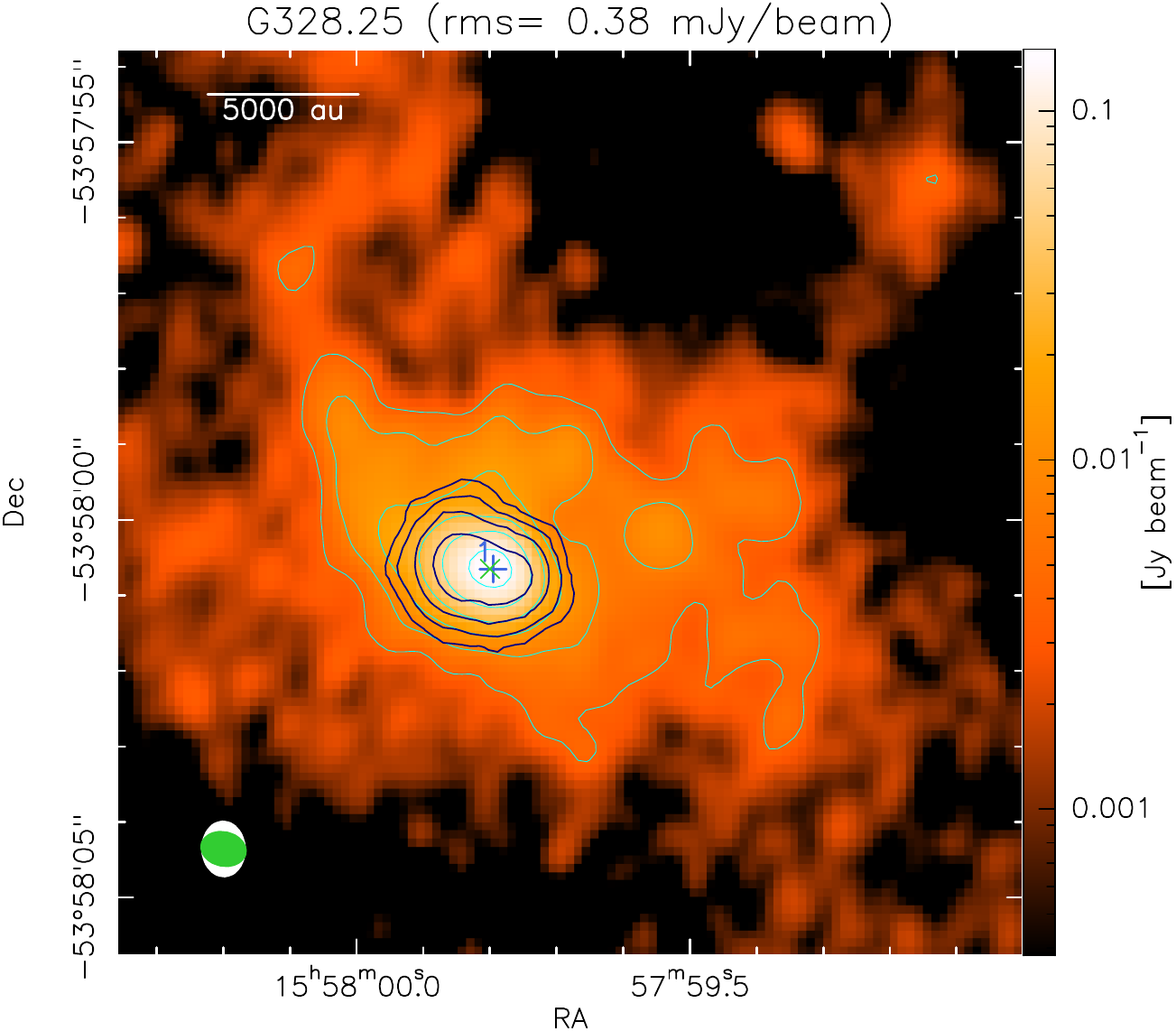}  &
       \includegraphics[width=\hsize]{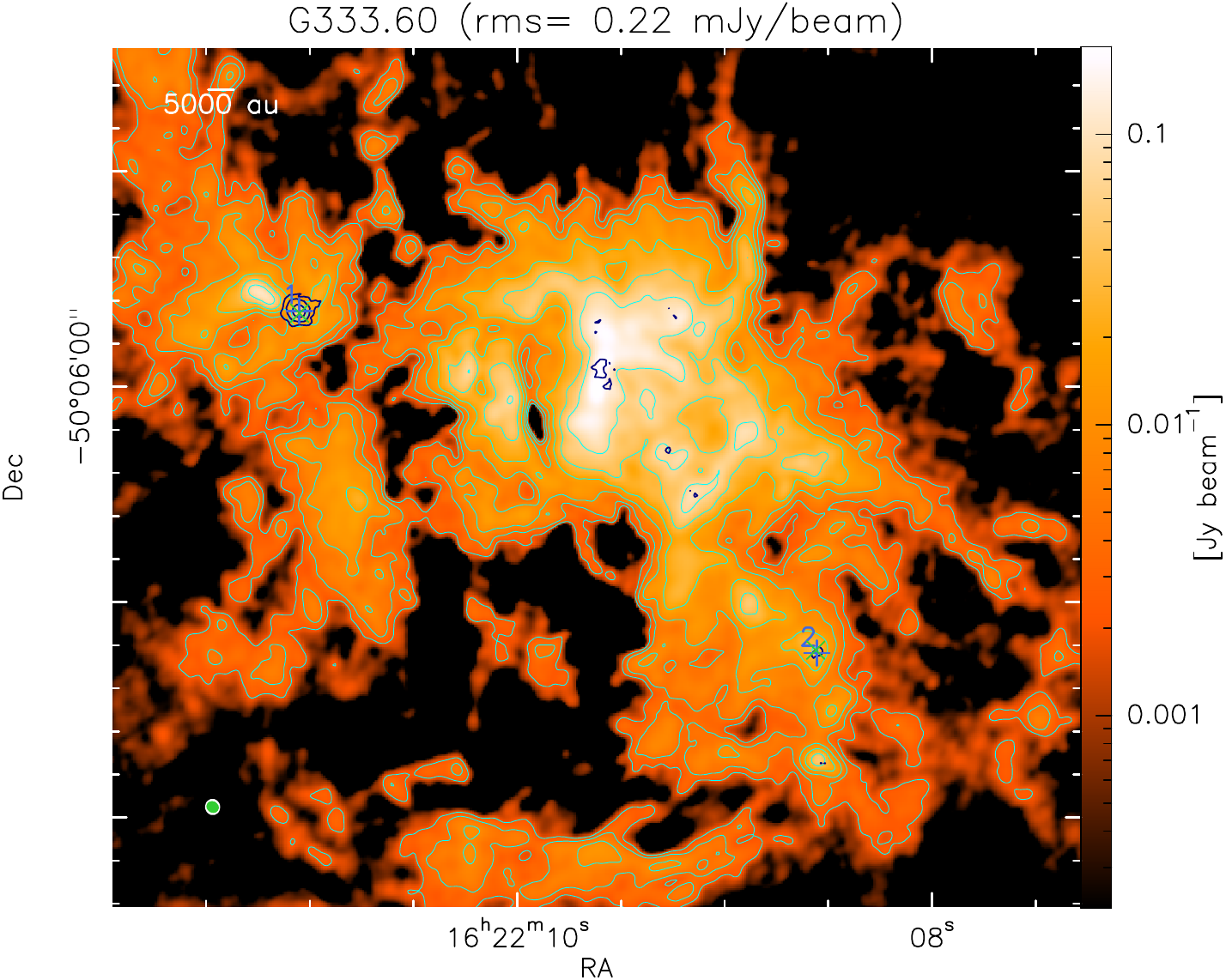}  \\
       \includegraphics[width=\hsize]{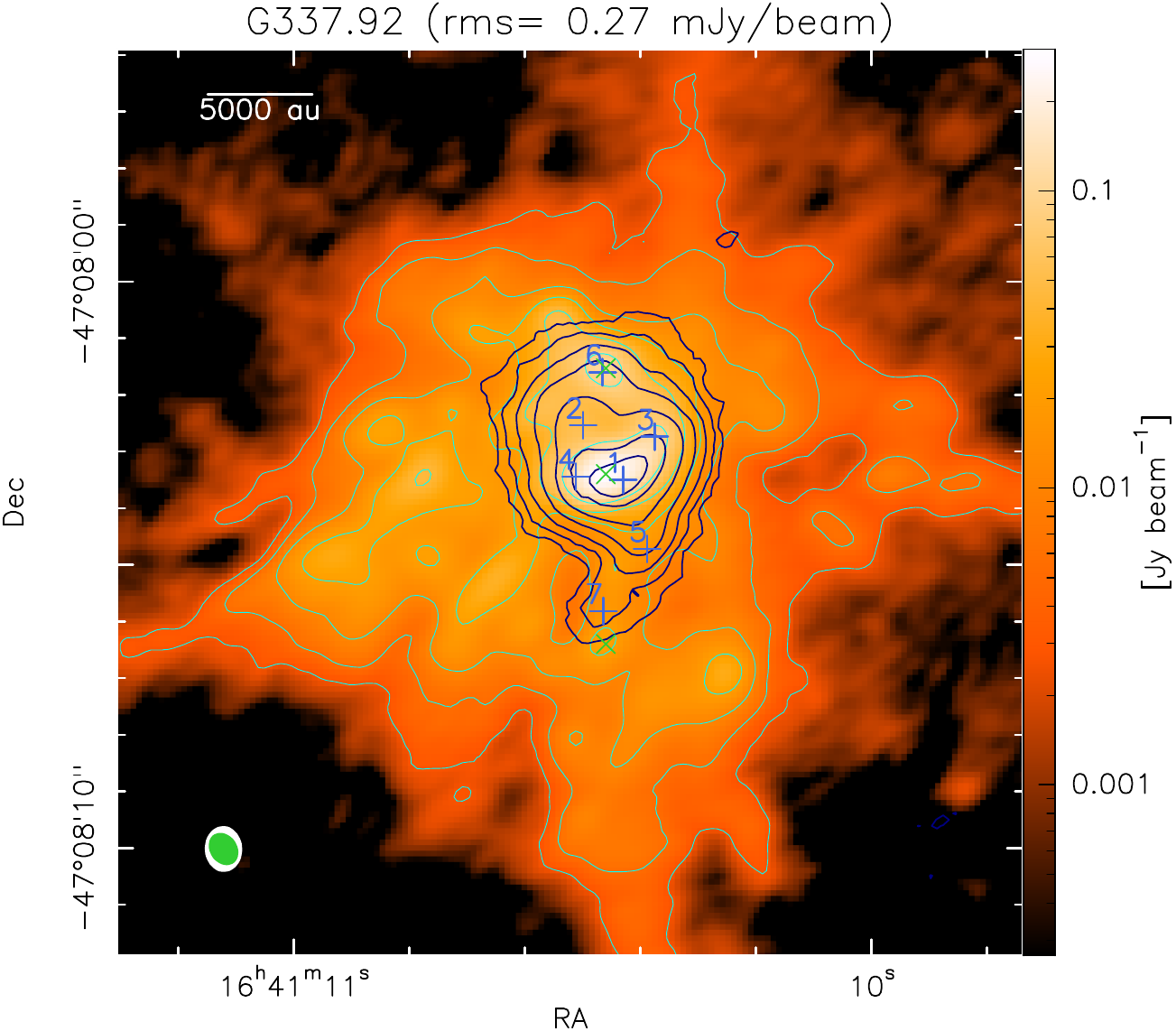} &
       \includegraphics[width=\hsize]{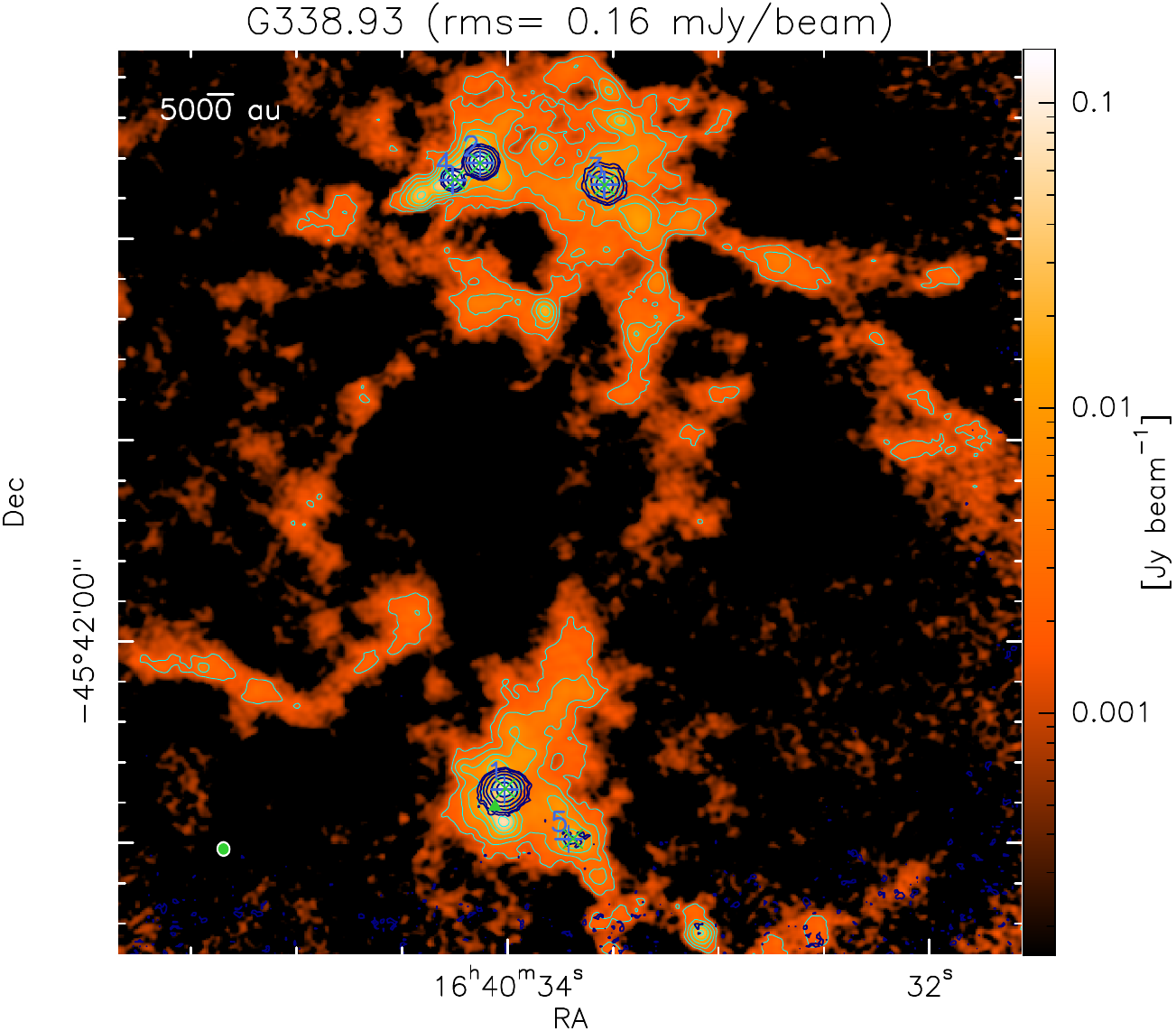}  \\
       \end{tabular}}
  \caption{\label{FIG-cont-mom0-maps2} Same as Fig.\,\ref{FIG-cont-mom0-maps}. The figure continues on the next page.} 
  \end{figure*}

\begin{figure*}[hbt!]
   \resizebox{\hsize}{!}
   {\begin{tabular}{cc}
        \includegraphics[width=\hsize]{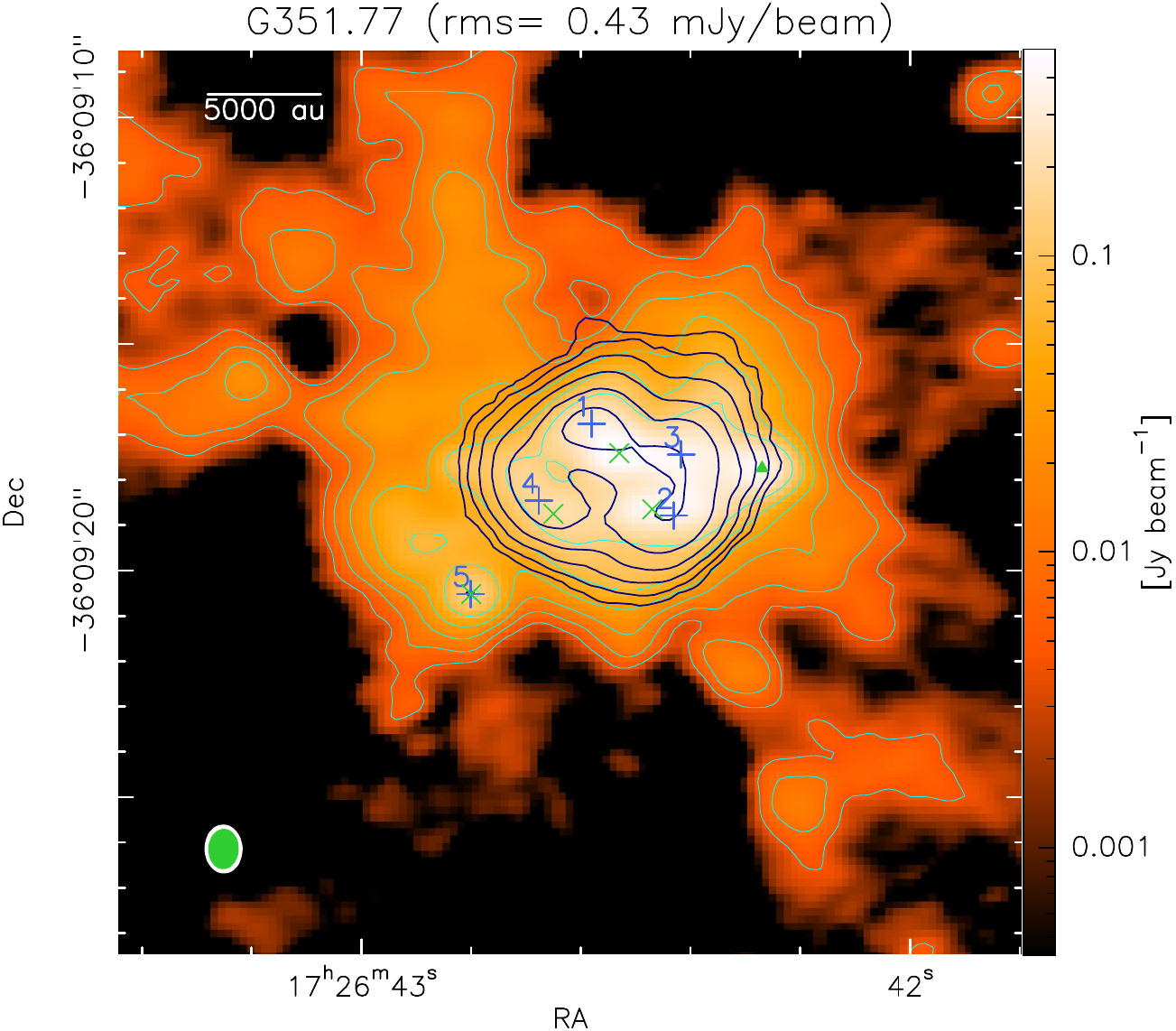}  &
        \includegraphics[width=\hsize]{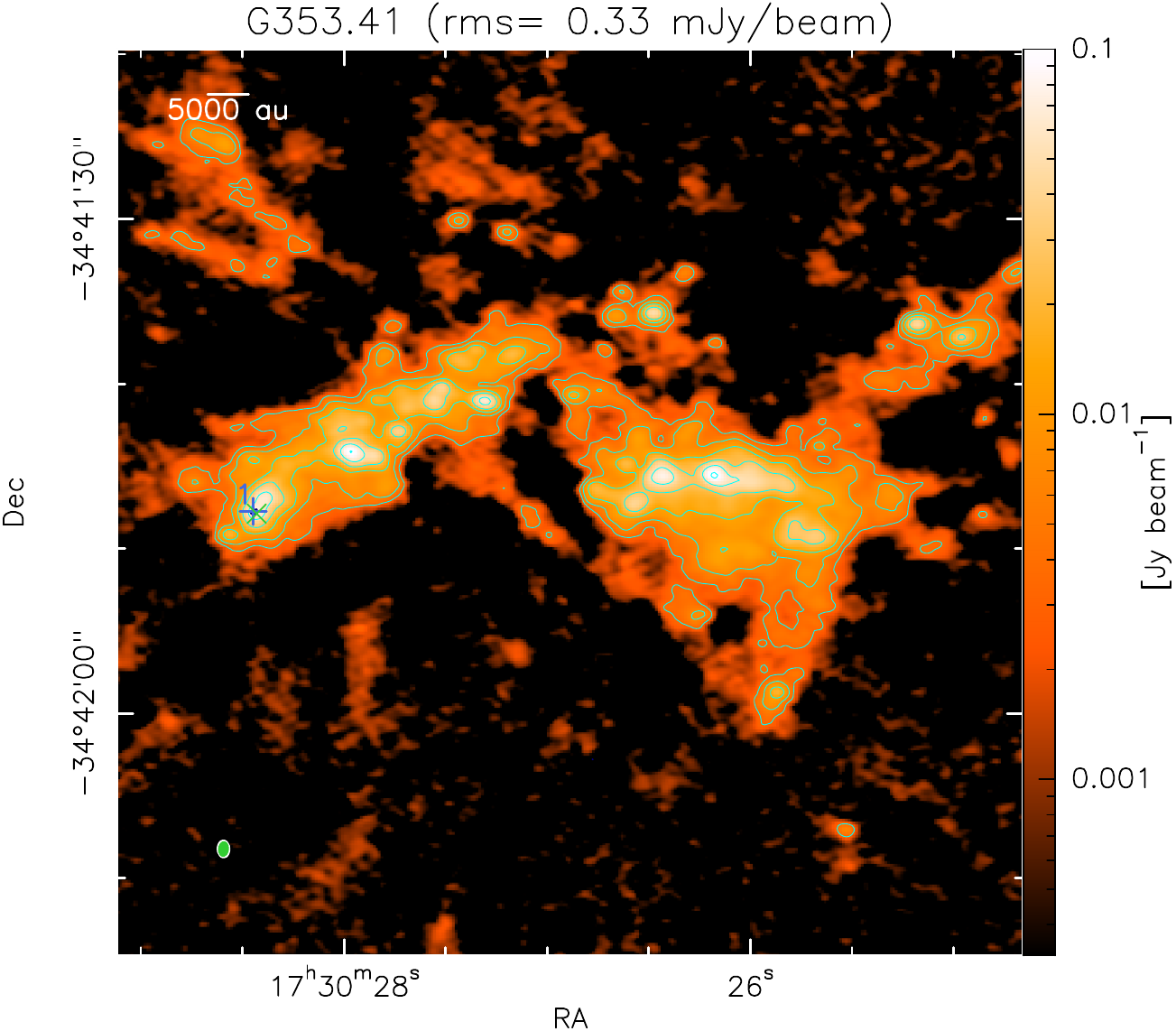}  \\
        \includegraphics[width=\hsize]{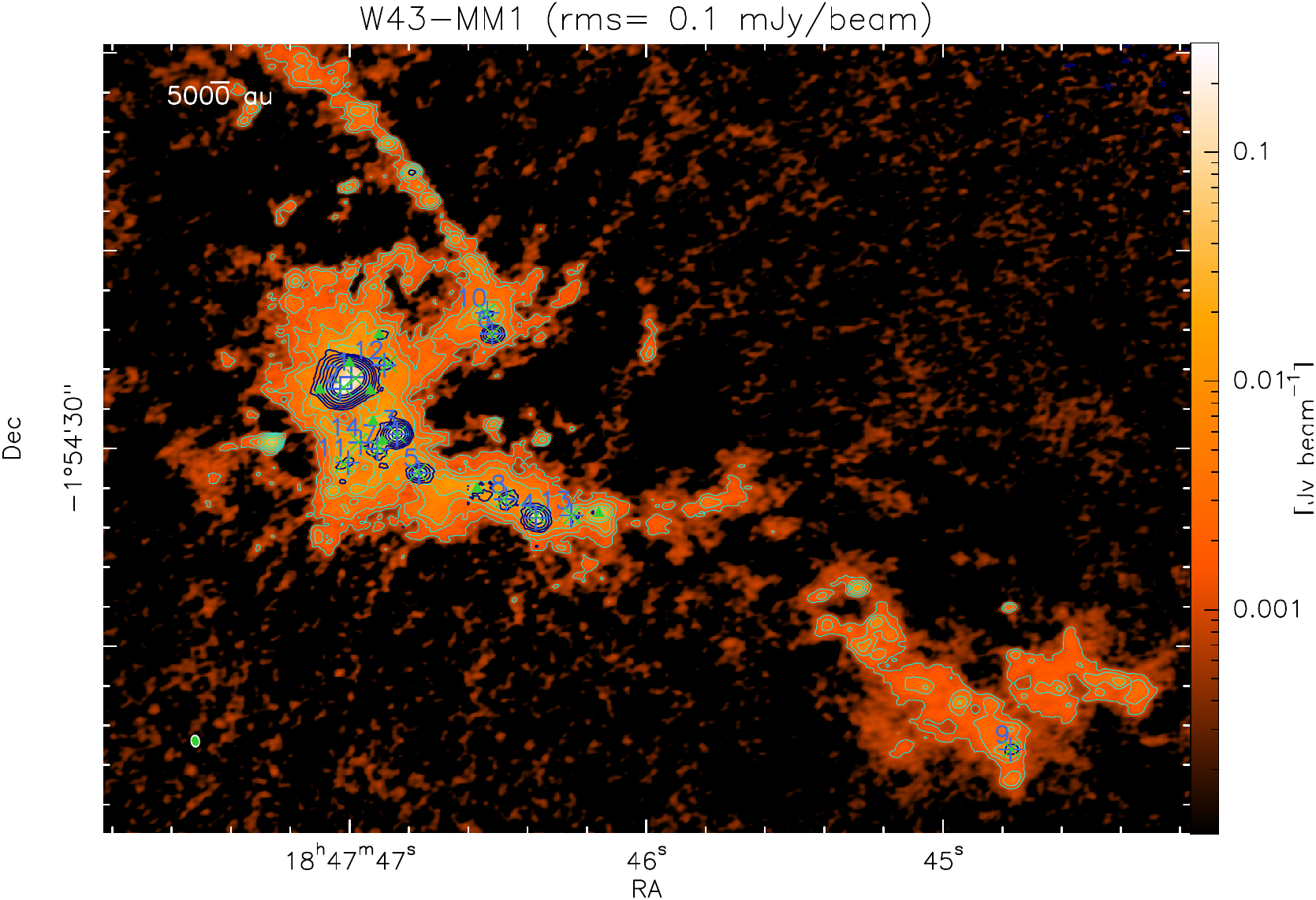}  &
        \includegraphics[width=\hsize]{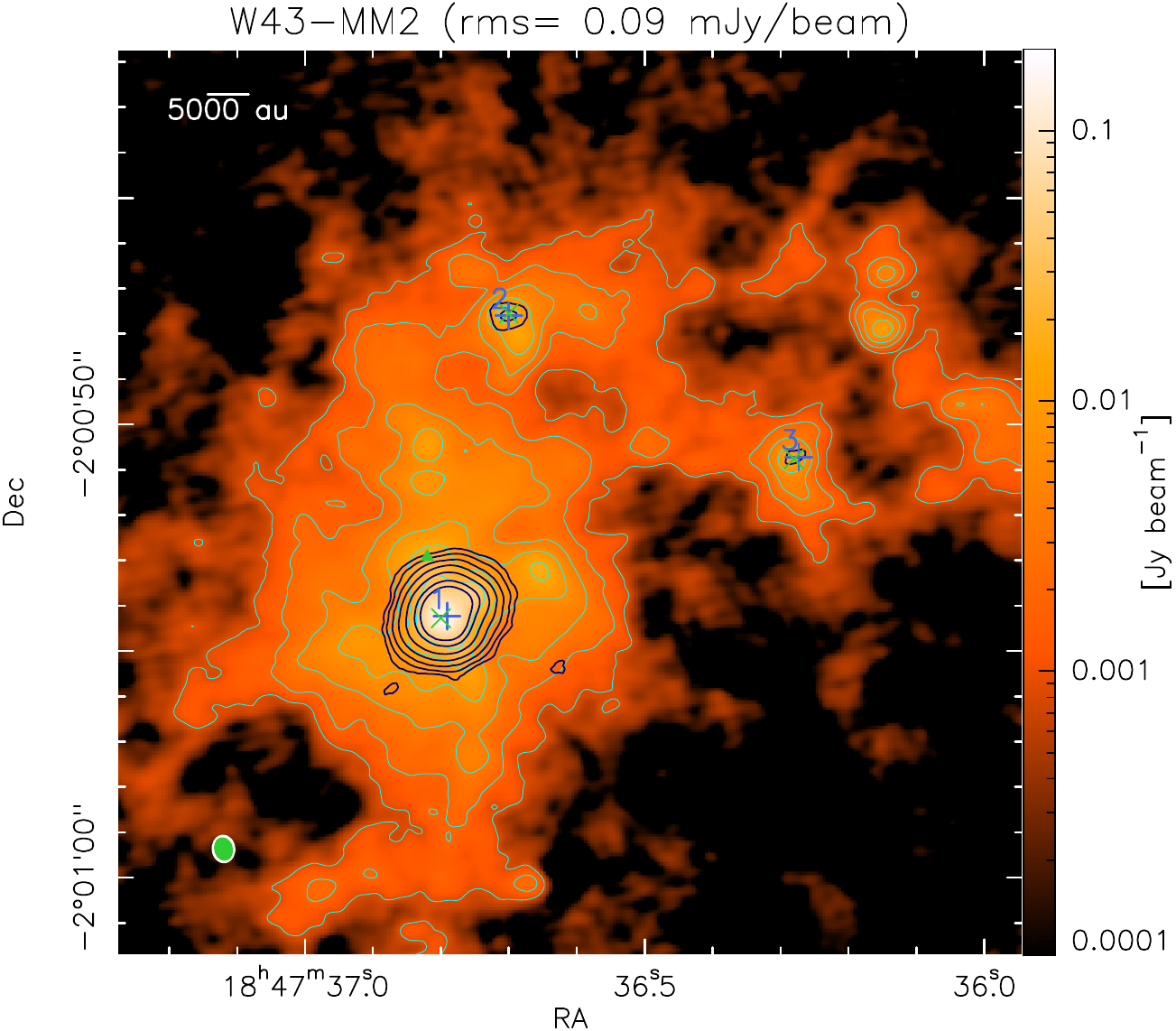}  \\
        \end{tabular}}
  \caption{\label{FIG-cont-mom0-maps3} Same as Fig.\,\ref{FIG-cont-mom0-maps}. The figure continues on the next page.}
  \end{figure*}

\begin{figure*}[hbt!]
   \resizebox{\hsize}{!}
   {\begin{tabular}{cc}
        \includegraphics[width=\hsize]{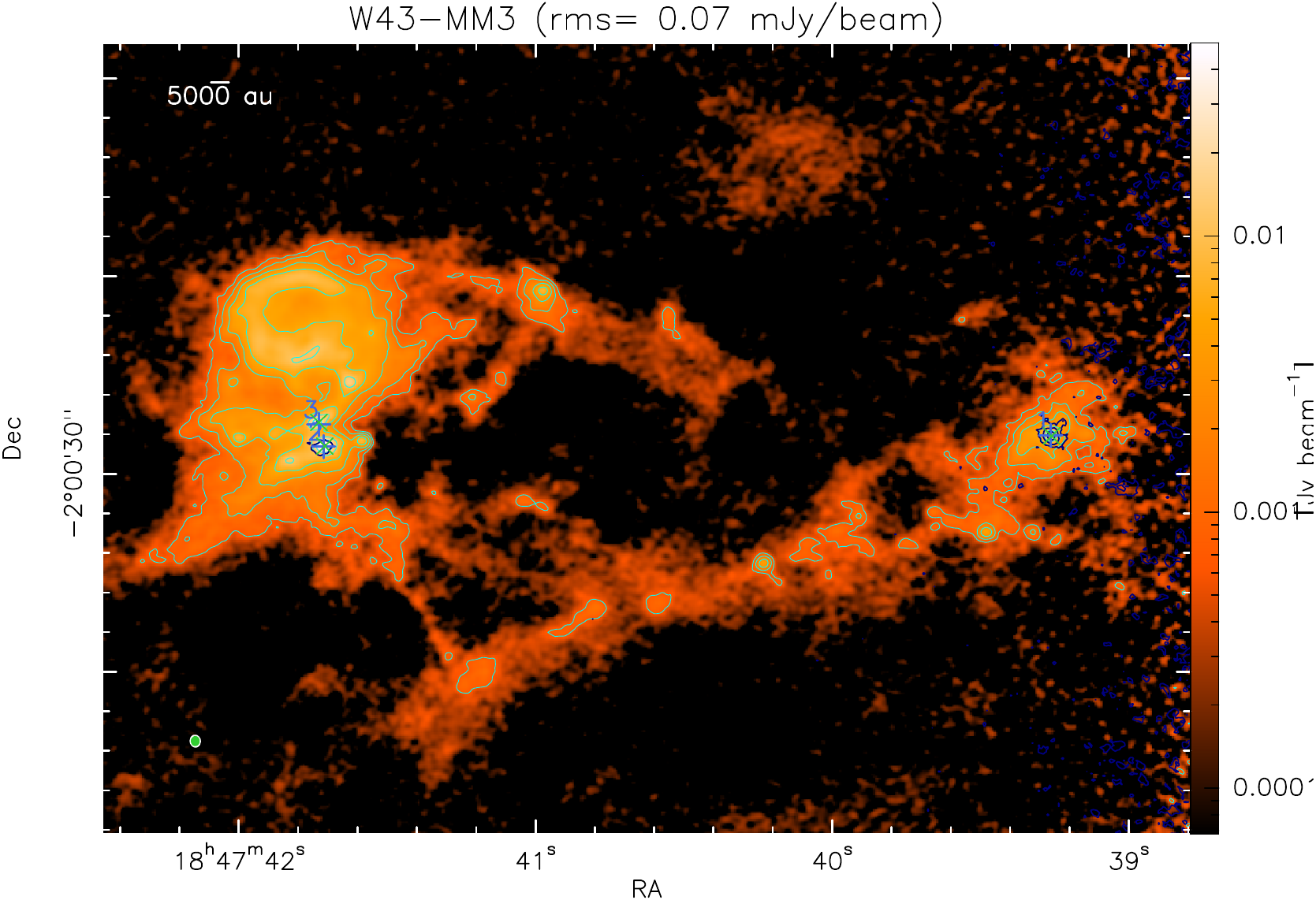}  &
        \includegraphics[width=\hsize]{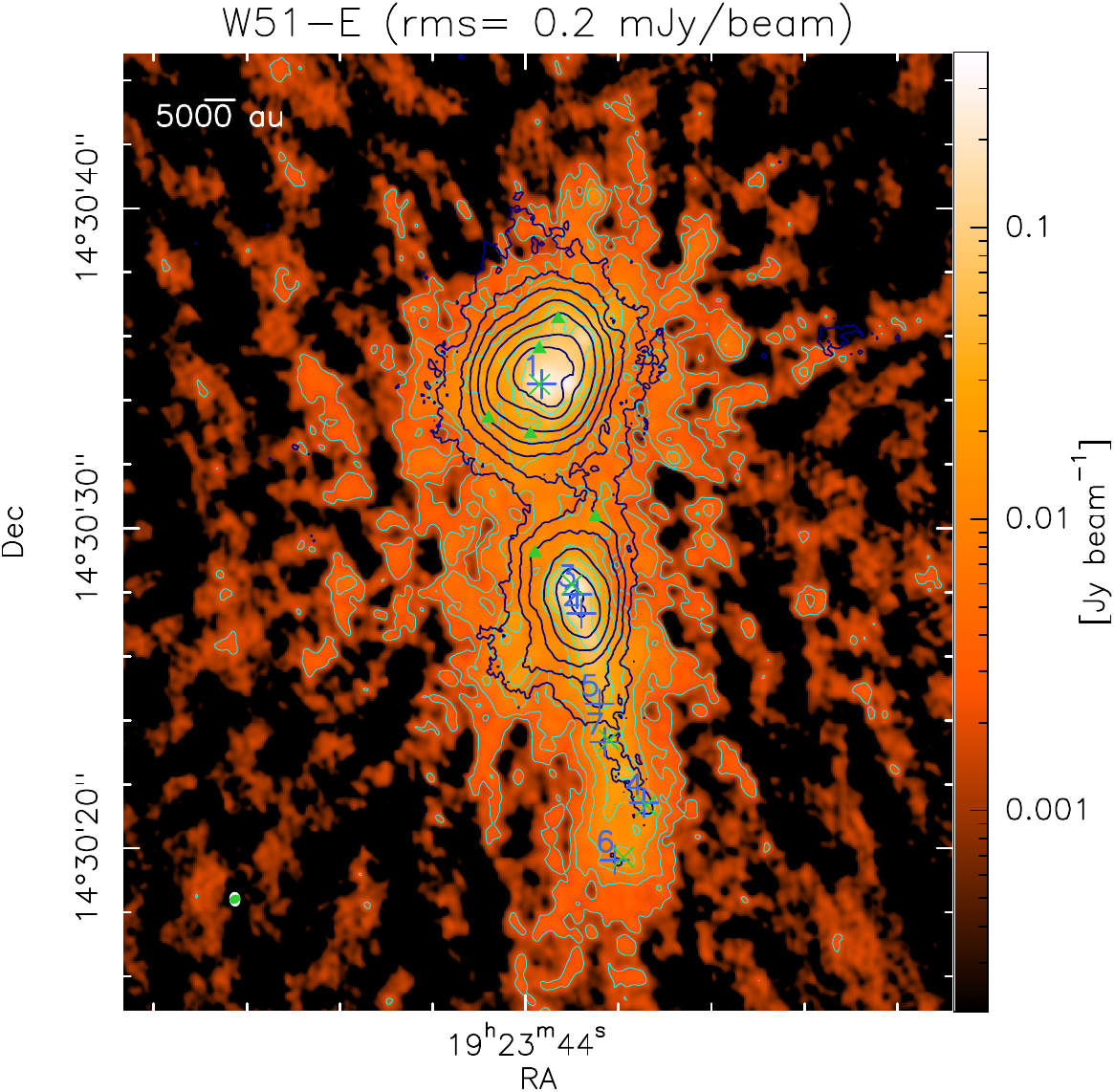}  \\
        \includegraphics[width=\hsize]{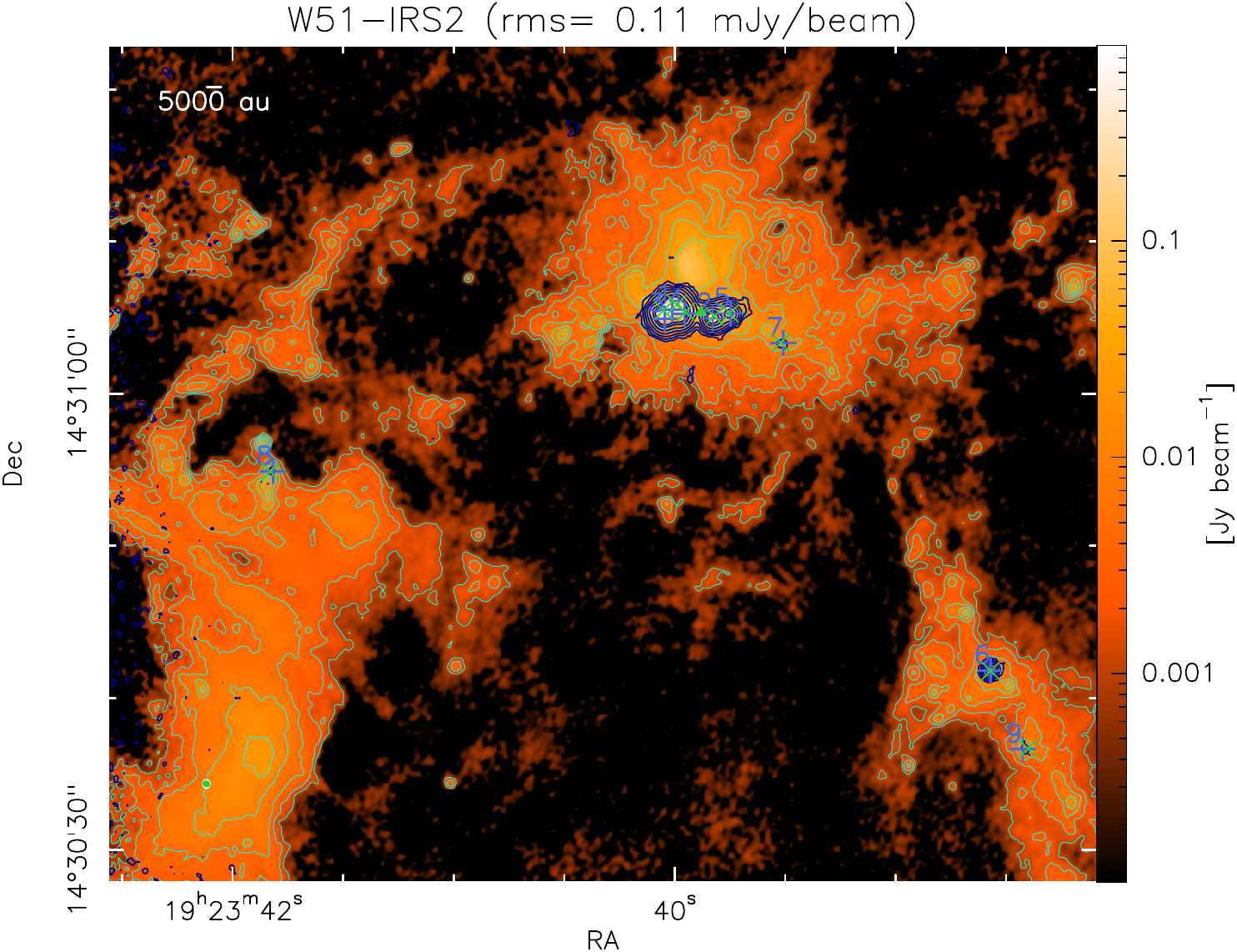}  &
        \end{tabular}}
  \caption{\label{FIG-cont-mom0-maps4} Same as Fig.\,\ref{FIG-cont-mom0-maps}.}
  \end{figure*}

In Figs.\,\ref{FIG-cont-mom0-maps}--\ref{FIG-cont-mom0-maps4} we present the 1.3\,mm ALMA-IMF continuum emission maps obtained towards the 15 protoclusters, compared to the contours of the methyl formate integrated emission. 

%======================
% FIGURE: CONT MAPS
%======================
  
\clearpage
\begin{landscape}

%------------------------------------------------------------------
\section{Catalog of continuum cores associated with methyl formate emission and their properties}
%------------------------------------------------------------------
\label{appendix-cont-core-cat}

\begin{table}[h!]
  \caption{\label{TAB-cont-cat-getsf} Properties of the continuum cores associated to the methyl formate sources towards the 15 ALMA-IMF protoclusters.} 
  \setlength{\tabcolsep}{0.6mm}
\begin{tabular}{lrrr|rrcrr|r|rrcrr}
    \hline
    \hline
    \multicolumn{4}{c|}{ } & \multicolumn{5}{c|}{1.3 mm} & \multicolumn{1}{c|}{3 mm} & \multicolumn{4}{c}{ }  \\
   \hline
    \multicolumn{1}{c}{ID$_{\mathrm{MF}}^{(a)}$} & \multicolumn{1}{c}{RA$^{(b)}$} & \multicolumn{1}{c}{Dec$^{(b)}$} & \multicolumn{1}{c|}{d$^{(c)}$}  & \multicolumn{1}{c}{$S^{\mathrm{peak} (d)}_{\mathrm{1.3 mm}}$} & \multicolumn{1}{c}{$S^{\mathrm{int} (d)}_{\mathrm{1.3 mm}}$} & \multicolumn{1}{c}{$\theta_{\mathrm{maj}}^{\mathrm{dec}} \times \theta_{\mathrm{min}}^{\mathrm{dec} (e)}$} &  \multicolumn{1}{c}{PA$^{\mathrm{dec} (e)}$} & \multicolumn{1}{c|}{FWHM$_{\mathrm{cont}}^{\mathrm{dec} (f)}$} &  \multicolumn{1}{c|}{$S^{\mathrm{int} (g)}_{\mathrm{3 mm}}$} & \multicolumn{1}{c}{$\tau^{(h)}$} & \multicolumn{1}{c}{Mass range$^{(i)}$} & \multicolumn{1}{c}{M$^{(j)}$} &  \multicolumn{1}{c}{$\alpha^{(k)}$}  & \multicolumn{1}{c}{frac$_{\mathrm{ff}}^{(l)}$} \\
    
     \multicolumn{1}{c}{} & \multicolumn{2}{c}{[J2000]} & \multicolumn{1}{c|}{[$\arcsec$]} & \multicolumn{1}{c}{[mJy beam$^{-1}$]}  & \multicolumn{1}{c}{[mJy]}  & \multicolumn{1}{c}{[$\arcsec$ $\times$ $\arcsec$]}  & \multicolumn{1}{c}{[deg]} & \multicolumn{1}{c|}{[au]}   & \multicolumn{1}{c|}{[mJy]}  &  & \multicolumn{1}{c}{[$\Msun$]} & \multicolumn{1}{c}{[$\Msun$]} &   & \multicolumn{1}{c}{[\%]}     \\
         \hline
    \multicolumn{15}{c}{G008.67} \\
     \hline
    G008--MF1 & 18:06:19.02 & -21:37:32.2 & 0.19 & 200.4$\pm$7.6 & 736.7$\pm$12.2 & 1.16$\times$1.0 & 0.4 & 3675.4 & 557.0$\pm$1.4 & 0.04 & 21.7--5.9 & 9.4 & 0.3 & 65 \\
    G008--MF2 & 18:06:23.48 & -21:37:10.5 & 0.02 & 64.4$\pm$1.5  & 84.5$\pm$1.4 & 0.37$\times$0.31 & -38.1 & 1166.2 & 9.1$\pm$0.2 & 0.04 & 6.7--2.0 & 3.1 & 2.7 & 0 \\
 \hline
    \multicolumn{15}{c}{G010.62} \\
     \hline  
G010--MF1  & 18:10:28.66  & -19:55:49.6  & 0.36 & 316.6$\pm$13.6 & 1537.0$\pm$23.8 & 1.18$\times$0.77 & -58.7 & 4727.2 & 2448.8$\pm$14.2 & \_ & \_  & \_ & -0.5 & 100 \\
G010--MF2  & 18:10:28.69  & -19:55:50.2   & 0.17 & 68.2$\pm$19.6  & 129.5$\pm$16.0 & 0.55$\times$0.52 & 67.1 & 2658.1 & 185.8$\pm$7.1 & \_ & \_ & \_ & -0.4 & 100 \\
G010--MF3$^{*}$  & \_           & \_            & \_   & 281.3$\pm$40.1 & 281.3$\pm$40.1 & \_  & \_ & 2306.7 & 232.6$\pm$47.2 & 0.09 & 15.4--3.9 & 6.2 & 0.2 & 72 \\
G010--MF4$^{*}$  & \_           & \_            & \_   & 50.7$\pm$19.2  & 50.7$\pm$19.2  & \_ & \_ & 2306.7 & 43.0$\pm$17.0 & 0.01 & 2.3--0.6 & 1.0 & 0.2 & 74 \\
G010--MF5$^{*}$  & \_           & \_            & \_   & 50.2$\pm$9.2   & 50.2$\pm$9.2   & \_ & \_ & 2306.7 & 44.4$\pm$11.7 & 0.01 & 2.0--0.5 & 0.8 & 0.1 & 77 \\
G010--MF6$^{*}$  & \_           & \_            & \_   & 24.3$\pm$11.4  & 24.3$\pm$11.4  & \_ & \_ & 2306.7 & 16.2$\pm$10.2 & 0.01 & 1.7--0.5 & 0.8 & 0.4 & 57 \\
G010--MF7  & 18:10:29.24  & -19:55:40.95  & 0.06 & 13.9$\pm$0.6   & 20.4$\pm$0.8   & \_ & \_ & 1153.3 & 3.0$\pm$0.2 & 0.02 & 3.3--1.0 & 1.5 & 2.3 & 0 \\
G010--MF8  & 18:10:28.82  & -19:55:51.17  & 0.42 & 22.9$\pm$7.0   & 45.0$\pm$6.3   & 0.6$\times$0.44 & 86.6 & 2569.0 & 7.3$\pm$2.7 & 0.02 & 6.8--2.0 & 3.1 & 2.2 & 9 \\
G010--MF9$^{*}$  & \_           & \_            & \_   & 38.3$\pm$10.5  & 38.3$\pm$10.5  & \_ & \_ & 2306.7 & 16.1$\pm$7.7 & 0.03 & 4.3--1.2 & 1.9 & 1.0 & 34 \\
G010--MF10 & 18:10:29.10  & -19:55:45.23  & 0.21 & 18.3$\pm$3.2   & 19.6$\pm$2.5   & \_ & \_ & 1153.3 & 2.4$\pm$0.4 & 0.02 & 3.2--0.9 & 1.5 & 2.5 & 0 \\
\hline
    \multicolumn{15}{c}{G012.80} \\
     \hline  
G012--MF1 & 18:14:11.84 & -17:55:32.5 & 0.05 & 195.9$\pm$5.7 & 310.3$\pm$6.0 & 0.78$\times$0.46 & 40.4 & 1442.3 & 16.0$\pm$0.3 & 0.07 & 12.7--3.7 & 5.8 & 3.6 & 0 \\
G012--MF2 & 18:14:13.77 & -17:55:21.0 & 0.59 & 107.6$\pm$4.9 & 229.9$\pm$5.7 & 1.07$\times$0.72 & 22.2 & 2114.4 & 8.3$\pm$2.2 & 0.04 & 9.1--2.7 & 4.2 & 4.0 & 0 \\
G012--MF3 & 18:14:13.14 & -17:55:40.4 & 0.16 & 91.0$\pm$16.0 & 166.5$\pm$15.4 & 0.87$\times$0.77 & 37.7 & 1980.0 & 10.3$\pm$4.7 & 0.03 & 6.5--1.9 & 3.0 & 3.3 & 0 \\
G012--MF4 & 18:14:11.66 & -17:55:34.2 & 0.40 & 61.0$\pm$6.2 & 159.3$\pm$7.0 & 1.72$\times$0.94 & -86.5 & 3069.6 & 11.5$\pm$0.6 & 0.02 & 6.2--1.9 & 2.9 & 3.2 & 0 \\
\hline
    \multicolumn{15}{c}{G327.29} \\
     \hline  
G327--MF1 & 15:53:07.79 & -54:37:06.4 & 0.01 & 669.3$\pm$4.8 & 3160.0$\pm$12.4 & 1.43$\times$0.84 & 54.6 & 2755.0 & 171.5$\pm$1.4 & 0.13 & 33.1--15.5 & 21.1 & 3.5 & 0 \\
G327--MF2 & 15:53:09.48 & -54:37:00.8 & 0.28 & 79.3$\pm$4.0 & 311.2$\pm$6.0 & 1.46$\times$1.00 & -30.2 & 3040.0 & 33.1$\pm$0.4 & 0.05 & 13.5--4.0 & 6.2 & 2.7 & 0 \\
G327--MF3 & 15:53:10.93 & -54:36:46.1 & 0.44 & 37.6$\pm$2.8 & 106.3$\pm$3.6 & 0.94$\times$0.89 & -89.0 & 2307.5 & 10.4$\pm$0.3 & 0.02 & 4.5--1.3 & 2.1 & 2.8 & 0 \\
\hline
    \multicolumn{15}{c}{G328.25} \\
     \hline  
G328--MF1 & 15:57:59.80 & -53:58:00.7 & 0.04 & 147.7$\pm$1.6 & 406.1$\pm$2.6 & 0.83$\times$0.44 & 71.3 & 1522.5 & 27.7$\pm$0.2 & 0.14 & 19.8--5.5 & 8.6 & 3.3 & 0 \\
\hline
    \multicolumn{15}{c}{G333.60} \\
     \hline  
G333--MF1 & 16:22:11.05 & -50:05:56.6 & 0.10 & 56.6$\pm$3.8 & 78.1$\pm$4.0 & \_ & \_ & 1163.4 & 8.6$\pm$0.1 & 0.04 & 9.1--2.7 & 4.2 & 2.6 & 4 \\
G333--MF2 & 16:22:08.56 & -50:06:12.2 & 0.14 & 25.0$\pm$3.3 & 87.6$\pm$4.2 & 1.07$\times$0.83 & 36.6 & 3990.0 & 13.1$\pm$2.1 & 0.02 & 9.6--2.9 & 4.4 & 2.3 & 8 \\
\hline
\end{tabular}
%\end{center}
\tablefoot{$^{(a)}$ ID of the methyl formate sources. $^{(b)}$ Position of the associated compact continuum core from the \textsl{getsf} unsmoothed catalog presented in \citepalias{louvet2024}. The sources marked with a $^*$ in the first column are not associated with any compact continuum core. $^{(c)}$ Angular offset between the peak position of the methyl formate source and its associated compact continuum core. $^{(d)}$ Peak and integrated intensities at 1.3 mm from the unsmoothed core catalog \citepalias{louvet2024}. For the sources that are not associated with compact continuum cores, peak intensities have been measured in the 1.3 mm continuum maps (see Sect. \ref{section-contassociation}). $^{(e)}$ Major $\times$ minor axis and position angle of the compact continuum core from the unsmoothed core catalog \citepalias{louvet2024}. These values are deconvolved from the continuum map beam size. $^{(f)}$ Deconvolved mean size (FWHM) of the continuum cores. $^{(g)}$ 3\,mm integrated intensity from the unsmoothed core catalog \citepalias{louvet2024}. $^{(h)}$ Opacity computed at $T_{\mathrm d}$ = 100\,K (except for the six most extreme sources for which we used 300\,K, see Sect.\ref{section-MFTd}). $^{(i)}$ Core mass range computed with $T_{\mathrm d}$ ranging from 50\,K to 150\,K (except for the six most extreme sources for which we used 200 -- 400\,K, see Sect.\ref{section-MFTd}). $^{(j)}$ Core mass used to plot Figs.\,\ref{FIG-histo-mass-distribution}, \ref{FIG-mass-to-size} and \ref{FIG-mHC-Mclump}, computed at $T_{\mathrm d}$ = 100\,K for all methyl formate sources, except for the six most extreme sources for which we used 300\,K (see Sect.\ref{section-MFTd})). $^{(k)}$ Spectral index (see Sect. \ref{section-freefree}). $^{(l)}$ Fraction of the fluxes measured at 1.3\,mm and 3\,mm that is due to free-free emission and that are subtracted to the fluxes given in $^{(k)}$ and $^{(g)}$ to obtain the mass estimates given in $^{(i)}$ and $^{(j)}$. The table continues on the next page.}
\end{table}
\end{landscape}

\begin{landscape}
\begin{table}
\ContinuedFloat
  \caption{continued.} 
  \setlength{\tabcolsep}{0.6mm}
\begin{tabular}{lrrr|rrcrr|r|rrcrr}
    \hline
    \hline
     \multicolumn{4}{c|}{ } & \multicolumn{5}{c|}{1.3 mm} & \multicolumn{1}{c|}{3 mm} & \multicolumn{4}{c}{ }  \\
   \hline
    \multicolumn{1}{c}{ID$_{\mathrm{MF}}^{(a)}$} & \multicolumn{1}{c}{RA$^{(b)}$} & \multicolumn{1}{c}{Dec$^{(b)}$} & \multicolumn{1}{c|}{d$^{(c)}$}  & \multicolumn{1}{c}{$S^{\mathrm{peak} (d)}_{\mathrm{1.3 mm}}$} & \multicolumn{1}{c}{$S^{\mathrm{int} (d)}_{\mathrm{1.3 mm}}$} & \multicolumn{1}{c}{$\theta_{\mathrm{maj}}^{\mathrm{dec}} \times \theta_{\mathrm{min}}^{\mathrm{dec} (e)}$} &  \multicolumn{1}{c}{PA$^{\mathrm{dec} (e)}$} & \multicolumn{1}{c|}{FWHM$_{\mathrm{cont}}^{\mathrm{dec} (f)}$} &  \multicolumn{1}{c|}{$S^{\mathrm{int} (g)}_{\mathrm{3 mm}}$} & \multicolumn{1}{c}{$\tau^{(h)}$} & \multicolumn{1}{c}{Mass range$^{(i)}$} & \multicolumn{1}{c}{M$^{(j)}$} &  \multicolumn{1}{c}{$\alpha^{(k)}$}  & \multicolumn{1}{c}{frac$_{\mathrm{ff}}^{(l)}$} \\
    
     \multicolumn{1}{c}{} & \multicolumn{2}{c}{[J2000]} & \multicolumn{1}{c|}{[$\arcsec$]} & \multicolumn{1}{c}{[mJy beam$^{-1}$]}  & \multicolumn{1}{c}{[mJy]}  & \multicolumn{1}{c}{[$\arcsec$ $\times$ $\arcsec$]}  & \multicolumn{1}{c}{[deg]} & \multicolumn{1}{c|}{[au]}   & \multicolumn{1}{c|}{[mJy]}  &  & \multicolumn{1}{c}{[$\Msun$]} & \multicolumn{1}{c}{[$\Msun$]} &   & \multicolumn{1}{c}{[\%]}     \\
         \hline
         \multicolumn{15}{c}{G337.92} \\
     \hline  
G337--MF1 & 16:41:10.46 & -47:08:03.4 & 0.32 & 234.3$\pm$13.6  & 549.1$\pm$17.5  & 0.91$\times$0.42 & -83.4 & 1682.1 & 86.5$\pm$2.5 & 0.22 & 35.0--8.9 & 14.2 & 2.2 & 0 \\
G337--MF2$^{*}$ & \_          & \_           & \_   & 110.0$\pm$110.0 & 110.0$\pm$110.0 & \_ & \_ & 1460.7 & 8.6$\pm$8.6 & 0.1 & 6.0--1.7 & 2.6 & 3.1 & 0 \\
G337--MF3$^{*}$ & \_          & \_           & \_   & 240.0$\pm$65.0  & 240.0$\pm$65.0  & \_ & \_ & 1460.7 & 15.8$\pm$5.0 & 0.23 & 15.4--3.9 & 6.2 & 3.3 & 0 \\
G337--MF4$^{*}$ & \_          & \_           & \_   & 136.0$\pm$136.0 & 136.0$\pm$136.0 & \_ & \_ & 1460.7 & 5.6$\pm$5.6 & 0.12 & 7.6--2.1 & 3.3 & 3.9 & 0 \\
G337--MF5$^{*}$ & \_          & \_           & \_   & 18.4$\pm$18.4   & 18.4$\pm$18.4   & \_ & \_ & 1460.7 & 1.1$\pm$1.1 & 0.02 & 0.9--0.2 & 0.4 & 3.4 & 0 \\
G337--MF6 & 16:41:10.46 & -47:08:01.5 & 0.08 & 71.3$\pm$17.7   & 151.9$\pm$19.5  & 0.74$\times$0.41 & 61.5 & 1503.9 & 21.9$\pm$1.8 & 0.06 & 7.9--2.3 & 3.6 & 2.4 & 0 \\
G337--MF7 & 16:41:10.46 & -47:08:06.4 & 0.58 & 13.6$\pm$3.5    & 16.3$\pm$2.7    & \_ & \_ & 731.7 & 2.0$\pm$0.1 & 0.01 & 0.8--0.2 & 0.3 & 2.6 & 0 \\
\hline
    \multicolumn{15}{c}{G338.93} \\
     \hline  
G338--MF1 & 16:40:34.01 & -45:42:07.3 & 0.06 & 103.2$\pm$1.8 & 148.6$\pm$1.8 & 0.39$\times$0.32 & 67.4 & 1396.2 & 8.8$\pm$0.3 & 0.09 & 16.6--4.7 & 7.4 & 3.4 & 0 \\
G338--MF2 & 16:40:34.13 & -45:41:36.3 & 0.01 & 154.1$\pm$1.3 & 280.7$\pm$2.4 & 0.45$\times$0.27 & -86.4 & 1380.6 & 21.3$\pm$0.4 & 0.14 & 33.2-9.1 & 14.3 & 3.1 & 0 \\
G338--MF3 & 16:40:33.54 & -45:41:37.3 & 0.01 & 70.8$\pm$1.1 & 158.8$\pm$2.3 & 0.47$\times$0.28 & 9.4 & 1450.8 & 13.2$\pm$0.3 & 0.06 & 17.1--5.0 & 7.8 & 3.0 & 0 \\
G338--MF4 & 16:40:34.25 & -45:41:37.1 & 0.10 & 141.5$\pm$1.3 & 237.4$\pm$1.4 & 0.53$\times$0.25 & -41.2 & 1443.0 & 31.9$\pm$0.3 & 0.13 & 27.7--7.7 & 12.0 & 2.4 & 0 \\
G338--MF5 & 16:40:33.69 & -45:42:09.8 & 0.22 & 26.9$\pm$1.4 & 65.0$\pm$2.0 & 0.67$\times$0.49 & 57.7 & 2250.2 & 4.9$\pm$0.2 & 0.02 & 6.7--2.0 & 3.1 & 3.1 & 0 \\
\hline
    \multicolumn{15}{c}{G351.77} \\
     \hline  
G351--MF1 & 17:26:42.53 & -36:09:17.4 & 0.89 & 339.3$\pm$35.3 & 897.8$\pm$47.2 & 1.37$\times$0.7 & 53.0 & 1972.0 & 35.0$\pm$1.2 & 0.15 & 28.6--7.8 & 12.2 & 3.9 & 0 \\
G351--MF2 & 17:26:42.47 & -36:09:18.7 & 0.49 & 271.7$\pm$31.9 & 1199.0$\pm$42.6 & 1.86$\times$1.4 & -66.5 & 3230.0 & 54.8$\pm$2.7 & 0.12 & 36.7--10.3 & 16.1 & 3.7 & 0 \\
G351--MF3$^{*}$ & \_     & \_           & \_ & 930.0$\pm$130.0 & 930.0$\pm$130.0 & \_ & \_ & 1544.0 & 68.4$\pm$10.0 & 0.49 & 52.1--8.9 & 14.9 & 3.1 & 0 \\
G351--MF4 & 17:26:42.65 & -36:09:18.8 & 0.43 & 94.3$\pm$34.8 & 254.0$\pm$35.9 & 1.31$\times$0.81 & 35.6 & 2074.0 & 26.5$\pm$0.01 & 0.04 & 7.1--2.1 & 3.2 & 2.7 & 0 \\
G351--MF5 & 17:26:42.80 & -36:09:20.5 & 0.01 & 106.9$\pm$32.1 & 175.7$\pm$30.8 & 0.7$\times$0.43 & -16.1 & 1102.0 & 5.0$\pm$1.7 & 0.05 & 4.9--1.4 & 2.2 & 4.3 & 0 \\
\hline
    \multicolumn{15}{c}{G353.41} \\
     \hline  
G353--MF1 & 17:30:28.43 & -34:41:47.9 & 0.25 & 53.5$\pm$3.0 & 127.0$\pm$3.3 & 1.06$\times$0.74 & -5.1 & 1780.0 & 5.0$\pm$0.4 & 0.02 & 3.4--1.0 & 1.6 & 3.9 & 0 \\
\hline
    \multicolumn{15}{c}{W43-MM1} \\
     \hline  
W43-MM1--MF1 & 18:47:46.98  & -01:54:26.5 & 0.21 & 118.6$\pm$3.8 & 361.6$\pm$4.3 & 0.72$\times$0.62 & 27.7 & 3685.0 & 23.8$\pm$0.4 & 0.18 & 89.0--23.6 & 37.3 & 3.2 & 0 \\
W43-MM1--MF2 & 18:47:47.02  & -01:54:26.9 & 0.20 & 311.6$\pm$3.9 & 640.9$\pm$5.0 & 0.45$\times$0.39 & 42.2 & 2332.0 & 29.2$\pm$0.3 & 0.56 & 356.0--46.6 & 79.5 & 3.7 & 0 \\
W43-MM1--MF3 & 18:47:46.84  & -01:54:29.3 & 0.01 & 187.8$\pm$5.8 & 297.2$\pm$6.8 & 0.38$\times$0.16 & -1.2 & 1358.5 & 20.5$\pm$0.5 & 0.3 & 86.8--20.1 & 32.5 & 3.2 & 0 \\
W43-MM1--MF4 & 18:47:46.37  & -01:54:33.4 & 0.11 & 96.8$\pm$1.9 & 227.0$\pm$2.8 & 0.55$\times$0.29 & 24.2 & 2205.5 & 17.4$\pm$0.2 & 0.15 & 53.5--14.7 & 23.0 & 3.0 & 0 \\
W43-MM1--MF5 & 18:47:46.77  & -01:54:31.2 & 0.07 & 45.5$\pm$3.1 & 72.5$\pm$3.1 & 0.4$\times$0.25 & 51.5 & 1771.0 & 5.4$\pm$0.2 & 0.07 & 15.5--4.5 & 7.0 & 3.1 & 0 \\
W43-MM1--MF6 & 18:47:46.52  & -01:54:24.2 & 0.01 & 15.8$\pm$1.4 & 21.1$\pm$1.3 & 0.31$\times$0.20 & 89.2 & 1413.5 & 1.4$\pm$0.1 & 0.02 & 4.3--1.3 & 2.0 & 3.2 & 0 \\
W43-MM1--MF7 & 18:47:46.90  & -01:54:30.0 & 0.09 & 26.3$\pm$5.0 & 33.2$\pm$3.9 & 0.26$\times$0.25 & 31.0 & 1424.5 & 1.9$\pm$0.3 & 0.04 & 6.9--2.0 & 3.2 & 3.4 & 0 \\
W43-MM1--MF8 & 18:47:46.48  & -01:54:32.6 & 0.12 & 30.4$\pm$3.4 & 46.2$\pm$3.2 & 0.33$\times$0.26 & 45.5 & 1644.5 & 3.1$\pm$0.2 & 0.04 & 9.6--2.9 & 4.4 & 3.2 & 0 \\
W43-MM1--MF9 & 18:47:44.77  & -01:54:45.2 & 0.04 & 12.2$\pm$0.9 & 20.7$\pm$0.9 & 0.37$\times$0.34 & 1.8 & 1980.0 & 0.9$\pm$0.1 & 0.02 & 4.2--1.2 & 1.9 & 3.7 & 0 \\
W43-MM1--MF10 & 18:47:46.54 & -01:54:23.1 & 0.06 & 43.0$\pm$1.0 & 74.9$\pm$1.3 & 0.41$\times$0.26 & -84.0 & 1826.0 & 5.9$\pm$0.1 & 0.06 & 16.0--4.7 & 7.3 & 3.0 & 0 \\
W43-MM1--MF11 & 18:47:47.02 & -01:54:30.8 & 0.22 & 20.1$\pm$5.0 & 41.6$\pm$4.9 & 0.61$\times$0.32 & -19.7 & 2447.5 & 2.9$\pm$0.3 & 0.03 & 8.5--2.6 & 3.9 & 3.1 & 0 \\
W43-MM1--MF12 & 18:47:46.87 & -01:54:25.7 & 0.20  & 14.5$\pm$3.5 & 17.5$\pm$2.9 & \_ & \_ & 1149.5 & 1.3$\pm$0.2 & 0.02 & 3.5--1.0 & 1.6 & 3.1 & 0 \\
W43-MM1--MF13 & 18:47:46.25 & -01:54:33.4 & 0.05 & 17.2$\pm$1.6 & 25.9$\pm$1.3 & 0.46$\times$0.26 & -22.4 & 1936.0 & 2.5$\pm$0.1 & 0.02 & 5.3--1.6 & 2.4 & 2.8 & 0 \\
W43-MM1--MF14 & 18:47:46.97 & -01:54:29.7 & 0.11 & 18.2$\pm$5.3 & 18.8$\pm$4.1 & \_ & \_ & 1149.5 & 2.0$\pm$0.3 & 0.03 & 3.8--1.1 & 1.8 & 2.6 & 0 \\

\hline
    \multicolumn{15}{c}{W43-MM2} \\
     \hline  
W43-MM2--MF1 & 18:47:36.80 & -02:00:54.3 & 0.14 & 144.2$\pm$1.1 & 432.7$\pm$3.2 & 0.61$\times$0.43 & -15.0 & 2832.5 & 30.3$\pm$0.3 & 0.18 & 108.8--28.8 & 45.5 & 3.2 & 0 \\
W43-MM2--MF2 & 18:47:36.70 & -02:00:47.6 & 0.04 & 14.8$\pm$0.7 & 28.2$\pm$0.9 & 0.45$\times$0.3 & -49.3 & 2029.5 & 2.3$\pm$0.1 & 0.02 & 5.8--1.7 & 2.7 & 3.0 & 0 \\
W43-MM2--MF3 & 18:47:36.28 & -02:00:50.8 & 0.11 & 10.7$\pm$0.4 & 19.8$\pm$0.5 & \_ & \_ & 1270.5 & 2.1$\pm$0.2 & 0.01 & 4.0--1.2 & 1.9 & 2.7 & 0 \\
\hline
\end{tabular}
%\end{center}
\tablefoot{The table continues on the next page.}
\end{table}
\end{landscape}

\begin{landscape}
\begin{table}
\ContinuedFloat
%\begin{center}
  \caption{continued.} 
  \setlength{\tabcolsep}{0.6mm}
\begin{tabular}{lrrr|rrcrr|r|rrcrr}
    \hline
    \hline
     \multicolumn{4}{c|}{ } & \multicolumn{5}{c|}{1.3 mm} & \multicolumn{1}{c|}{3 mm} & \multicolumn{4}{c}{ }  \\
   \hline
    \multicolumn{1}{c}{ID$_{\mathrm{MF}}^{(a)}$} & \multicolumn{1}{c}{RA$^{(b)}$} & \multicolumn{1}{c}{Dec$^{(b)}$} & \multicolumn{1}{c|}{d$^{(c)}$}  & \multicolumn{1}{c}{$S^{\mathrm{peak} (d)}_{\mathrm{1.3 mm}}$} & \multicolumn{1}{c}{$S^{\mathrm{int} (d)}_{\mathrm{1.3 mm}}$} & \multicolumn{1}{c}{$\theta_{\mathrm{maj}}^{\mathrm{dec}} \times \theta_{\mathrm{min}}^{\mathrm{dec} (e)}$} &  \multicolumn{1}{c}{PA$^{\mathrm{dec} (e)}$} & \multicolumn{1}{c|}{FWHM$_{\mathrm{cont}}^{\mathrm{dec} (f)}$} &  \multicolumn{1}{c|}{$S^{\mathrm{int} (g)}_{\mathrm{3 mm}}$} & \multicolumn{1}{c}{$\tau^{(h)}$} & \multicolumn{1}{c}{Mass range$^{(i)}$} & \multicolumn{1}{c}{M$^{(j)}$} &  \multicolumn{1}{c}{$\alpha^{(k)}$}  & \multicolumn{1}{c}{frac$_{\mathrm{ff}}^{(l)}$} \\
    
     \multicolumn{1}{c}{} & \multicolumn{2}{c}{[J2000]} & \multicolumn{1}{c|}{[$\arcsec$]} & \multicolumn{1}{c}{[mJy beam$^{-1}$]}  & \multicolumn{1}{c}{[mJy]}  & \multicolumn{1}{c}{[$\arcsec$ $\times$ $\arcsec$]}  & \multicolumn{1}{c}{[deg]} & \multicolumn{1}{c|}{[au]}   & \multicolumn{1}{c|}{[mJy]}  &  & \multicolumn{1}{c}{[$\Msun$]} & \multicolumn{1}{c}{[$\Msun$]} &   & \multicolumn{1}{c}{[\%]}     \\
     \hline
    \multicolumn{15}{c}{W43-MM3} \\
     \hline  
W43-MM3--MF1 & 18:47:39.26 & -02:00:28.1 & 0.06 & 16.8$\pm$0.4 & 36.6$\pm$0.6 & 0.51$\times$0.33 & 50.3 & 2277.0 & 1.3$\pm$0.01 & 0.02 & 7.5--2.2 & 3.5 & 4.0 & 0 \\
W43-MM3--MF2 & 18:47:41.71 & -02:00:28.6 & 0.07 & 51.1$\pm$2.0 & 78.3$\pm$2.5 & 0.38$\times$0.21 & -88.0 & 1589.4 & 6.6$\pm$0.4 & 0.06 & 16.7--4.9 & 7.6 & 3.0 & 0 \\
W43-MM3--MF3 & 18:47:41.73 & -02:00:27.4 & 0.10 & 22.5$\pm$1.7 & 25.8$\pm$1.5 & \_ & \_ & 1303.5 & 1.9$\pm$0.1 & 0.03 & 5.3--1.6 & 2.4 & 3.1 & 0 \\
         \hline
    \multicolumn{15}{c}{W51-E} \\
     \hline      
W51-E--MF1 & 19:23:43.97 & 14:30:34.5  & 0.06 & 363.0$\pm$8.8 & 2004.0$\pm$19.0 & 0.68$\times$0.51 & -0.2 & 3218.3 & 358.7$\pm$7.5 & 0.32 & 102.5--43.6 & 61.0 & 2.1 & 11 \\
W51-E--MF2$^{*}$ & \_          & \_          & \_ & 245.6$\pm$32.6 & 245.6$\pm$32.6 & \_ & \_ & 1636.2 & 48.3$\pm$9.3 & 0.24 & 13.2--5.8 & 8.1 & 2.0 & 0 \\
W51-E--MF3 & 19:23:43.90 & 14:30:28.2  & 0.33 & 250.8$\pm$17.8 & 853.7$\pm$25.1 & 0.49$\times$0.46 & -1.2 & 2597.4 & 215.8$\pm$5.2 & 0.19 & 36.0--16.3 & 22.4 & 1.6 & 18 \\
W51-E--MF4 & 19:23:43.74 & 14:30:21.3  & 0.11 & 17.9$\pm$4.9 & 30.0$\pm$4.4 & 0.32$\times$0.24 & 81.2 & 1528.1 & 4.7$\pm$0.6 & 0.05 & 6.1--1.8 & 2.8 & 2.2 & 0 \\
W51-E--MF5$^{*}$ & \_          & \_          & \_ & 55.5$\pm$9.2 & 55.5$\pm$9.2 & \_ & \_ & 1636.2 & 16.8$\pm$1.4 & 0.12 & 9.6--2.6 & 4.1 & 1.4 & 22 \\
W51-E--MF6 & 19:23:43.79 & 14:30:19.7  & 0.27 & 13.9$\pm$3.3 & 30.6$\pm$3.5 & 0.39$\times$0.29 & 57.1 & 1836.0 & 10.5$\pm$0.9 & 0.04 & 6.1--1.8 & 2.8 & 1.3 & 0 \\
W51-E--MF7 & 19:23:43.82 & 14:30:23.4  & 0.15 & 17.7$\pm$6.3 & 44.4$\pm$6.9 & 0.43$\times$0.36 & 4.8 & 2165.4 & 10.3$\pm$1.0 & 0.05 & 9.0--2.7 & 4.1 & 1.7 & 0 \\
\hline
    \multicolumn{15}{c}{W51-IRS2} \\
     \hline  
W51-IRS2--MF1 & 19:23:39.99 & 14:31:05.9 & 0.43 & 252.7$\pm$13.4 & 506.4$\pm$12.4 & 0.55$\times$0.52 & 60.6 & 2910.6 & 49.7$\pm$6.4 & 0.09 & 23.3--11.1 & 15.0 & 2.8 & 3 \\
W51-IRS2--MF2 & 19:23:39.82 & 14:31:05.1 & 0.18 & 223.5$\pm$14.9 & 393.5$\pm$13.0 & 0.56$\times$0.39 & 46.4 & 2554.2 & 104.0$\pm$2.7 & 0.22 & 81.2--20.8 & 32.5 & 1.6 & 19 \\
W51-IRS2--MF3 & 19:23:40.05 & 14:31:05.5 & 0.53 & 847.2$\pm$15.2 & 1789.0$\pm$20.7 & 0.52$\times$0.39 & 75.1 & 2457.0 & 368.0$\pm$12.6 & 0.3 & 87.1--37.6 & 52.4 & 1.9 & 14 \\
W51-IRS2--MF4 & 19:23:39.95 & 14:31:05.4 & 0.14 & 333.2$\pm$13.7 & 488.3$\pm$11.9 & 0.39$\times$0.27 & 85.6 & 1771.2 & 128.4$\pm$7.0 & 0.35 & 123.0--26.3 & 43.0 & 1.6 & 19 \\
W51-IRS2--MF5 & 19:23:39.75 & 14:31:05.3 & 0.04 & 141.1$\pm$14.3 & 196.0$\pm$11.1 & 0.46$\times$0.25 & 63.5 & 1857.6 & 17.3$\pm$2.8 & 0.16 & 44.8--12.0 & 19.0 & 2.9 & 2 \\
W51-IRS2--MF6 & 19:23:38.57 & 14:30:41.8 & 0.01 & 68.2$\pm$1.0 & 197.3$\pm$2.6 & 0.49$\times$0.36 & -32.4 & 2284.2 & 21.7$\pm$0.6 & 0.08 & 41.8--12.1 & 18.8 & 2.7 & 0 \\
W51-IRS2--MF7 & 19:23:39.52 & 14:31:03.4 & 0.17 & 17.7$\pm$2.5 & 29.9$\pm$2.9 & 0.43$\times$0.14 & 67.7 & 1360.8 & 1.1$\pm$1.7 & 0.02 & 5.9--1.8 & 2.7 & 4.0 & 0 \\
W51-IRS2--MF8 & 19:23:41.83 & 14:30:54.9 & 0.17 & 12.2$\pm$1.1 & 15.9$\pm$0.9 & 0.39$\times$0.17 & 9.8 & 1409.4 & 4.2$\pm$0.4 & 0.01 & 3.1--0.9 & 1.4 & 1.6 & 0 \\
W51-IRS2--MF9 & 19:23:38.40 & 14:30:36.7 & 0.36 & 11.5$\pm$1.1 & 35.7$\pm$1.5 & 0.80$\times$0.52 & 49.9 & 3515.4 & 2.5$\pm$0.1 & 0.01 & 7.0--2.1 & 3.3 & 3.2 & 0 \\
\hline
\end{tabular}
%\end{center}
%\tablefoot{}
\end{table}
\end{landscape}

\clearpage
\twocolumn

%------------------------------------------------------------------
\section{Correction for free-free contribution}
%------------------------------------------------------------------
\label{appendix-freefree}

\begin{figure*}[!ht]
   \resizebox{\hsize}{!}
   {\begin{tabular}{cc}  
       \includegraphics[width=\hsize]{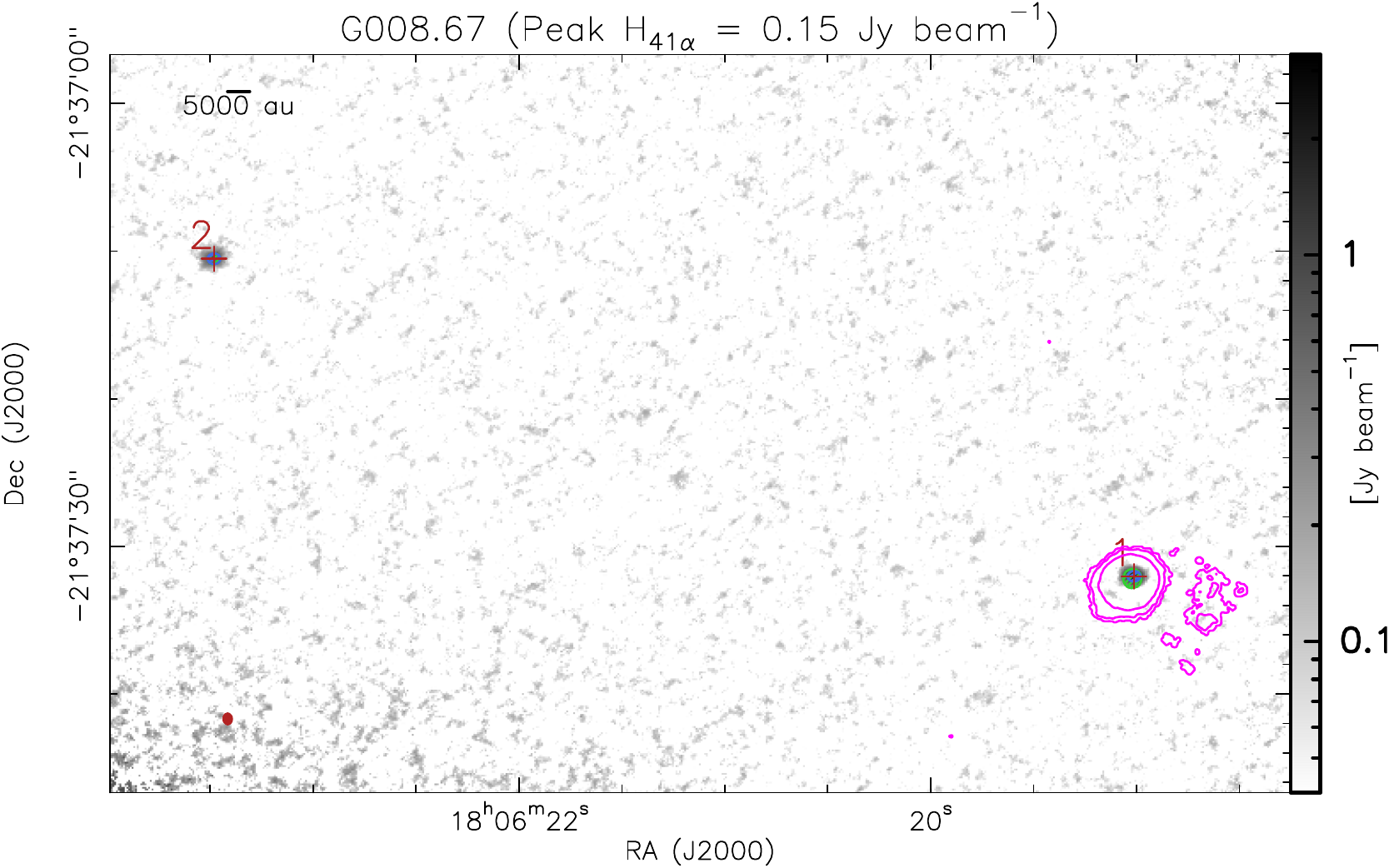}  &
       \includegraphics[width=\hsize]{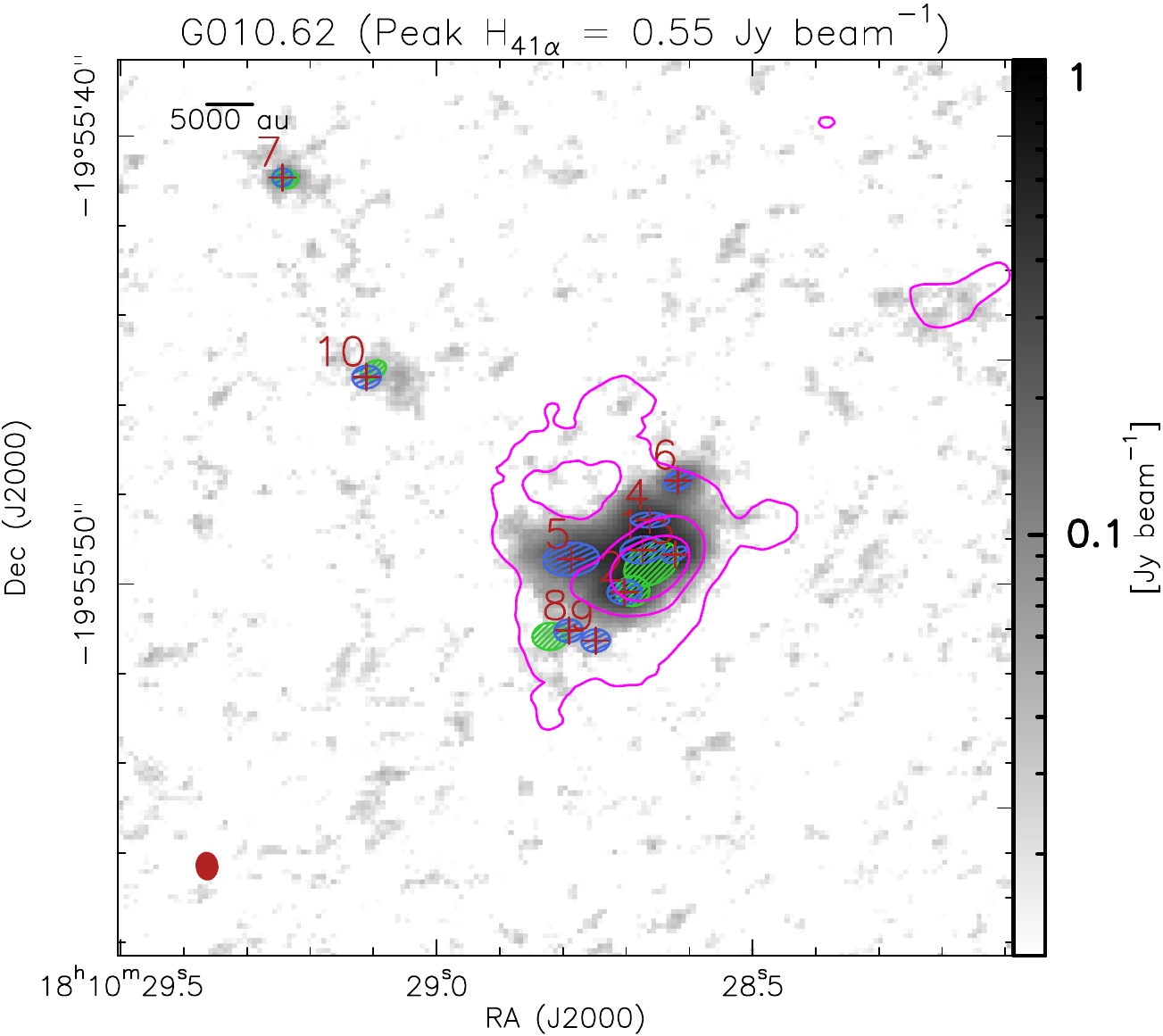}  \\
       \includegraphics[width=\hsize]{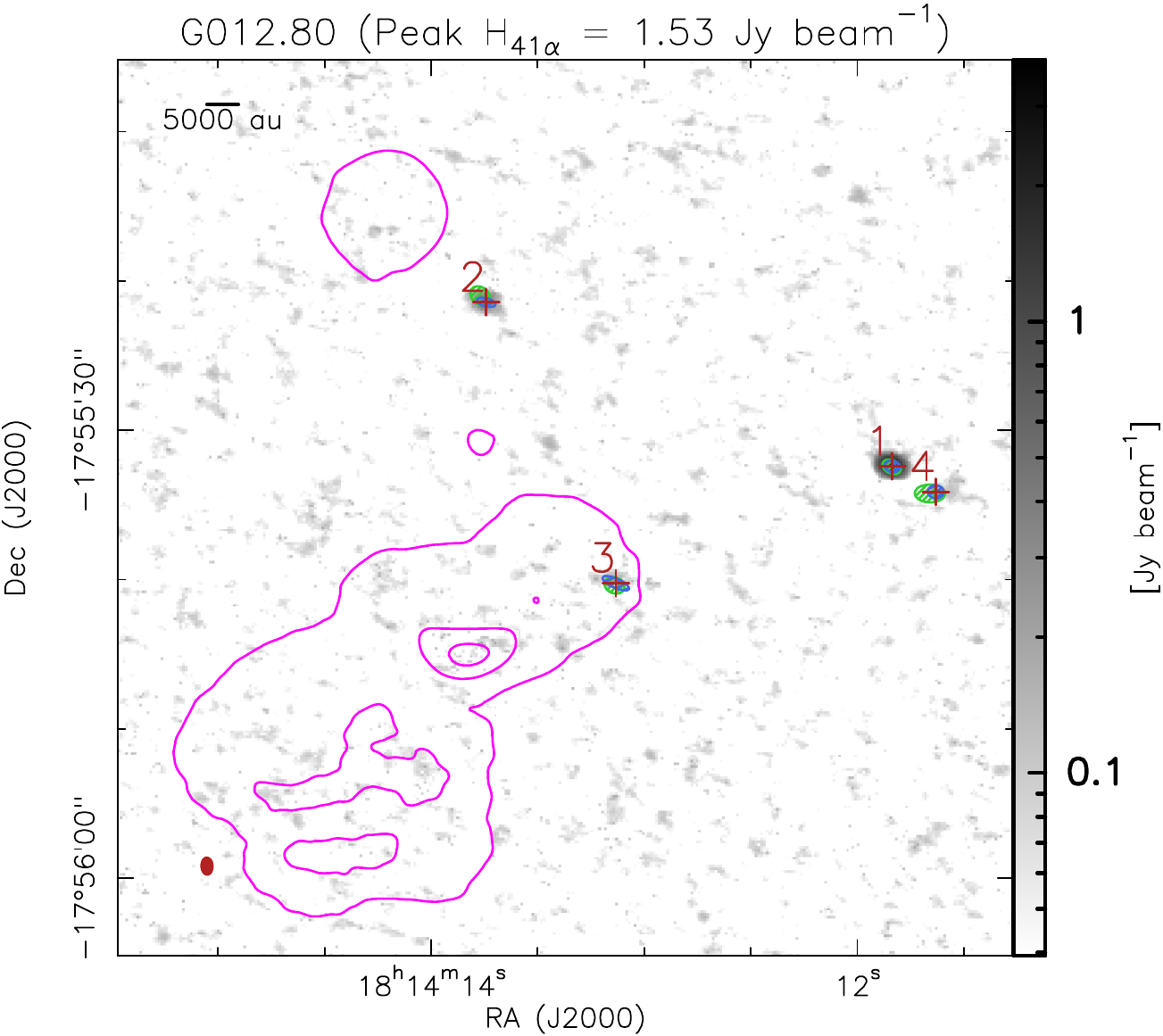}  &
       \includegraphics[width=\hsize]{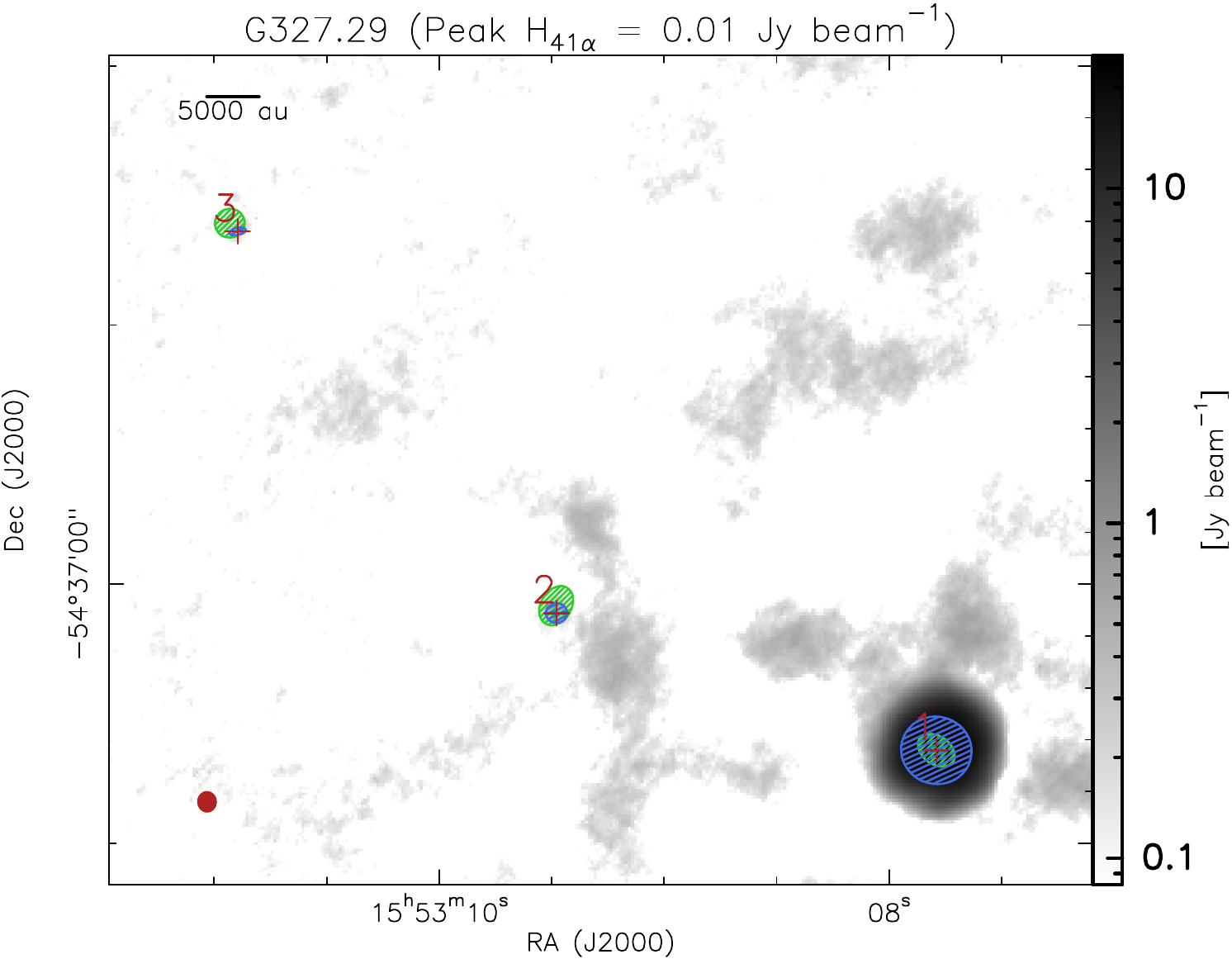}  \\
       \end{tabular}}
\caption{\label{FIG-ellipses-h41alpha} Methyl formate moment 0 maps (background image) as shown in Fig. \ref{FIG-mom0-maps}. The red crosses indicate the peak positions of the methyl formate sources, while the blue ellipses show the deconvolved emission sizes. The green ellipses show the deconvolved source sizes of the associated compact continuum cores. The synthesized beam size on the ALMA-IMF B6-SPW0 line cubes are shown with a red ellipse in the bottom left corner of each panel. Contours of the H$_{41 \alpha}$ emission are overlaid in magenta on top of the moment 0 maps of methyl formate, showing 2, 20, and 50\% of the peak intensity, indicated on top of each panel. The figure continues on the next page.} 
       \end{figure*}

\begin{figure*}[!ht]
   \resizebox{\hsize}{!}
   {\begin{tabular}{cc}  
       \includegraphics[width=\hsize]{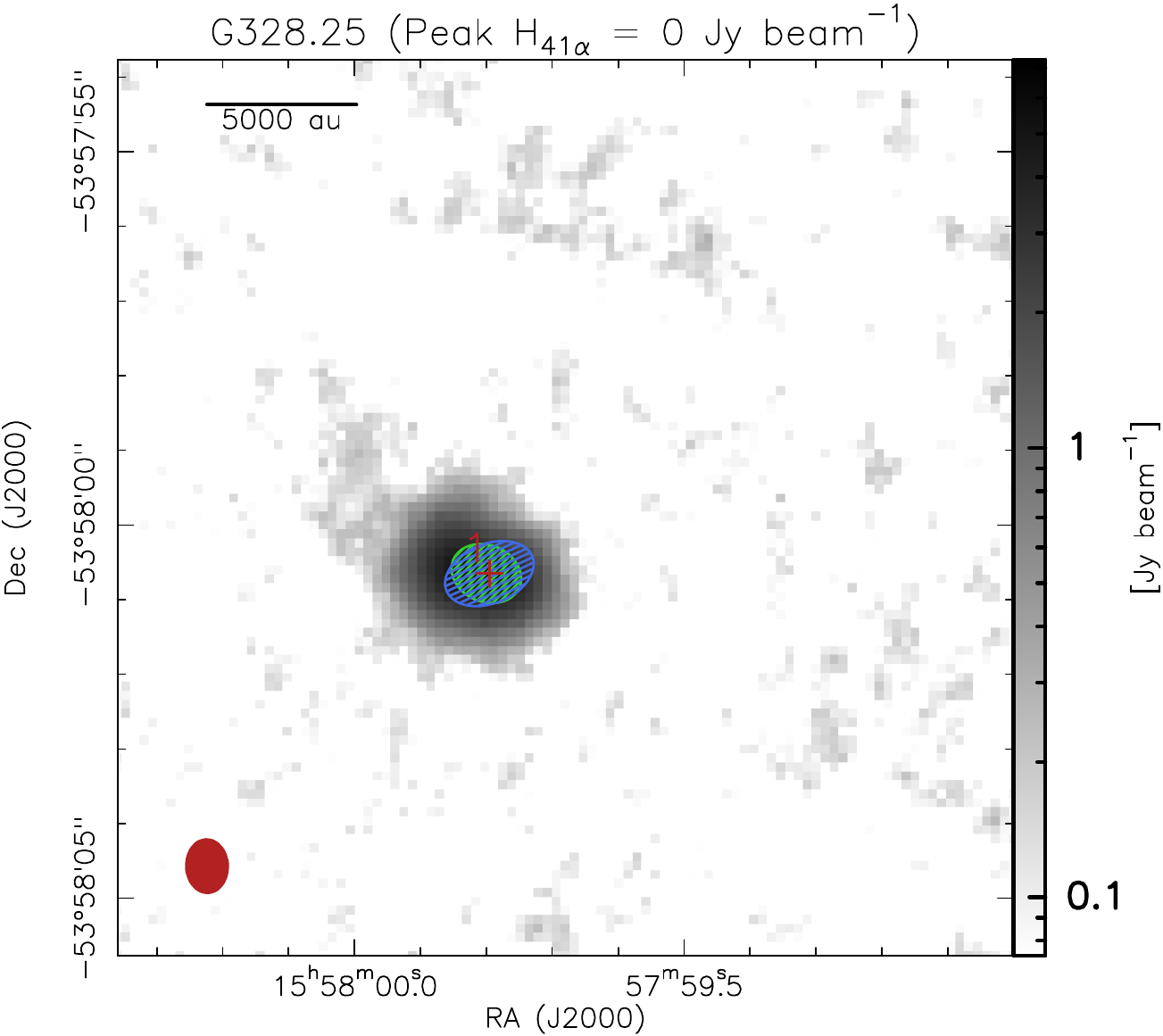}  &
       \includegraphics[width=\hsize]{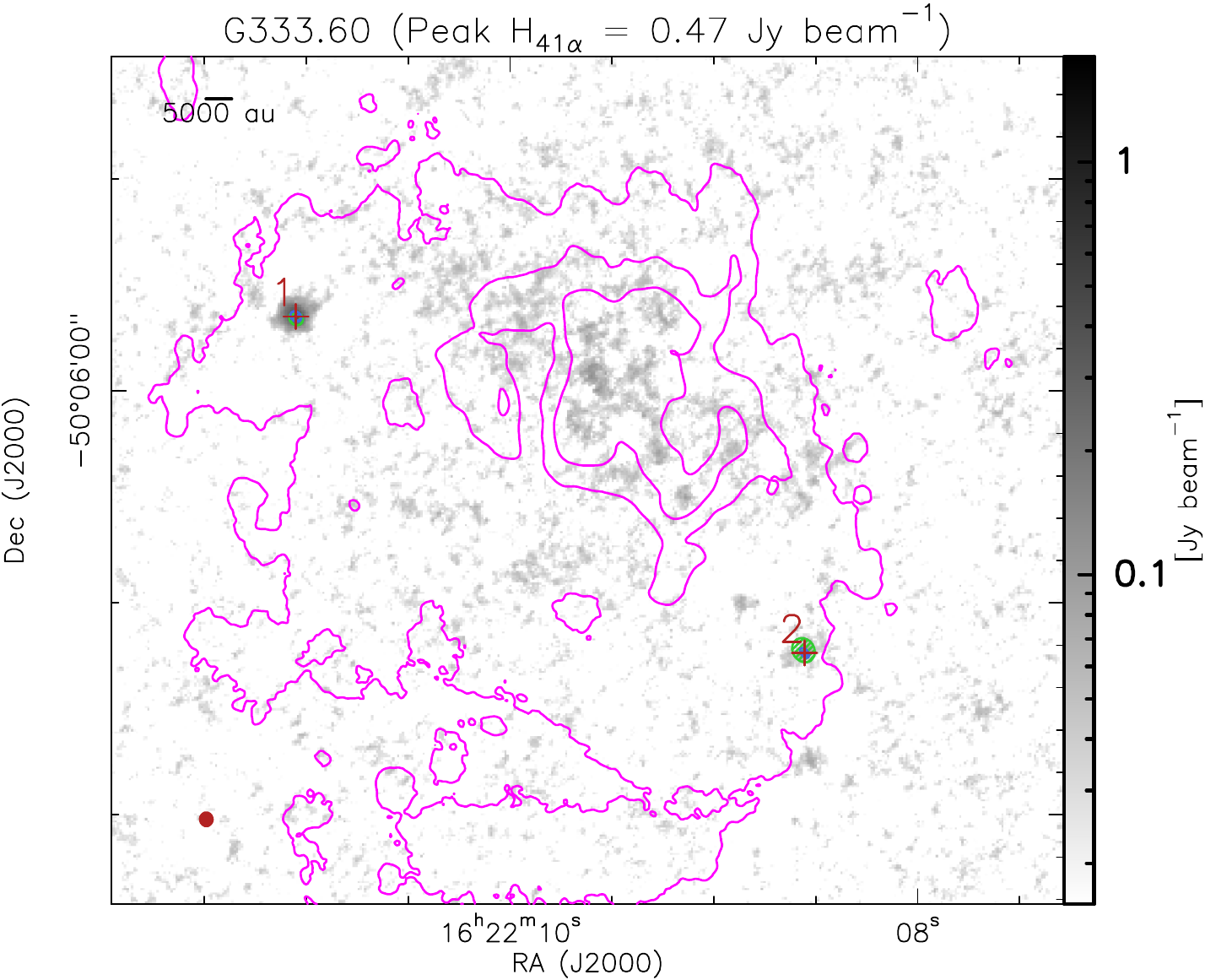}  \\
       \includegraphics[width=\hsize]{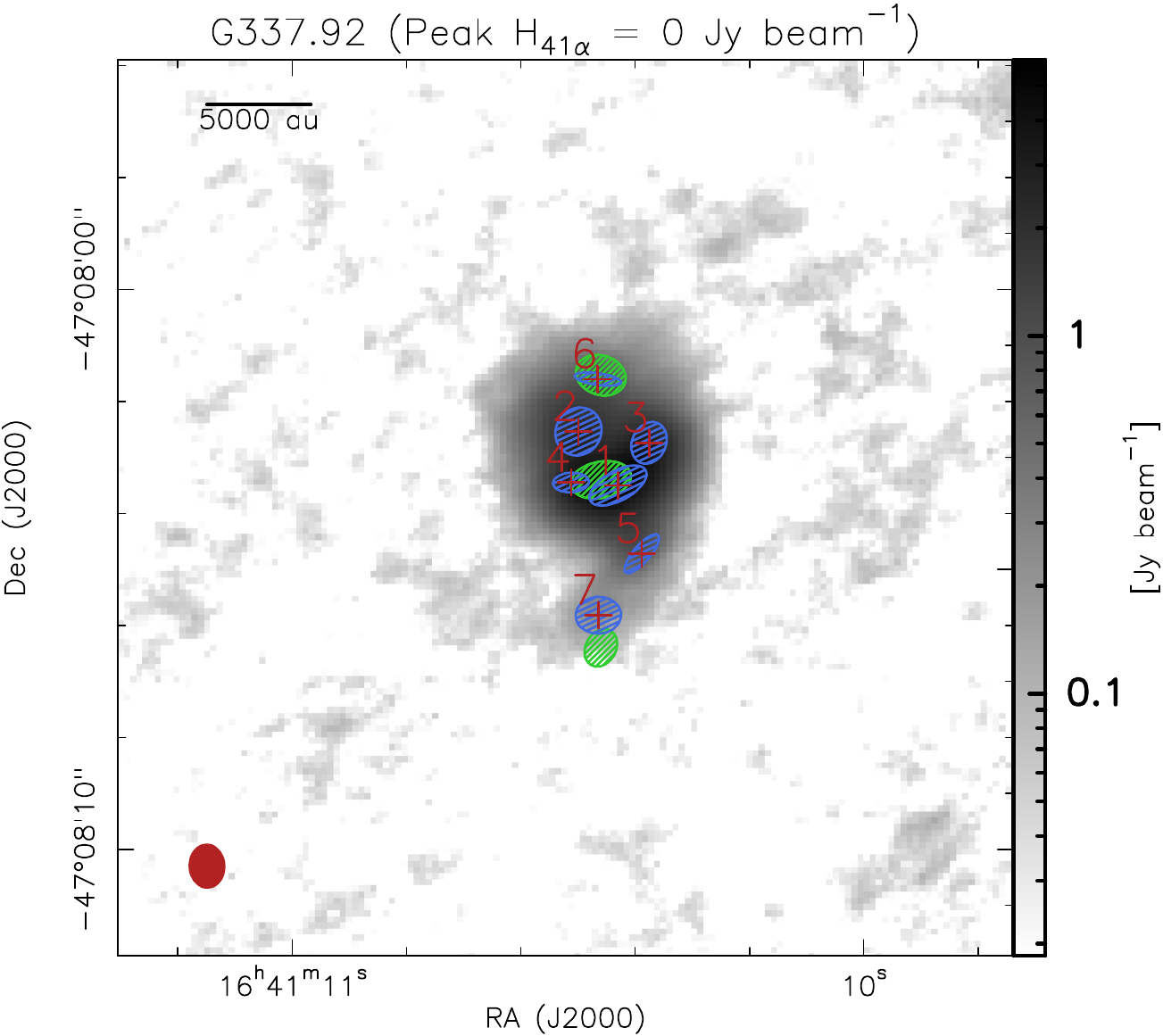}  &
       \includegraphics[width=\hsize]{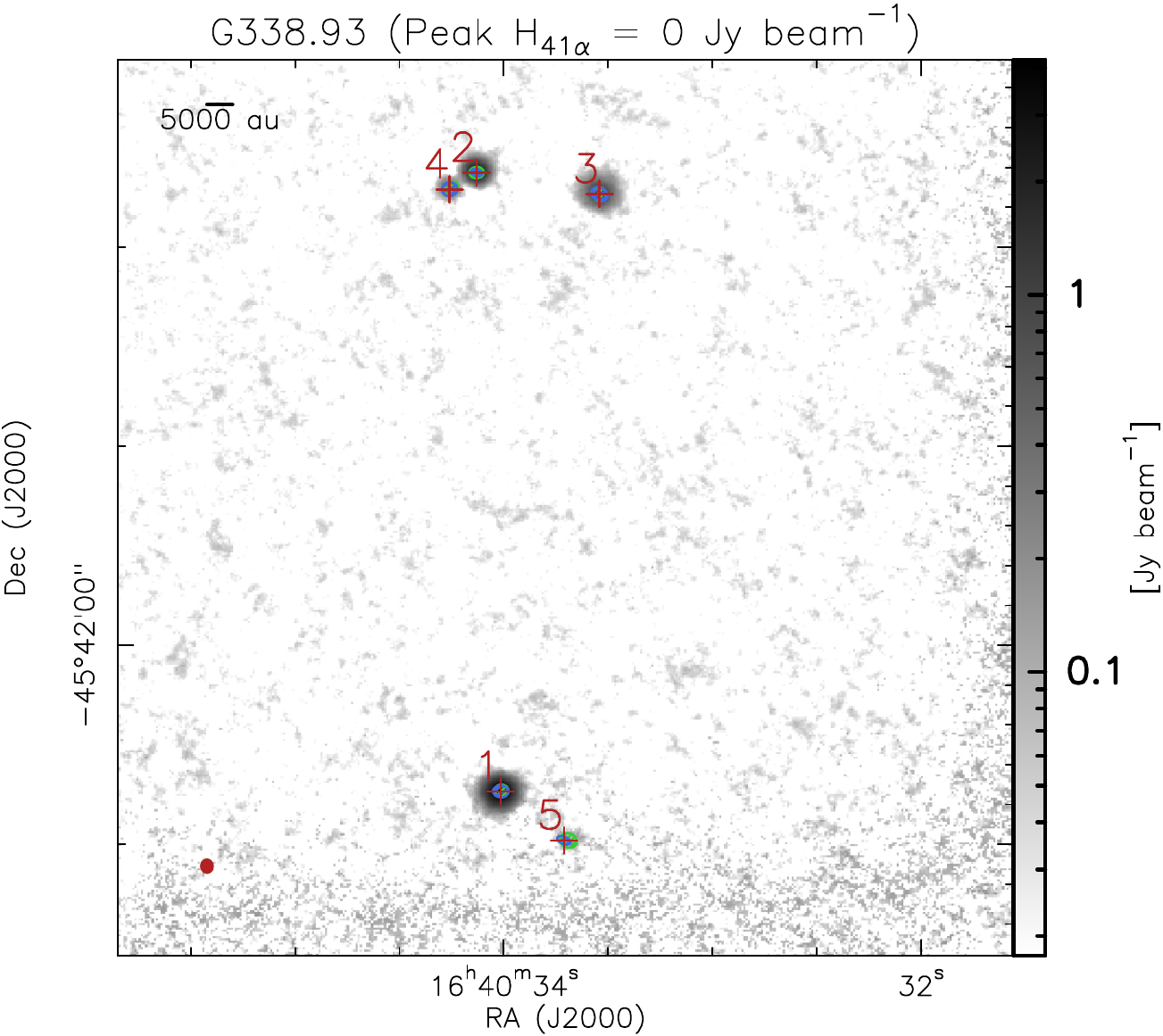}  \\
       \end{tabular}}
    \caption{\label{FIG-ellipses-h41alpha2} Same as Fig.\,\ref{FIG-ellipses-h41alpha}. The figure continues on the next page.}
       \end{figure*}

\begin{figure*}[!ht]
   \resizebox{\hsize}{!}
   {\begin{tabular}{cc}  
       \includegraphics[width=\hsize]{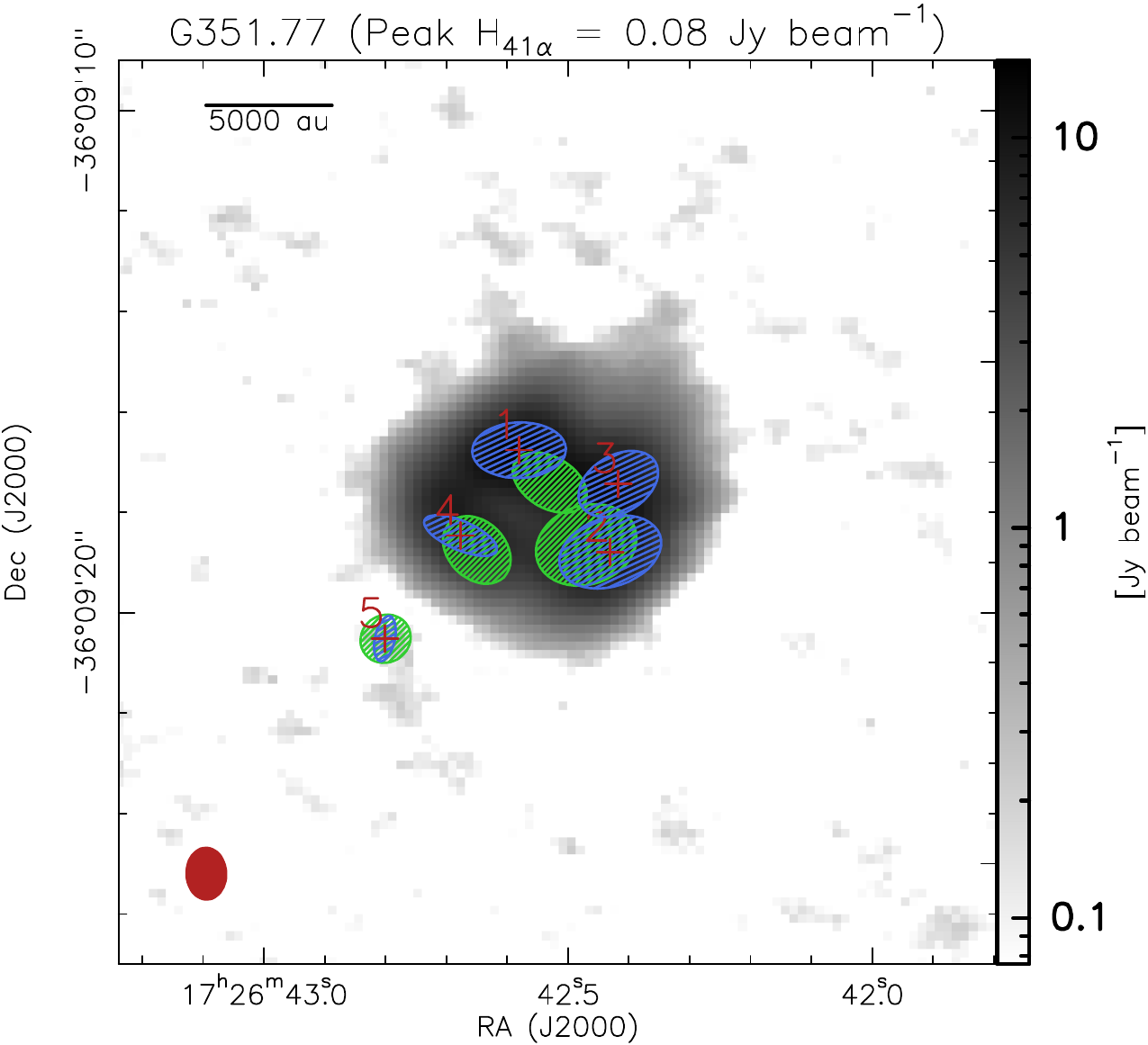}  &
       \includegraphics[width=\hsize]{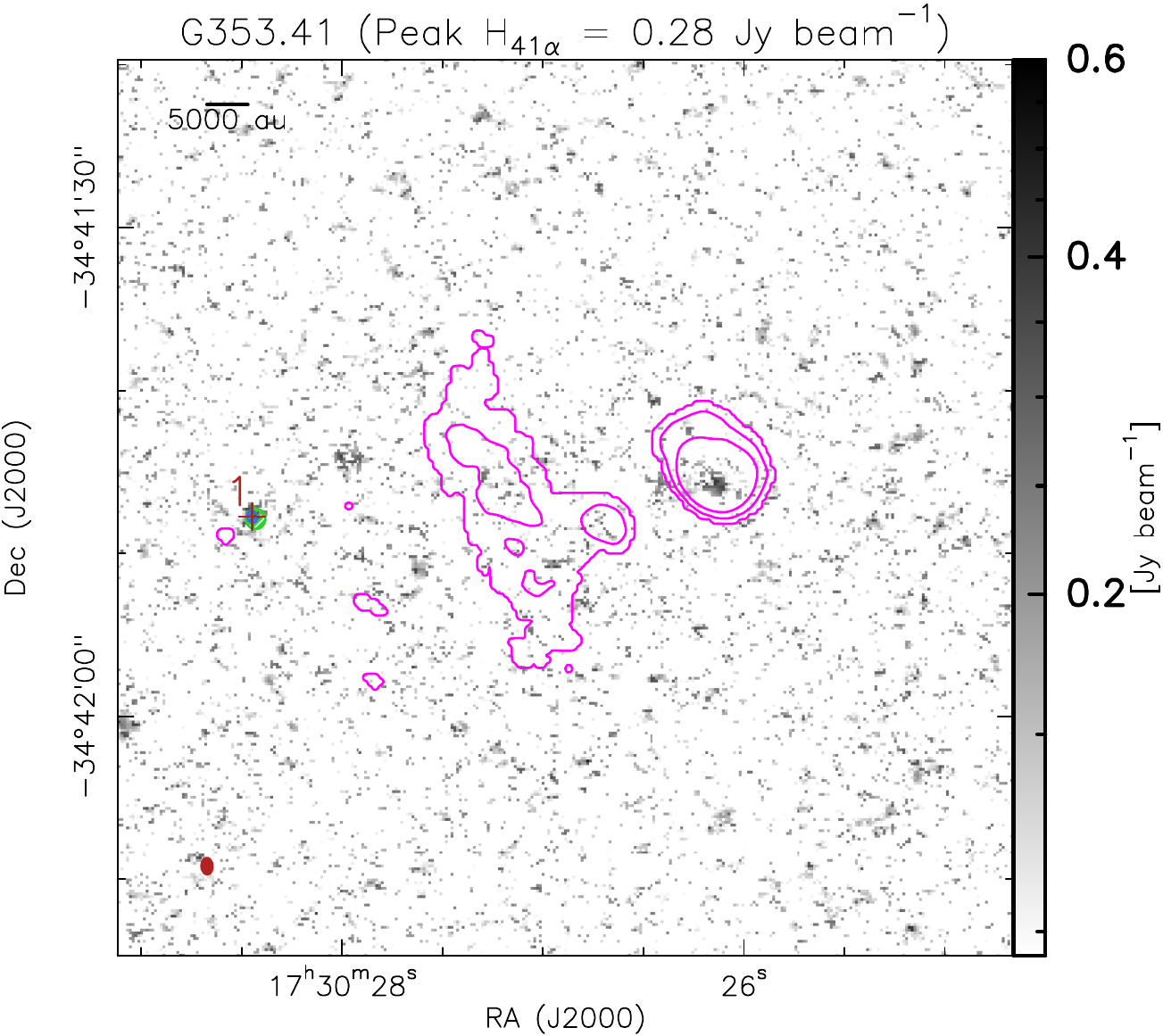}  \\
       \includegraphics[width=\hsize]{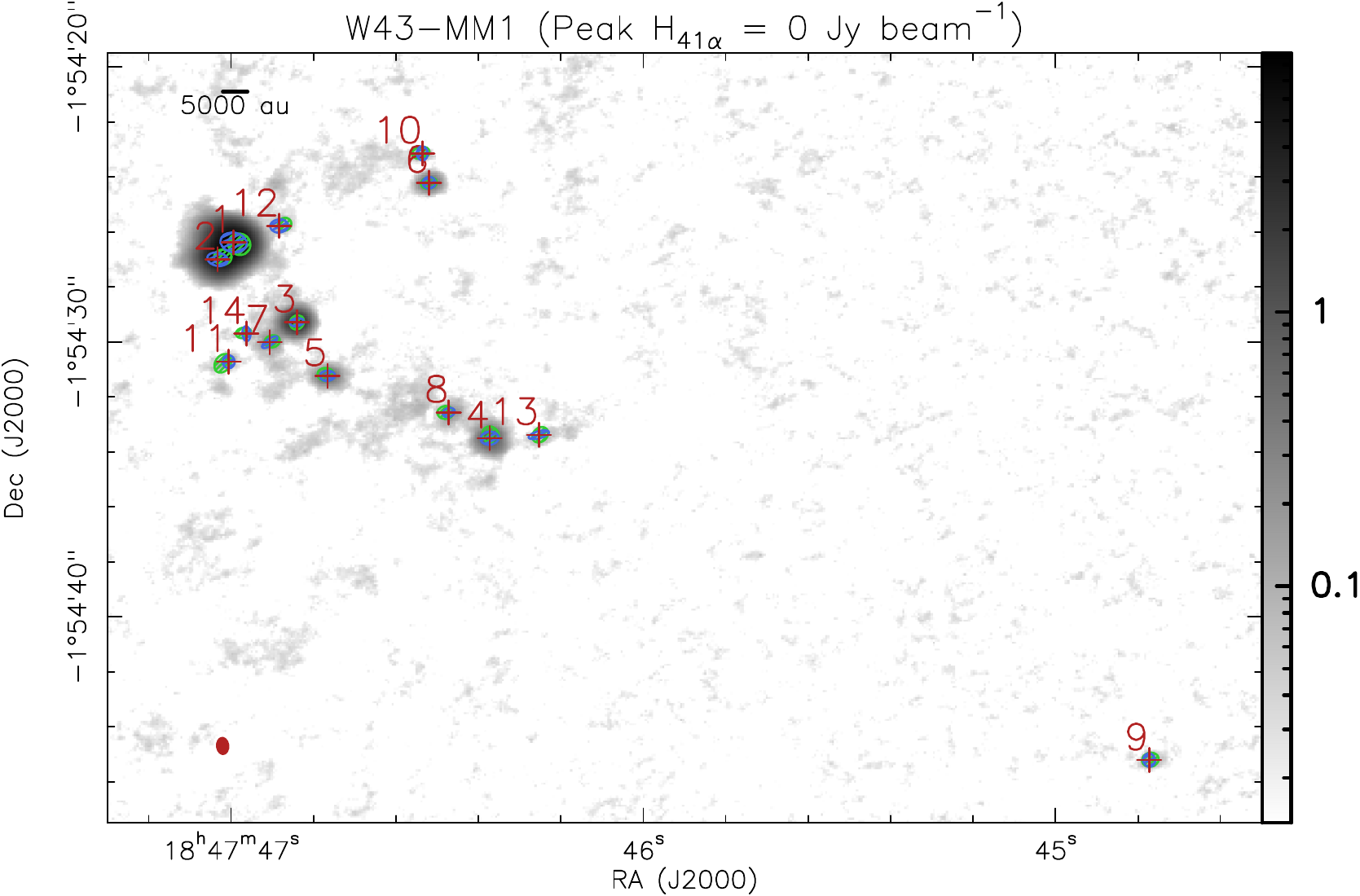}  &
       \includegraphics[width=\hsize]{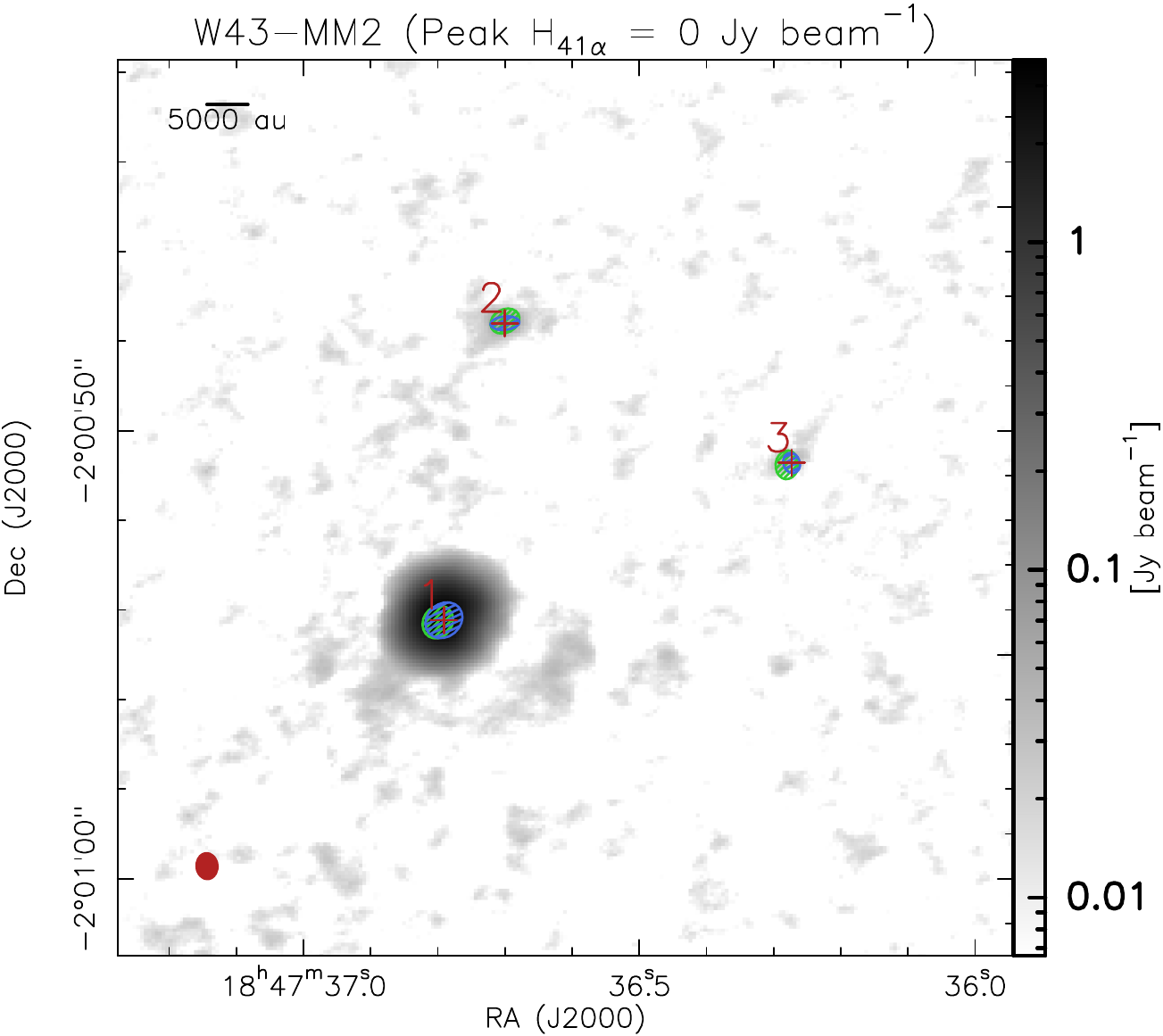}  \\
       \end{tabular}}
    \caption{\label{FIG-ellipses-h41alpha3} Same as Fig.\,\ref{FIG-ellipses-h41alpha}. The figure continues on the next page.}
       \end{figure*}

\begin{figure*}[!ht]
   \resizebox{\hsize}{!}
   {\begin{tabular}{cc}  
       \includegraphics[width=\hsize]{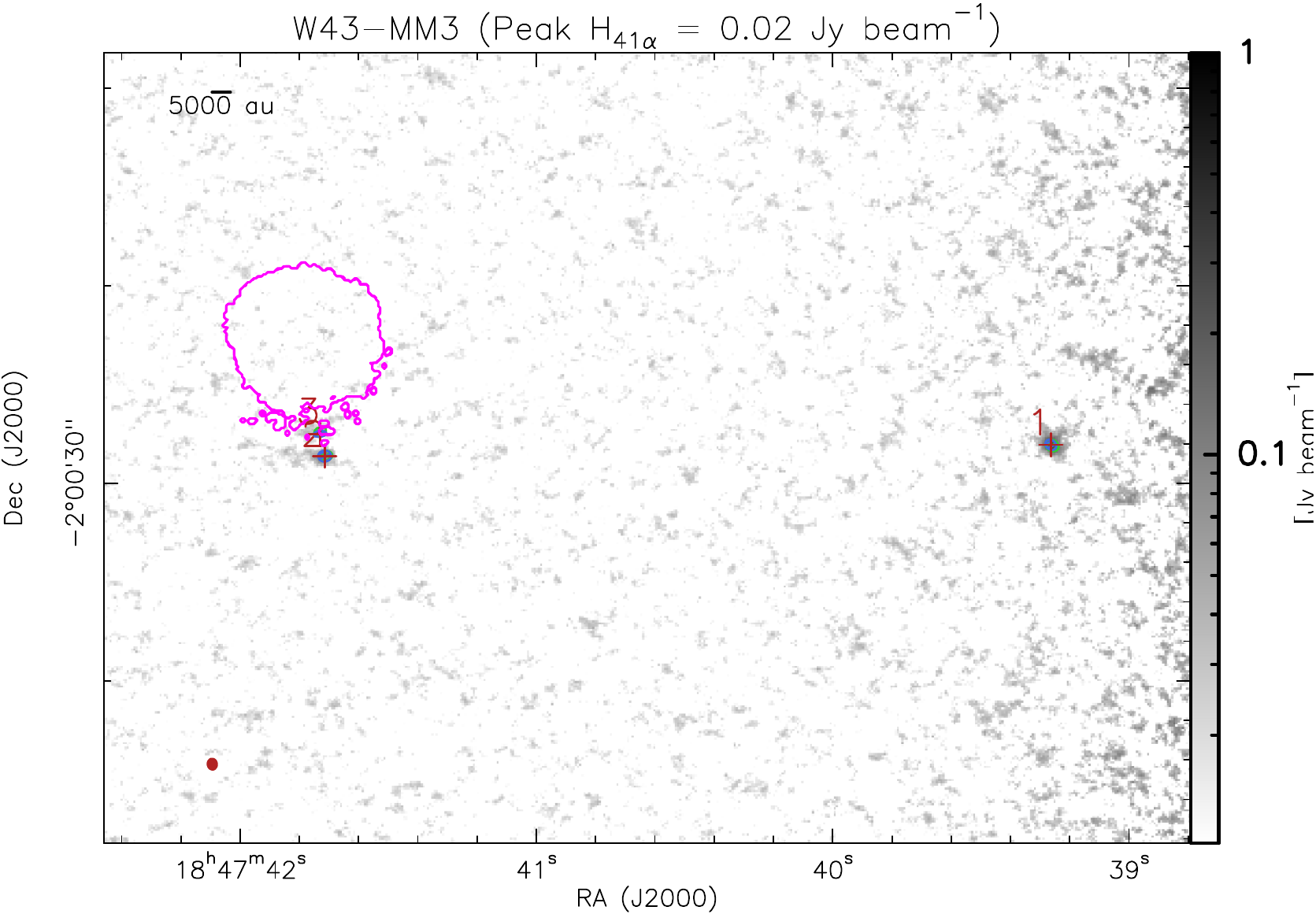}  &
       \includegraphics[width=\hsize]{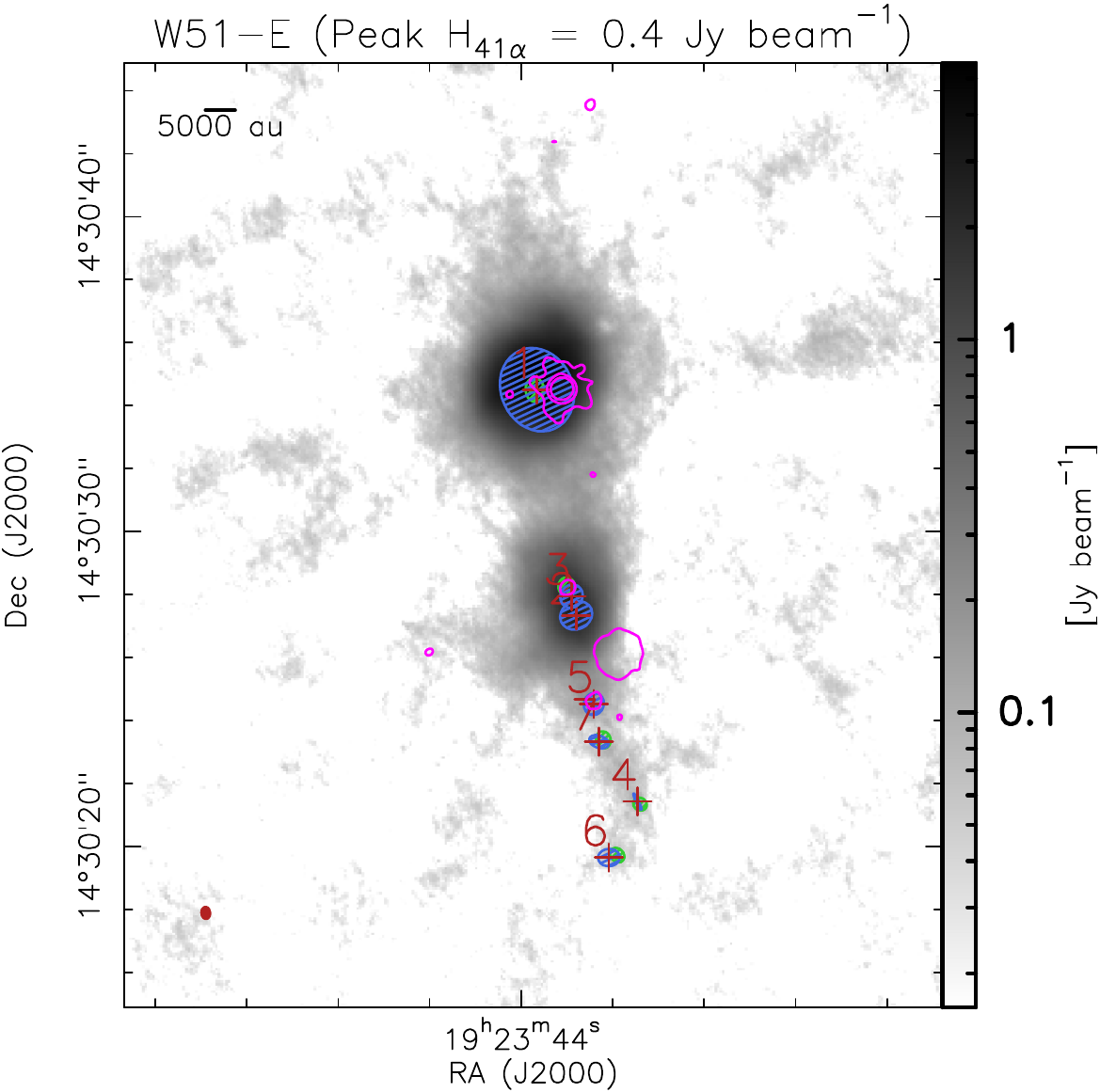}  \\
       \includegraphics[width=\hsize]{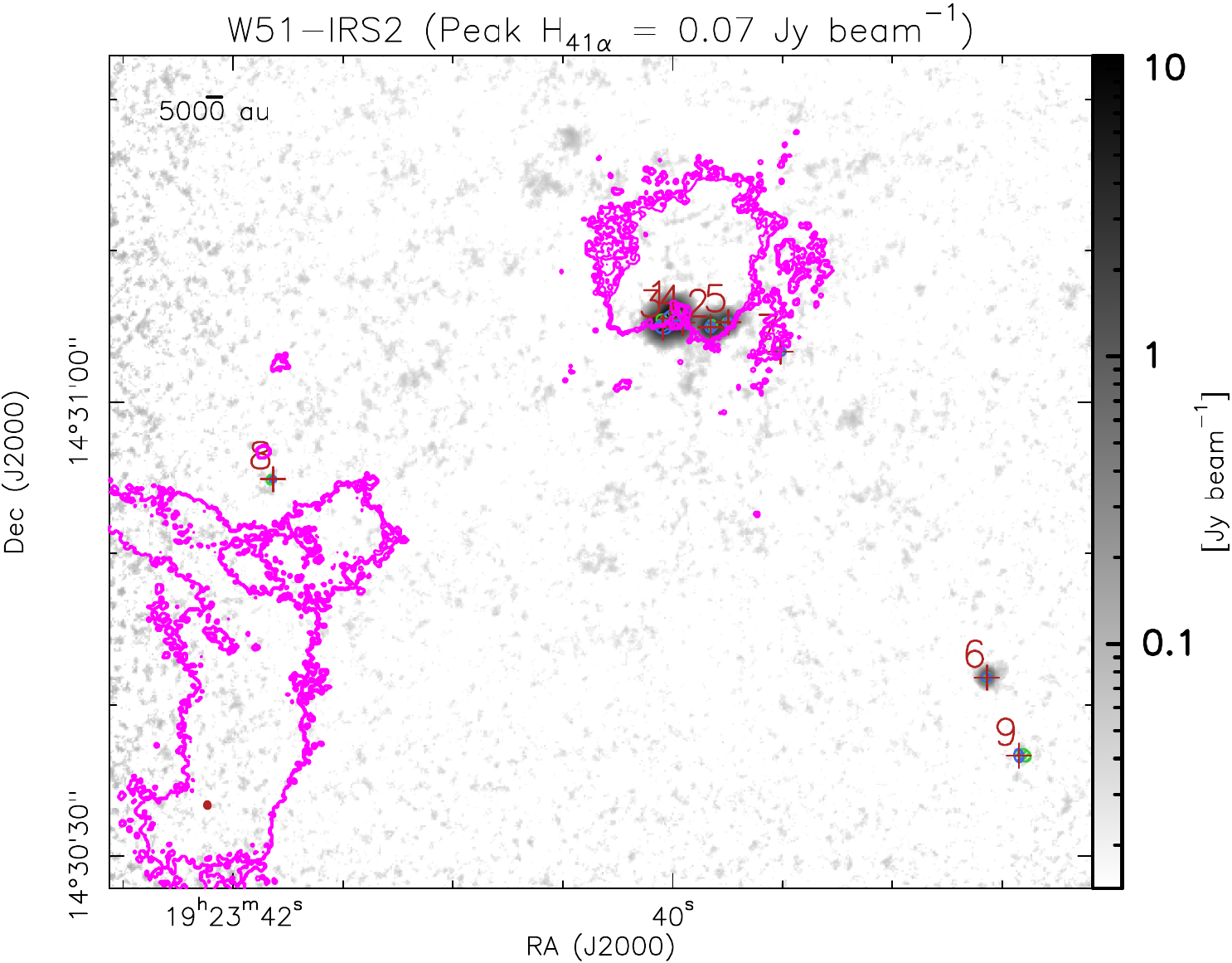}  \\
       \end{tabular}}
    \caption{\label{FIG-ellipses-h41alpha4} Same as Fig.\,\ref{FIG-ellipses-h41alpha}.}
       \end{figure*}

As mentioned already in Sect.\,\ref{section-freefree}, the ALMA-IMF dataset covers the H$_{41 \alpha}$ recombination line at 92.0 GHz that we use here to trace ionized gas coming from {\hii} \,regions. Figures\,\ref{FIG-ellipses-h41alpha}--\ref{FIG-ellipses-h41alpha4} show the regions that are expected to be contaminated by free-free emission based on the emission contours of H$_{41 \alpha}$ (see pink contours). We identify 17 methyl formate sources, in G008.67, G010.62, G012.80, G333.60, W51-E and W51-IRS2, that lie in regions containing free-free emission. In order to estimate the contribution of free-free emission to the 1.3 mm flux densities measured for these 17 cores, we proceed as follows. 

First the 3 mm integrated fluxes are rescaled to the 1.3 mm sizes to allow a direct comparison of these fluxes as described in \citetalias{pouteau2022}. Then we compute the theoretical flux ratio expected for thermal dust emission ($\gamma^{\mathrm{dust}}_{\mathrm{th}}$) under the optically thin assumption and considering that the fluxes arise from the same area:
\begin{align}
\label{eq-th-flux-ratio-dust}
 \gamma^{\mathrm{dust}}_{\mathrm{th}} &= \frac{S_{\mathrm{1.3mm}}^{\mathrm{int}}}{S_{\mathrm{3mm}}^{\mathrm{int}}}  \\
  &= \frac{\kappa_{\mathrm{1.3mm}}}{\kappa_{\mathrm{3mm}}} \frac{B_{\mathrm{1.3mm}}(T_{\mathrm d})}{B_{\mathrm{3mm}}(T_{\mathrm d})} = \frac{\kappa_{\mathrm{1.3mm}}}{\kappa_{\mathrm{3mm}}}  \times \left( \frac{\nu_{3mm}}{\nu_{1.3mm}} \right)^3 \frac{e^{h\nu_{\mathrm{3mm}}/k_{\mathrm b}T_{\mathrm d}}-1}{e^{h\nu_{\mathrm{1.3mm}}/k_{\mathrm b}T_{\mathrm d}}-1} 
\end{align}
where $B_{\mathrm{1.3mm}}(T_{\mathrm d})$ and $B_{\mathrm{3mm}}(T_{\mathrm d})$ are the Planck function for the dust temperature $T_d$ at 1.3 mm and 3 mm, respectively, $k_b$ is the Boltzmann constant, $h$ is the Planck constant, $\nu_{\mathrm{1.3mm}}$ and $\nu_{\mathrm{3mm}}$ the central frequencies of the continuum maps at 1.3 mm and 3 mm, respectively (see Table \ref{TAB-linecubes}). Following \citetalias{pouteau2022}, we adopted a dust opacity per unit of mass (gas + dust) $\kappa_{\mathrm{1.3mm}}$ = 0.01 cm g$^{-1}$. The dust mass opacity at 3 mm, $\kappa_{\mathrm{3mm}}$ is computed as follows:
\begin{equation}
\label{eq-kappa}
  \kappa_{\mathrm{3mm}} = \kappa_{\mathrm{1.3mm}} \times \left( \frac{\nu_{3mm}}{\nu_{1.3mm}} \right)^\beta
\end{equation}
where $\beta$ = $\alpha$ - 2 \citep[][]{terebey1993}. Following \citetalias{pouteau2022}, we assume $\alpha_{\mathrm{dust}}$ = 3.5, which corresponds to $\beta_{\mathrm{dust}}$ = 1.5, suitable for optically thin dense gas at the core scale \citep{ossenkopf1994}. Then by combining Eqs. \ref{eq-kappa} and \ref{eq-th-flux-ratio-dust}, we obtain: 
\begin{equation}
\label{eq-th-flux-ratio-dust2}
 \gamma^{\mathrm{dust}}_{\mathrm{th}} = \left(\frac{\nu_{\mathrm{1.3mm}}}{\nu_{\mathrm{3mm}}}\right)^{\alpha_{\mathrm{dust}}+1}  \times \frac{e^{h\nu_{\mathrm{3mm}}/k_{\mathrm b}T_{\mathrm d}}-1}{e^{h\nu_{\mathrm{1.3mm}}/k_{\mathrm b}T_{\mathrm d}}-1}, 
\end{equation}
where $\gamma^{\mathrm{dust}}_{\mathrm{th}}$ = 16.62 -- 17.34, for $T_{\mathrm d}$ = 50 -- 150K. This is valid if the integrated flux measured at 1.3\,mm ($S_{\mathrm{1.3mm}}^{\mathrm{int}}$) and 3\,mm ($S_{\mathrm{3mm}}^{\mathrm{int}}$) are only due to thermal dust emission. However, at 3 mm we expect that a non-negligible fraction of the measured flux is due to free-free emission, such that: 
\begin{equation}
\label{eq-Sintcorr-3mm}
 S_{\mathrm{3mm}}^{\mathrm{int}} = S_{\mathrm{3mm}}^{\mathrm{int-corr}} - S_{\mathrm{3mm}}^{\mathrm{int-freefree}} \,,
\end{equation}
where $S_{\mathrm{3mm}}^{\mathrm{int-corr}}$ is the 3\,mm integrated flux corrected from the free-free contribution, and $S_{\mathrm{3mm}}^{\mathrm{int-freefree}}$ the flux due to free-free emission, such that: 
\begin{equation}
\label{eq-th-flux-ratio-freefree}
 \gamma^{\mathrm{freefree}}_{\mathrm{th}} = \frac{S_{\mathrm{1.3mm}}^{\mathrm{int-freefree}}}{S_{\mathrm{3mm}}^{\mathrm{int-freefree}}}\,,
\end{equation}
with $\gamma^{\mathrm{freefree}}_{\mathrm{th}}$ = 0.86 -- 0.90, for $T_{\mathrm d}$ = 50 -- 150\,K, and for a spectral index $\alpha_{\mathrm{freefree}}$ = -0.1 that corresponds to optically thin free-free emission \citep[see e.g.][]{keto2008}. \\

In order to estimate integrated fluxes corrected from the free-free contribution, we assume in first approximation that the flux measured at 1.3 mm is optically thin and is only due to dust thermal emission, such that: 
\begin{equation}
\label{eq-Sintcorr-3mm-2}
 S^{\mathrm{int-corr}}_{\mathrm{3mm}} = \frac{S^{\mathrm{int}}_{\mathrm{1.3mm}}}{\gamma^{\mathrm{dust}}_{\mathrm{th}}}
\end{equation}
However, for the sources contaminated by free-free emission, we expect $S^{\mathrm{int-corr}}_{\mathrm{1.3mm}} \ne S^{\mathrm{int}}_{\mathrm{1.3mm}}$, which implies that the integrated flux measured at 1.3 mm is not fully due to dust thermal emission like assumed above. From Eqs. \ref{eq-Sintcorr-3mm}, \ref{eq-th-flux-ratio-freefree}, and \ref{eq-Sintcorr-3mm-2}, we derive the 1.3 mm flux corrected for the free-free contribution: 
\begin{equation}
\label{eq-Sintcorr-13mm}
 S^{\mathrm{int-corr}}_{\mathrm{1.3mm}} = S^{\mathrm{int}}_{\mathrm{1.3mm}} - \gamma^{\mathrm{freefree}}_{\mathrm{th}} (S^{\mathrm{int}}_{\mathrm{3mm}} - S^{\mathrm{int-corr}}_{\mathrm{3mm}}) 
\end{equation}
For more consistency, those calculations are performed in an iterative way, by replacing $S^{\mathrm{int}}_{\mathrm{1.3mm}}$ in Eq. \ref{eq-Sintcorr-3mm} by its corrected value, $S^{\mathrm{int-corr}}_{\mathrm{1.3mm}}$. Then Eqs. \ref{eq-Sintcorr-3mm-2} and \ref{eq-Sintcorr-13mm} are computed until the calculations converge onto a final value for $S^{\mathrm{int-corr}}_{\mathrm{1.3mm}}$. The resulting values are given in the last column of Table\,\ref{appendix-cont-core-cat}, as a correction factor, frac$_{ff}$, that indicates the fraction of the flux initially measured that is due to free-free emission for each continuum core. This correction factor must be applied to both the peak and integrated flux measured at 1.3\,mm and 3\,mm for the 17 methyl formate sources that are contaminated by free-free emission.

\clearpage
\onecolumn

%======================
% FIGURE: H41alpha emission + ellipse source sizes
%======================

%======================

\clearpage
\twocolumn

%------------------------------------------------------------------
\section{Source size deconvolution}
%------------------------------------------------------------------
\label{appendix-deconvolve-source-size}

The source sizes derived for the 76 hot core candidates are presented in Sect.\,\ref{section-MFsizes}. They were obtained by fitting 2D Gaussians to the methyl formate moment 0 maps of the 15 ALMA-IMF protoclusters (Sect.\,\ref{section-MFextraction}), using the source extraction algorithm \textsl{GExt2D}. In order to to retrieve the actual source sizes, we deconvolve the source sizes inferred from the Gaussian fits, from the beam size of the linecubes as follow (Pety, priv. comm.). First we define:
\begin{multline}
    a = [\theta_{\rm maj} \times cos(\phi_{\rm s})]^2 + [\theta_{\rm min} \times sin(\phi_{\rm s})]^2 \\
    - [\theta_{\rm maj}^{\rm beam} \times cos(\phi_{\rm beam})]^2 - [\theta_{\rm min}^{\rm beam} \times sin(\phi_{\rm beam})]^2 \, ,
\end{multline}
\begin{multline}
    b = [\theta_{\rm maj} \times sin(\phi_{\rm s})]^2 + [\theta_{\rm min} \times cos(\phi_{\rm s})]^2 \\
    - [\theta_{\rm maj}^{\rm beam} \times sin(\phi_{\rm beam})]^2 - [\theta_{\rm min}^{\rm beam} \times cos(\phi_{\rm beam})]^2 \, ,
\end{multline}
\begin{multline}
    c = 2 \times [(\theta_{\rm min}^2 - \theta_{\rm maj}^2) \times sin(\phi_{\rm s}) \times cos(\phi_{\rm s}) \\
    - (\theta_{\rm min}^{\rm beam}\,^2 - \theta_{\rm min}^{\rm beam}\,^2) \times sin(\phi_{\rm beam}) \times cos(\phi_{\rm beam})] \, ,
\end{multline}
where 
\begin{equation}
\phi_{\mathrm{s}} = \rm PA \times \frac{\pi}{180}, 
\end{equation}
and 
\begin{equation}
\phi_{\rm beam} = \rm PA_{\rm beam} \times \frac{\pi}{180}, 
\end{equation}
are expressed in radians. The major ($\theta_{\mathrm{maj}}$) and minor ($\theta_{\mathrm{min}}$) axes of the measured source size (FWHM$_{\mathrm{MF}}$), as well as the major ($\theta_{\mathrm{maj}}^{\mathrm{beam}}$) and minor ($\theta_{\mathrm{min}}^{\mathrm{beam}}$) axes of the synthesized beam size of the line cubes, are expressed in arcseconds. The beam parameters ($\theta_{\mathrm{min}}^{\mathrm{beam}} \times \theta_{\mathrm{maj}}^{\mathrm{beam}}$, PA$_{\mathrm{beam}}$) are taken from Table\,\ref{TAB-linecubes}, while the parameters for the methyl formate source sizes ($\theta_{\mathrm{min}} \times \theta_{\mathrm{maj}}$, PA) are taken from Table\,\ref{TAB-MFcat}. We also define: 
\begin{equation}
d = a + b,
\end{equation}
and
\begin{equation}
e = \sqrt{(a-b)^2+c^2},
\end{equation}
such that the deconvolved source sizes parameters ($\theta_{\mathrm{min}}^{\mathrm{dec}} \times \theta_{\mathrm{maj}}^{\mathrm{dec}}$, PA$_{\mathrm{dec}}$) are given by: 
\begin{equation}
\theta_{\mathrm{maj}}^{\mathrm{dec}} = \sqrt{\frac{|d+e|}{2}},
\end{equation}
\begin{equation}
\theta_{\mathrm{min}}^{\mathrm{dec}} = \sqrt{\frac{|d-e|}{2}},
\end{equation}
in arcseconds, and
\begin{equation}
\rm PA^{\rm dec} = \frac{180}{\pi} \times \frac{ \textit{atan2}(-c,a-b)}{2}.
\end{equation}
in degrees, which uses the 2-argument arctangent (\textit{atan2}) function of the GILDAS\footnote{\href{https://www.iram.fr/IRAMFR/GILDAS/}{https://www.iram.fr/IRAMFR/GILDAS/}} software. 

Finally, the deconvolved source size is given by: 
\begin{equation}
\label{EQ-deconvolved-MF}
\theta_{\rm MF}^{\rm dec} = \sqrt{\theta_{\mathrm{maj}}^{\mathrm{dec}} \times \theta_{\mathrm{min}}^{\mathrm{dec}}},
\end{equation}
which may also be expressed in astronomical units, as the physical size of the source ($FWHM_{\rm MF}^{\rm dec}$) at the distance of the protocluster. 

As mentioned already in Sect.\, \ref{section-MFsizes}, we have set a minimum deconvolved size for each region that is equal to half the FWHM of the synthesized beam of the linecube, in order to limit deconvolution effects that may give excessively small and thus unrealistic sizes. 

The resulting methyl formate deconvolved source sizes are shown in Figs. \ref{FIG-ellipses-h41alpha}--\ref{FIG-ellipses-h41alpha4} together with the deconvolved continuum core sizes for the comparison.

\begin{figure}[!ht]
  \begin{center}
       \includegraphics[width=\hsize]{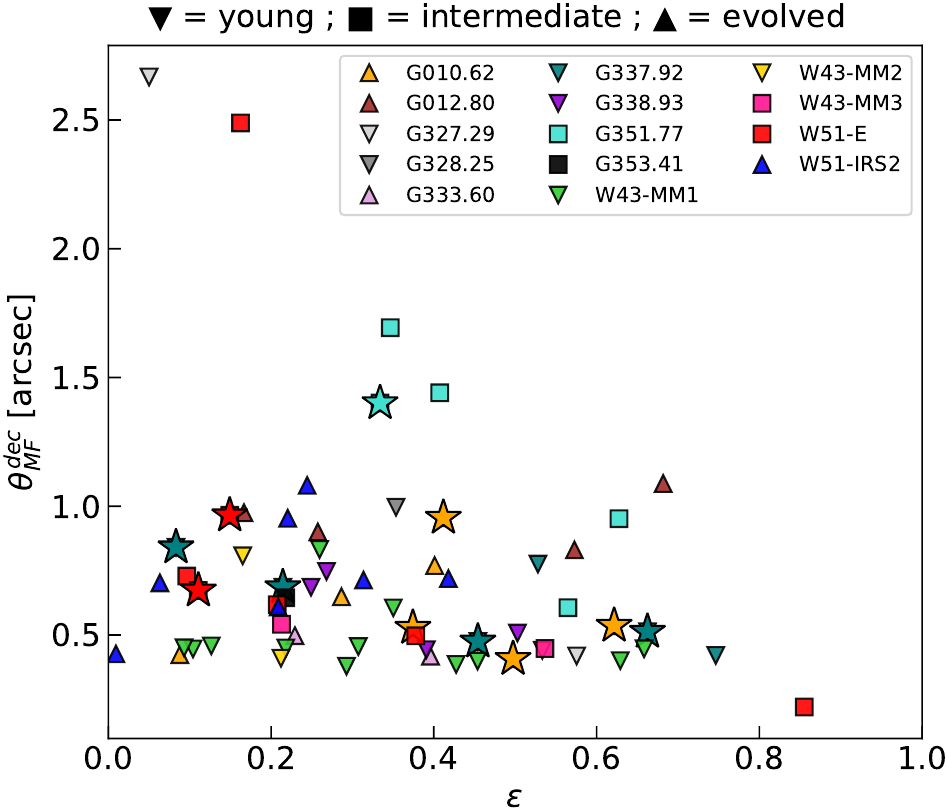}  \\
  \caption{\label{FIG-deconvolve-ellipticity} Methyl formate deconvolved source sizes ($\theta_{\rm MF}^{\rm dec}$) as a function of their ellipticity ($\epsilon$). The different symbols represent the different evolutionary stages of the protoclusters: young, intermediate, or evolved, as shown on top of the figure. The methyl formate sources that are not associated with compact continuum cores are shown with stars.The sources for which $\theta_{\rm MF}^{\rm dec}$ is lower than half of the linecube beam size are not shown on the figure.} 
  \end{center}
\end{figure}

Figure\,\ref{FIG-deconvolve-ellipticity} shows the deconvolved source sizes ($\theta_{\rm MF}^{\rm dec}$, Eq.\,\ref{EQ-deconvolved-MF}), plotted as a function of the source ellipticity ($\epsilon$), which is given by: 
\begin{equation}
\epsilon = 1 - \frac{\theta_{\mathrm{min}}^{\mathrm{dec}}}{\theta_{\mathrm{maj}}^{\mathrm{dec}}}.
\end{equation}
It shows that most methyl formate sources have an ellipticity $\epsilon$ $<$ 0.5.

%======================
% FIGURE: DECONVOLVED SIZES VS. ELLIPTICITY
%======================
%======================

% ------------------------------------

% ------------------------------------

\end{appendix}

\end{document}